\documentclass[rmp,aps,nofootinbib,floatfix]{revtex4}

\usepackage{graphics}
\usepackage{epsfig}

\input myBabarsym
\input definitions

\begin{document}
\title{The Physics of Hadronic Tau Decays}

\author{Michel Davier$^a$}
\email{davier@lal.in2p3.fr}
\author{Andreas H\"ocker$^b$}
\email{andreas.hocker@cern.ch}
\author{Zhiqing Zhang$^a$}
\email{zhangzq@lal.in2p3.fr}
\affiliation{$^a$Laboratoire de l'Acc\'el\'erateur Lin\'eaire, IN2P3/CNRS
	     and Universit\'e de Paris-Sud, 
             BP-34, F-91898 Orsay, France \\ 
	     $^b$CERN, CH-1211 Geneva, Switzerland}  
\begin{abstract}  
Hadronic $\tau$ decays provide a clean laboratory for the precise 
study of quantum chromodynamics (QCD). Observables based 
on the spectral functions of hadronic $\tau$ decays can be related 
to QCD quark-level calculations to determine fundamental 
quantities like the strong coupling constant, parameters of
the chiral Lagrangian, $|V_{us}|$, the mass of 
the strange quark, and to simultaneously test the concept of 
quark-hadron duality. Using the best available measurements and a
revisited analysis of the theoretical framework, the value
$\asm = 0.345 \pm 0.004_{\rm exp} \pm 0.009_{\rm th}$ is obtained.
Taken together with the determination of \asZ from the global electroweak
fit, this result leads to the most accurate test of asymptotic freedom:
the value of the logarithmic slope of $\alpha_s^{-1}(s)$ is found to
agree with QCD at a precision of 4\%.
The $\tau$ spectral functions 
can also be used to determine hadronic quantities that, due to the 
nonperturbative nature of long-distance QCD, cannot be computed 
from first principles. An example for this is 
the contribution from hadronic vacuum polarization to loop-dominated 
processes like the anomalous magnetic moment of the muon. This 
article reviews the measurements of nonstrange and strange $\tau$
spectral functions and their phenomenological applications.
\end{abstract}                                                                 

%\date{May 2001}
\maketitle
\tableofcontents
\vfill\newpage

\section{INTRODUCTION}
\label{sec:introduction}

Because of its relatively large mass and the simplicity of its decay
mechanism, the $\tau$ lepton offers many interesting, and sometimes unique,
possibilities for testing and improving the Standard Model (SM). 
These studies involve the leptonic and hadronic sectors and encompass 
a large range of topics, from the measurement of the leptonic couplings 
in the weak charged current and the search for lepton flavor violation,
providing precise lepton universality tests, to a complete investigation 
of hadronic production from the Quantum Chromodynamics (QCD) vacuum. 
For the latter case, the $\tau$
decay results have proven to be complementary to those from $e^+e^-$ data,
allowing us to perform detailed studies at the fundamental level through
the determination of the {\em spectral functions}, which embody both the rich
hadronic structure seen at low energy, and the quark behavior relevant
in the higher energy regime. The spectral functions play an important 
role in the understanding of hadronic dynamics in the intermediate
energy range. They represent the basic input for QCD studies and for 
evaluating low-energy contributions from hadronic vacuum polarization. 
This is required for the calculation of the muon anomalous magnetic 
moment and the running of the electromagnetic coupling constant.
The topics of interest also include: studying isospin violation
between the weak charged and electromagnetic hadronic currents, evaluation 
of chiral sum rules making use of the separate determination of vector and
axial-vector components, and a global QCD analysis including perturbative 
and nonperturbative contributions, offering the possibility for a precise
extraction of the strong coupling at a relatively low energy scale, and 
hence providing a sensitive test of the running when compared to the value
obtained at the $M_Z$ scale. The spectral functions into
strange ($|\Delta S|=1$) final states, though Cabibbo-suppressed, can be used 
to determine the $s$-quark mass and the CKM element $|V_{us}|$. In addition,
the study of $\tau^+ \tau^-$ pair production at LEP has proven to be an
invaluable tool for investigating the neutral current sector of the electroweak
theory, particularly through the measurement of the $\tau$ polarization.

This article reviews the measurements of nonstrange and strange $\tau$
spectral functions and their phenomenological applications.
It is organized as follows. We begin in Section~\ref{sec:tauphysics}
with an overview on the experimental conditions leading to
the measurement of the $\tau$ branching fractions 
(Section~\ref{sec:tauspecfun_brs}) and hadronic spectral functions 
(Section~\ref{sec:tauspecfun}) by the ALEPH and OPAL experiments 
at LEP as well as CLEO at CESR. 
%We also describe the measurements of the leptonic and hadronic 
%$\tau$ branching fractions, a crucial ingredient to the computation 
%of the spectral functions and the QCD analysis. 
Using isospin symmetry, the $\tau$ vector spectral functions are compared 
to the corresponding \ee annihilation data in Section~\ref{sec:cvc}.
We discuss in this context isospin-breaking and radiative corrections.
The use of spectral functions in dispersion relations to compute
hadronic vacuum polarization is reviewed in Section~\ref{sec:vacpol}.
A comprehensive review of hadronic $\tau$ decays within QCD is presented
in Sections~\ref{sec:qcd} and \ref{sec:strangesf}. The main topics here
are the determinations of the strong coupling constant and the mass of the 
strange quark at the scale of the $\tau$ mass together with $|V_{us}|$. 
We discuss nonperturbative contributions to the $\tau$ hadronic width and 
present results from chiral QCD sum rules. 

For all these topics we critically examine experimental and theoretical 
limitations, and provide detailed comparisons between experiments 
as well as perspectives for the future.

\section{THE EXPERIMENTAL STAGE}
\label{sec:tauphysics}

\subsection{Different experimental conditions}
\label{subsec:experiments}

Tau physics in the nineties has been dominated by two rather 
different and in many views complementary experimental facilities:
on one hand the LEP experiments ALEPH, DELPHI, L3 and OPAL, operating 
at the $Z$ resonance at center-or-mass (CM) energies of 91.2\gev,
and on the other hand CLEO at CESR running at the \FourS resonance
(10.6\gev). The $\tau$ pair production and decay characteristics 
at LEP were:
\bei

\item	a mean $\tau$ flight length of $2.2\mm$.

\item	a strong boost of the produced $\tau$'s leading to a
	collimated back-to-back event topology; hence tracks and photon 
	clusters overlap in the detector, requiring excellent double
	hit resolution in the tracking system and a highly granular
	electromagnetic calorimeter.

\item	a large track reconstruction efficiency and low nuclear
	interactions and multiple scattering due to the high energetic
	particles in the final state.

\item	since hadronic background from $Z$ decays occurs with 
        significantly larger particle multiplicity, the selection of
	$\tau$ pairs is almost background free ($<1-2\%$), and the 
	efficiency is large (up to $96\%$ at ALEPH, depending on 
	the final state).

\item	limited statistics: up to $1.6\times10^5$ $\tau$ pairs have been 
	recorded by each of the LEP experiments between 1990 and 1995.

\eei
For CLEO, one has
\bei

\item	a mean $\tau$ flight length of $0.25\mm$.

\item	well separated $\tau$ decay products in the detector.

\item	significant hadronic background from $\epem\to\q\qbar$ continuum
	events due to the relatively low average multiplicity in the 
 	final state of these events. This requires $\tau$ tagging, \ie,
	one $\tau$ is reconstructed (``tagged'') in a leptonic decay, or 
	a hadronic mode with less than three hadrons in the final state.
	The recoil side of the tagged event is then used as an unbiased
	sample for physics studies.

\item	large statistics: for example, the $\tau^-\to\pim\piz\nut$ 
	spectral function analysis~\cite{cleotaurho} uses
	a data sample that contains $3.2\times10^6$ produced $\tau$
	pairs.

\eei

Due to its superior statistics, CLEO discovered a number of rare
modes, and performed studies of their hadronic substructures.
Examples for these are $\tau^-\to\eta K^-\nut$~\cite{cleo_etak} and 
$\tau^-\to\eta \pi^-\piz\nut$~\cite{cleo_etapipi}, the latter one
proceeding through a vector current that is due to the Wess-Zumino 
chiral anomaly~\cite{pich:1987,Decker:1993}. The anomaly violates the rule
that the weak vector and axial-vector currents produce an even and odd
number of pseudoscalars, respectively.
The $\tau^-\to\eta \pi^-\nut$ mode is a $G$-parity
suppressed {\em second-class current}, which has not been observed
yet.

For the $\tau$ decays with a branching fraction greater than $0.1\%$, 
the cleaner LEP events provided the more precise branching fraction
measurements, many of them with a sub-percent relative accuracy, 
which are a crucial ingredient to the QCD study of hadronic $\tau$ decays.
The Cabibbo-suppressed $\tau$ decays with $|\Delta S|=1$
have been studied by all experiments, albeit with limited success
due to the restricted particle identification capabilities (except for
DELPHI and CLEO III). Much progress is expected in this area from 
the $B$-Factory experiments  \babar\  and Belle, which have dedicated 
Cherenkov particle-identification devices.

\subsection{Investigating the resonance structure in $\tau$ decays}
\label{sec:structure}

Most of this review will be concerned with the interplay between
measurements involving hadronic (or more precisely semileptonic) 
$\tau$ decays and QCD. This may appear
a bit paradoxical in view of the relatively low energy scale of $\tau$ 
physics. Indeed, $\tau$ decays to hadrons ($+~\nu_\tau$) reveal a rich
structure of resonances, the study of which opens an interesting
field of research on meson dynamics. The leptonic environment provides
a significant advantage for isolating clean hadronic systems and measuring
their parameters.
Many $\tau$ decay modes have been investigated for their resonance 
structure. The phenomenologically important $\pim\piz$ vector state
is analyzed in detail in Section~\ref{sec:cvc}.

\begin{figure}[t]
   \centerline{\epsfxsize6.9cm\epsffile{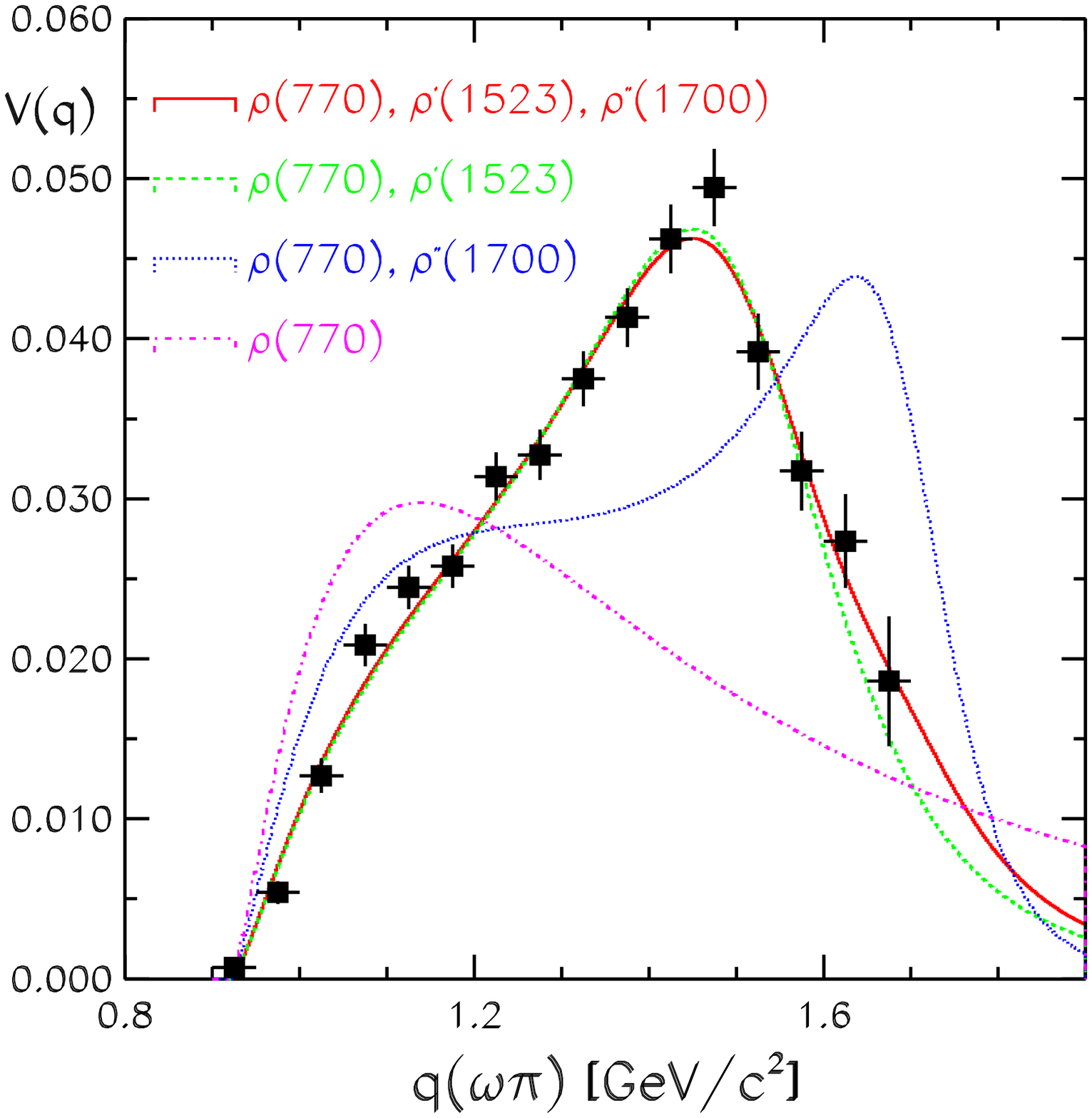}
 	       \epsfxsize7.3cm\epsffile{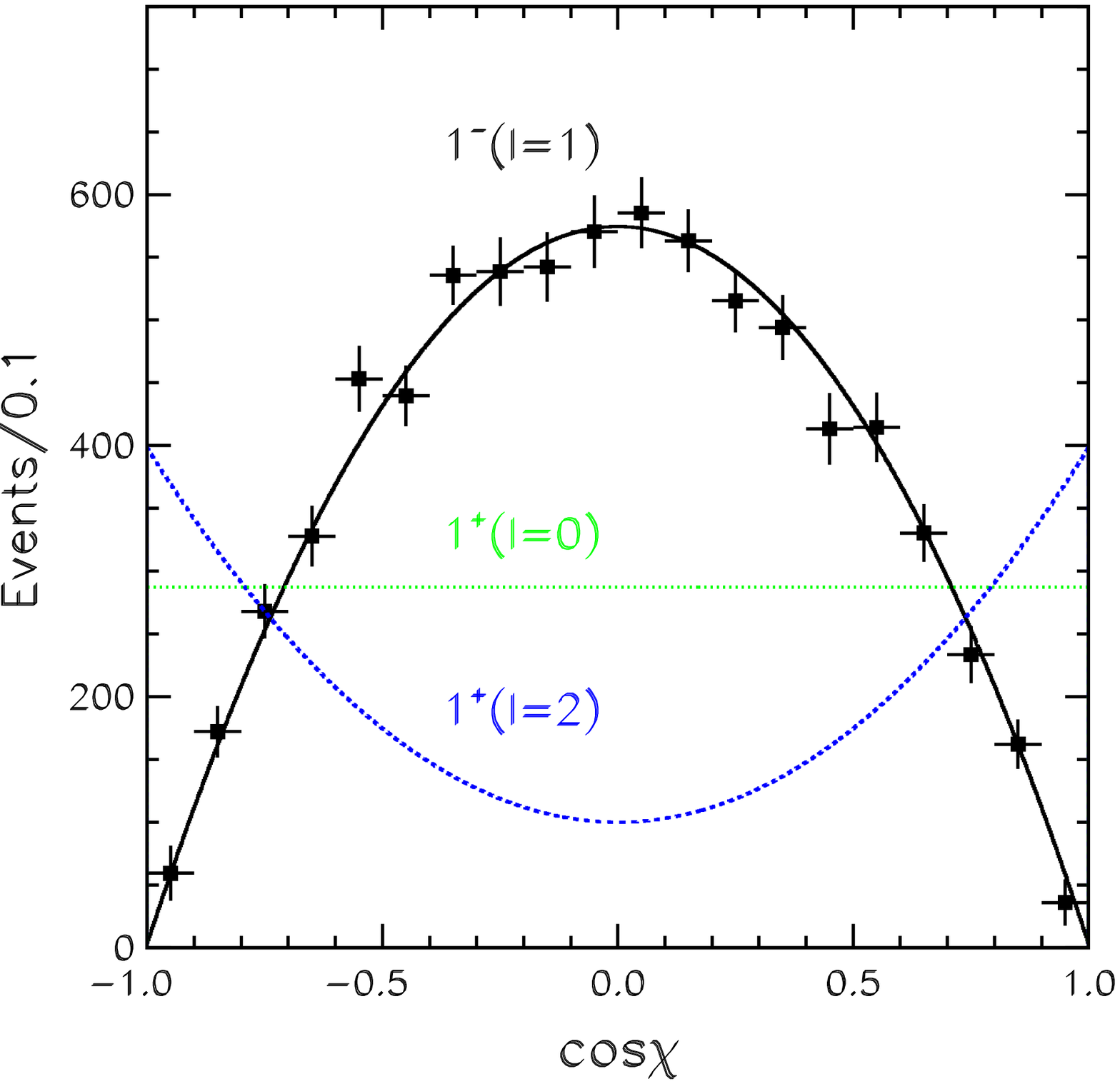}}
   \vspace{0.3cm}
   \caption{\underline{Left:} the spectral function (see Section~\ref{sec:tauspecfun}
     for the definition) for the decay $\taum \to \omega\pi^-\nu_\tau$
     vs. the hadronic mass from the CLEO analysis~\cite{cleo_4pi}. 
     The curves are the results of fits to various combinations of 
     $\rho$, $\rho(1450)$, $\rho(1700)$ resonance contributions. 
     \underline{Right:} the angular distribution in the $\omega\pi$
     center-of-mass frame compared to predictions for different ($l$)
     partial waves. Second-class currents would reveal themselves as
     $l=0,2$ waves.}
\label{cleo_omegapi}
\end{figure}
Taking advantage of its large statistics sample, the CLEO Collaboration
was able to perform partial-wave analyses of multi-hadron final states.
In particular, the $(3\pi)^-\pi^0$ mode is expected to proceed through the
vector current dominated by $\rho$, $\rho(1450)$, $\rho(1700)$ isovector 
mesons. Resonances are also produced in two-body decays, such as $\omega\pi$,
$\eta\pi$, $a_1\pi$. CLEO measured~\cite{cleo_4pi} the $\omega\pi$ 
contribution, which they successfully modeled with interfering $\rho$, 
$\rho(1450)$, $\rho(1700)$ resonances as shown in Fig.~\ref{cleo_omegapi}. At
least the first two contributions are needed to reproduce the experimental 
mass spectrum. The helicity angle distribution in the $\omega\pim$ 
center-of-mass is found to be consistent with the expected $l=1$ wave
(Fig.~\ref{cleo_omegapi}), whereas a wrong $G$-parity contribution 
arising from a weak second-class current, such as $\tau\to b_1^- \nu_\tau$, 
$b_1^- \to \omega\pim$, would lead to an even $l$ value. Such a contribution
has been bound to be less than 5.4\% of the $\omega\pim$ rate.
Besides $\omega\pim$, the total $(3\pi)^-\pi^0$ mode is shown to be 
dominated by $\rho\pi\pi$, consistent with originating from $a_1 \pi$, 
$a_1 \to \rho\pi$.

The decay $\tau \to \nu_\tau (3\pi)^-$ is the cleanest mode to study
axial-vector resonance structure. The spectrum is dominated by the 
$J^P=1^+$ $a_1$ state, known to decay essentially through $\rho \pi$. The 
hadronic structure of this decay has been studied in a model-independent
analysis by the OPAL collaboration, where no evidence for vector or 
scalar currents has been found~\cite{opal_a1a,opal_a1b}. A comprehensive 
analysis of the $\pi^- 2\pi^0$ channel has been presented by CLEO. 
First, a model-independent determination of the hadronic structure 
functions gave no evidence for non-axial-vector contributions 
($<17\%$ at $90\%$ CL)~\cite{cleo_3pisf}. Second, a partial-wave
amplitude analysis was performed~\cite{cleoa1}: while the dominant 
$\rho \pi$ mode has been confirmed, it came as a surprise that an 
important contribution ($\sim 20\%$) from scalars 
($\sigma$, $f_0(1470)$, $f_2(1270)$) was found in the $2\pi$ subsystem.
The precisely determined $a_1^- \to \pi^- 2\pi^0$ line shape 
shows the opening of the $K^*K$ decay channel.
Some interference with a higher mass state ($a'_1$) was advocated
in~\cite{delphi_a1p}. However, no conclusive evidence was found in the 
CLEO analysis, with a sensitivity about tenfold higher. 

Evidence for the decay $\tau \rar f_1(1285) \pi^- \nu_\tau$ has been 
obtained, with $f_1$ decay into $\eta \pi^+ \pi^-$ and 
$\eta \pi^0 \pi^0$~\cite{cleoeta3pi}, and recently into 
$2\pi^+ 2\pi^-$~\cite{babar_5h}, the latter decay mode being consistent with
$\rho^0 \pi^+ \pi^-$. Thus many spectroscopy issues can be tackled with
$\tau$ leptons in a clean environment for single resonance studies.

\section{TAU BRANCHING FRACTIONS}
\label{sec:tauspecfun_brs}

Tau hadronic spectral functions (Section~\ref{sec:tauspecfun}) are 
obtained from the measured vector and axial-vector invariant mass spectra,
normalized to the corresponding inclusive branching fractions. 
These are derived from the sum of exclusive branching fractions.
Before discussing the spectral functions, we shall thus briefly review 
the measurement of the $\tau$ branching fractions.

\subsection{The measurement of branching fractions}
\label{sec:brs_general}

In principle, the measurement of the branching fractions of exclusive 
$\tau$ decays relies on the identification of the corresponding final 
state and the knowledge of the number of the produced $\tau$ leptons 
in the sample under study. The latter number can be derived from the 
production cross section, known to high precision in \ee annihilation, and 
the measured luminosity. The measurement of the final state however 
requires excellence in many aspects of detector technology and analysis 
techniques: charged particle tracking and identification for $e/\mu/\pi/K$ 
separation, photon detection and $\pi^0$ ($\eta$) reconstruction.

Since very clean and high-efficiency $\tau^+\tau^-$ samples have been
produced at LEP, another approach can be considered where all decay
channels are simultaneously measured. This global method was pioneered
by CELLO at PETRA~\cite{cello_tau} with much smaller statistics and less 
favorable conditions than those encountered at LEP. The detector performance
achieved by ALEPH, in particular the excellent granularity of its
electromagnetic calorimeter, made it possible to implement this method
and take full advantage of the LEP conditions on the $Z$ 
peak~\cite{aleph01,aleph13_l,aleph13_h,aleph_taubr}.

The ALEPH global analysis of $\tau$ branching fractions proceeds from the
selection of the $\taup\taum$ sample, based on the rejection of non-$\tau$
backgrounds: $Z$ decays into $\epem$, $\mup\mun$ and $q\qbar$
pairs, $\gamma\gamma$-induced processes $e^+e^- \to e^+e^-~X$
(where $X$ can be any fermion-antifermion pair), and cosmic-ray events. 
The kinematic properties of these backgrounds and their characteristic
detector response permit to achieve a high $\taup\taum$ selection
efficiency of 78.9\% (91.7\% when the $\taup\taum$ direction points into the
active area of the detector) and a low background fraction of 1.2\%.
Each $\tau$ decay is classified on the basis of charged particle
and photon/$\pi^0$ multiplicities, and the nature of the charged particles.
Charged particle identification is achieved through the response of 
electromagnetic and hadronic calorimeters, and the ionization loss measurement.
Photons need to be identified among the neutral electromagnetic clusters,
some of which are induced by charged hadron interactions or by spatial
fluctuations of nearby showers (fake photons). 
Fourteen classes are thus defined in~\cite{aleph_taubr} 
for reconstructed decays, 
where the last one collects the fraction (3.6\%)
of rejected candidates corresponding to the cases where particle 
identification is not reliable. The first 13 classes correspond to the
following final states produced with $\nut$: $\en \nueb$,
$\mun \numb$, $h^-$, $h^- \pi^0$, $h^- 2\pi^0$, $h^- 3\pi^0$, 
$h^- 4\pi^0$, $2h^-h^+$, $2h^-h^+\pi^0$, $2h^-h^+2\pi^0$, 
$2h^-h^+ 3\pi^0$, $3h^-2h^+$, $3h^-2h^+ \pi^0$,
where $h$ stands for a pion or kaon.

The branching fractions are determined using
\beqn
\label{eq:lin_br}
 n^{\rm obs}_i - n^{\rm bkg}_i &=& \sum_{j} \eff_{ji} N^{\rm prod}_j~,\\
 \BR_j &=& \frac {N^{\rm prod}_j} {\sum_{j} N^{\rm prod}_j}~,
\eeqn
where $n^{\rm obs}_i (n^{\rm bkg}_i)$ is the number of observed $\tau$ 
candidates (the non-$\tau$ background events)
that are reconstructed in class $i$, $\eff_{ji}$ the cross-talk
efficiency of events that are produced in class $j$ and reconstructed in 
class $i$, and $N^{\rm prod}_j$ the number of produced events in class $j$.
The efficiency matrix $\eff_{ji}$ is determined from Monte Carlo
simulation, based on the event generator KORALZ07~\cite{was} and including
the full detector response. The simulation has been calibrated using data 
for a large number of observables. Among these are particle identification
efficiencies, $\taup\taum$ selection efficiency and the number of fake 
photon candidates. The selection efficiency remains constant within 
$\pm\sim5\%$ for all 13 topologies considered. The linear system of 
equations~(\ref{eq:lin_br}) is solved for the $N^{\rm prod}_j$ unknown,
hence the branching fractions $\BR_j$.

Extensive systematic studies are performed to assess possible biases
in the decay assignment. These studies are conducted on data samples in
order not to rely on the detector simulation, however detailed it can be.
They include: event selection, non-$\tau$ background, particle 
identification, photon detection efficiency, converted photons, photon
identification, fake photons, $\pi^0$ reconstruction, bremsstrahlung and
radiative photons, $\pi^0$ Dalitz decays, secondary nuclear interactions,
tracking of charged particles, dynamics in the Monte Carlo generator.
The dominant systematic uncertainties for the hadronic modes stem from
the photon/$\pi^0$ treatment and secondary interactions.

While measurements of branching fractions larger than 0.1\% are more 
precise at LEP because of its small systematic uncertainties, the larger 
statistics accumulated by CLEO makes it more competitive for smaller 
branching fractions. The different experimental conditions also provide 
complementary information and valuable cross checks.  

The ALEPH global analysis assumes a standard $\tau$ decay description. 
However, unknown decay modes not included in the simulation may exist.
Since large detection efficiencies are achieved in the $\taum\taup$ 
selection, which is therefore robust, these decays would be
difficult to pass unnoticed. An independent measurement of the
branching fraction for undetected decay modes, \ie, modes not passing
the selection cuts, using a direct search with
a one-sided $\tau$ tag, was done by ALEPH~\cite{aleph_undetect},
limiting this branching fraction to less than 0.11\% at 95\% CL. This
result justifies the assumption that the sum of the branching fractions 
for visible $\tau$ decays is equal to one. The consistency between the
results and the $\tau$ decay description in the simulation can be 
investigated further by verifying that the ``branching fraction'' for a
$14^{\rm th}$ class in the decay classification, which collects rejected 
events, turns out to be zero within the measurement errors. The result, 
$\BR_{14}= (0.066 \pm (0.042)\%$, provides a nontrivial test of the 
overall procedure at the 0.1\% level for branching fractions.

The global method at LEP reaches its limit for decay modes with small 
branching fractions, which compete with leading channels with fractions 
that could be more than two orders of magnitude larger. This limit is 
also determined by the overall $\tau$ statistics. It turns out that the 
two factors contribute at about the same level. Thus final states with more
than five hadrons are poorly determined at LEP, whereas CLEO can use its
larger statistics to apply stricter cuts for these higher-multiplicity
decays with low branching ratios.

A final check of the global analysis can be performed by comparing the 
sums of the relevant exclusive decay fractions to the topological branching 
fractions $\BR_i$, where $i$ is the charged particle multiplicity in 
the decay. Even though the latter branching fractions have 
only limited direct physics implications, their determination represents
a valuable cross check as the measurement does not depend on photon 
reconstruction in a calorimeter. Assuming a negligible contribution from 
charged-particle multiplicities larger than five, in agreement
with the 90\% CL limit achieved by CLEO 
($\BR_7 < 2.4 \times 10^{-6}$~\cite{cleo7pr}), 
ALEPH finds:
\beqn
 \BR_3 &=& (14.652 \pm 0.067 \pm 0.086) \%~, \\
 \BR_5 &=& ( 0.093 \pm 0.009 \pm 0.012) \%~, 
\eeqn
in agreement with the values from a dedicated topological
analysis performed by DELPHI~\cite{delphitopol},
\beqn
 \BR_3^{\rm top} &=& (14.569 \pm 0.093 \pm 0.048) \%~, \\ 
 \BR_5^{\rm top} &=& (0.115 \pm 0.013 \pm 0.006) \%~.
\eeqn

\subsection{Leptonic branching fractions}
\label{sec:brs_lepton}

The measurement of the two leptonic decay modes of the $\tau$ is of great 
importance in our 
context, since they are complementary to the hadronic modes, the fraction 
of which can be determined by a simple difference. The leptonic branching 
ratios also turn out to be the most precisely determined $\tau$ decay 
fractions, not affected by systematic uncertainties that are specific to 
the presence of hadrons in the detector.

Before examining the experimental situation, we shall recall the theoretical
expectations within the standard $V-A$ theory with leptonic coupling $g_\l$ 
at the $W \l \nub_\l$ vertex. The $L=\mu,\tau$ leptonic ($\l=e,\mu$)
partial widths can be computed, including radiative 
corrections~\cite{Marciano:1988} and safely neglecting neutrino masses:
\beq
\Gamma(L \to \nu_L \l \nub_\l (\gamma)) =
\frac {G_L G_\l m^5_L}{192 \pi^3}\, f\!\left(\frac {m^2_\l}{m^2_L}\right)
 \delta^L_W \delta^L_\gamma~,
\eeq
where
\beqn
 G_\l &=& \frac {g^2_\l}{4 \sqrt{2} M^2_W}  \nonumber \\
 \delta^L_W &=& 1 + \frac {3}{5} \frac {m^2_L}{M^2_W} \nonumber \\
 \delta^L_\gamma &=& 1+\frac {\alpha(m_L)}{2\pi}
 \left(\frac {25}{4}-\pi^2\right) \nonumber \\
 f(x) &=& 1 -8x +8x^3 -x^4 -12x^2 {\rm ln}x 
\eeqn
Numerically, the radiative and $W$ propagator corrections are small:
$\delta^\tau_W =1+2.9\times 10^{-4}$,
$\delta^\tau_\gamma=1-43.2\times 10^{-4}$,
$\delta^\mu_W =1+1.0\times 10^{-6}$, and 
$\delta^\mu_\gamma=1-42.4\times 10^{-4}$.

\begin{figure}[tp]
   \centerline{\epsfxsize8.1cm\epsffile{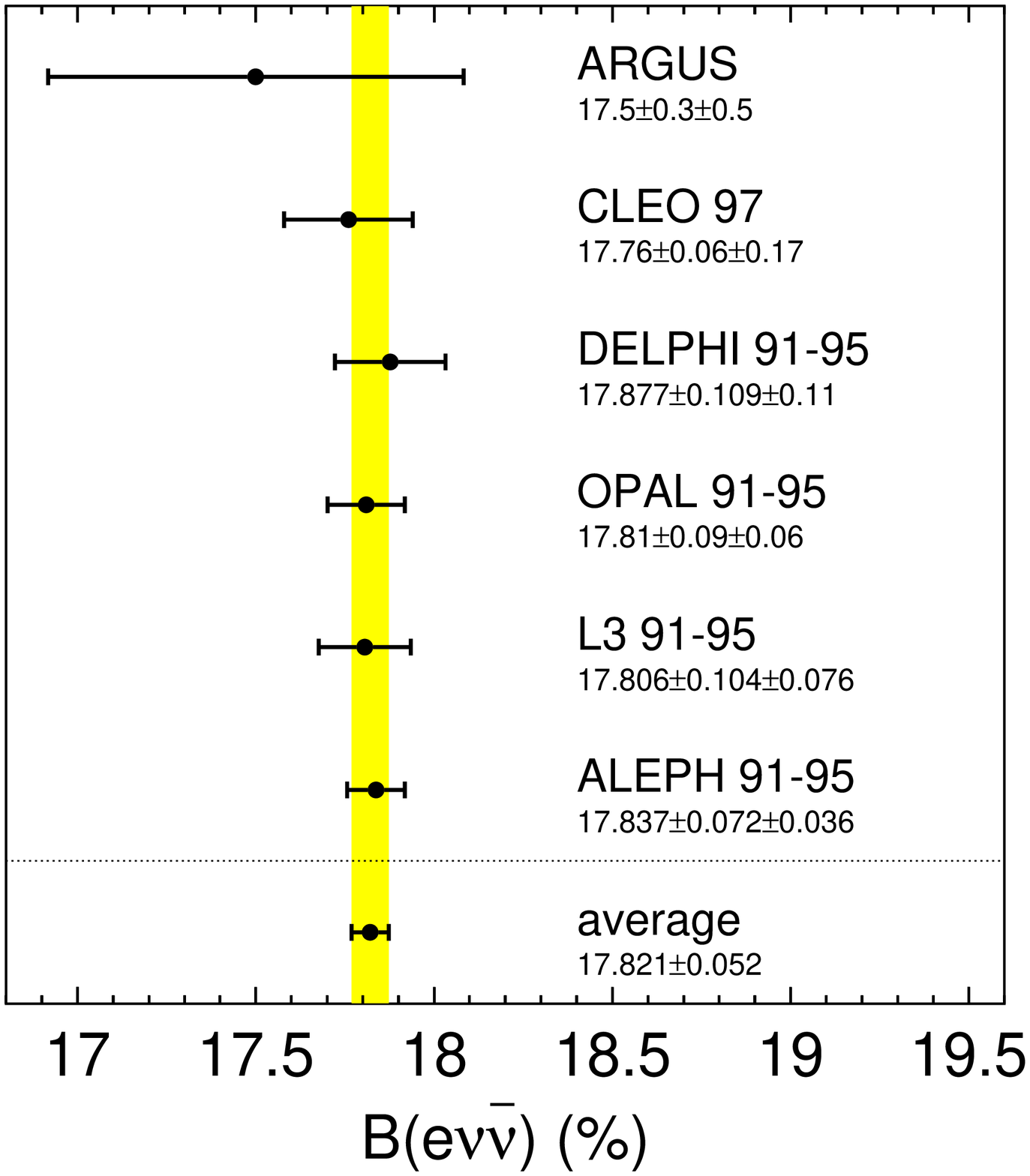}\hspace{0.7cm}
               \epsfxsize8.1cm\epsffile{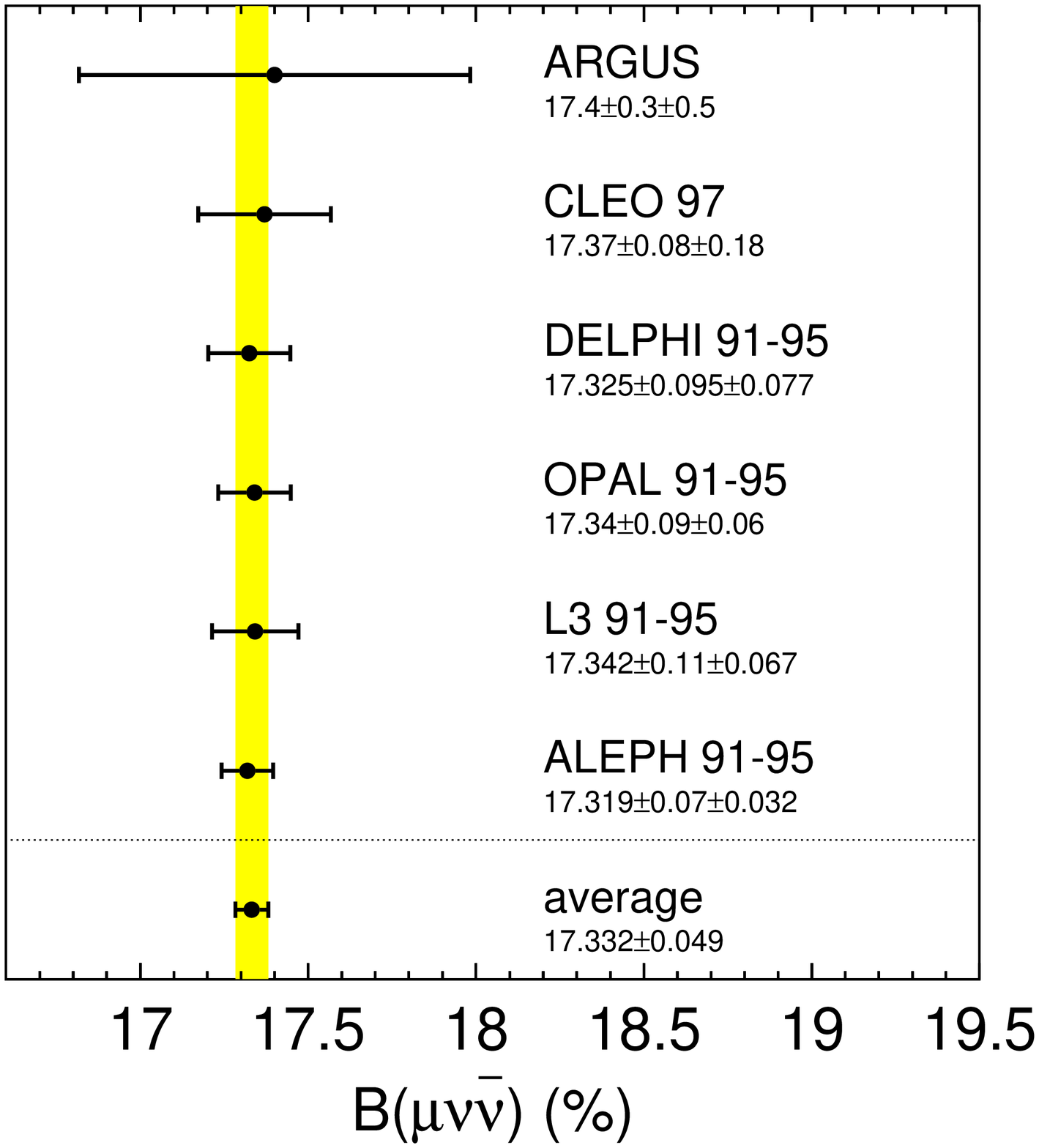}}
   \vspace{-0.1cm}
   \caption{Measurements of the branching fraction for
    $\taum \to \en \nueb \nut$ (left) and $\taum \to \mun \numb \nut$ (right). 
    References for experiments are~\cite{argus_be, cleo_be, delphi_be,
    opal_be, opal_bmu, l3_be, aleph_taubr}.
    The world averages are indicated by the shaded vertical bands.}
\label{br_leptons}
\end{figure}

The experimental situation on the electronic and muonic branching fractions,
$\BR_e$ and $\BR_\mu$, is given in Fig.~\ref{br_leptons}. The results for 
their average are
\beqn
\label{bebmu}
 \BR_e     &=& (17.821 \pm 0.052)\%~, \\
 \BR_\mu   &=& (17.332 \pm 0.049)\%~,
\eeqn
with a relative precision of 0.3\%. The values for $\BR_e$ and $\BR_\mu$
agree with the universality assumption of equal leptonic couplings, 
$g_e=g_\mu=g_\tau$. Their ratio
\beq
 \frac {\BR_\mu} {\BR_e} = 0.9726 \pm 0.0041
\eeq 
is consistent with the predicted value equal to $f(m_\mu^2/m_\tau^2) / 
f(m_e^2/m_\tau^2) = 0.972565\pm0.000009$, using lepton masses 
from~\cite{Eidelman:2004}. The uncertainty arises essentially from the
$\tau$ mass determination, $m_\tau=(1777.03 ^{+0.30} _{-0.26})\mev$, 
dominated by the BES result~\cite{besmtau}.

The result for $\BR_e$ 
can also be compared with the value obtained from the $\tau$ and $\mu$ 
masses and lifetimes $\tau_{\tau,\mu}$, under the assumption of 
$\tau$--$\mu$ universality, according to 
\beqn
  \BR_e &=& \frac {\tau_\tau}{\tau_\mu} \left(\frac {m_\tau}{m_\mu}\right)^5
            \frac {f(m_e^2/m_\tau^2)}{f(m_e^2/m_\mu^2)}
   \frac {\delta_W^\tau \delta_\gamma^\tau}{\delta_W^\mu \delta_\gamma^\mu}
   \\ \nonumber
        &=& \frac{\tau_\tau}{(1\,631.9\pm1.4)\fs}~.
\eeqn
Using the $\tau$ lifetime~\cite{Eidelman:2004},
$\tau_\tau = (290.6 \pm 1.1)\fs$, as well as the $\mu$ lifetime, 
one obtains $\BR_e = (17.807 \pm 0.067)\%$, 
in excellent agreement with the direct measurements.

The direct value for $\BR_e$ and the two derived values
from $\BR_\mu$ and $\tau_\tau$ are in good agreement with each other,
providing a consistent and precise combined result for the electronic 
branching fraction,
\beq
\label{eq:uni_be} 
  	\BR_e^{\rm uni}=(17.818 \pm 0.032)\%~.
\eeq
Hence we find for the branching fraction of 
$\tau \to  \nut\,{\rm hadrons}$, $\BR_{\rm had}$, assuming 
lepton universality and unitarity
\beq
 \BR_{\rm had} 
	= 1-\BR_e-\BR_\mu
	= 1-1.97257\,\BR_e^{\rm uni} = (64.853 \pm 0.063)\%~.
\eeq

\subsection{Decay modes with kaons}
\label{sec:brs_k}

Measurements of $\tau$ decay modes with kaons in the final state are
important on at least two counts: {\it (i)} to determine the 
Cabibbo-suppressed component of the hadronic decay width, leading to 
interesting physics applications (see Section~\ref{sec:strangesf}), and
{\it (ii)} to perform the necessary corrections to the measured 
topological branching fractions, based on charged particles and 
$\pi^0$ multiplicities. This
latter step is made necessary by the difficulty separating charged
pions and kaons, and identifying $\KL$'s at the level of an event.
The subtraction of strange final states is then performed on a
statistical basis. 

The most complete measurements of $\tau$ branching fractions involving 
kaons are from ALEPH~\cite{3prong,k0decay,1prong,keta,ALEPH:1999}.
From these results a comprehensive study of the strange sector 
has been performed, as well as some aspects of 
the dynamics in the final states with a $K\Kb$ pair. The
experimental difficulty resides in the fact that both charged and neutral
kaons have to be identified into multi-particle final states. The analyses 
involve the detection of charged kaons through the measurement of specific 
ionization losses and neutral kaons through either their decay into 
$\pi^+\pi^-$ ($\KS$) or their interactions in a calorimeter ($\KL$).
Assuming \CP invariance the $\KS$ and $\KL$ branching fractions
can be averaged. The higher efficiency achieved for $\KL$ detection
leads to smaller statistical uncertainties. 
Although the strange final states are Cabibbo-suppressed, 
the precision achieved in these measurements is sufficient for 
various significant tests, ranging from $\tau-\mu$ lepton
universality to hadron dynamics, resonance production, QCD analyses,
and the measurement of the Cabibbo angle.
In addition, the consistency of the results is checked through the
verification of isospin relations between the rates observed for
different charge configurations of the same final states.   

New experimental results have been released by CLEO~\cite{CLEOK99,
CLEOK04,CLEOK05,cleoksteta} and OPAL~\cite{OPAL04} achieving
significant progress in the $\Kb\pi\pi$ modes. 
The new measurements tend to favor larger branching fractions.
In particular, the branching fraction for the $K^- \pi^+ \pi^-$
mode, $(2.14 \pm 0.47)\times 10^{-3}$ for ALEPH, becomes in the average
$(3.30 \pm 0.28)\times 10^{-3}$ with a $\chi^2$ of 9.6 for 2 DF (about 
1\% probability). Also, the branching fraction for
$\Kzb \pi^- \pi^0$, $(3.27 \pm 0.51)\times 10^{-3}$ for ALEPH,
becomes $(3.60 \pm 0.40)\times 10^{-3}$ in the average. A compilation of the 
world average $S=1$ branching fractions is given in Table~\ref{brs}.
\begin{table}[t]
\caption[.]{World average branching fractions ($10^{-3}$) for $\tau$ decays
            into strange final states and $\nut$ 
            from~\cite{ALEPH:1999,OPAL04,CLEOK04,CLEOK05,Eidelman:2004}.
            The value for the $(\Kb 3\pi)^- $ branching fraction
            takes into account measurements from ALEPH and CLEO, as well 
            as estimates for unseen final states using isospin relations.
            The estimate for the very small $(\Kb 4\pi)^-$ 
            fraction, in addition to the measured $\Kstarm \eta$ 
            rate~\cite{cleoksteta}, is obtained from an empirical
            observation of pionic modes and Cabibbo suppression.}
\label{brs}
\begin{center}
\setlength{\tabcolsep}{0.0pc}
\begin{tabular*}{\textwidth}{@{\extracolsep{\fill}}lc} 
\hline\noalign{\smallskip}
 Mode & \BR $(10^{-3})$ \\
\noalign{\smallskip}\hline\noalign{\smallskip}
  $K^-$                         & $6.81 \pm 0.23$ \\
  $K^- \pi^0$                   & $4.54 \pm 0.30$ \\
  $\Kzb \pi^-$        & $8.78 \pm 0.38$ \\
  $K^- \pi^0 \pi^0$             & $0.58 \pm 0.24$ \\
  $\Kzb \pi^- \pi^0$  & $3.60 \pm 0.40$ \\
  $K^- \pi^+ \pi^-$             & $3.30 \pm 0.28$ \\
  $K^- \eta$                    & $0.27 \pm 0.06$ \\
  $(\Kb 3\pi)^-$ (estimated)       & $0.74 \pm 0.30$ \\
  $K_1(1270)^- \to K^- \omega$        
                                & $0.67 \pm 0.21$ \\
  $(\Kb 4\pi)^-$ (estimated) and $\Kstarm \eta$     
                                & $0.40 \pm 0.12$ \\
\noalign{\smallskip}\hline\noalign{\smallskip}
  Sum                           &$29.69 \pm 0.86$ \\
\noalign{\smallskip}\hline
\end{tabular*}
\end{center}
\end{table}

The branching fraction for the decay $\taum \to \Km \nu$, $\BR_K$,
\beq
\label{k_exp}
 \BR_K            = (0.681 \pm 0.023)\%~, \\
\eeq
can be compared to the theoretical prediction assuming $\tau$--$\mu$  
universality in the charged weak current, $\BR_K^{\rm uni}$, using 
the values~\cite{Eidelman:2004} for the $\tau$ and $K$ lifetimes and 
masses, the $\Km \to \mu^- \numb$ branching fraction, and a small radiative 
correction~\cite{decker-fink}, $\delta_{\tau/K}=1.0090\pm0.0022$,
\beqn
\label{k_uni}
 \BR_K ^{\rm uni} &=& \frac {\tau_\tau \BR_{K \to \mu \nu}}{\tau_K}
      \frac {m_\tau^3}{2 m_K m_\mu^2}
      \left(\frac {1-m_K^2/m_\tau^2}{1-m_\mu^2/m_K^2}\right)^2 
      \delta_{\tau/K}     \\ \nonumber 
                  &=&\frac {\tau_\tau} {(40\,636 \pm 162)\fs}
                  \;=\; (0.715 \pm 0.003)\%~.
\eeqn

\subsection{Nonstrange hadronic branching fractions}

Nonstrange hadronic branching fractions represent the largest part of
$\tau$ decays and the physics goals require the achievement of small
systematic uncertainties. The main experimental 
difficulty for this task is the photon/$\pi^0$ reconstruction.
The high collimation of $\tau$ decays at LEP energies makes
this task quite often difficult, since these photons are close to one
another or close to the showers generated by charged hadrons. Of particular
relevance is the rejection of fake photons, which are due to 
hadronic interactions, fluctuations of electromagnetic showers, or an
overlap of several showers. These problems reach a tolerable level in
ALEPH thanks to the fine granularity of its electromagnetic calorimeter, 
comprising about 70,000 cells covering the detector angular acceptance,
each cell being segmented threefold along the photon direction in order to 
discriminate between hadronic and electromagnetic showers. Even then,
proper and reliable reconstruction methods had to be developed 
to correctly identify photon candidates. In the ALEPH analysis,
a likelihood method is used for discriminating between genuine 
and fake photons, taking advantage of the detailed calorimeter information
for each detected cluster of energy not associated to a charged track.
Photons converted in the detector material are also included.
Then $\piz$'s are reconstructed: resolved $\piz$'s from two-photon 
pairing, unresolved high-energy $\piz$'s from merged clusters, and residual
$\piz$'s from the remaining single photons after removing radiative,
bremsstrahlung and fake photons with a likelihood method. 

The 13 $\tau$ decay classes defined in Section~\ref{sec:brs_general} are
not fully exclusive as they contain contributions
from final states involving kaons. They are coming from charged
kaons included in the class definition using hadrons ($h$) and from
neutral kaons not taken into account in the classification. These
contributions are subtracted on a statistical basis using the
branching fractions measured separately (Section~\ref{sec:brs_k}).

The $\tau$ decays involving $\eta$ or $\omega$ mesons also require 
special attention because of their electromagnetic
decay modes. Indeed the final state classification relies in part 
on the $\piz$ multiplicity, thereby assuming that 
all photons---except those specifically identified as bremsstrahlung or 
radiative---originate from $\piz$ decays. 
Therefore the non-$\piz$ photons from
$\eta$ and $\omega$ decays are treated as $\piz$ candidates in the
ALEPH analysis and the systematic bias introduced by this effect must be
evaluated. The corrections are based on specific measurements by ALEPH 
of $\tau$ decay modes containing those mesons~\cite{keta}. Hence
the final results correspond to exclusive branching fractions obtained
from the values measured in the topological classification, 
corrected by the contributions from $K$, $\eta$ and $\omega$  
modes measured separately, taking into account through the simulation
their specific selection and reconstruction efficiencies.
This delicate bookkeeping takes into account all the
major decay modes of the considered mesons~\cite{Eidelman:2004}, including
the isospin-violating $\omega \to \pi^+\pi^-$ decay mode.
The main decay modes considered are $\pim \omega$, $\pim \pi^0 \omega$
and $\pim \pi^0 \eta$ with branching fractions of 
$(2.26 \pm 0.18)\times 10^{-2}$, $(4.3 \pm 0.5)\times 10^{-3}$, 
and $(1.80 \pm 0.45)\times 10^{-3}$~\cite{keta}, 
respectively. The first two values are derived from 
the branching fractions for the $2\pim\pip \pi^0$ and $2\pim\pip 2\pi^0$ 
modes obtained in the global analysis and the measured $\omega$ 
fractions of $0.431 \pm 0.033$ from~\cite{keta} 
and the average value, $0.78 \pm 0.06$, from~\cite{keta} 
and~\cite{cleoomega}, respectively.

Small contributions with $\eta$ and $\omega$ have been identified and 
measured in~\cite{cleoeta3pi,cleoeta3pia} with the decay modes 
$\taum \to \nut \eta\pi^-\pi^+\pi^-$ 
$(2.4 \pm 0.5)$, $\taum \to \nut \eta\pi^-2\piz$ $(1.5 \pm 0.5)$,
$\taum \to \nut \omega\pi^-\pi^+\pi^-$ $(1.2\pm 0.2)$,
and $\taum \to \nut \omega\pi^-2\piz$ $(1.5 \pm 0.5)$, where
the branching fractions are given in units of $10^{-4}$.
For the sake of completeness these modes have been included. 
Another small correction has been applied to take into account the $a_1$ 
radiative decay into $\pi \gamma$ with a branching fraction of 
$(2.1 \pm 0.8)\times 10^{-3}$~\cite{zielinski}.

\begin{table}[t]
\caption{Results~\cite{aleph_taubr} on exclusive hadronic branching fractions
         for modes without kaons. The contributions from channels with 
         $\eta$ and $\omega$ are given separately, the latter one only for 
         the electromagnetic $\omega$ decays. The measured branching ratio 
         for \phyx\  is consistent with zero and the value listed in the
         table has been estimated with the help of CLEO 
         results~\cite{cleo_b3h2-3pi0}.}
\setlength{\tabcolsep}{0.0pc}
\begin{center}
\begin{tabular*}{\textwidth}{@{\extracolsep{\fill}}lrc} 
\hline\noalign{\smallskip} 
 Mode & $\BR$ $\pm$ $\sigma_{\hbox{stat}}$ $\pm$ $\sigma_{\hbox{syst}}$
  (in \%) \\
\noalign{\smallskip}\hline\noalign{\smallskip}
 \phyiii  &     10.828 $\pm$  0.070 $\pm$  0.078 &\\
 \phyiv   &     25.471 $\pm$  0.097 $\pm$  0.085 &\\
 \phyv    &      9.239 $\pm$  0.086 $\pm$  0.090 &\\
 \phyvi   &      0.977 $\pm$  0.069 $\pm$  0.058 &\\
 \phyxiii &      0.112 $\pm$  0.037 $\pm$  0.035 &\\
 \phyvii  &      9.041 $\pm$  0.060 $\pm$  0.076 &\\
 \phyviii &      4.590 $\pm$  0.057 $\pm$  0.064 &\\
 \phyix   &      0.392 $\pm$  0.030 $\pm$  0.035 &\\
 \phyx    &      0.013 $\pm$  0.000 $\pm$  0.010 & estimate\\
 \phyxi   &      0.072 $\pm$  0.009 $\pm$  0.012 &\\
 \phyxii  &      0.014 $\pm$  0.007 $\pm$  0.006 &\\
$\pi^- \pi^0 \eta$ & 0.180 $\pm$ 0.040 $\pm$ 0.020 & \cite{keta}\\
$\pi^- 2\pi^0 \eta$ & 0.015 $\pm$ 0.004 $\pm$ 0.003 & \cite{cleoeta3pi,cleoeta3pia}\\
$\pi^- \pi^- \pi^+ \eta$ & 0.024 $\pm$ 0.003 $\pm$ 0.004 & \cite{cleoeta3pi,cleoeta3pia}\\
$a_1^- (\to \pi^- \gamma)$ & 0.040 $\pm$ 0.000 $\pm$ 0.020 & estimate \\
$\pi^- \omega (\to \pi^0 \gamma, \pi^+ \pi^-)$ & 0.253 $\pm$ 0.005 $\pm$ 0.017 & \cite{keta}\\
$\pi^- \pi^0 \omega (\to \pi^0 \gamma, \pi^+ \pi^-)$ & 0.048 $\pm$ 0.006 $\pm$ 0.007 & \cite{keta,cleoomega}\\
$\pi^- 2\pi^0 \omega (\to \pi^0 \gamma, \pi^+ \pi^-)$ & 0.002 $\pm$ 0.001 $\pm$ 0.001 & \cite{cleoeta3pi,cleoeta3pia}\\
$\pi^- \pi^- \pi^+\omega (\to \pi^0 \gamma, \pi^+ \pi^-)$ & 0.001 $\pm$ 0.001 $\pm$ 0.001 & \cite{cleoeta3pi,cleoeta3pia}\\
\noalign{\smallskip}\hline
\end{tabular*}
\label{finalBR}
\end{center}
\end{table}
The final ALEPH results on the hadronic branching fractions for $\tau$
decay modes not including kaons are given in Table~\ref{finalBR}. 
The branching fractions obtained for the different channels are
correlated with each other. On one hand the statistical fluctuations
in the data and the Monte Carlo samples are driven by the multinomial
distribution of the corresponding events, producing well-understood
correlations. On the other hand the systematic effects also induce
significant correlations between the different channels.
All the systematic studies were done keeping track of the correlated
variations in the final branching fraction results, thus allowing a proper
propagation of errors. The full covariance matrices from statistical 
and systematic origins are given in~\cite{aleph_taubr}.

The branching fraction $\BR_\pi$ for the $\pim \nu$ decay mode, 
\beq
 \BR_\pi            = (10.83 \pm 0.11)\%~,
\eeq
agrees well 
with a theoretical prediction only depending on the assumption of 
$\tau$--$\mu$ universality, $\BR_\pi^{\rm uni}$, using 
the values~\cite{Eidelman:2004} for the $\tau$ and $\pi$ lifetimes and 
masses, the $\pim \to \mu^- \numb$ branching fraction, and a small radiative 
correction~\cite{decker-fink}, $\delta_{\tau/\pi}=1.0016\pm0.0014$,
\beqn
 \BR_\pi ^{\rm uni} &=& \frac {\tau_\tau \BR_{\pi \to \mu \nu}}{\tau_\pi}
      \frac {m_\tau^3}{2 m_\pi m_\mu^2}
      \left(\frac {1-m_\pi^2/m_\tau^2}{1-m_\mu^2/m_\pi^2}\right)^2 
      \delta_{\tau/\pi}     \\ \nonumber 
                    &=& \frac {\tau_\tau} {(2\,663.7 \pm 4.0)\fs}
                    \;=\; (10.910 \pm 0.044)\%~,
\eeqn
where the error in the denominator of the prediction is dominated by
the uncertainty on the structure-dependent radiative correction.

Concerning the $(3\pi)^- \nu$ decay mode, which is dominated by the $a_1$ 
resonance, it is interesting to compare the rates in the $2\pim\pip$ and 
$\pim 2\pi^0$ channels. Dominance of the $\rho \pi$ intermediate state 
in the $a_1$ decay predicts equal rates, but (small) isospin-breaking 
effects are expected from different charged and neutral $\pi$ (and 
possibly $\rho$) masses, slightly favoring the $\pim 2 \pi^0$ channel.
However, a partial-wave analysis of the $\pim 2 \pi^0$ final state performed 
by CLEO~\cite{cleoa1} revealed
that this decay is in fact rather complicated involving many intermediate 
states, in particular isoscalars, which amount to about 20\% of the total 
rate and produce strong interference effects. A good description of the 
$a_1$ decays was achieved in this study, which can be applied to the $(3\pi)^-$ 
final state. Including known isospin-breaking effects from the pion masses,
it predicts a ratio of the $2 \pim\pip$ to the $\pim 2 \pi^0$ rate of
0.985. This value is in agreement with the ALEPH measurement~\cite{aleph_taubr}
\beq
 \frac{\BR_{2 \pim\pip}}{\BR_{\pim 2 \pi^0}} = 0.979 \pm 0.018~,
\eeq
supporting the expected trend.

\begin{figure}[t]
  \centerline{
        \epsfxsize7.7cm\epsffile{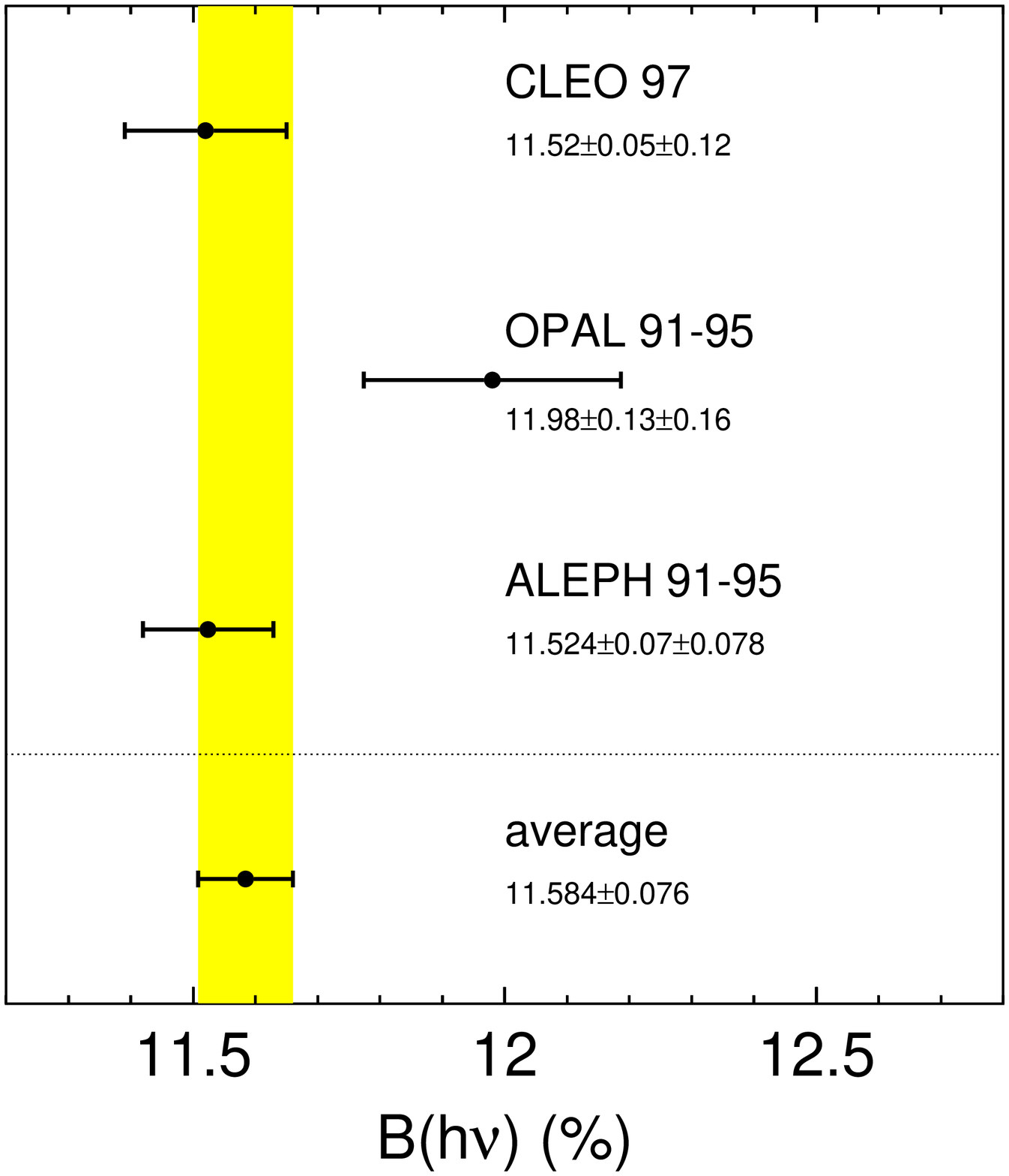}\hspace{0.7cm}
        \epsfxsize7.7cm\epsffile{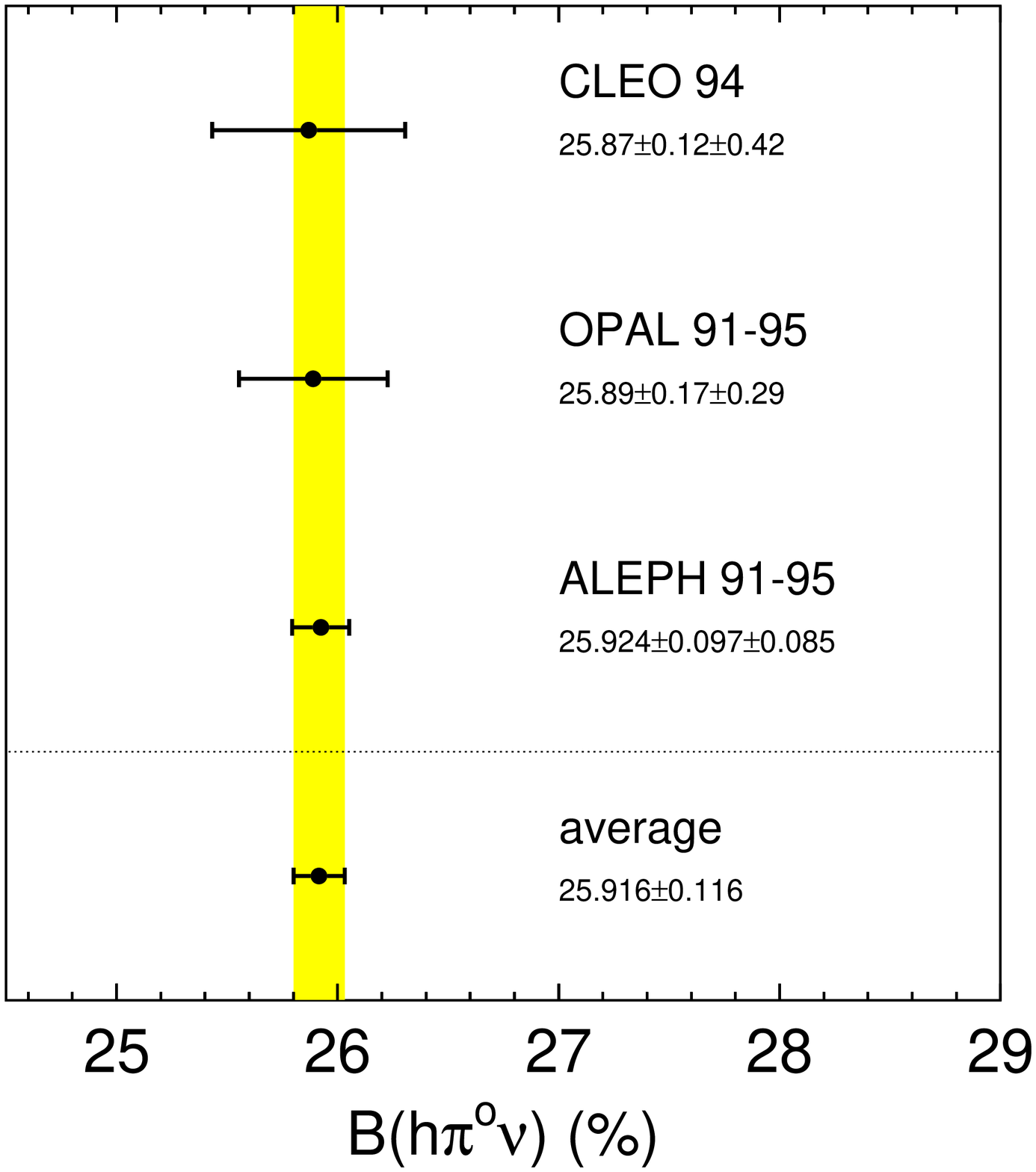}} 
  \centerline{
        \epsfxsize7.7cm\epsffile{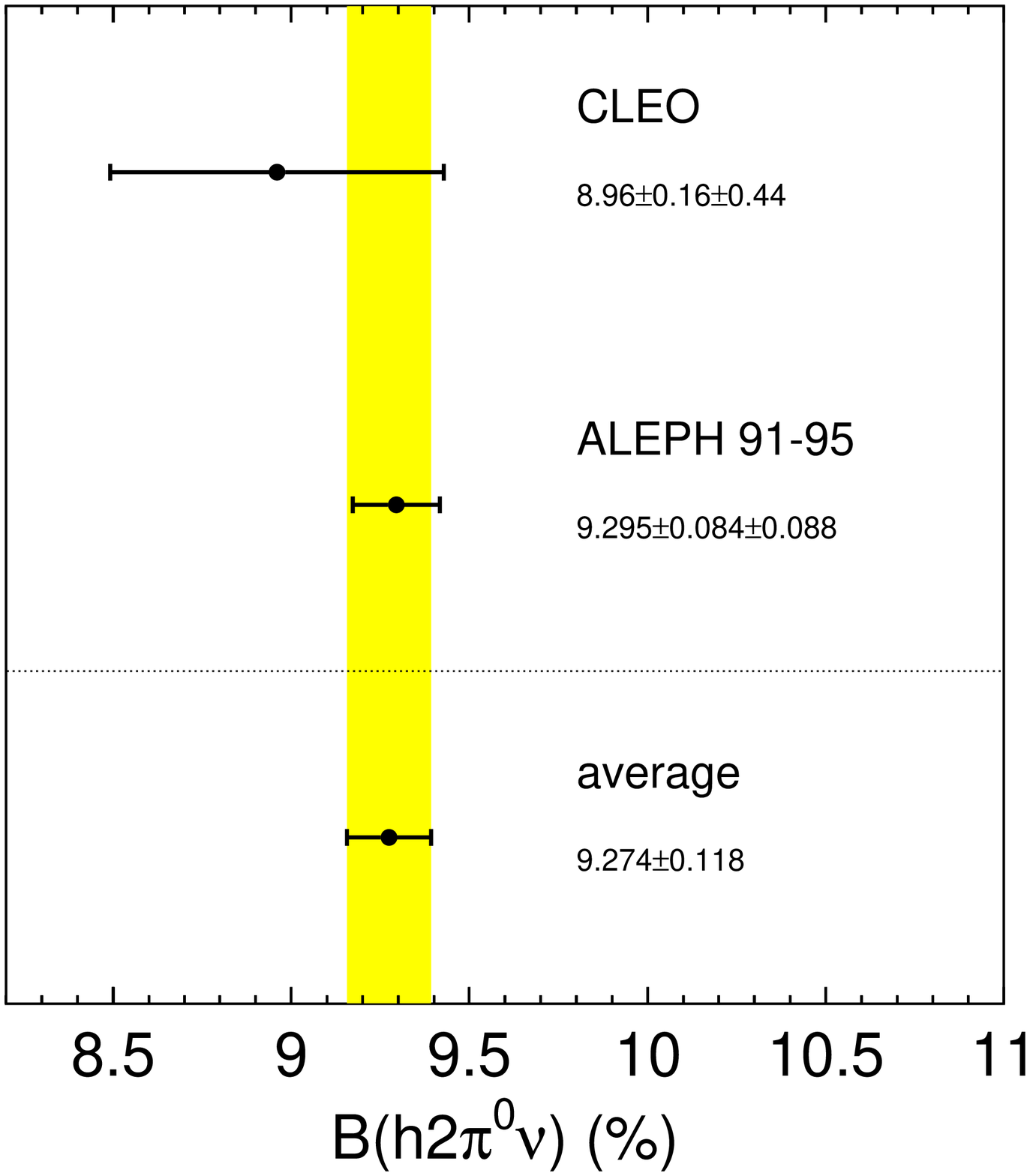}\hspace{0.7cm}
        \epsfxsize7.7cm\epsffile{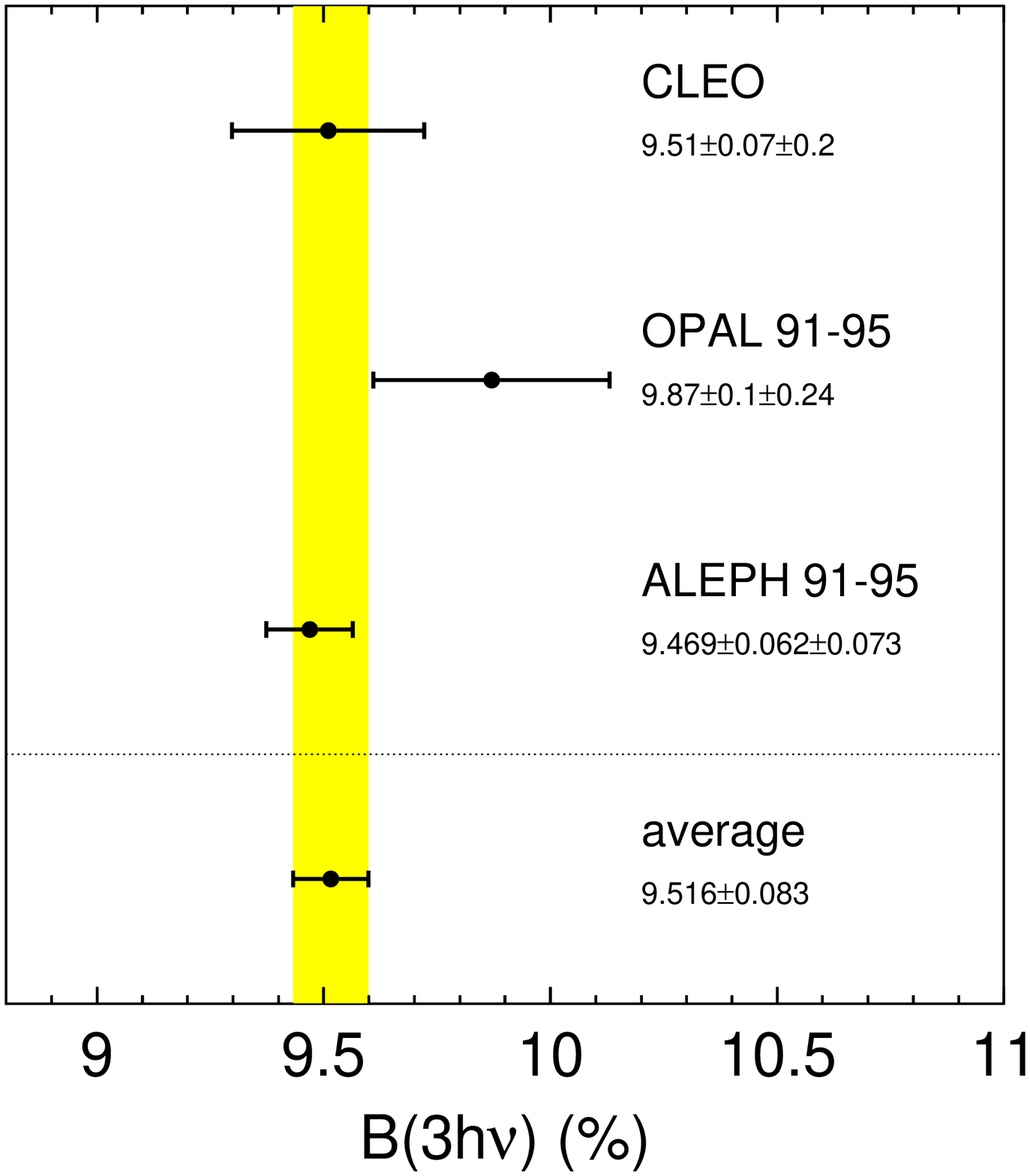}}
  \centerline{
        \epsfxsize7.7cm\epsffile{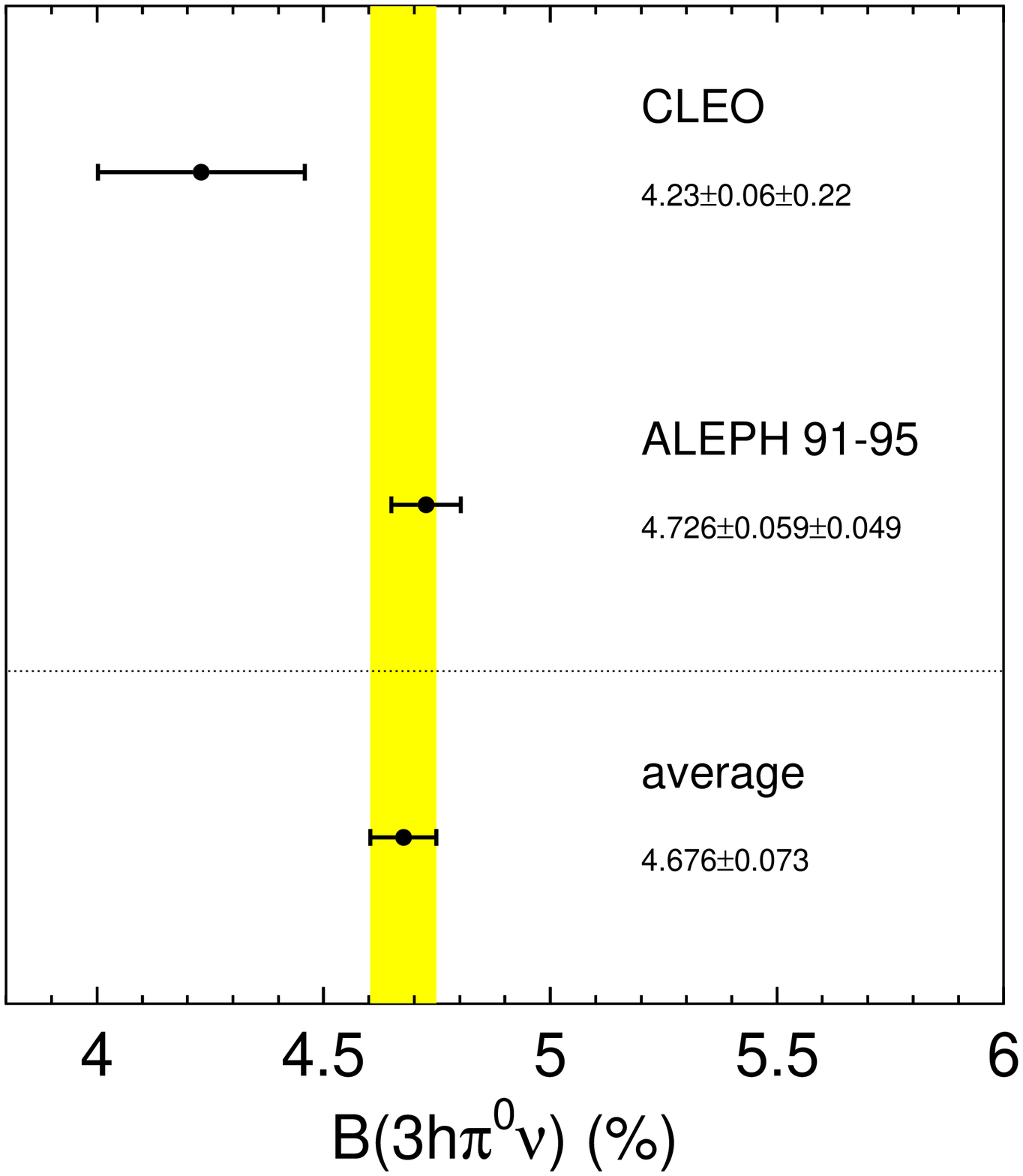}\hspace{0.7cm}
        \epsfxsize7.7cm\epsffile{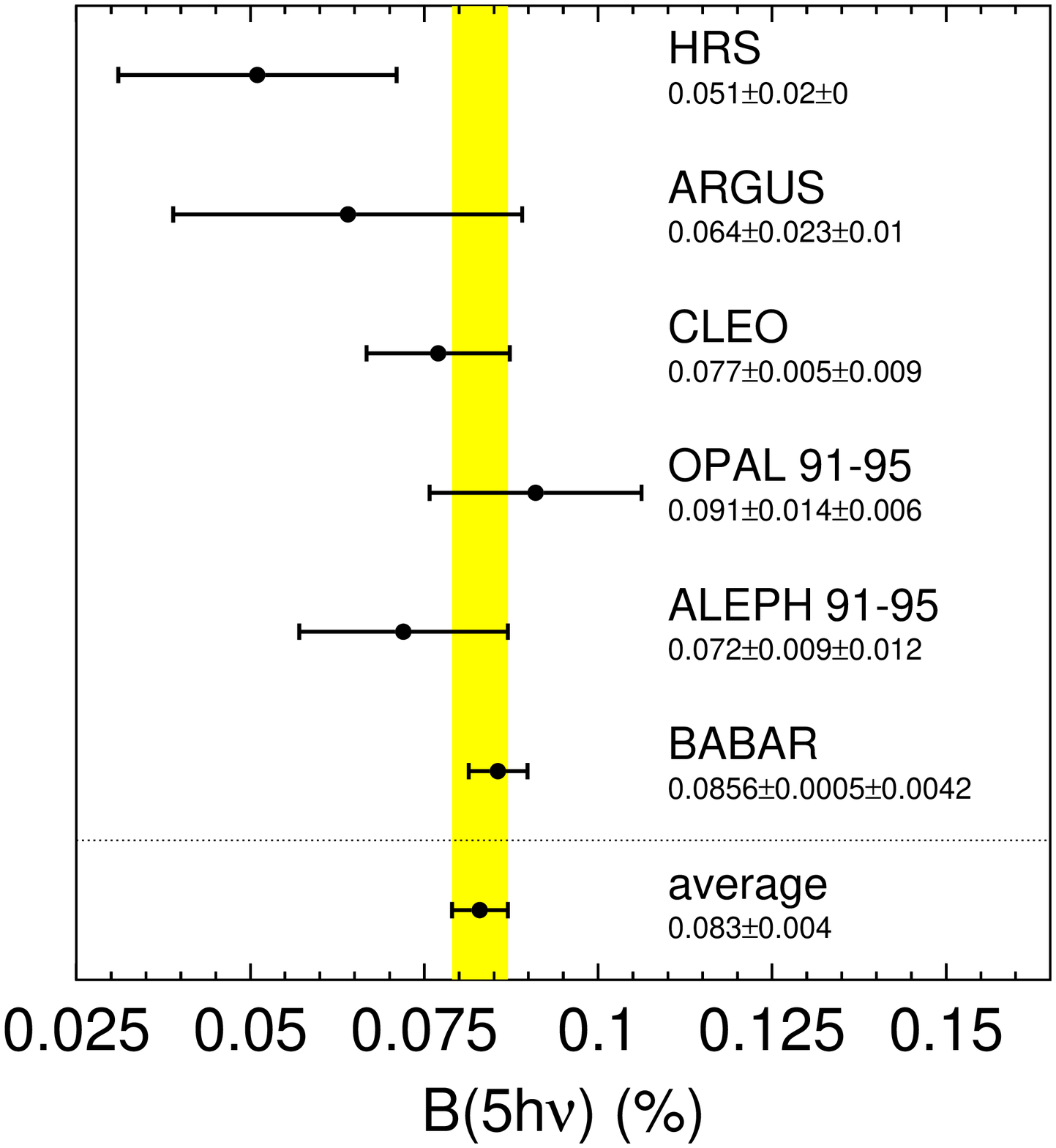}}
  \vspace{-0.1cm}
  \caption[.]{\label{fig:comp_had_1}
   Comparison of results on the main branching fractions for $\tau$ hadronic
   modes~\cite{aleph_taubr, cleo_be, cleo_bhpi0, cleo_bh2pi0, cleo_b3h, 
   cleo_b5h, opal_bh, opal_b5h, argus_b5h, hrs_b5h, babar_5h}. 
   The world averages are indicated by the shaded vertical bands.}
\end{figure}

A comparison of recent measurements of the dominant hadronic 
branching fractions is provided in Figs.~\ref{fig:comp_had_1}.
To compare with other experiments, 
results are given there without charged hadron identification, \ie,
the branching fractions correspond to the sum of all branching 
fractions of the same charged particle topology, whether pions or kaons. 

\subsection{Separation of vector and axial-vector contributions}
\label{sec:vasep}

From the complete analysis of the $\tau$ branching fractions presented 
in~\cite{aleph_taubr} it is possible to determine the nonstrange vector 
($V$) and axial-vector ($A$) contributions to the total $\tau$ hadronic 
width. It is conveniently expressed in terms of their ratios to the 
leptonic width, denoted $\RtauV$ and $\RtauA$, respectively,
while the inclusive strange counterpart is denoted $\RtauS$.
Only limited attempts to separate vector and axial-vector $|\Delta S|=1$ 
currents have been performed by the experiments so far~\cite{ALEPH:1999}.

The ratio \Rtau for the total hadronic width is calculated from 
the leptonic \brs alone, and eventually from the (massless) electronic 
\br only~\footnote
{\label{foot:1}
	The branching ratios $\BR_e^{\rm uni}({\rm ALEPH})=(17.810\pm0.039)\%$ 
        and $\BR_S({\rm ALEPH})=(2.85\pm0.11)\%$ used
	in this section are from the ALEPH analysis only~\cite{aleph_taubr} 
	and differ slightly from the world average values given in 
	Sections~\ref{sec:brs_lepton} and \ref{sec:brs_k}.
},
\beq
\label{eq:rtau}
  \Rtau	= \frac{\Gamma(\taum\to\nut{\rm hadrons}^-)}{\Gamma(\taum\to \nut e^-\nueb)}
	=\frac{1-\BR_e-\BR_\mu}{\BR_e}
           =\frac{1}{\BR_e^{\rm uni}}-1.9726 
       	= 3.642 \pm 0.012~, 
\eeq
for ALEPH, assuming universality in the leptonic weak current. All $R$ values 
given below are rescaled so that the sum of the hadronic and the 
leptonic branching fractions, the latter one computed using the ``universal'' 
value $\BR_e^{\rm uni}({\rm ALEPH})$, add up to one. The deviations 
introduced in this way are small, less than 10\%
of the experimental error, but this procedure guarantees the consistency 
of all values. Using the ALEPH measurement of the strange branching 
fractions~\cite{ALEPH:1999}, supplemented by a (small) contribution from 
the $\Kstarm \eta$ channel measured by CLEO~\cite{cleoksteta}, one finds
(see Footnote~\ref{foot:1})
\beqn
   \BR_S       	&=& (2.85 \pm 0.11)\%~,\nonumber \\
   \RtauS	&=& 0.1603\pm0.0064~,
\eeqn
and consequently
\beqn
\label{eq:rtauvpa}
    \BR_{V+A}        &=& (62.01 \pm 0.14)\%~,\nonumber \\
    R_{\tau,V+A}   &=& 3.482 \,\pm\, 0.014~,
\eeqn
for the nonstrange component.

Separation of $V$ and $A$ components in hadronic final states with only pions
is straightforward. Even number of pions has $G=1$ corresponding to
vector states, while an odd number of pions has $G=-1$, which tags axial-vector 
states. This property places a strong requirement on the knowledge of the $\piz$
reconstruction efficiency in the detector.

Modes with a $K \Kb$ pair are not in general eigenstates 
of $G$-parity and contribute to both $V$ and $A$ channels. 
While the decay to $K^- K^0$ is pure vector, the situation is {\em a priori} 
not clear in the $K \Kb \pi$ channel, observed in three charged modes:
$K^- K^+ \pi^-$, $K^- K^0 \pi^0$ and $K^0 \Kzb \pi^-$.
Three sources of information exist on the possible $V$/$A$ content in these
decays:
\begin{enumerate}

\item 	In the ALEPH analysis of $\tau$ decay modes with 
      	kaons~\cite{ALEPH:1999}, an estimate of the vector contribution was 
      	obtained using the available $e^+e^-$ annihilation data in the  
      	$K \Kb \pi$ channel, extracted in the $I=1$ state. 
      	This contribution was found to be small, yielding 
      	the branching fraction
      	$\BR_{\rm CVC} (\taum \to \nut (K \Kb \pi)^-_V) = 
      	(0.26 \pm 0.39)\times10^{-3}$, which corresponds to an axial fraction
      	of $f_{A,{\rm CVC}}= 0.94^{+0.06}_{-0.08}$.

\item 	The CVC result is corroborated by a partial-wave and 
      	lineshape analysis of the $a_1$ resonance from $\tau$ decays in the 
      	$\nut \pi^- 2\pi^0$ mode in~\cite{cleoa1}. 
      	The observation through unitarity of the opening of the $\Kstar K$ 
      	decay mode of the $a_1$ is postulated and the branching fraction 
      	$\BR(a_1 \to \Kstar K) = (3.3 \pm 0.5)\%$ is derived. With the known
      	decay rate of $\taum \to \nut a_1^-$, it is easy to see 
      	that such a result saturates, and even exceeds, 
      	the total rate for the $K \Kb \pi$ channel. 
      	The corresponding axial fraction is $f_{A,3\pi} = 1.30 \pm 0.24$.

\item 	A new piece of information, also contributed by CLEO, but 
      	conflicting with the two previous results, was recently 
      	published~\cite{CLEO:2004}. It is based on a partial-wave analysis 
      	in the $K^- K^+ \pi^-$ channel using two-body resonance production and 
      	including many possibly contributing channels. A much smaller axial 
      	contribution is found here: $f_{A,K\Kb\pi} = 0.56 \pm 0.10$.

\end{enumerate}

Since the three determinations are inconsistent, a conservative value 
of $f_A = 0.75 \pm 0.25$ is assumed. It encompasses the range allowed 
by the previous results and still represents some progress over earlier 
analyses~\cite{aleph_vsf,aleph_asf}, where a value of $0.5 \pm 0.5$ was used. 
For the decays into $K \Kb\pi\pi$ no information is available 
in this respect, and a fraction of $0.5 \pm 0.5$ is used.

The total nonstrange vector and axial-vector contributions 
obtained in~\cite{aleph_taubr} are
\beqn
\label{eq:rtauv}
   \BR_V      	&=& (31.82 \pm 0.18 \pm 0.12)\%~,\\
  \RtauV	&=& 1.787 \pm 0.011 \pm 0.007~,\\
\label{eq:rtaua}
   \BR_A	&=& (30.19 \pm 0.18 \pm 0.12)\%~,\\
  \RtauA	&=& 1.695 \pm 0.011 \pm 0.007~,
\eeqn
where the second errors reflect the uncertainties in the $V/A$ 
separation in the channels with $K \Kb$ pairs. Taking care 
of the (anti)correlations between the respective uncertainties, 
one obtains the difference between the vector and axial-vector 
components, which is physically related to the amount of 
nonperturbative contributions in the nonstrange hadronic $\tau$ 
decay width (see Section~\ref{sec:qcd})
\beqn
    \BR_{V-A}     &=& (1.63 \pm 0.34  \pm 0.24)\%~,\\
    R_{\tau,V-A}  &=& 0.092 \pm 0.018 \pm 0.014~,
\label{eq:rtauvma}
\eeqn
where the second error has the same meaning 
as in Eqs.~(\ref{eq:rtauv}) and (\ref{eq:rtaua}). The ratio
\beq
\label{eq:rvmavpa}
    \frac {R_{\tau,V-A}} {R_{\tau,V+A}}     = 0.026 \pm 0.007
\eeq
is a measure of the relative importance of nonperturbative contributions.
Its smallness, in spite of our subtraction of observables with 
very different spectral contributions from hadronic resonances,
represents a proof of the remarkable property that is referred 
to as {\em quark-hadron duality}$\,$\footnote
{
	Reference~\cite{shifman} gives the following definition of quark-hadron
	duality. ``In a nutshell, a {\em truncated} OPE is analytically
	continued, term by term, from the Euclidean to the Minkowski
	domain. A smooth quark curve obtained in this way is supposed
	to coincide at high energies (energy releases) with the actual 
	hadronic cross section.''
}~\cite{svz}. 
It is the basis of the QCD analyses of hadronic $\tau$ decays 
presented in this review.

The inclusive vector and axial-vector branching fractions provide 
the normalization of the corresponding spectral functions, which 
will be discussed below and which are the ingredients for the vacuum 
polarization calculations and the QCD analysis of hadronic $\tau$ decays.

\section{TAU HADRONIC SPECTRAL FUNCTIONS}
\label{sec:tauspecfun}

The $\tau$ is the only lepton of the three-generation SM
that is heavy enough to decay into hadrons. It is therefore 
an ideal laboratory for studying the charged weak hadronic 
currents and QCD. 
In analogy to the purely leptonic $\tau$ decays, the invariant 
amplitude for the hadronic decays can be
written in form of a factorized current-current interaction
\beq
   {\cal M}(\tau^-\to {\rm hadrons}^-\nu_\tau) =
      \frac{\GF}{\sqrt{2}}|V_{\rm CKM}|\l_\mu h^\mu~,
\eeq
with the corresponding nonstrange or strange CKM matrix element 
$|V_{\rm CKM}|$, and where $\l_\mu$ describes 
the leptonic $V-A$ current
\beq
   \l_\mu = \nutb\gamma_\mu(1-\gamma_5)\tau~
\eeq
of the weak interaction.
The hadronic transition current $h_\mu$ is the piece of 
interest here. It probes the matrix element of the 
left-handed charged current between the vacuum and the 
hadronic final state. Restricting to a $V-A$ structure, 
one can write
\beq
   h_\mu = \langle{\rm hadrons}|V_\mu(0)-A_\mu(0)|0\rangle~.
\eeq
The differential $\tau$ hadronic width can be expressed 
as follows
\beq
\label{eq:tensor}
   d\Gamma(\tau^-\to {\rm hadrons}^-\nu_\tau) =
      \frac{\GF^2}{4M_\tau}|V_{\rm CKM}|^2L_{\mu\nu}H^{\mu\nu}\,d{\rm PS}~,
\eeq
with the leptonic (hadronic) tensor $L_{\mu\nu}$ ($H_{\mu\nu}$)
and the Lorentz invariant phase space element $d$PS. The hadronic tensor 
obeys a Lorentz invariant decomposition, the dynamical structure being 
embedded in so-called structure functions~\cite{Kuehn:1992}. 

For the simplest hadronic decay modes $\tau^-\to h^-\nut$
($h= \pi,K$), the structure functions reduce to 
$\delta$-distributions, whereas resonances appear
in $\tau$ decays into multiple hadrons. The hadronic structure 
is described by spectral functions. 
The quantum number corresponding to the (conserved) vector and
axial-vector currents is the isotopic $G$-parity. It is a 
generalized multiplicative symmetry of multi-pion 
systems under the successive operations $C$ and $R$, where $C$ 
conjugates the charge and $R$ rotates the system around the 
second axis in the $I=1$ isospace.
The measurement of the $\tau$ vector and axial-vector current \sfs\ 
requires the selection and identification of $\tau$ decay modes with 
a defined $G$-parity $G=+$1 and $G=-1$, and hence
hadronic channels with an even and odd number of neutral {\it or} 
charged pions, respectively. Since hadronic final states 
of different $G$-parity differ also in their $J^P$ quantum
numbers, there is no interference between these two states.
Hence the total hadronic width separates into
$\Gamma_{\rm tot} = \Gamma_V+\Gamma_A$.

\subsection{Definitions}
\label{sec:tauspecfun_def}

The \sf\ $v_1$ ($a_1$, $a_0$), where the subscript 
refers to the spin $J$ of the hadronic system, is defined
for a nonstrange ($|\Delta S|=0$) or strange ($|\Delta S|=1$) vector 
(axial-vector) hadronic $\tau$ decay channel ${V^-}\nut$ (${A^-}\nut$). 
The \sf\  is obtained by dividing the normalized invariant mass-squared 
distribution $(1/N_{V/A})(d N_{V/A}/d s)$ for a given hadronic mass 
$\sqrt{s}$ by the appropriate kinematic factor
\beqn
\label{eq:sf}
   v_1(s)/a_1(s) 
   &=&
           \frac{m_\tau^2}{6\,|V_{\rm CKM}|^2\,\Sew}\,
              \frac{\BR(\tau^-\to {V^-/A^-}\,\nut)}
                   {\BR(\tau^-\to e^-\,\nueb\nut)} \nonumber \\
   & & 
              \times\,\frac{d N_{V/A}}{N_{V/A}\,ds}\,
              \left[ \left(1-\frac{s}{m_\tau^2}\right)^{\!\!2}\,
                     \left(1+\frac{2s}{m_\tau^2}\right)
              \right]^{-1}\hspace{-0.3cm}, \\[0.2cm]
   a_0(s) 
   &=& 
           \frac{m_\tau^2}{6\,|V_{\rm CKM}|^2\,\Sew}\,
              \frac{\BR(\tau^-\to {\pi^-(K^-)}\,\nut)}
                   {\BR(\tau^-\to e^-\,\nueb\nut)}
              \frac{d N_{A}}{N_{A}\,ds}\,
              \left(1-\frac{s}{m_\tau^2}\right)^{\!\!-2}\,
              \hspace{-0.3cm},
\label{eq:spect_fun}
\eeqn
where $\Sew$ accounts for electroweak radiative 
corrections~\cite{Marciano:1988} (see Section~\ref{sec:cvc_isobreak} for
the relevant values of $\Sew$). 
Since CVC is a very good approximation for the nonstrange sector,
the $J=0$ contribution to the nonstrange vector \sf\ is put to zero, 
while the main contributions to $a_0$ are from the pion or kaon poles,
with $d N_{A}/N_{A}\,ds = \delta (s-m_{\pi,K}^2)$.
They are connected through partial conservation of the axial-vector 
current (PCAC) to the corresponding decay 
constants, $f_{\pi,K}$.
The \sfs\  are normalized by the ratio of the vector/axial-vector 
\bfr\ $\BR(\tau^-\to {V^-/A^-}\nut)$ to the \bfr\ of 
the massless leptonic, \ie, electron, channel (discussed in 
Section~\ref{sec:brs_lepton}).

The values for the CKM matrix elements $V_{ud}$ and $V_{us}$ have been 
recently much debated, because the determinations from nuclear $\beta$ 
decays and semileptonic kaon decays did not satisfy the
unitarity relation in the first row of the CKM matrix, that is
$|V_{ud}|^2+|V_{us}|^2+|V_{ub}|^2=1$ (for this comparison, 
the $V_{ub}$ element has a negligible contribution). The situation has
improved with the availability of new results from $K_{\l3}$
and $K_{\mu2}$ decays with revisited corrections from chiral perturbation
theory on one hand, and more precise radiative corrections for the nuclear
$\beta$ decays~\cite{marciano_ckm05} on the other hand. The current
results~\cite{isidori_ckm05}, which are still preliminary\footnote
{
	We thank V.~Cirigliano and G.~Isidori for their advice on 
	the current best values for $|V_{ud}|$ and $|V_{us}|$.
},
\beqn
 |V_{ud}| &=& 0.9739 \pm 0.0003 ~~~~~~~~{\rm nuclear}~\beta~{\rm decays}~, \\
 |V_{us}| &=& 0.2248 \pm 0.0016 ~~~~~~~~K_{\l3} + K_{\mu2}~,
\eeqn
are consistent with CKM unitarity and provide a new level of precision
in this sector. The $|V_{ud}|$ value gives $|V_{us}|=0.2269 \pm 0.0013$ 
if CKM unitarity is used.

Using unitarity and analyticity, the \sfs\ 
are connected to the imaginary part of the two-point correlation (or 
hadronic vacuum polarization) functions
\beqn
\label{eq:correlator}
\Pi_{ij,U}^{\mu\nu}(q) &\equiv& i\int d^4x\,e^{iqx}
	\langle 0|T(U_{ij}^\mu(x)U_{ij}^\nu(0)^\dag)|0\rangle \nonumber\\
	&=&
	\left(-g^{\mu\nu}q^2+q^\mu q^\nu\right)
	\,\Pi^{(1)}_{ij,U}(q^2)+q^\mu q^\nu\,\Pi^{(0)}_{ij,U}(q^2)
\eeqn
of vector $(U_{ij}^\mu= V_{ij}^\mu=\qbar_j\gamma^\mu q_i)$
or axial-vector ($U_{ij}^\mu= A_{ij}^\mu=\qbar_j\gamma^\mu\gamma_5 q_i$) 
color-singlet quark currents, and
for time-like momenta-squared $q^2>0$. Lorentz decomposition 
is used to separate the correlation function into its $J=1$ and $J=0$ 
parts. The polarization function $\Pi_{ij,U}^{\mu\nu}(s)$ have a branch 
cut along the real axis in the complex $s=q^2$ plane. Their imaginary 
parts give the spectral functions defined in~(\ref{eq:sf}), for nonstrange 
(strange) quark currents
\beq
\label{eq:imv}
   \Im\Pi^{(1)}_{\ubar d(s),V/A}(s)
   = \frac{1}{2\pi}v_1/a_1(s)~,\hspace{1cm}
   \Im\Pi^{(0)}_{\ubar d(s),A}(s)
   = \frac{1}{2\pi}a_0(s)~,
\eeq
which provide the basis for comparing short-distance theory with 
hadronic data.

The analytic vacuum polarization function $\Pi_{ij,U}^{(J)}(q^2)$ 
obeys the dispersion relation
\beq
  \label{eq:dispersion}
	   \Pi_{ij,U}^{(J)}(q^2)  =
	     \frac{1}{\pi} \intl_{0}^{\infty}
	     ds\,\frac{{\rm Im}\Pi_{ij,U}^{(J)}(s)}{s-q^2-i\varepsilon} \,+\,
	     {\rm subtractions}~,
\eeq
where the unknown but in general irrelevant subtraction constants 
can be removed by taking the derivative of $\Pi_{ij,U}(q^2)$. The 
dispersion relation allows one to connect the experimentally accessible
spectral functions to the correlation functions $\Pi_{ij,U}^{(J)}(q^2)$, 
which can be derived from theory (QCD).

\subsection{The physical mass spectra and their unfolding}
\label{sec:tauspecfun_meas}

The measurement of the $\tau$ \sfs\  defined in Eq.~(\ref{eq:sf})
requires the determination of the invariant mass-squared 
distributions, obtained from the experimental distributions after 
unfolding from the effects of measurement distortion. The unfolding 
procedure used by the ALEPH collaboration~\cite{aleph_vsf,aleph_asf}
follows a method based on the regularized inversion of the 
simulated detector response matrix using the Singular Value Decomposition 
(SVD) technique~\cite{unfold}. The regularization function applied\footnote
{
	Regularization is necessary since in certain cases the unfolding
	produces results with unphysical behavior: statistically 
	insignificant components in the detector response matrix
	can lead to large oscillations in the unfolded distribution.
	The regularization is achieved by applying a smooth damping 
	function to the unfolded distribution. 
} 
minimizes the average curvature of the distribution. The optimal choice 
of the regularization strength is found by means of the Monte Carlo 
simulation where the true distribution is known and chosen to be close to 
the expected physical distribution. The OPAL collaboration~\cite{opal_vasf}
employs a similar regularized unfolding technique~\cite{blobelunf},
which is however based on cubic splines instead of the binned histograms 
used by ALEPH. Due to the superior mass resolution, the CLEO 
collaboration considers it to be sufficient to only apply a bin-to-bin 
migration correction~\cite{cleotaurho}.

To measure exclusive \sfs, individual unfolding procedures 
with specific detector response matrices and regularization parameters 
are applied for each $\tau$ decay channel considered. An iterative
procedure is followed to correct the Monte Carlo spectral functions
used to subtract the cross-feed among the modes. 

Before unfolding the mass distributions, the $\tau$ and non-$\tau$ 
backgrounds are subtracted. In the case of $\tau$ feed-through the
Monte Carlo distributions normalized to the measured branching 
fractions are used. When the \sfs\  are measured for the nonstrange 
exclusive final states, the
contributions from strange modes classified in the same topology are
subtracted using their Monte Carlo \sfs\  normalized by the measured
branching fractions.

The systematic uncertainties affecting the decay classification 
of the exclusive modes are contained in the systematic errors on 
the branching fractions. Additional systematic uncertainties related 
to the shape of the unfolded mass-squared distributions, 
and not its normalization, are included. They are dominated by the
photon and $\pi^0$ reconstruction.

\subsection{Inclusive nonstrange spectral functions}
\label{sec:tauspecfun_inclusiveresults}

\subsubsection{Vector and axial-vector spectral functions}

\begin{figure}[t]
 \centerline{
        \epsfxsize8.3cm\epsffile{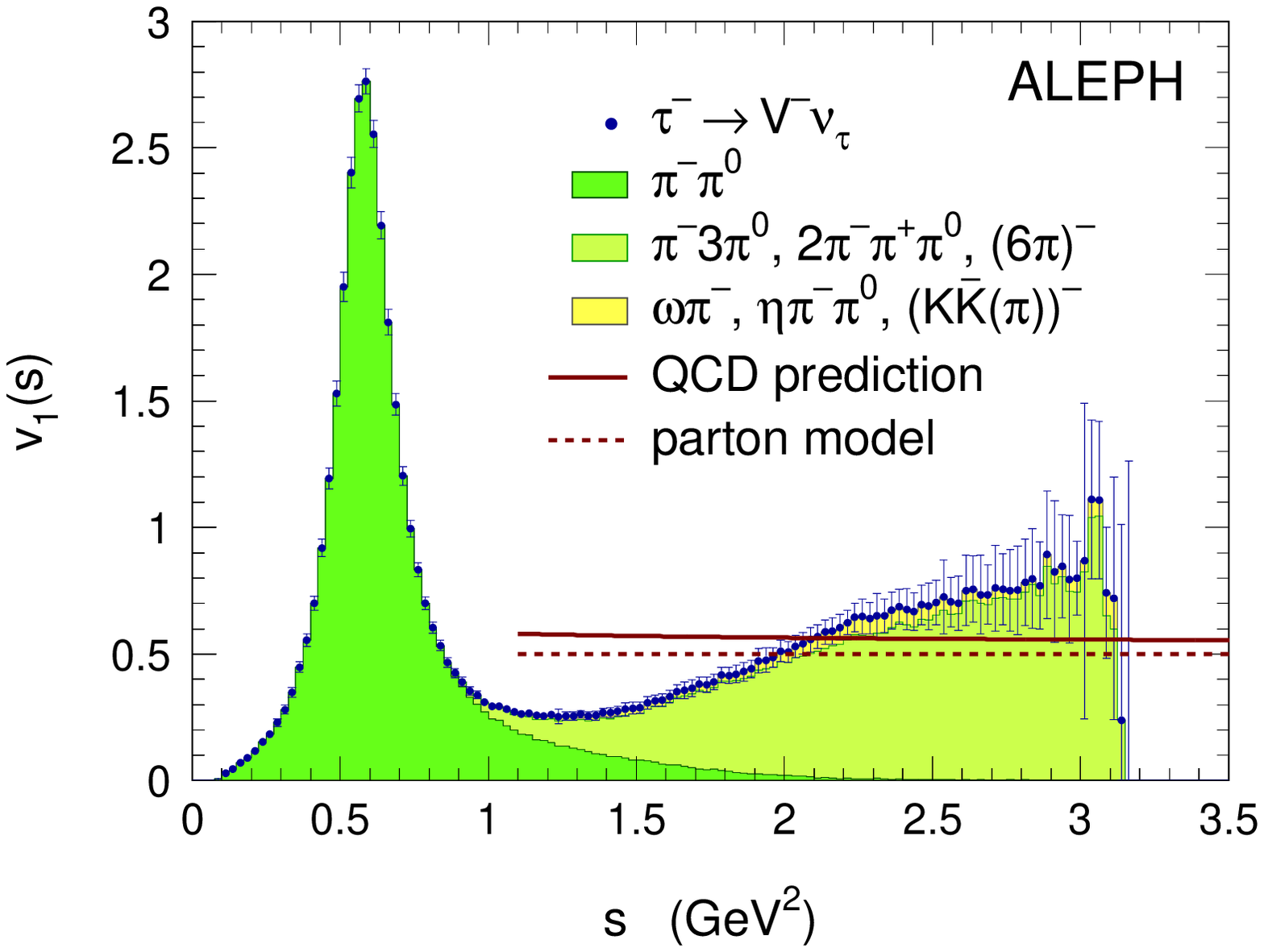}
        \epsfxsize8.3cm\epsffile{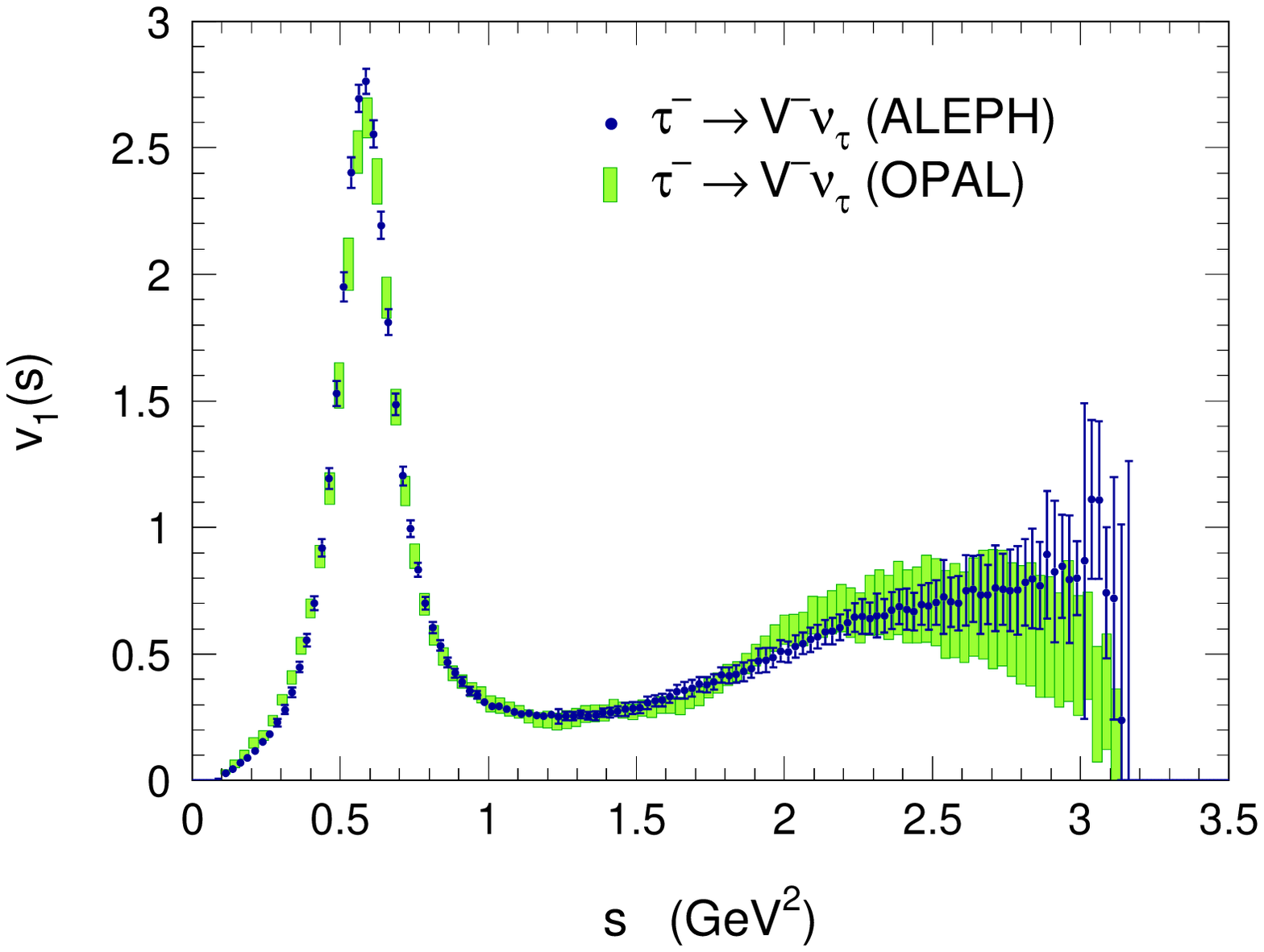}}
  \vspace{0.0cm}
  \centerline{
        \epsfxsize8.3cm\epsffile{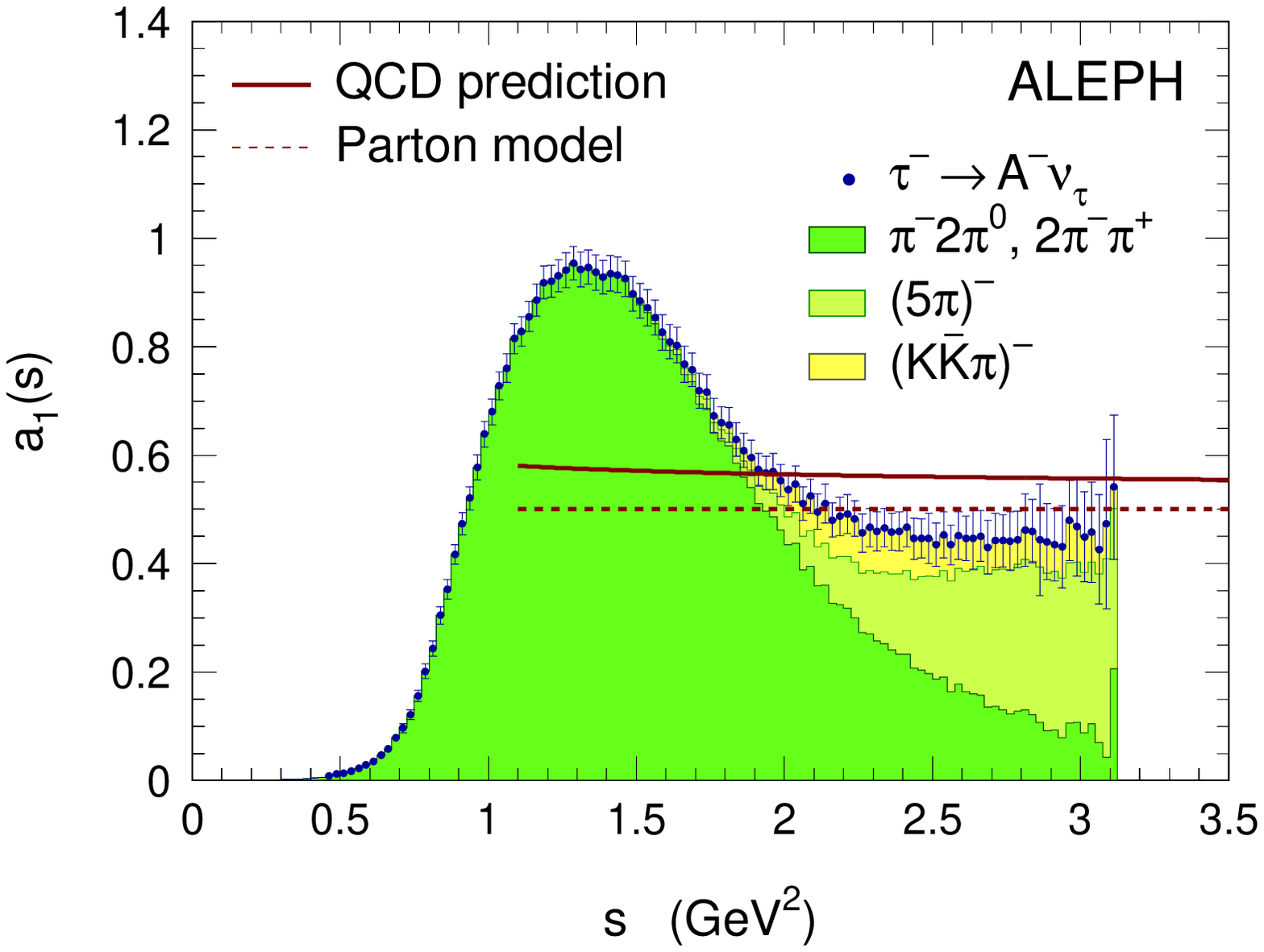}
        \epsfxsize8.3cm\epsffile{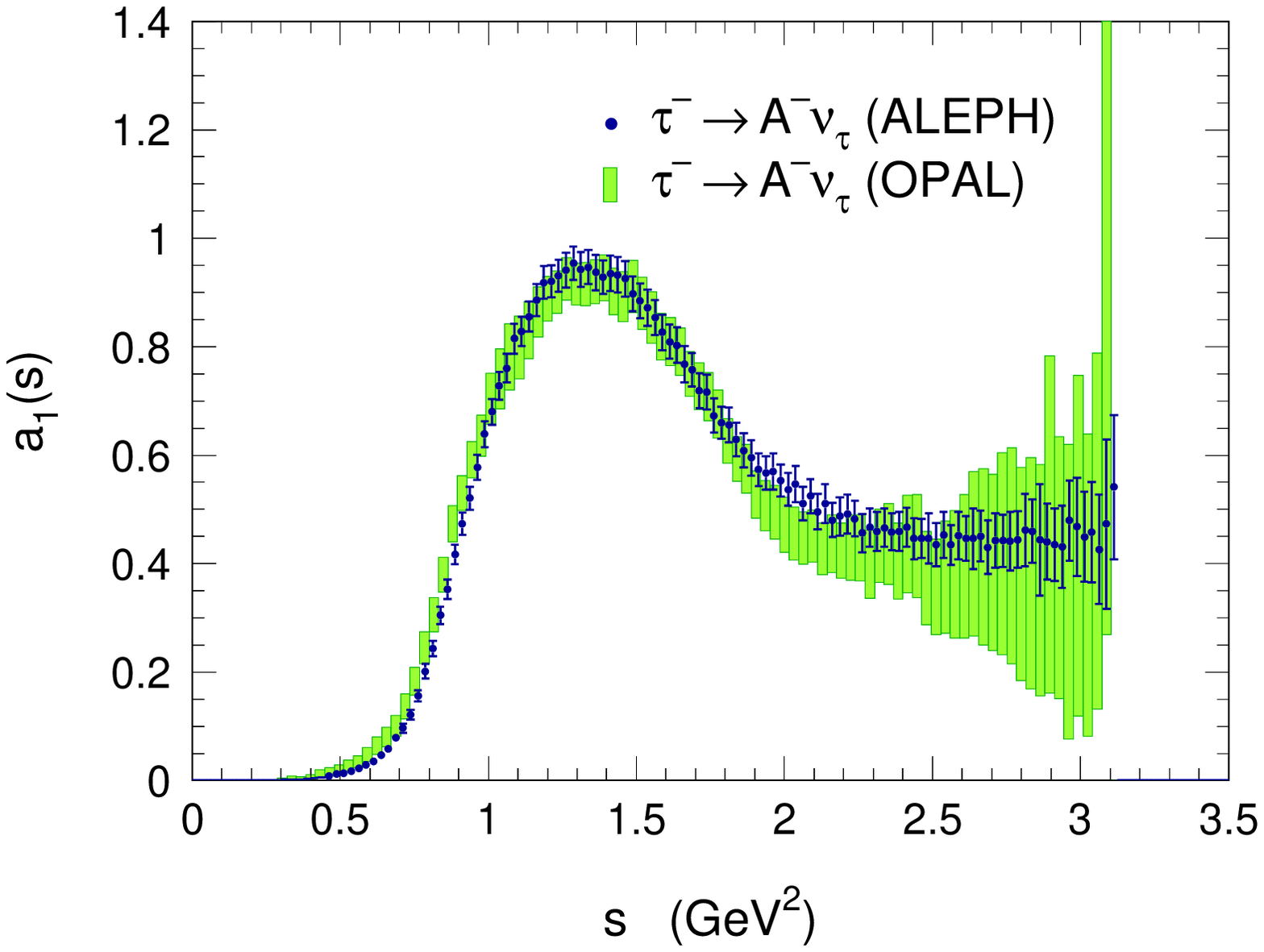}}
  \vspace{-0.3cm}
  \caption[.]{\label{fig:vasf}
        \underline{Left hand plots:} the inclusive vector (upper) and 
	axial-vector (lower) \sfs\   as measured 
	in~\cite{aleph_taubr}. The shaded areas 
	indicate the contributing 
	exclusive $\tau$ decay channels. The curves show the 
	predictions from the parton model (dotted) and from massless 
        perturbative QCD using $\as(M_Z^2)=0.120$ (solid).
	\underline{Right hand plots:} comparison of the inclusive 
	vector (upper) and axial-vector (lower) \sfs\
	obtained by ALEPH and OPAL~\cite{opal_vasf}.}
\end{figure}

The inclusive $\tau$ vector and axial-vector \sfs\   are shown 
in the upper and lower plots of Fig.~\ref{fig:vasf}, respectively.
The left hand plots give the ALEPH 
result~\cite{aleph_taubr,aleph_vsf,aleph_asf} 
together with its most important exclusive contributions, 
and the right hand plots
compare ALEPH with OPAL~\cite{opal_vasf}. The agreement between the
experiments is satisfying with the exception of the $\pim\piz$ (vector)
mode, where some discrepancies occur. We note that the branching fraction 
used for this mode to obtain the inclusive \sf\  is not the same in
both experiments. The \br used by ALEPH~\cite{aleph_taubr} exceeds 
the one used by OPAL~\cite{opal_vasf} by 0.9\% (relative).

The curves in the left hand plots of Fig.~\ref{fig:vasf} represent 
the parton model (leading-order QCD) prediction (dotted) and the improved 
(massless) perturbative QCD prediction (solid), assuming the relevant 
physics to be governed by short distances (Section~\ref{sec:qcd}). 
The difference between 
the two curves is due to higher order terms in the strong coupling 
$(\as(s)/\pi)^n$ with $n=1,2,3$. At high energies the 
spectral functions are assumed to be dominated by continuum production,
which locally agrees with perturbative QCD. This asymptotic region
is not yet reached at $s=m_\tau^2$ for the vector and axial-vector
\sfs.

To build the vector \sf\  ALEPH has measured the two- and four-pion 
final states exclusively, while only parts of the six-pion state.
The total six-pion \br has been determined using isospin 
symmetry~\cite{aleph_vsf}. 
However, one has to account for the fact that the six-pion channel 
is contaminated by isospin-violating $\tauto\eta(3\pi)^-\nut$
decays, as reported by CLEO~\cite{cleoeta3pi,cleoeta3pia}.
A small fraction of the $\omega\pim\nut$ decay channel that is not
reconstructed in the four-pion final state is added using the simulation.
Similarly, one corrects for the $\eta \pi^-\piz\nut$ decay mode where
the $\eta$ decays into pions. For the $\eta\to2\gamma$ mode, the $\tau$ 
decay is classified in the $h^- 3\piz\nut$ 
final state, since the two-photon 
mass is inconsistent with the $\piz$ mass and consequently each photon is 
reconstructed as a $\piz$.
The $\Km\Kz\nut$ mass distribution is taken entirely from the 
simulation (recall that this mode is pure vector). The \sfs\  for the 
final states $K\Kb\pi$ and $K\Kb\pi\pi$ (see Section~\ref{sec:vasep} for 
the discussion on the $G$-parity of these modes) are obtained from the
Monte Carlo simulation, where large systematic 
errors are applied to cover this approximation~\cite{aleph_vsf}.

In analogy to the vector \sf, the inclusive axial-vector
\sf\ is obtained by summing up the exclusive axial-vector \sfs\ 
with the addition of small unmeasured modes taken from the 
Monte Carlo simulation. The small fraction of the 
$\omega \pi^-\piz\nut$ decay channel that is not accounted for 
in the $2\pim\pip2\piz\nut$ final state is added
from the simulation. Also considered are the axial-vector 
$\eta(3\pi)^-\nut$ final states~\cite{cleoeta3pi,cleoeta3pia}. 
CLEO observed that the dominant part of this mode issues from the 
$\tauto f_1(1285)\pi^-\nut$ intermediate state, with
$\BR(\tauto f_1\pi^-\nut)=(0.068\pm0.030)\%$, measured in the 
$f_1\to\eta\pi^+\pi^-$ and $f_1\to\eta 2\piz$ 
decay modes~\cite{cleoeta3pi,cleoeta3pia}. Since the $f_1$ meson is isoscalar, 
the \brs relate as 
$\BR(\tau^-\to\eta\tpi\nut)=2\times\BR(\tau^-\to\eta\pidpiz\nut)$. 
The distributions are taken 
from the ordinary six-pion phase space simulation accompanied 
by large systematic errors. The $K\Kb\pi$ 
final states contribute mostly to the inclusive axial-vector 
\sf, with errors that are fully anticorrelated with the inclusive 
vector \sf. The invariant mass distributions for these channels are 
taken from the simulation.

\subsubsection{Inclusive $V \pm A$ spectral functions}

\begin{figure}[t]
  \centerline{
        \epsfxsize8.3cm\epsffile{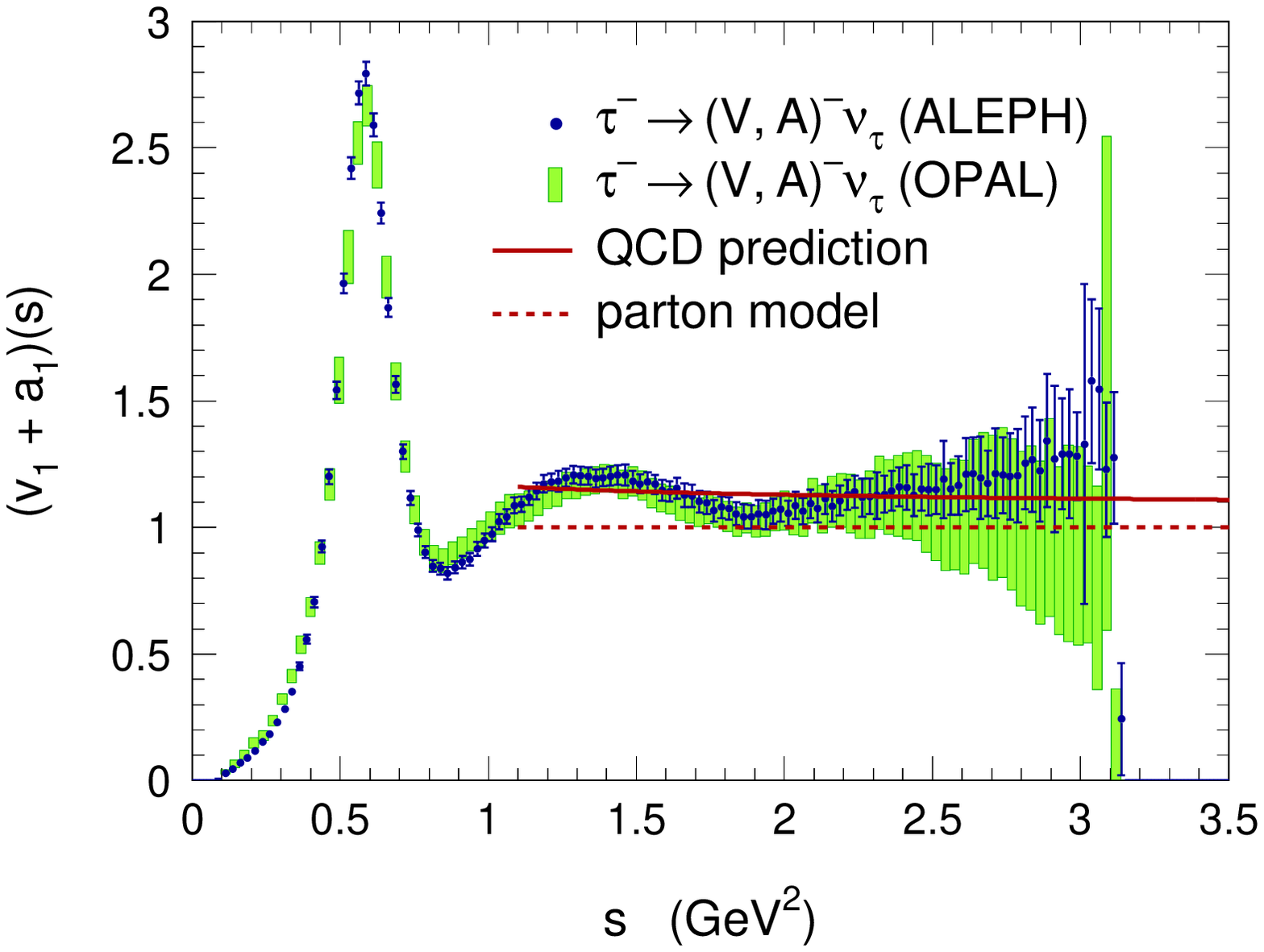}
        \epsfxsize8.3cm\epsffile{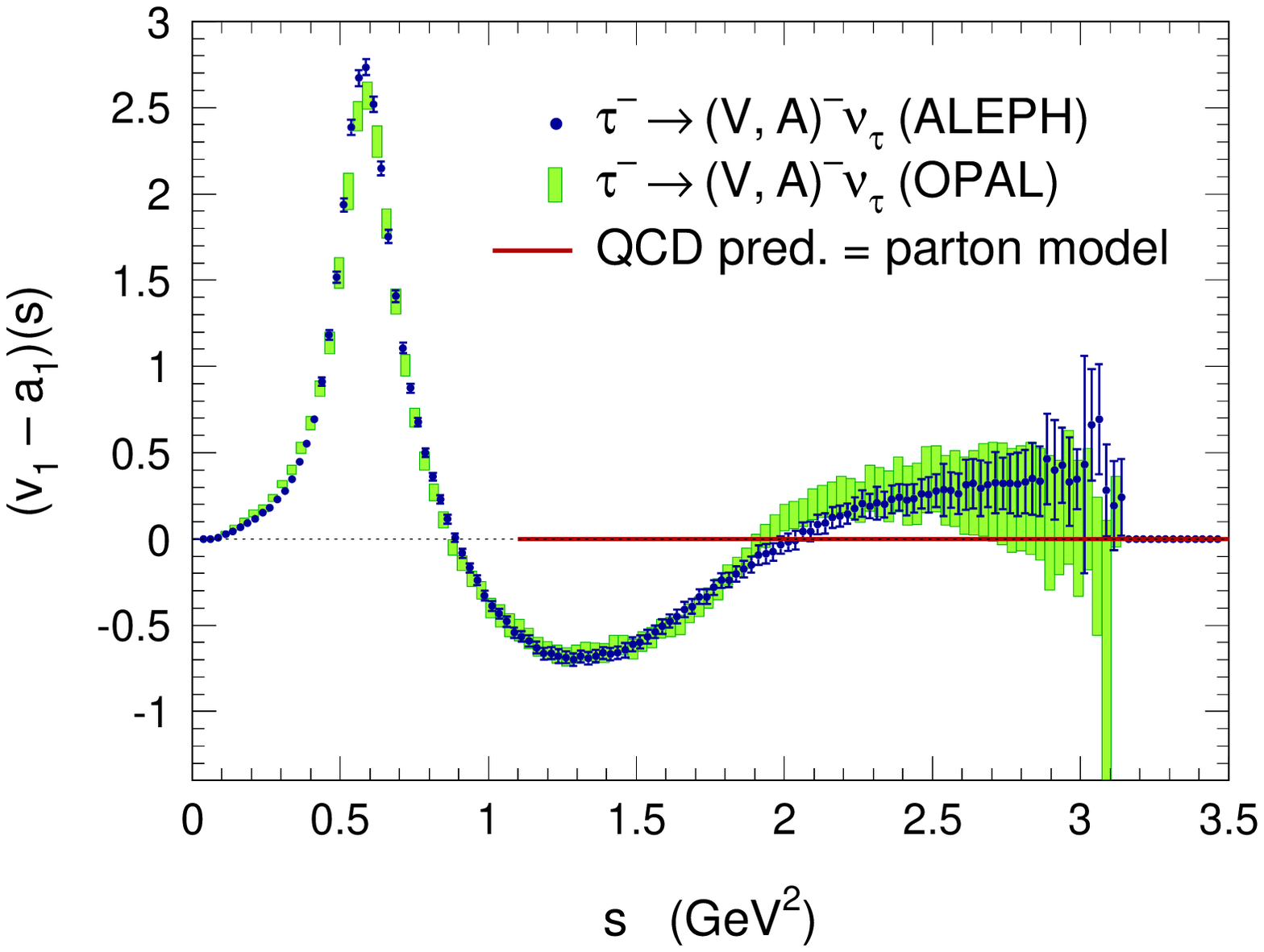}
  }
  \vspace{-0.3cm}
  \caption[.]{\label{fig:vpmasf}
        Inclusive vector plus axial-vector (left) and 
	vector minus axial-vector \sf\  (right) as measured 
	in~\cite{aleph_taubr} (dots with errors bars) 
	and~\cite{opal_vasf} (shaded one standard deviation 
	errors). The lines show the predictions from the 
	parton model (dotted) and from massless 
        perturbative QCD using $\as(M_Z^2)=0.120$ (solid).
	They cancel to all orders in the difference.}
\end{figure}

For the total $v_1+a_1$ hadronic \sf\  it is not necessary to 
distinguish in the experiment whether a given event belongs to
one or the other current.
The one, two and three-pion final states dominate and their exclusive 
measurements are added with proper accounting for the (anti)correlations. 
The remaining contributing topologies are treated inclusively, \ie, without 
separation of the vector and axial-vector decay modes. This reduces the
statistical uncertainty. The effect of the feed-through between $\tau$
final states on the invariant mass spectrum is described by the Monte 
Carlo simulation and resolution effects are corrected by the data unfolding. 
In this procedure the simulated mass distributions are iteratively corrected
using the exclusive vector/axial-vector unfolded mass spectra. 
Also, one does not have to separate the vector/axial-vector currents 
of the $K\Kb\pi$ and $K\Kb\pi\pi$ modes. The $v_1+a_1$ \sfs\  for ALEPH
and OPAL are plotted in the left hand plot of Fig.~\ref{fig:vpmasf}. 
The improvement in precision when comparing to a sum of the two parts 
in Fig.~\ref{fig:vasf} is significant at higher mass-squared values.

One nicely identifies the oscillating behavior of the \sf\  and it is
interesting to observe that, unlike 
the vector/axial-vector \sfs, it does approximately reach the asymptotic 
limit predicted by perturbative QCD at $s\to m_\tau^2$. Also,
the $V+A$ \sf, including the pion pole,
exhibits the features expected from global quark-hadron duality: 
despite the huge oscillations due to the prominent $\pi$, $\rho(770)$, 
$a_1$ and $\rho(1450)$ resonances, the \sf\  qualitatively averages out to 
the quark contribution from perturbative QCD. 

In the case of the $v_1-a_1$ \sf, uncertainties on the $V/A$ 
separation are reinforced due to their anticorrelation.
In addition, anticorrelations in the branching fractions
between $\tau$ final states with adjacent numbers of pions 
increase the errors. The $v_1-a_1$ \sfs\  for ALEPH and OPAL are 
shown in the right hand plot of Fig.~\ref{fig:vpmasf}. The oscillating 
behavior of the respective $v_1$ and $a_1$ \sfs\  is emphasized and the
asymptotic regime is not reached at $s=m_\tau^2$.
However again, the strong oscillation generated by the hadron resonances
mostly averages out to zero, as predicted by perturbative QCD.

\subsection{Strange spectral functions}

\begin{figure}[t]
  \centerline{\epsfxsize9cm\epsffile{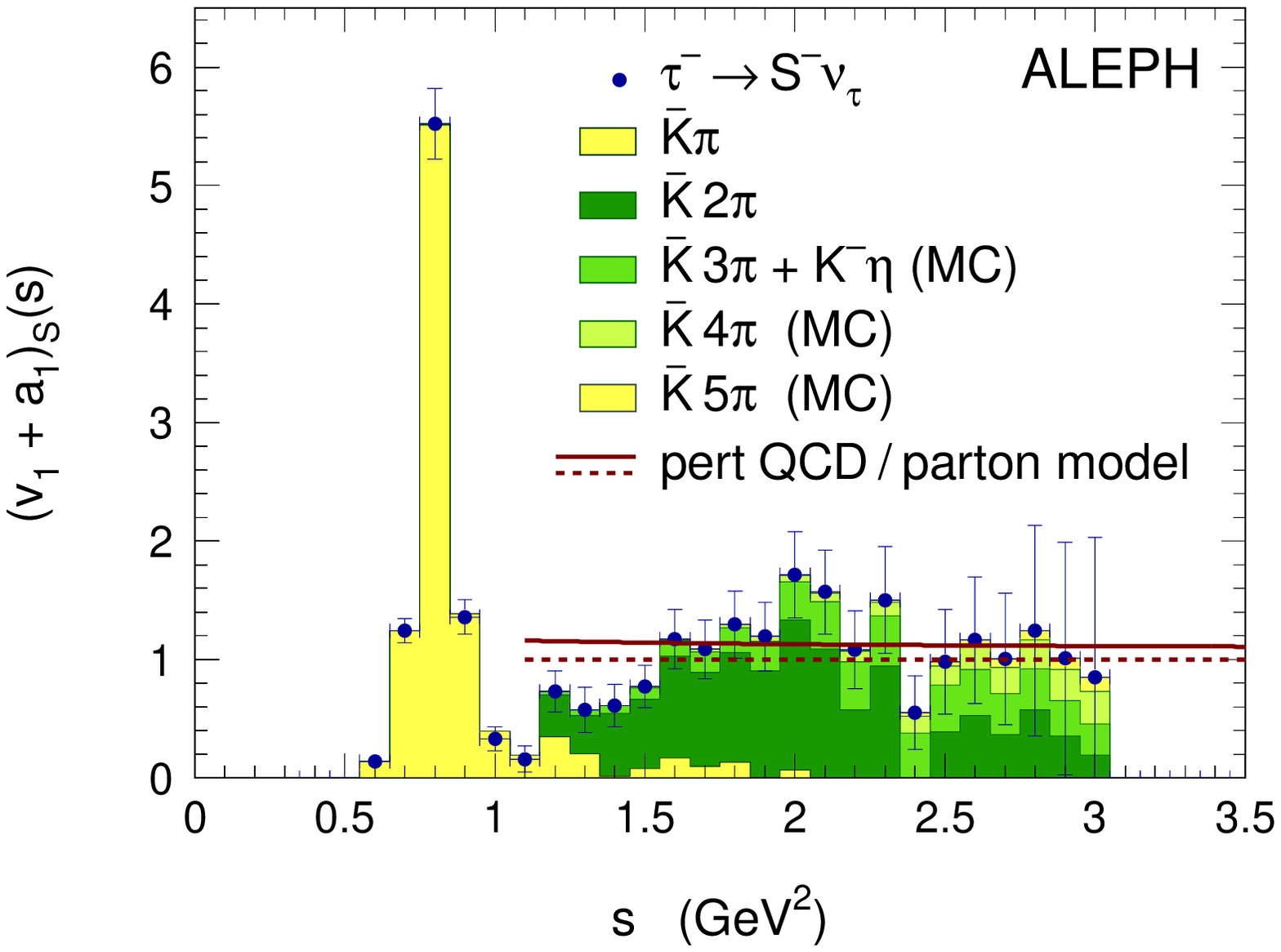}}
  \centerline{\epsfxsize9cm\epsffile{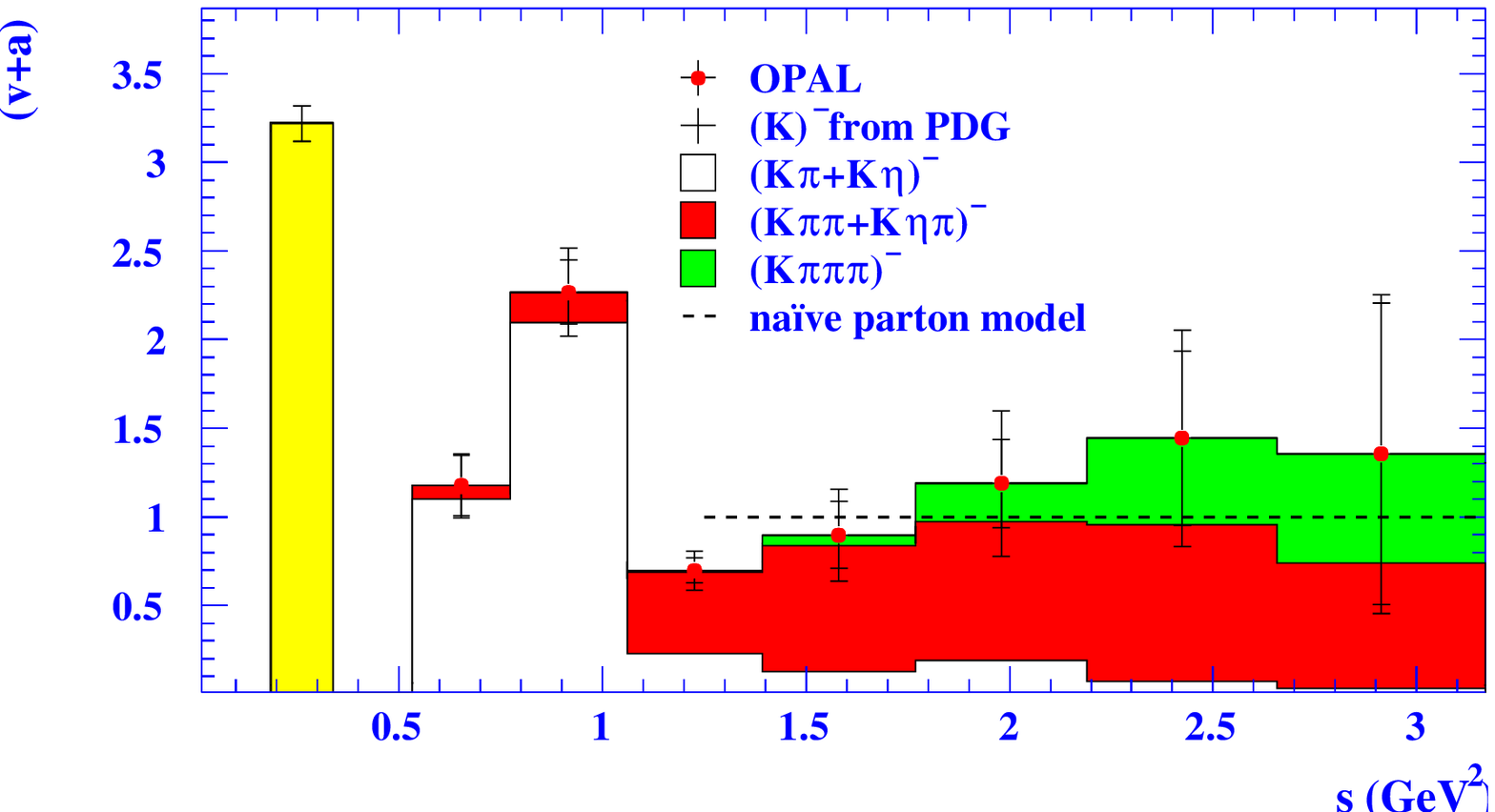}}
  \vspace{0.3cm}
\caption{Inclusive hadronic vector plus axial-vector spectral
	function from $\tau$ decays into $S=\pm1$ final states
	with its different contributions as indicated 
        from~\cite{ALEPH:1999} and~\cite{OPAL04}. 
	The kaon pole is not included in the ALEPH plot.
	The parton model prediction is given by the dashed 
        line.}
\label{specfun_aleph}
\end{figure}

The total $|\Delta S|=1$ spectral functions from the ALEPH and OPAL 
analyses are shown in Fig.~\ref{specfun_aleph}. 
For ALEPH, the spectra for the $\Kb\pi$ and 
$\Kb\pi\pi$ modes are obtained from the
corrected mass spectra, taking into account acceptance and bin migration
corrections. The spectrum for the $\Kb\pi$ mode is taken from the
$\Km\pi^0$ and $\KS \pim$ final states because of their superior mass
resolution as compared to the higher statistics $\KL \pi$ state where the
$\KL$ detection suffers from poor energy resolution. For the other modes 
a phase-space generator is used to simulate the mass 
distributions. The contributions from the respective 
channels are normalized to the corresponding branching fractions.

Since OPAL normalizes each channel using the corresponding world average 
branching fraction, which is dominated by the ALEPH measurement, the added
information is on the shape, however limited by the relatively low
statistics. As a consequence the OPAL and ALEPH spectral functions are 
correlated.

The $\Kstar(892)^-$ resonance stands 
out clearly on the low mass side, while within a large uncertainty 
(dominated by background subtraction in the $(\Kb\pi\pi)^-$ modes)
the higher energy part is consistent with the  parton model (or 
perturbative QCD) expectation.

\section{\boldmath COMPARING TAU SPECTRAL FUNCTIONS WITH ELECTRON-POSITRON ANNIHILATION DATA VIA THE CONSERVED VECTOR CURRENT}
\label{sec:cvc}

The vacuum polarization functions~(\ref{eq:correlator}) for 
the various types of quark currents $ij$ carry the dynamical information
for the theoretical evaluation of total cross sections and decay widths.
In the limit of isospin invariance the vector current is conserved
(CVC), so that the \sf\  of a vector $\tau$ decay mode $X^-\nut$
in a given isospin state for the hadronic system is related 
to the \ee  annihilation cross section of the corresponding 
isovector final state $X^0$
\beq
\label{eq:cvc}
  \sigma_{e^+e^-\to X^0}^{I=1}(s) \:=\:
         \frac{4\pi\alpha^2}{s}\,v_{1,\,X^-}(s)~,
\eeq
where $\alpha$ is the electromagnetic fine structure constant.

\subsection{Isospin breaking and radiative corrections}
\label{sec:cvc_isobreak}

Since the breaking of isospin symmetry is expected at some level, 
in particular from electromagnetic effects, it is useful to 
carefully write down all the factors involved in the 
comparison of $\tau$ and \ee  spectral functions in order to make
explicit the possible sources causing the breakdown of CVC. For the 
dominant $\pi\pi$ spectral functions, we have on the \ee  side
\beq
\label{eq:ee_ff}
	\sigma (\epem\to \pi^+\pi^-)
	=	\frac{4\pi\alpha^2}{s} v_0(s)~,
\eeq
where the spectral function $v_0(s)$ is related to the pion form 
factor $F^0_\pi(s)$ by
\beq
    	v_0(s)=	\frac {\beta_0^3(s)} {12} |F^0_\pi(s)|^2~,
\eeq
and where $\beta_0^3(s)$ is the threshold kinematic factor. Similarly,
on the $\tau$ side the relation between spectral function~(\ref{eq:sf}) 
and the charged pion form factor reads
\beq
\label{eq:tau_ff}
      	v_-(s) = \frac {\beta_-^3(s)} {12} |F^-_\pi(s)|^2~.
\eeq
Isospin symmetry implies $v_-(s) =  v_0(s)$. The threshold functions 
$\beta_{0,-}$ are defined by
\beq
 	\beta_{0,-} = \beta(s,m_{\pi^-},m_{\pi^{+,0}})~,
\eeq
where
\beq
 \beta(s,m_1,m_2)=\left[\left(1-\frac{(m_1+m_2)^2}{s}\right)
                     \left(1-\frac{(m_1-m_2)^2}{s}\right)\right]^{1/2}~.
\eeq

The experiments measure $\tau$ decays inclusively with respect to radiative 
photons, \ie, for $\taum \to \nu_\tau \pim \piz (\gamma)$. The measured spectral 
function is hence $v_-^*(s) =  v_-(s)~G(s)$, where $G(s)$ is a radiative 
correction.

Several levels of ${\rm SU}(2)$ breaking can be identified:
\bei
\item 	{\em Electroweak radiative corrections to $\tau$ decays} 
	are contained in the $\Sew$ 
	factor~\cite{Marciano:1988,braaten}, which is dominated by 
	short-distance effects. It is expected to be weakly 
	dependent on the specific hadronic final state, as verified 
	for the $\taum \to (\pim, \Km) \nut$ decays~\cite{decker-fink}. 
	Detailed calculations have been performed for the 
	$\pim \pi^0$ channel~\cite{ecker,ecker2}, which also confirm the 
	relative smallness of the long-distance contributions. The 
	total correction is
	$\Sew = \Sew^{\rm had} \Sem^{\rm had}/\Sem^{\rm lep}$,
	where $\Sew^{\rm had}$ is the leading-log short-distance electroweak
	factor (which vanishes for leptons) and $\Sem^{\rm had,lep}$ are the
	nonleading electromagnetic corrections. The latter corrections
	have been calculated at the quark level~\cite{braaten}, 
	at the hadron level for the $\pim \pi^0$ decay mode~\cite{ecker},
	and for leptons~\cite{Marciano:1988,braaten}. The 
	total correction amounts to~\cite{Davier:2003a} 
	$\Sew^{\rm incl} = 1.0198 \pm 0.0006$ 
	for the inclusive hadron decay rate and 
	$\Sew^{\pi \pi^0} = (1.0232 \pm 0.0006)\cdot\Gem^{\pi \pi^0}(s)$ 
	for the $\pi \pi^0$ decay mode, where $\Gem^{\pi \pi^0}(s)$ is an
	$s$-dependent long-distance radiative correction~\cite{ecker}.

\item 	{\em The pion mass splitting} breaks isospin symmetry in 
	the spectral functions~\cite{adh,czyz} since $\beta_-(s) \neq \beta_0(s)$.

\item 	Isospin symmetry is also broken in {\em the pion form factor} due
	to the $\pi$ mass splitting~\cite{adh,ecker}.

\item 	A similar effect is expected from {\em the $\rho$ mass splitting}. 
	The theoretical expectation~\cite{bijnens} gives a limit ($<0.7$~MeV), 
	but this is only a rough estimate. Hence the question must be 
	investigated experimentally, the best approach being the explicit 
	comparison of $\tau$ and $\epem$ $2\pi$ spectral functions, after 
	correction for the other isospin-breaking effects. No correction 
	for $\rho$ mass splitting is applied initially.

\item 	Explicit {\em electromagnetic decays} such as $\pi \gamma$, 
	$\eta \gamma$, $\ell^+\ell^-$ and $\pi \pi \gamma$ introduce
	small differences between the widths of the charged and neutral $\rho$'s.

\item 	Isospin violation in the strong amplitude through the {\em mass 
	difference between $u$ and $d$ quarks} is expected to be negligible.

\item 	When comparing $\tau$ with $e^+e^-$ data, an obvious and
        locally large correction must be applied to the $\tau$ \sf\
	to introduce the effect of {\em$\rho$--$\omega$ mixing}, 
	only present in the
	neutral channel. This correction is computed using the parameters 
	determined by the \ee experiments in their form factor fits to the 
	$\pi^+ \pi^-$ lineshape modeling $\rho$--$\omega$ 
	interference~\cite{cmd2_new}. 

\eei

\subsection{Tau spectral functions and $e^+e^-$ data}
\label{sec:cvceetau}

\begin{figure}[t]
   \centerline{\epsfxsize8.3cm\epsffile{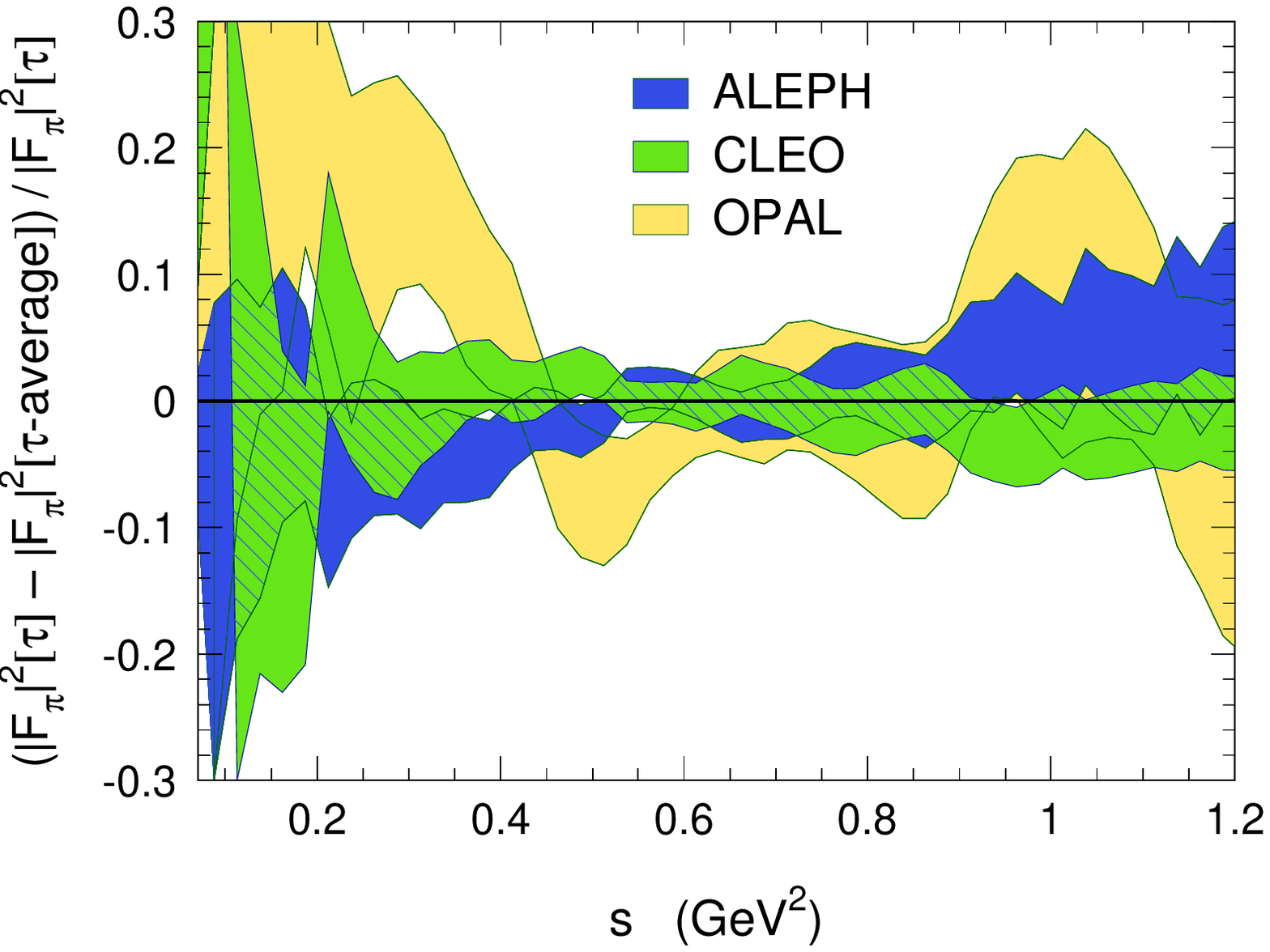}
	       \epsfxsize8.3cm\epsffile{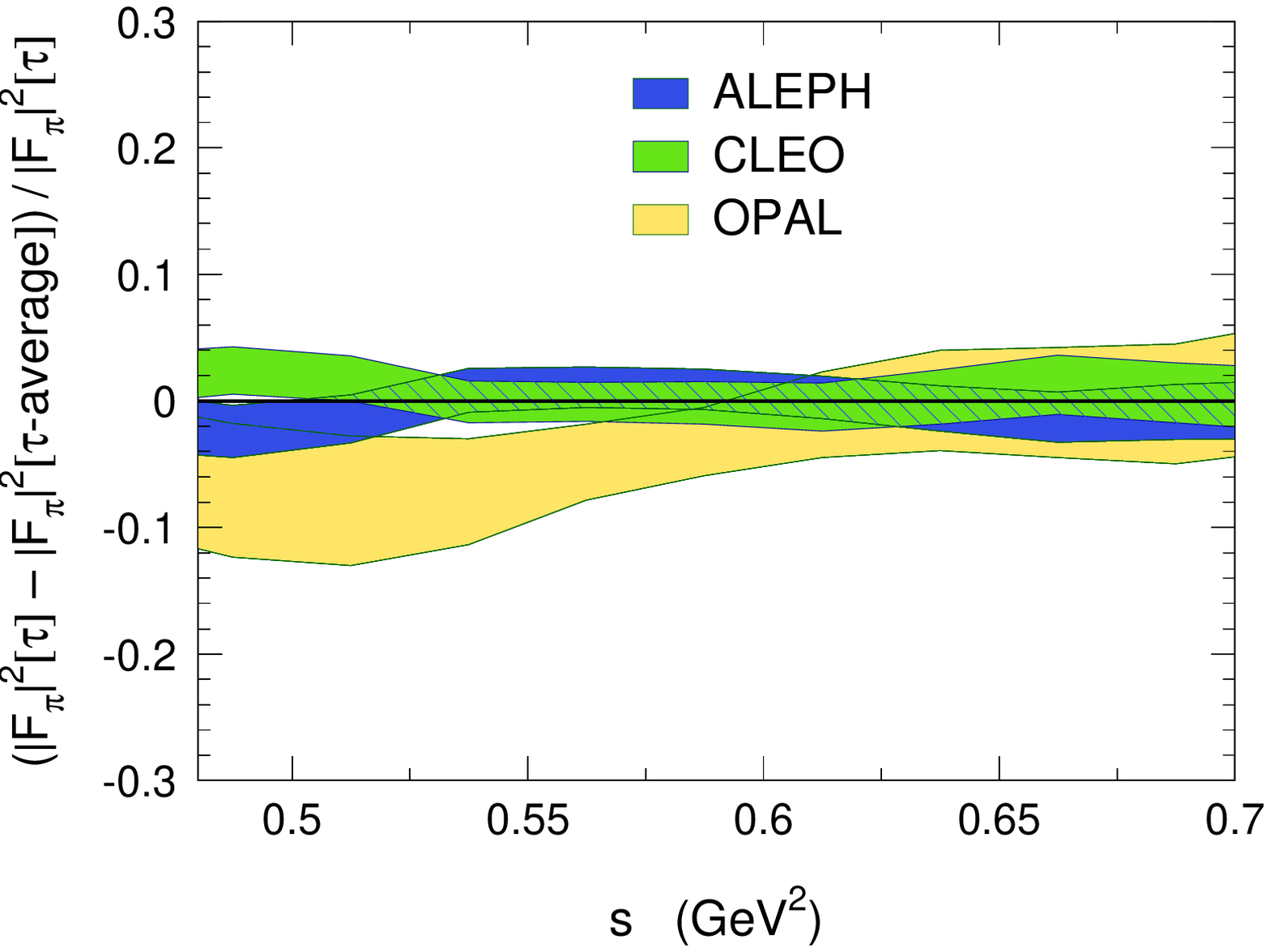}}
   \vspace{-0.3cm}
   \caption[.]{The left figure shows a relative comparison of 
        the $\pi^+\pi^-$ \sfs\
    	extracted from $\tau$ data from the different experiments, 
	expressed as a ratio to the average $\tau$ \sf. 
        The right figure emphasizes 
        the $\rho$ region. For CLEO only the statistical errors are shown.} 
\label{comp_tau_2pi}
\end{figure}

Data from $\tau$ decays into two- and four-pion final states
\tauto\nut\pipiz, \tauto\nut\pitpiz\ and \tauto\nut\tpipiz,
are available from ALEPH~\cite{aleph_taubr}, 
CLEO~\cite{cleotaurho,cleo_4pi} and OPAL~\cite{opal_vasf}.
For the two-pion final states, they are compared in 
Fig.~\ref{comp_tau_2pi} in the region around the $\rho$ resonance.
For this comparison, each data set is normalized to the world average
branching fraction and plotted with respect to their weighted average.
The two most precise results from ALEPH and CLEO are in agreement.
The statistics are comparable in the two cases, however due to a flat
acceptance in ALEPH and a strongly increasing one (with the mass-squared)
in CLEO, ALEPH data are more precise below the $\rho$ peak, while CLEO 
has the better precision above. In the following we will use the 
$\taum\to\pim\piz\nut$ \sf\  averaged over the three experiments.

The exclusive low-energy \ee cross sections have been mainly measured by 
experiments running at \ee colliders in Novosibirsk and Orsay. 
The most precise $\ee\to\pi^+\pi^-$ measurements come from 
CMD-2, which are now available in their final form~\cite{cmd2_new}, after 
a significant revision where problems related to radiative corrections 
have been removed. 
%The results are corrected for leptonic and hadronic 
%vacuum polarization, and for photon radiation by the pions (final state 
%radiation, FSR), so that the measured final state corresponds to 
%$\pi^+\pi^-$ including pion-radiated photons and virtual final state QED 
%effects. 
The overall systematic error of the final data is quoted to be 
0.6\% and is dominated by the uncertainties in the radiative corrections 
(0.4\%). 
 
%The comparison between the cross section results from CMD-2 and from 
%previous experiments (corrected for vacuum polarization and FSR,
%according to the procedure discussed in Section~\ref{sec:rad}) shows
%agreement within the much larger uncertainties (2--10\%) quoted 
%by the older experiments. But the new CMD-2 results only cover the mass
%range from 0.61 to 0.96\gev, so the older data must still be relied 
%upon below and above these values. See Section~\ref{sec:cvc} for 
%detailed numerical and graphical comparisons.

Recently, new data on the $\pip\pim$ spectral function
in the mass region between $0.60$ and $0.97\gev$ were presented by
the KLOE collaboration~\cite{kloe_2pi}, using the---for the purpose
of precision measurements---innovative technique 
of the radiative return~\cite{isr1,isr2}. The statistical precision 
of these data by far outperforms the Novosibirsk sample, but the
systematic errors are about twice as large as those assigned by CMD-2.

\subsection{Comparing $\tau$ with $e^+e^-$ data}
\label{sec:comp_eetau_2pi}

\subsubsection{The $2\pi$ spectral function}

\begin{figure}[t]
  \centerline{\epsfxsize8.3cm\epsffile{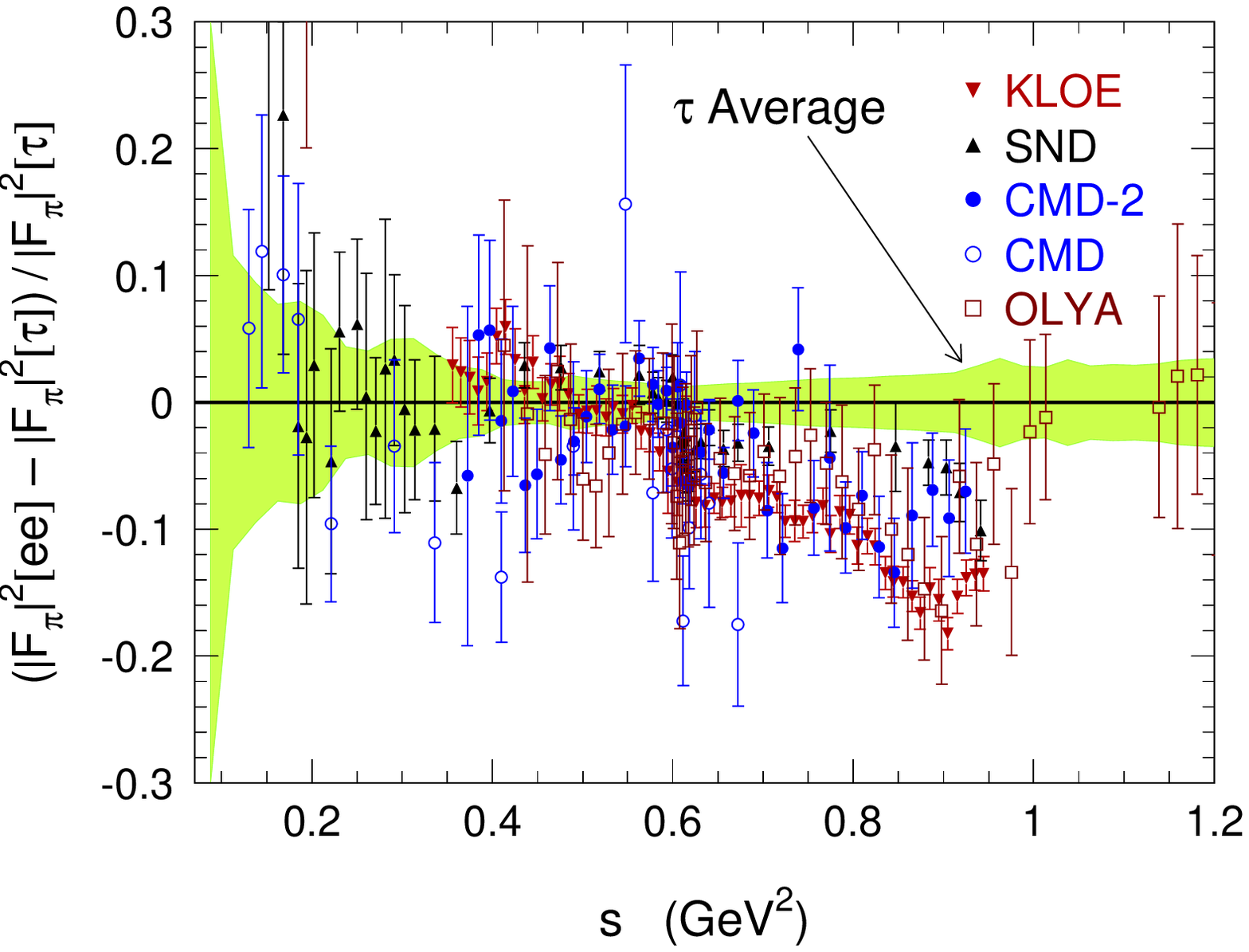}
	      \epsfxsize8.3cm\epsffile{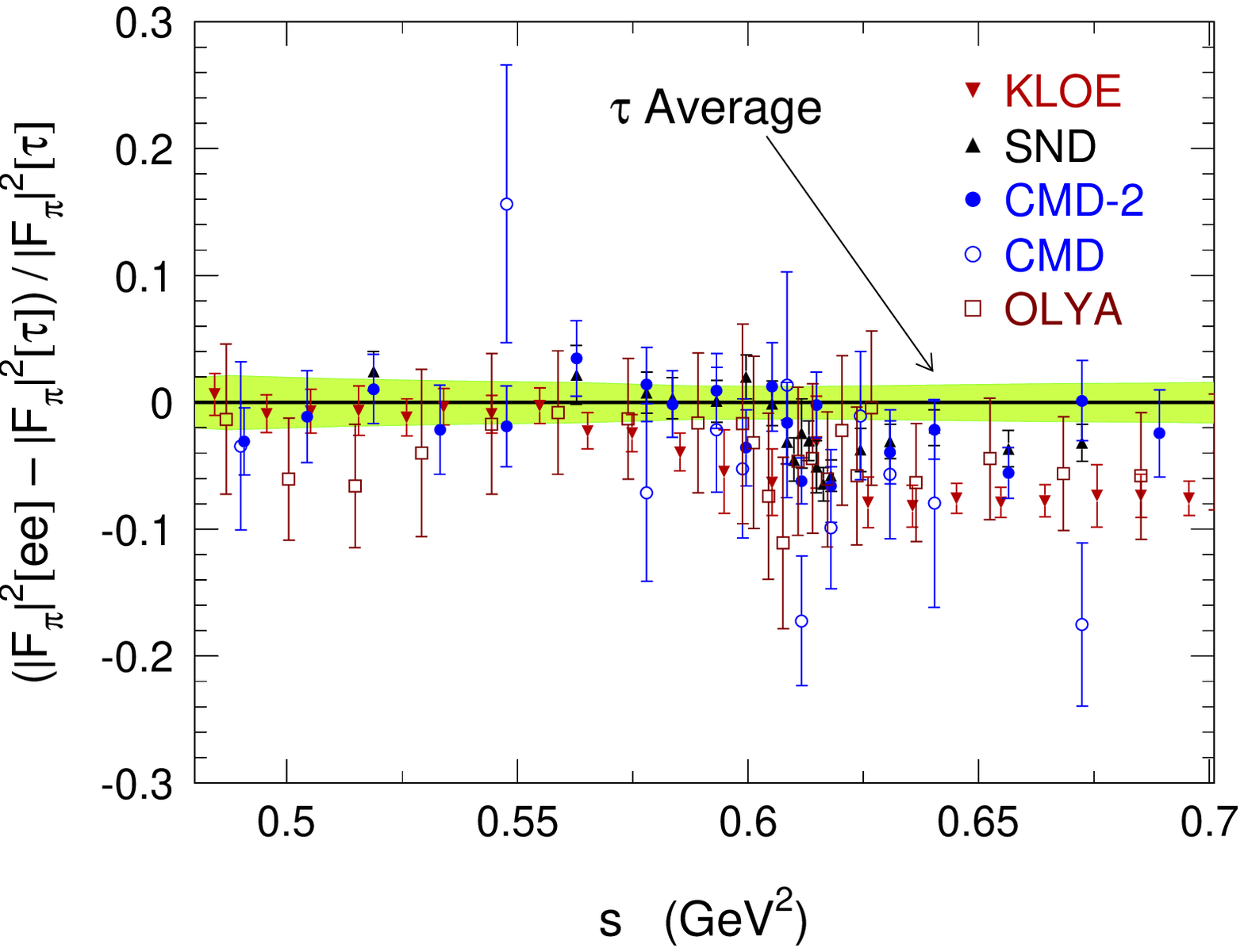}}
  \vspace{-0.3cm}
  \caption[.]{Relative comparison of the isospin-breaking corrected $\tau$
        data (world average) and $\pi^+\pi^-$ \sfs\ from \ee annihilation,
	expressed as a ratio to the $\tau$ \sf. The shaded band gives the 
        uncertainty on the $\tau$ spectral function. The right hand plot 
	zooms into the $\rho(770)$ peak region where the data density is 
	high. The $\rho$--$\omega$ mixing has been phenomenologically 
	corrected using a fit to the CMD-2 data. The larger mixing observed
	in the SND data, expressed as a $40\%$ increase in the value for the 
	$\omega\to\pip\pim$ \br~\cite{snd:2005}, is responsible for the 
	residual effect after correction seen in the right hand plot.
	}
\label{fig:comp_eetau_2pi}
\end{figure}

The $2\pi$ \sfs\  extracted from $\tau$ decays and $\ee$ annihilation data 
are compared in Fig.~\ref{fig:tau_ee_2pi}. The \ee  data
are taken from~\cite{tof_2pi, olya_2pi, cmd_2pi, cmd2_new, dm1_2pi, dm2_2pi}. 
All error bars shown contain statistical and systematic errors.
Visually, the agreement appears to be satisfactory, however the large 
dynamical range involved does not permit an accurate test. To do so, 
the \ee  data are plotted as a point-by-point ratio to the $\tau$  
\sf\  in Fig.~\ref{fig:comp_eetau_2pi}. In these latter plots we also show
more recent \epem-annihilation results from the SND collaboration~\cite{snd:2005}, 
and from the radiative return analysis performed by the KLOE 
collaboration~\cite{kloe_2pi}. Several observations can be made:
\begin{figure}[t]
  \centerline{\epsfxsize14.3cm\epsffile{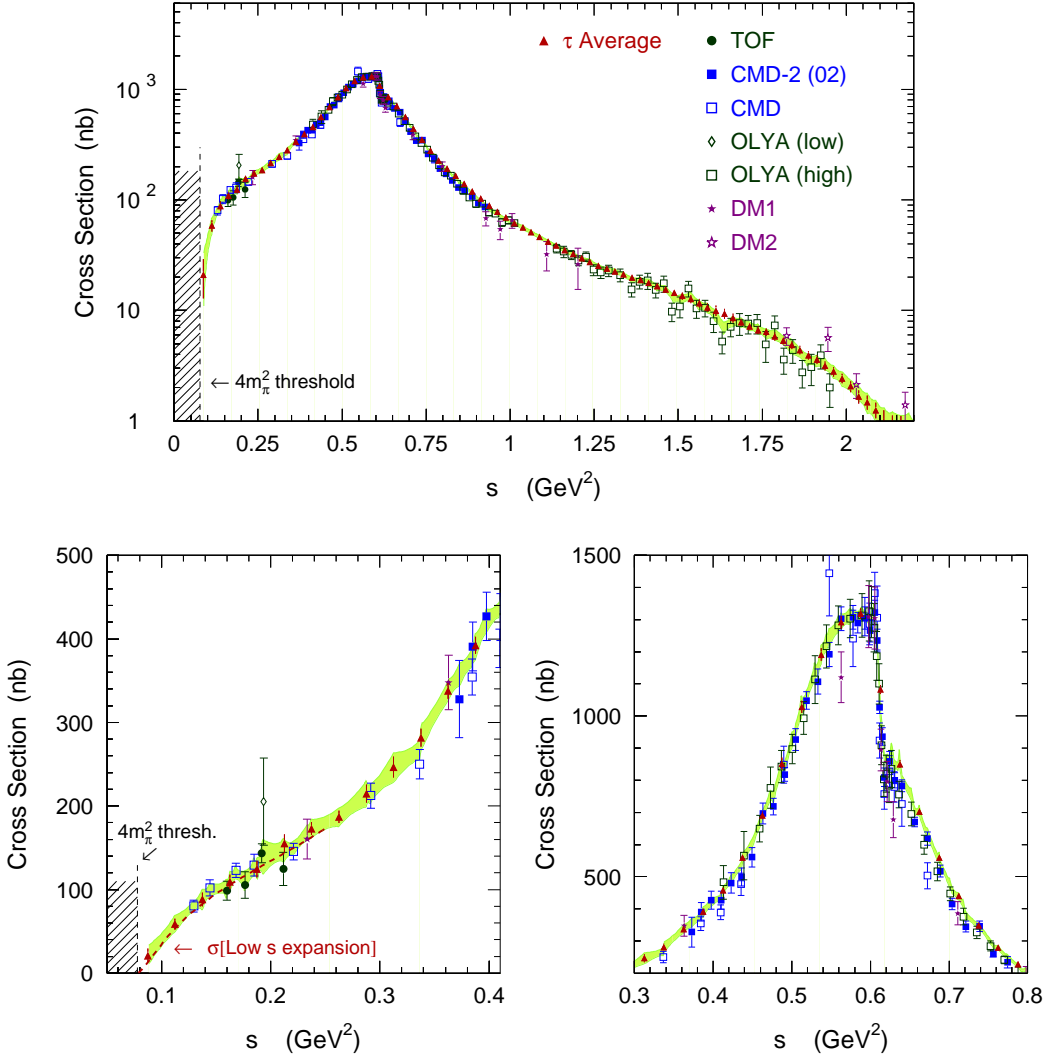}}
  \vspace{0.1cm}
   \caption[.]{Comparison of the $\pi^+\pi^-$ \sfs\ from 
        isospin-breaking corrected $\tau$ data (world average) 
        expressed as \ee cross sections, and \ee  annihilation. 
        The band indicates 
	the combined $\tau$ and \ee result within $1\sigma$ errors.
	It is given for illustration purpose only. A compendium of 
	references for
        the \ee data is given in~\cite{Davier:2003a, Davier:2003b}.}
\label{fig:tau_ee_2pi}
\end{figure}
\bei

\item 	A significant discrepancy, mainly above the $\rho$ peak
	is found between $\tau$ and the \epem data from CMD-2 as well	
	as older data from OLYA.

\item	Overall, the KLOE data seem to confirm the trend exhibited by the 
	other (older) \epem data.

\item	Some disagreement between KLOE and CMD-2
	occurs on the low mass side (KLOE data are large),
  	on the $\rho$ peak (KLOE below CMD-2) as well as
	on the high mass side (KLOE data are low).

\item 	A significant discrepancy between the (most recent) SND data and 
	KLOE is observed in the energy domain above the $\rho(770)$ peak.
	The SND data largely dissolve the discrepancy observed
	with the $\tau$ data.

\eei

At this stage, the $\tau$ spectral function has not been corrected 
for a possible $\rho^-$--$\rho^0$ mass and width 
splitting~\cite{gj,davier_pisa_g-2,aleph_taubr}.
In contrast to earlier experimental~\cite{aleph_vsf} and 
theoretical results~\cite{bijnens}, a combined pion form 
factor fit to the new precise data on $\tau$ spectral 
functions and \epem leads to $m_{\rho^-}-m_{\rho^0}=(2.4\pm0.8)\mev$
(Section~\ref{sec:eetau_combinedfit}), while no significant width 
splitting$^{\,}$ is observed within the fit error of $1.0\mev$.

Considering the mass splitting in the isospin-breaking correction 
of the $\tau$ spectral function tends to locally improve though 
not restore the agreement between $\tau$ and CMD-2 data, leaving 
an overall normalization discrepancy. Increasing the
$\Gamma_{\rho^-}-\Gamma_{\rho^0}$ width splitting by $+3\mev$
improves the agreement between $\tau$ and KLOE data in the 
peak region, while it cannot correct the discrepancies in
the tails. Note that a correction of the mass splitting 
alone would {\em increase} the discrepancy between the $\tau$ 
and \epem-based  results for \amuhadLO (Section~\ref{sec:g-2com}).

%No convincing theoretical solution for a $\tau$-vs-$\ee$ discrepancy
%has been suggested to date. A possible $S$-wave contribution to the $\tau$ 
%decay amplitude \via\  exchange of a charged Higgs boson~\cite{morse} has 
%been found too small to explain the observation~\cite{daonenggao}.
%The possibility of contributions from tensor couplings introduced 
%by a new force has been outlined in~\cite{chizhov:2003}.
\begin{figure}[t]
  \vspace{-1.1cm}
  \centerline{\epsfxsize8.3cm\epsffile{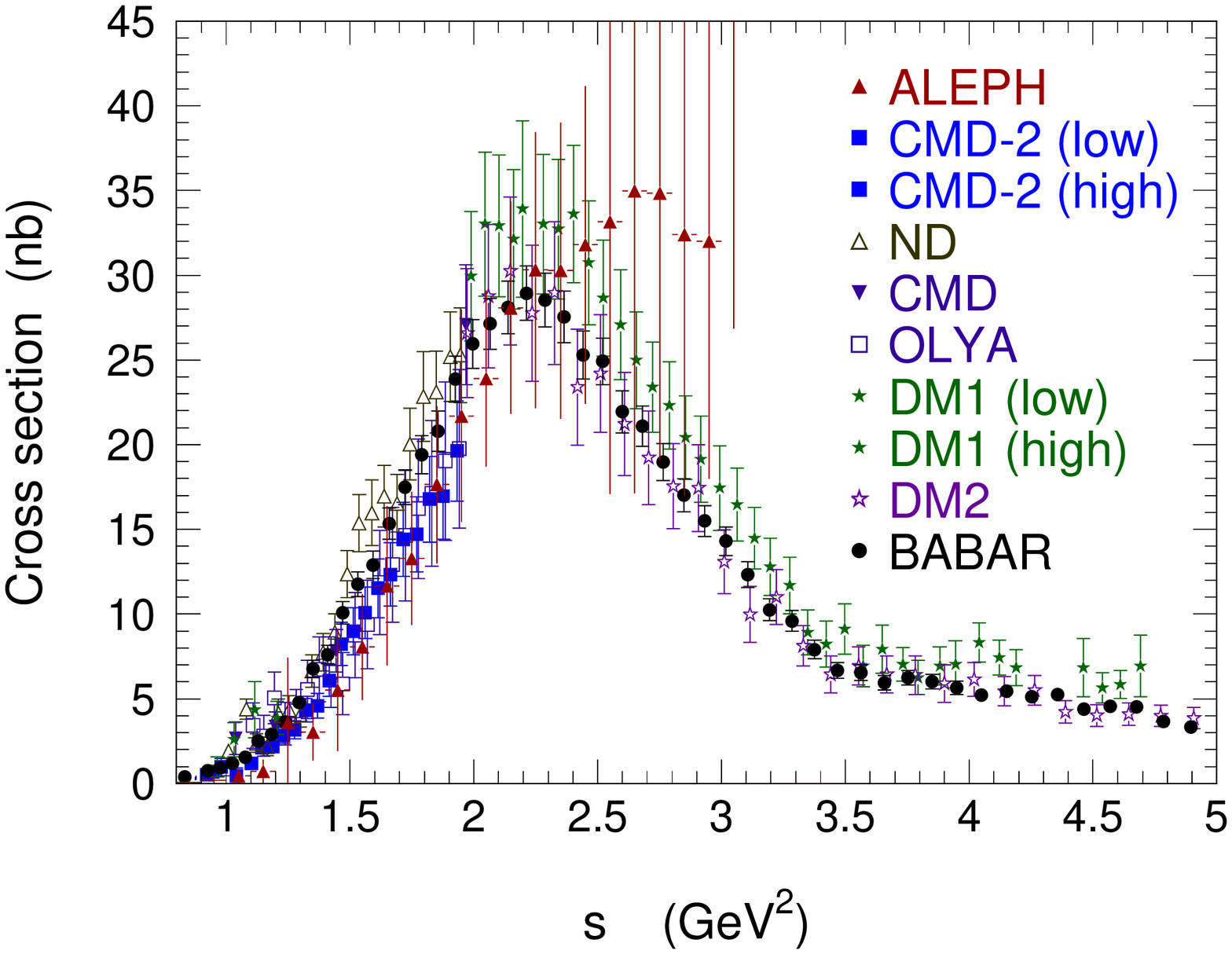}
              \epsfxsize8.3cm\epsffile{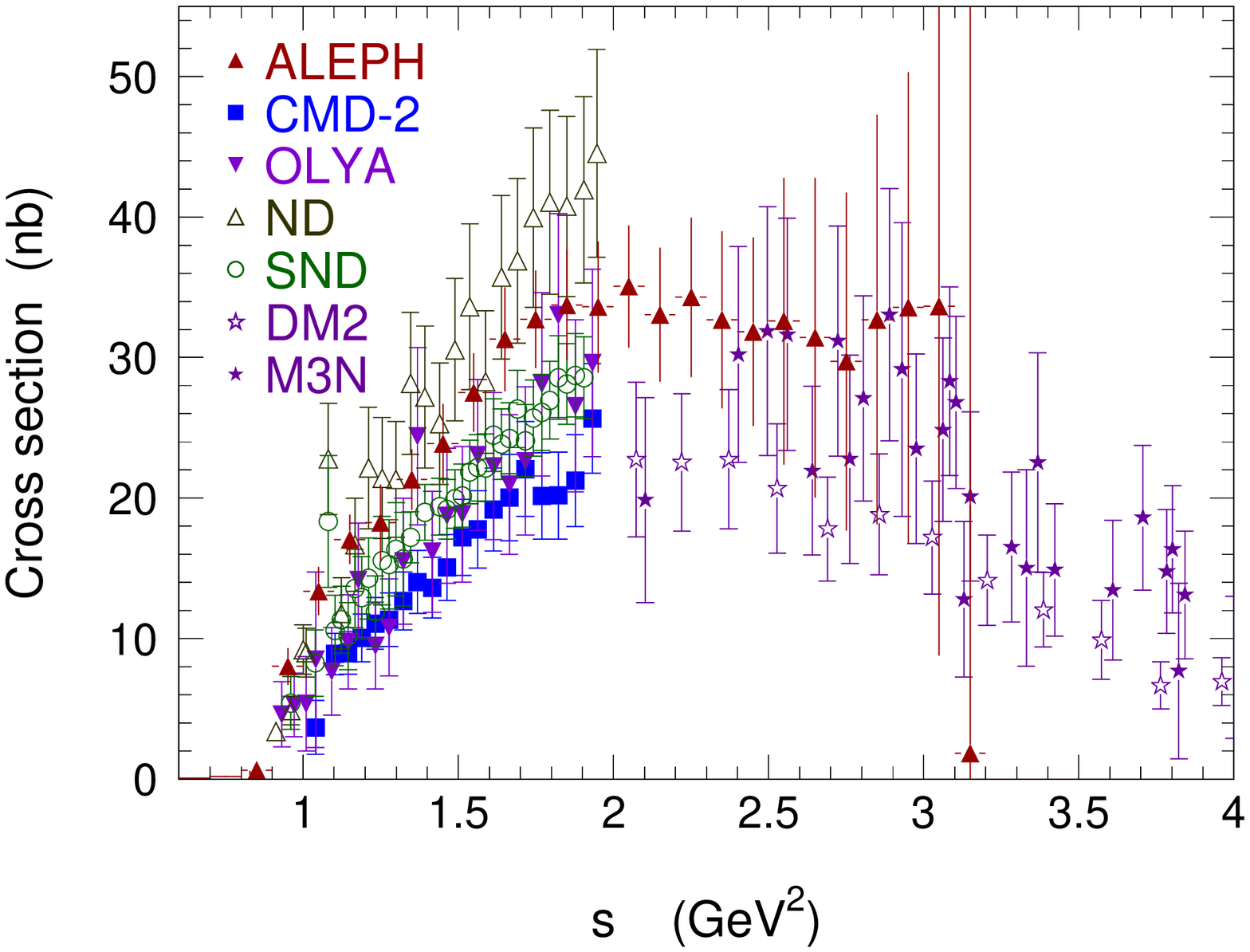}}
  \vspace{-0.3cm}
  \caption{Comparison of the $2\pi^+2\pi^-$ (left) and $\pi^+\pi^- 2\pi^0$
	(right) \sfs\  from \ee and isospin-breaking corrected $\tau$ 
	data (ALEPH), expressed as \ee cross sections. 
	References for the \ee data are given in~\cite{Davier:2003a, Davier:2003b}.
        The \babar\  results are taken from~\cite{babar_4pi}.}
\label{fig:comp_4pi_eetau}
\end{figure}

In light of the new SND data~\cite{snd:2005}, it seems appropriate
to consider the possibility of a bias in the KLOE results, and one may
also question the very small systematic uncertainties claimed by CMD-2.
We also point out that the application of the $\tau$ 
long-distance correction $\Gem^{\pi \pi^0}(s)$~\cite{ecker} worsens 
the agreement with the SND data in the energy domain above the $\rho(770)$ 
peak.

\subsubsection{The $4\pi$ spectral functions}
\label{sec:4pisf}

The \sf\ measurements of the $\tau$ vector-current final states \pitpiz\  
and \tpipiz\  are compared to the cross sections of the corresponding \ee 
annihilation into the isovector states $2\pi^- 2\pi^+$ and $\pi^- \pi^+ 2\pi^0$. 
Using Eq.~(\ref{eq:spect_fun}) and isospin invariance, the following relations 
hold
\begin{eqnarray}
\label{eq:cvc_4pi}
 \sigma_{e^+e^-\rightarrow\pi^+\pi^-\pi^+\pi^-}^{I=1} 
        & = &
             2\cdot\frac{4\pi\alpha^2}{s}\,
             v_{1,\,\pi^-\,3\pi^0\,\nut}~, \\[0.3cm]
\label{eq:cvc_2pi2pi0}
 \sigma_{e^+e^-\rightarrow\pi^+\pi^-\pi^0\pi^0}^{I=1} 
        & = &
             \frac{4\pi\alpha^2}{s}\,
             \left[v_{1,\,2\pi^-\pi^+\pi^0\,\nut} 
                  \:-\:
                     v_{1,\,\pi^-\,3\pi^0\,\nut}
             \right]~.
\end{eqnarray}
The comparison of these cross sections is given in 
Fig.~\ref{fig:comp_4pi_eetau} for $2\pi^+2\pi^-$ (left hand 
plot) and $\pi^+\pi^-2\pi^0$ (right hand plot). The latter mode
suffers from discrepancies between the results from the various \ee
experiments. The $\tau$ data, combining two measured \sfs\ according to 
Eq.~(\ref{eq:cvc_2pi2pi0}), and (coarsely) corrected for isospin breaking 
originating from the pion mass splitting~\cite{czyz}, 
lie somewhat in between, with large uncertainties above $1.4\gev$ because 
of the lack of statistics and a large feed-through background in the 
mode with three neutral pions. In spite of these difficulties the  
$\pi^-3\pi^0$ \sf\  is in agreement with \ee  data as can be seen in 
Fig.~\ref{fig:comp_4pi_eetau}. Due to the inconsistencies among the \ee 
experiments a quantitative test of CVC in the $\pi^+\pi^-2\pi^0$ channel 
is premature.

\subsubsection{Branching fractions in $\tau$ decays and CVC}
\label{sec_brcvc}

It is instructive to compare the $\tau$ and \ee spectral functions 
in a more quantitative way by calculating weighted integrals
over the mass range of interest up to the $\tau$ mass. 
One convenient choice is provided by the $\tau$ branching fractions,
which for the spin one part involve as a weight the kinematic factor 
$(1-s/m_\tau^2)^2(1+2s/m_\tau^2)$
coming from the $V-A$ charged current in $\tau$ decay. It is then 
possible to directly compare the measured $\tau$ \brs
to their prediction through isospin invariance (CVC), with 
the \ee  isovector \sfs\  as input.
 
Using the universality-improved branching fraction~(\ref{eq:uni_be}), 
one finds the results given for the dominant channels 
in Table~\ref{tab:brcvc}. The errors quoted for the CVC values are split
into uncertainties from ({\it i}) the experimental input (the \ee annihilation 
cross sections) and the numerical integration procedure,
({\it ii}) additional radiative corrections applied to some of the \ee 
data~\cite{Davier:2003a}, and ({\it iii}) the isospin-breaking corrections 
when relating $\tau$ and \ee  \sfs. 

\begin{figure}[t]
  \centerline{\epsfxsize8.6cm\epsffile{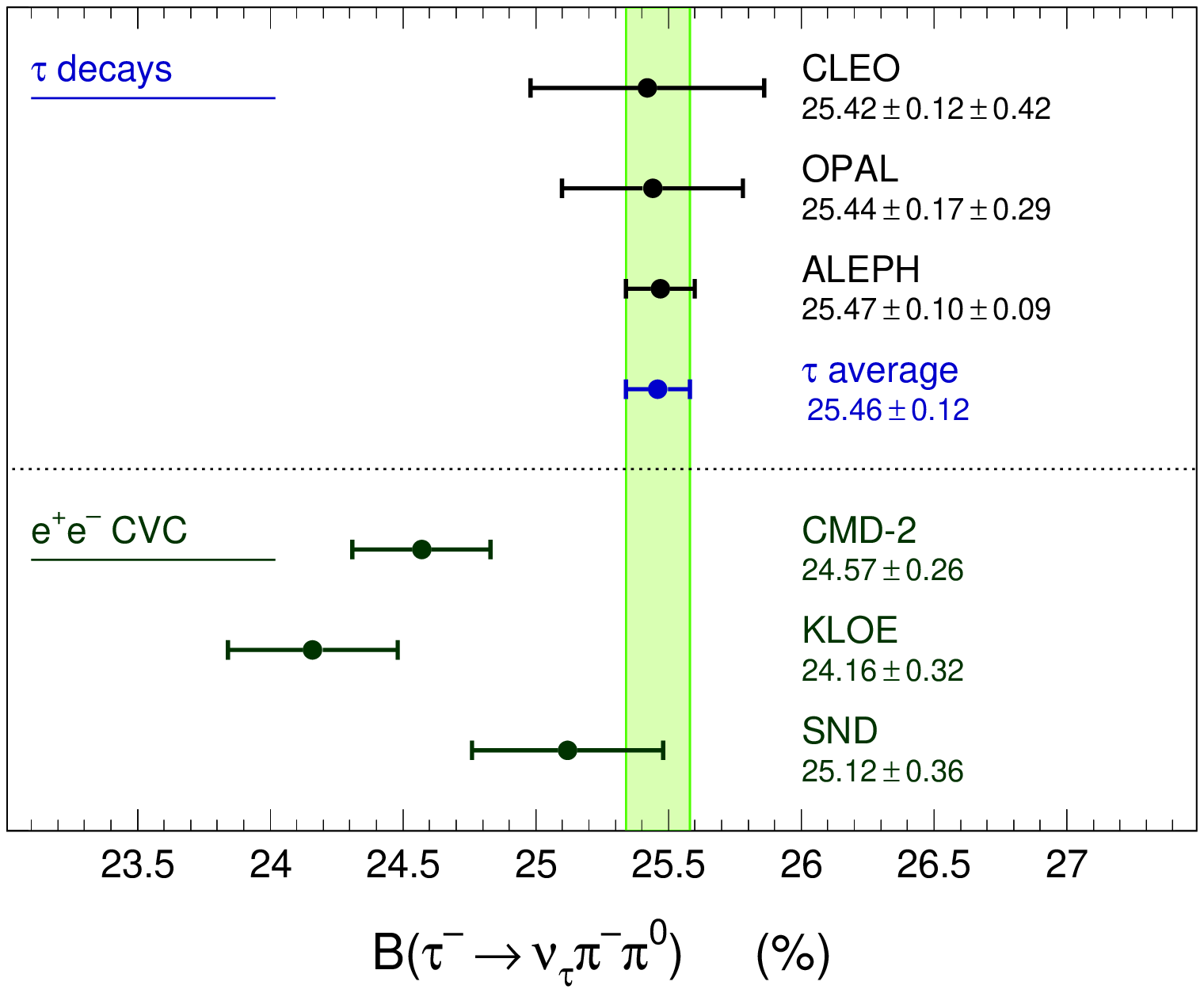}}
  \vspace{0.1cm}
  \caption[.]{The measured branching fractions for 
       	$\taum\to\pim\piz\nut$~\cite{aleph_taubr,cleo_bhpi0,opal_bh} compared 
	to the predictions from the $e^+e^-\to\pi^+\pi^-$ \sfs, applying the 
	isospin-breaking correction factors discussed in 
	Section~\ref{sec:cvc_isobreak}. 
	For the \epem results we have used only the data from the 
	indicated experiments where they are available ($0.61$--$0.96\gev$ 
	for CMD-2, $0.60$--$0.97\gev$ for KLOE, and $0.39$--$0.97\gev$
	for SND), and all data combined in the remaining energy 
	domains below $m_\tau$. Since the compatibility between 
	the \epem data is marginal (\cf\  Fig.~\ref{fig:comp_eetau_2pi}),
	we have refrained from taking their average.
	}
\label{fig:cvc_2pi}
\end{figure}

\begin{table}[t]
\caption{\label{tab:brcvc}
	Branching fractions of $\tau$ vector decays into 2 and 4 pions in
        the final state~\cite{aleph_taubr}. Second column: $\tau$ decay
	measurements. Third column: results inferred from \ee  spectral
	functions using the isospin 
	relations~(\ref{eq:cvc},\ref{eq:tau_ff},\ref{eq:cvc_4pi}) and
        correcting for isospin breaking~\cite{Davier:2003a, Davier:2003b}. 
	The 2 pion prediction does not include the recent KLOE and SND data.
	Experimental errors, including uncertainties on the integration 
        procedure, and theoretical errors (missing radiative corrections 
	for \ee, and isospin-breaking corrections and $|V_{ud}|$ for 
	$\tau$) are shown separately. Last column: differences between the
	direct measurements in $\tau$ decays and the CVC evaluations,
        where the various errors have been added in quadrature.}
\begin{center}
\setlength{\tabcolsep}{0.0pc}
\begin{tabular*}{\textwidth}{@{\extracolsep{\fill}}lrrr} \hline 
&&& \\[-0.3cm]
		& \mc{3}{c}{Branching fractions  (in \%)} \\
\rs{Mode} 	& \mc{1}{c}{$\tau$ } 	
		& \mc{1}{c}{$e^+e^-$ \via\  CVC} & $\Delta(\tau-e^+e^-)$ 
\\[0.15cm]
\hline
&&& \\[-0.3cm]
$~\tau^-\to\nut\pi^-\pi^0$
		& $25.47 \pm 0.13$ 
		& $24.52 \pm \underbrace{0.26_{\rm exp}	
			\pm 0.11_{\rm rad}\pm 0.12_{\rm SU(2)}}_{0.31}$ 
		& $+0.95 \pm 0.33$ 
	\\[0.0cm]
$~\tau^-\to\nut\pi^-3\pi^0$
		& $ 0.98 \pm 0.09$ 
		& $ 1.09 \pm \underbrace{0.06_{\rm exp}
		        \pm 0.02_{\rm rad}\pm 0.05_{\rm SU(2)}}_{0.08}$ 
		& $-0.11 \pm 0.12$ 
	\\[0.0cm]
$~\tau^-\to\nut2\pi^-\pi^+\pi^0$
		& $ 4.59 \pm 0.09$ 
		& $ 3.63 \pm \underbrace{0.19_{\rm exp}
			\pm 0.04_{\rm rad}\pm 0.09_{\rm SU(2)}}_{0.21}$ 
		& $+0.96 \pm 0.23$ 
	\\[0.5cm]
\noalign{\smallskip}\hline
\end{tabular*}
\end{center}
\end{table} 
As expected from the preceding discussion, a discrepancy is 
observed for the $\taum\to \pim\piz\nut$ \br, with a difference of 
$(0.95\pm0.13_\tau\pm0.26_{\rm ee}\pm0.11_{\rm rad}\pm0.12_{\rm SU(2)})\%$, 
where the uncertainties are from the $\tau$ \br, 
\ee  cross section, \ee  missing radiative corrections and isospin-breaking 
corrections (also including the uncertainty on $|V_{ud}|$), respectively. 
Adding all errors in quadrature gives a $2.9\sigma$ effect. Including
the new SND measurements~\cite{snd:2005} into the $\pip\pim$ data sample 
would decrease this discrepancy to approximately $2.5\sigma$. The improvement
would have been stronger, if we had discarded older data sets. In effect,
using only SND data in the region where these are available 
$[0.39$--$0.97\gev]$, and all other data elsewhere, one finds for
the predicted \br: $\BR_{\rm CVC}(\tau^-\to\nut\pi^-\pi^0)=(25.12\pm0.36)\%$, 
which is in agreement with the $\tau$ result\footnote
{
	However, we point out that the scarce SND data points in the tails 
	of the $\rho(770)$ resonance, where the line shape is concave, 
	may lead to an overestimate of the integral computed by the 
	trapezoidal rule as it is done here.
}. 
More information on this comparison is displayed in 
Fig.~\ref{fig:cvc_2pi}, where we also give the results for using
only CMD-2 and only KLOE data, respectively, in the available energy 
regions (and the average of all other data elsewhere). The discrepancy
between the \epem-based prediction using the KLOE data and 
the $\tau$ result is at the $3.8\sigma$ level.

The situation in the $4\pi$ channels is different. While there is agreement 
for the $\pi^-3\pi^0$ mode within a relative accuracy of $12\%$, 
the comparison is not satisfactory for $2\pi^-\pi^+\pi^0$. 
In the latter case, the relative difference is very large, 
$(23\pm6)\%$, compared to a reasonable level of isospin symmetry 
breaking. As such, it rather points to experimental problems that require
further investigation. These are emphasized by the scatter observed among 
the different \ee  results.

\subsection{Fits to the $\pi\pi$ spectral function}

Phenomenological fits to the pion form factor have been performed by the 
ALEPH and CLEO collaborations~\cite{aleph_taubr,aleph_vsf,cleotaurho}.
Appropriate parameterizations, \eg, the one proposed in~\cite{kuhnsanta},
employ relativistic $P$-wave Breit-Wigner propagators, and coherent summation 
of the resonance amplitudes using the isobar model.
The $\pi\pi$ spectral function is dominated by the broad $\rho$ resonance,
parameterized by both collaborations following Gounaris-Sakurai~\cite{gounarissak} 
(GS). The GS parameterization takes into account analyticity and unitarity 
properties. We will in the following discuss the fits performed in~\cite{aleph_taubr}.

Assuming vector dominance, the pion form factor is given by interfering
amplitudes from the known isovector meson resonances $\rho(770)$, 
$\rho(1450)$ and $\rho(1700)$ with relative strengths 1, $\beta$, and $\gamma$. 
Although one could expect from the quark model that $\beta$ and $\gamma$ are real
and respectively negative and positive, the phase of $\beta$, $\phi_\beta$, is left 
free in the fits, and only the much smaller $\gamma$ is assumed to be real for lack 
of precise experimental information at large masses. Taking into account 
$\rho$--$\omega$ mixing, one writes for the $\pi^+\pi^-$ form factor in 
\ee annihilation
\beq\label{fpi_vdm}
F^{I=1,0}_\pi(s) 
                 \:=\; 
                    \frac{{\rm BW}_{\rho(770)}(s)\,
                    \frac{1+\delta\,{\rm BW}_{\omega(783)}(s)}
                         {1+\delta} \:+\:                            
                             \beta\,{\rm BW}_{\rho(1450)}(s) \:+\: 
                             \gamma\,{\rm BW}_{\rho(1700)}(s)}
                         {1\:+\:\beta\:+\:\gamma}~,
\eeq
with the Breit-Wigner propagators
\beq\label{eq_BW}
{\rm BW}^{\rm GS}_{\rho(m_\rho)}(s) 
               = 
          \frac{m_\rho^2(\,1+d\cdot\Gamma_\rho/m_\rho)}
               {m_\rho^2\:-\:s\:+\:f(s)\:-\:i\sqrt{s}\,\Gamma_\rho(s)}~,
\eeq
where
\beq
f(s) = \Gamma_\rho \frac{m_\rho^2}{k^3(m_\rho^2)}\,
           \Bigg[ 
 \, k^2(s) \left( h(s)-h(m_\rho^2)\right) \:+\:
                 (\,m_\rho^2-s)\,k^2(m_\rho^2)\,
                   \frac{d h}{d s}\bigg|_{s=m_\rho^2}
         \,\Bigg] 
\eeq
and $\Gamma_\rho=\Gamma_\rho (m_\rho^2)$~.
The $P$-wave energy-dependent width is given by
\beq\label{width}
\Gamma_\rho(s) = \Gamma_\rho \,
                    \frac{m_\rho}{\sqrt{s}}
                    \left(\frac{k(s)}{k(m_\rho^2)}\right)^{\!\!3}~,
\eeq
where $k(s)=\frac{1}{2}\,\sqrt{s}\,\beta^-_\pi(s)$ and $k(m_\rho^2)$ are 
pion momenta in the $\rho$ rest frame. The function $h(s)$ is defined as
\beq
h(s) = \frac{2}{\pi}\,\frac{k(s)}{\sqrt{s}}\,
           {\rm ln}\frac{\sqrt{s}+2k(s)}{2m_\pi}~,
\eeq
with $dh/ds|_{m_\rho^2} = 
h(m_\rho^2)\left[(8k^2(m_\rho^2))^{-1}-(2m_\rho^2)^{-1}\right]
\,+\, (2\pi m_\rho^2)^{-1}$.
Since interference with the isospin-violating electromagnetic 
$\omega\rightarrow\pi^+\pi^-$ decay occurs only in \ee annihilation
$\delta$ is fixed to zero when fitting $\tau$ data.
The normalization BW$^{\rm GS}_{\rho(m_\rho)}(0)= 1$ fixes the
parameter $d=f(0)/(\Gamma_\rho m_\rho)$, which is found to 
be~\cite{gounarissak}
\beq
d = \frac{3}{\pi}\frac{m_\pi^2}{k^2(m_\rho^2)}\,
        {\ln}\frac{m_\rho+2k(m_\rho^2)}{2m_\pi} \:+\:
        \frac{m_\rho}{2\pi\,k(m_\rho^2)} \:-\: 
        \frac{m_\pi^2 m_\rho}{\pi\,k^3(m_\rho^2)}~.
\eeq

\subsubsection{Fit to $\tau$ data}

The fit of the GS model to the isovector $\tau$ data establishes the need
for the $\rho(1450)$ contribution to the weak pion form 
factor~\cite{aleph_taubr,aleph_vsf}. Only weak evidence is found for a 
$\rho(1700)$ contribution lying close to the $\tau$ end-point. 
Most of the fit parameters exhibit large correlations, which have
to be taken into account when interpreting the results.
The $\rho$ mass uncertainty is found to be dominated by systematic effects, 
the largest being the knowledge of the $\pi^0$ energy scale (calibration).

\subsubsection{Combined fit to $\tau$ and $e^+e^-$ data}
\label{sec:eetau_combinedfit}

The ALEPH collaboration has performed a combined fit using  
$\tau$ (ALEPH and CLEO) and \ee data (not yet including KLOE and SND)
in order to better constrain the lesser known parameters in the 
phenomenological form factor~\cite{aleph_taubr,aleph_vsf}. 
For this study, the $\tau$ \sf\  is duly corrected for the 
isospin-breaking effects identified in Section~\ref{sec:cvc_isobreak}. Also 
photon vacuum polarization contributions are removed in the \ee \sf\ 
since they are absent in the $\tau$ data. In this way, the mass and width 
of the dominant $\rho(770)$ resonance in the two isospin states can be 
determined. For the subleading amplitudes from the higher
vector mesons, isospin symmetry is assumed so that common masses 
and widths can be used in the fit.

The result of the combined fit is given in Table~\ref{tab:rho_common_fit}. 
The differences between the masses and widths of the charged 
and neutral $\rho(770)$'s are found to be~\cite{aleph_taubr}
\beqn
\label{eq:splitting}
   m_{\rho^-}-m_{\rho^0}		&=& (2.4 \pm 0.8)\mev~, \\
   \Gamma_{\rho^-}-\Gamma_{\rho^0}	&=& (0.2 \pm 1.0)\mev~.
\eeqn
\begin{figure}[t]
   \centerline{\epsfxsize8.3cm\epsffile{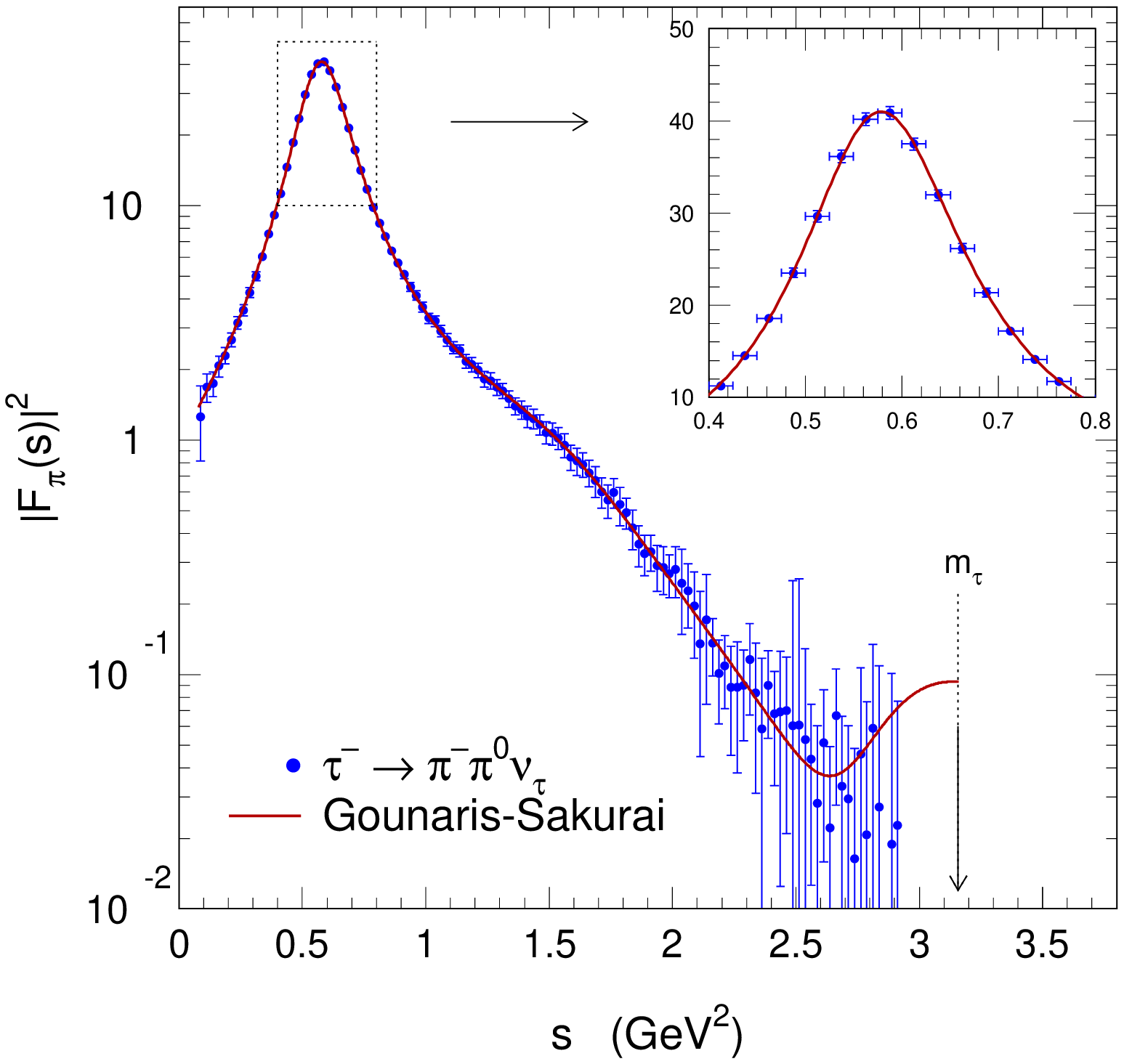}\hspace{0.1cm}
	       \epsfxsize8.3cm\epsffile{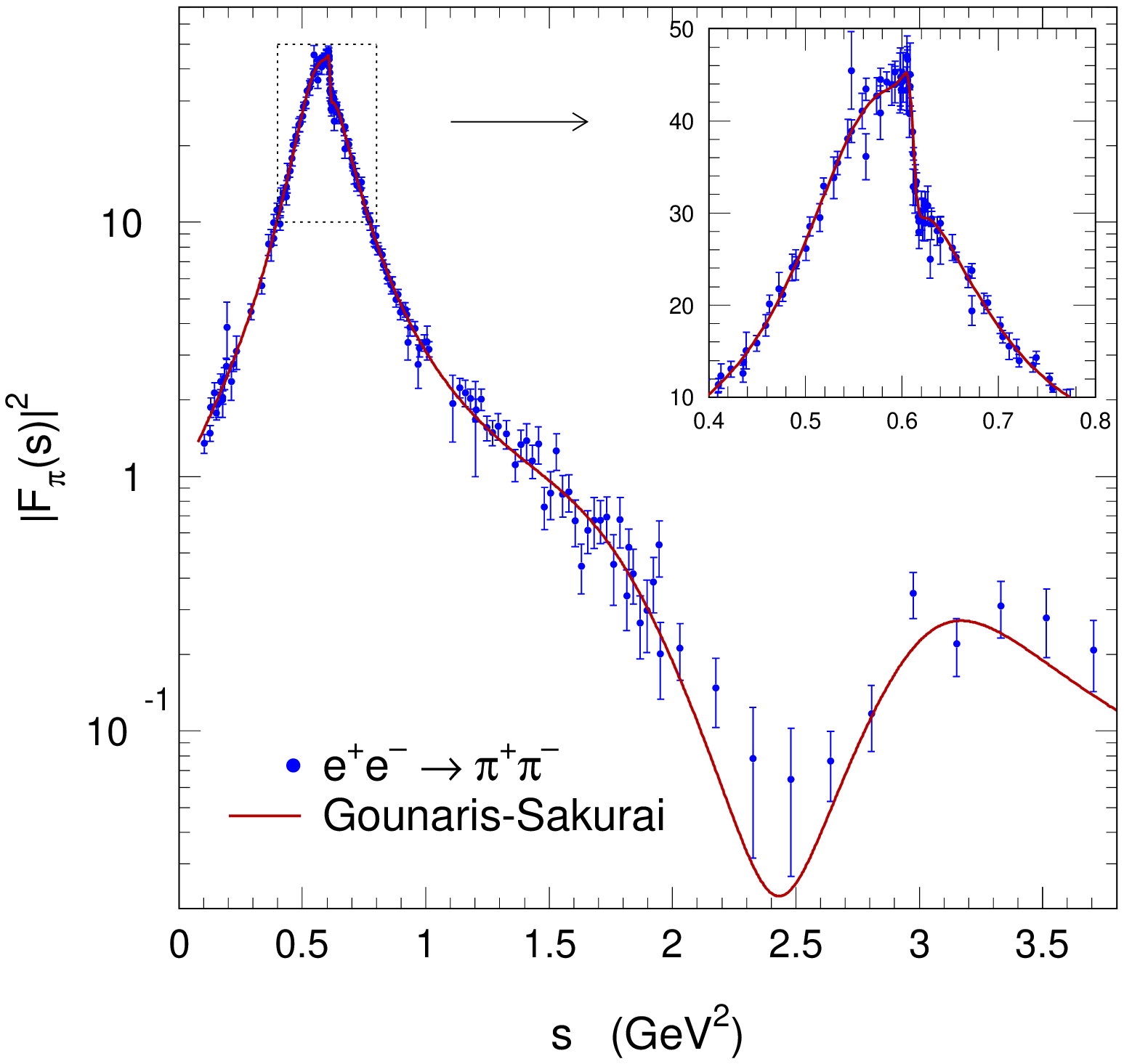}}
   \vspace{-0.3cm}
   \caption{Phenomenological fit to the ALEPH \pipiz\ spectral function 
	(left) and to \ee  annihilation data (right) using the 
	Gounaris-Sakurai parameterization. The fits are taken 
	from~\cite{aleph_taubr}.}
\label{rho_fit_aleph}
\end{figure}
The mass splitting is somewhat larger than the theoretical prediction 
($<0.7$ MeV)~\cite{bijnens}, but only at the $2\sigma$ level. The expected
width splitting, from known isospin breaking, but not taking into account any 
$\rho$ mass splitting, is $(0.7\pm0.3)\mev$~\cite{ecker,Davier:2003a}. 
However, if the mass difference is taken as an experimental fact, a larger 
width difference would be expected. From the chiral model of the $\rho$ 
resonance~\cite{pich-portoles1,pich-portoles2,ecker}, one expects
\begin{table}[t]
\caption{Combined fit to the pion form factor-squared to ALEPH, CLEO 
      	$\tau$ and all \ee data, where vacuum polarization and known 
	sources of isospin violation have been 
	excluded from the latter data. The parameterization of the $\rho(770)$, 
	$\rho(1450)$ and $\rho(1700)$ line shapes follows the Gounaris-Sakurai 
	formula. Separate masses and widths are fit for the $\rho(770)$, 
	while common values are kept for the higher vector mesons. All mass 
	and width values are in units of \mev and the phases are 
        in degrees. The value of the $\chi^2$ estimator per degree of 
	freedom (DF) is also quoted.}
\label{tab:rho_common_fit}
\begin{center}
\setlength{\tabcolsep}{0.5pc}
{\normalsize
\begin{tabular*}{\textwidth}{@{\extracolsep{\fill}}lcclcclc}
\hline\noalign{\smallskip}
\mc{8}{c}{Simultaneous fit to $\tau$ and \ee  \sfs\
($\chi^2/{\rm DF}=383/326$)} \\
\noalign{\smallskip}\hline\noalign{\smallskip}
$m_{\rho^-(770)}$      	& 775.5 $\pm$ 0.6  	&\vline&
$\Gamma_{\rho^-(770)}$ 	& 148.2 $\pm$ 0.8  	&\vline&
$\delta$ 		& $ (2.03 \pm 0.10)\times10^{-3}$  \\
$m_{\rho^0(770)}$      	& 773.1 $\pm$ 0.5  	&\vline&
$\Gamma_{\rho^0(770)}$ 	& 148.0 $\pm$ 0.9  	&\vline&
$\phi_\delta$         	& $ (13.0 \pm 2.3)$  	\\
$m_{\rho(1450)}$   	& 1409 $\pm$ 12     	&\vline&
$\Gamma_{\rho(1450)}$  	& 501 $\pm$ 37       	&\vline&
$\beta$   		& 0.166 $\pm$ 0.005 	\\
$m_{\rho(1700)}$      	&  1740 $\pm$ 20     	&\vline&
$\Gamma_{\rho(1700)}$   &  $\equiv235$       	&\vline&
$\phi_\beta$   		& 177.8 $\pm$ 5.2   	\\
			&			&&
			&			&\vline&
$\gamma$   		& 0.071 $\pm$ 0.006  	\\
			&			&&
			&			&\vline&
$\phi_\gamma$   	& $\equiv0$		\\
\noalign{\smallskip}\hline
\end{tabular*}
}
\end{center}
\end{table}
%\beq
%\label{eq:gamma_calc}
%\Gamma_{\rho^0} = \Gamma_{\rho^-}\left(\frac {m_{\rho^0}}{m_{\rho^-}}\right)^{\!\!3} 
%              \left(\frac {\beta_0}{\beta_-}\right)^{\!\!3}\;+\;\Delta \Gamma_{\rm EM}
%\eeq  
%where $\Delta \Gamma_{\rm EM}$ is the width difference from electromagnetic decays
%(as discussed above). Using~(\ref{eq:splitting}) leads to 
a total width difference 
of $(2.1 \pm 0.5)\mev$, which is only marginally consistent with the observedvalue.

%Since the $\tau$ results from ALEPH, CLEO and OPAL have been shown to be 
%consistent, the question of the correction of the observed $\rho$ mass 
%splitting when relating $\tau$ to \epem data is more relevant. However, a 
%combined fit using $\tau$ {\em and} \ee data, and requiring for consistency 
%the constraint from Eq.~(\ref{eq:gamma_calc}), has a $\chi^2$ probability of only 
%$0.6\%$~\cite{aleph_taubr}. In fact, correcting for different masses extracted 
%from the fit, and using the corresponding constrained widths, improves the 
%agreement between the $\tau$ and \ee line shapes, but at the expense of a 
%significant discrepancy in normalization\footnote
%{ 
%	It should be noted that a correction for this apparent 
%	$\rho$ mass splitting increases the present discrepancy for the muon 
%	anomalous magnetic moment between the estimates based on $\tau$ and 
%	\ee  \sfs~\cite{davier_pisa_g-2}. See the discussion in 
%	Section~\ref{sec:vacpol}.
%}.
As seen in the previous applications, the situation should improve with
the inclusion of the new SND data~\cite{snd:2005}. Also the observed $\rho$ 
mass difference~(\ref{eq:splitting}) is expected to decrease in this case.

\section{TAU DECAYS AND HADRONIC VACUUM POLARIZATION}
\label{sec:vacpol}

Hadronic vacuum polarization in the photon propagator plays an important 
role in precision tests of the SM. This is the case for the
evaluation of the electromagnetic coupling at the $Z$ mass scale,
$\alpha (M_Z^2)$, which receives a contribution 
$ \Delta\alpha_{\rm had}(M_{Z}^2)$ of the order of $2.8\times10^{-2}$.
This correction must be known to a relative accuracy of better than $1\%$ 
so that it does not limit the accuracy on the indirect determination of the 
Higgs boson mass from the measurement of ${\rm sin}^2\theta_W$. Another 
example is provided by the anomalous magnetic moment $\amu=(g_\mu -2)/2$ 
of the muon, where the lowest-order hadronic vacuum polarization component 
$a^{\rm had,LO}_\mu$ is the leading contributor to the uncertainty of the 
theoretical prediction.

Starting from~\cite{cabibbo,bouchiat} there is a long history 
of calculating the contributions from hadronic vacuum polarization 
in these processes. As they cannot be obtained 
from first principles because of the low energy scale
involved, the computation relies on analyticity and unitarity so that the
relevant integrals can be expressed in terms of an experimentally determined
spectral function, which is proportional to the cross section for \ee
annihilation into hadrons. The accuracy of the calculations has therefore
followed the progress in the quality of the corresponding data~\cite{eidelman}.
Because the data were not always suitable, it was deemed necessary to resort 
to other sources of information. One such possibility was the 
use of the vector spectral functions derived from the study of hadronic 
$\tau$ decays for the energy range less than $m_\tau$~\cite{adh}. 
Also, it was demonstrated that essentially perturbative QCD could be 
applied to energy scales as low as $1$--$2\gev$~\cite{aleph_asf,opal_vasf},
thus offering a way to replace poor \epem data in some energy regions 
by a reliable and precise theoretical 
prescription~\cite{martin,dh97,steinhauserkuhn,erler,groote,dh98}. %jobby 
Finally, without any further 
theoretical assumption, it was proposed to use QCD sum rules~\cite{groote,dh98}
in order to improve the evaluation in energy regions dominated by
resonances where one has to rely on experimental data. 
The complete theoretical prediction includes in addition 
quantum electrodynamics (QED), 
weak and higher order hadronic contributions.

Since in this review of hadronic $\tau$ decays we are mostly dealing with 
the low energy region\footnote
{
	It is the property of the dispersion relation describing 
	the hadronic contribution to $(g-2)_\mu$ that the $\pi\pi$ 
	spectral function provides the major part of the total hadronic
	vacuum polarization contribution. Hence the experimental 
	effort focuses on this channel.
}, 
common to both $\tau$ and \ee data, and because 
of the current interest in the muon magnetic moment prompted by the excellence
of the recent experimental results (see below), the emphasis will be on 
\amuhadLO rather than \daqedhZ. We note that the presently achieved accuracy 
on \daqedhZ  is meeting the goals for the LEP/SLC/FNAL global electroweak
fit. However the situation will change in the long run when very
precise determinations of ${\rm sin}^2\theta_{\rm W}$, as could be available 
from the beam polarization asymmetry at the International Linear Collider, 
require a significant increase in the accuracy of \daqedhZ~\cite{linear}.

%
% ----------------  Muon Magnetic Anomaly -------------------
%
\subsection{Muon magnetic anomaly}
\label{anomaly}

\begin{figure}[t]
   \newcommand\thisSize{2.8cm}
   \newcommand\thisHspace{\hspace{0.9cm}}
   \centerline{\epsfxsize\thisSize\epsffile{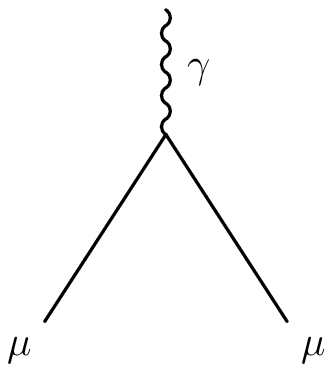}
               \thisHspace
               \epsfxsize\thisSize\epsffile{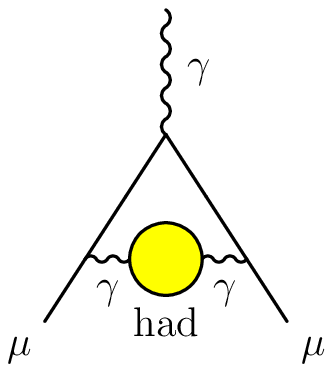}
               \thisHspace
               \epsfxsize\thisSize\epsffile{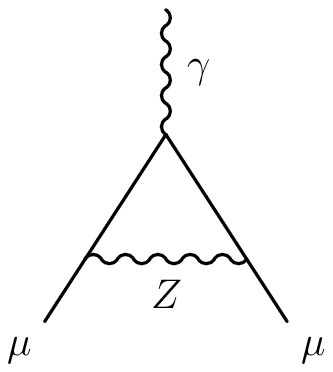}
               \thisHspace
               \epsfxsize\thisSize\epsffile{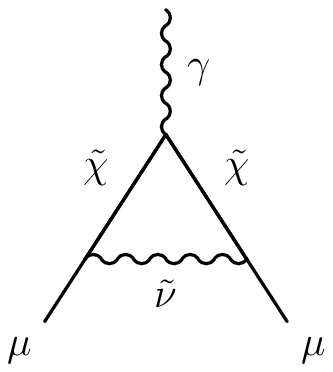}}
   \vspace{0.0cm}
   \centerline{\epsfxsize\thisSize\epsffile{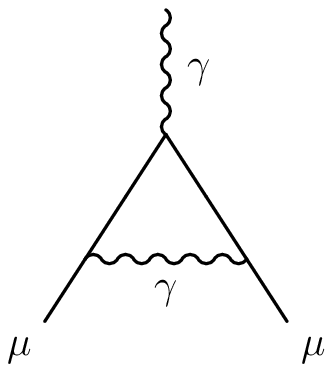}
               \thisHspace
               \epsfxsize\thisSize\epsffile{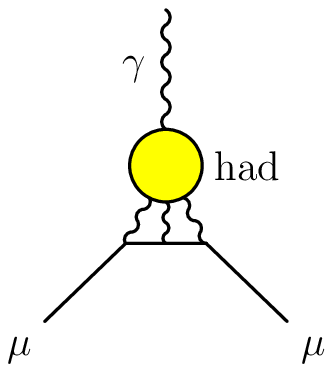}
               \thisHspace
               \epsfxsize\thisSize\epsffile{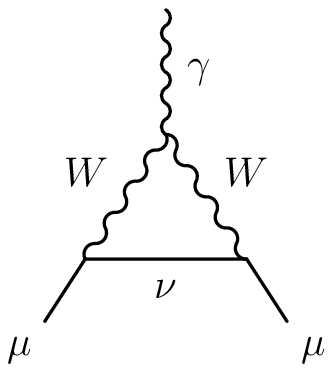}
               \thisHspace
               \epsfxsize\thisSize\epsffile{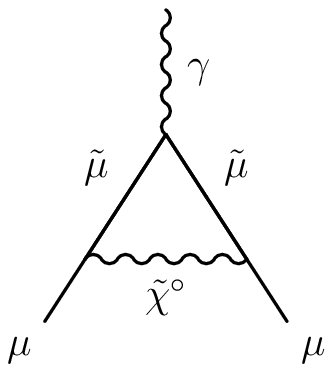}}               
   \vspace{0.3cm}
   \caption[.]{
	Representative diagrams contributing to $\amu$.
	First column: lowest-order diagram (upper) and first
	order QED correction (lower); second column: lowest-order 
        hadronic contribution (upper) and hadronic 
	light-by-light scattering (lower); third column: weak
	interaction diagrams; last column: possible 
	contributions from lowest-order supersymmetry.}
   \label{fig:g-2feyn}
\end{figure}

One of the great successes of the Dirac equation~\cite{dirac} was its 
prediction that the magnetic dipole moment, $\vec\mu$, of a spin 
$|\vec{s}|=1/2$ particle such as the electron (or muon) is given by
\beq
\label{eq:mul}
  \vec\mu_{\l} = g_{\l}\frac{e}{2m_{\l}}\vec{s}, \qquad \l=e,\mu\dots
\eeq
with gyromagnetic ratio $g_{\l}=2$, a value already implied by early atomic
spectroscopy.  Later it was realized that a relativistic quantum field theory
such as QED can give rise via quantum fluctuations to
a shift in $g_{\l}$
\beq
 \label{eq:ale}
   a_{\l}\equiv~\frac{g_{\l}-2}{2}~,
\eeq
called the magnetic anomaly.  In a now classic QED calculation,
Schwinger~\cite{schwinger} found the leading (one loop) effect 
(Fig.~\ref{fig:g-2feyn}, lower-left)
$   a_\l = \alpha/(2\pi)\simeq 0.00116$, 
with
$  \alpha\equiv e^2/(4\pi)\simeq 1/137.036
$,
which agreed beautifully with experiment~\cite{rabi,zacharias}, thereby 
providing strong confidence in the validity of perturbative QED.
Today, the tradition of testing QED and its 
${\rm SU}(3)_{C}\times {\rm SU}(2)_{L}\times U(1)_{Y}$ SM extension 
is continued (which includes strong and electroweak interactions) by measuring
$a_{\l}^{\rm exp}$ for the electron and muon even more precisely and comparing 
with $a_{\l}^{\rm SM}$ expectations, calculated to much higher order in 
perturbation theory.  Such comparisons test the validity of the SM and probe 
for new physics effects, which if present in quantum loop fluctuations 
should cause disagreement at some level. An experiment underway at 
Harvard~\cite{gabrielse} aims to improve the best present 
measurement~\cite{dehmelt} of $a_{e}$ by about a factor of 15.  
Combined with a much improved independent determination of $\alpha$, 
it would significantly test the validity of perturbative 
QED. It should be noted, however, 
that $a_{e}$ is in general not very sensitive to new physics
at a high mass scale $\Lambda$ because its effect on $a_{e}$ is
expected to be quadratic in $1/\Lambda$~\cite{czar2}
\beq
\label{eq:dla}
  \Delta a_{e}(\Lambda)\sim {\cal O}\left(\frac{m_{e}^{2}}{\Lambda^{2}}\right)
\eeq
and, hence, highly suppressed by the smallness of the electron mass.  
It would be much more sensitive if $\Delta a_{e}$ were linear in $1/\Lambda$;
but that is unlikely if chiral symmetry is present in the 
$m_{e}\to~0$ limit.

The muon magnetic anomaly has recently been measured for positive 
and negative muons with a relative 
precision of $5 \times 10^{-7}$ by the E821 collaboration at Brookhaven 
National Laboratory~\cite{bnl_2004}. Combined with the older, less precise 
results from CERN~\cite{bailey}, and averaging over charges, gives
\beq
 \label{eq:aeav}
    \amu^{\rm exp} \:=\: (11\,659\,208.0 \pm 5.8)\tmten~.
\eeq
Although the accuracy is 200 times worse than $a_{e}^{\rm exp}$, $\amu$ is 
about $m_\mu^2/m_e^2\simeq 40,000$ times more sensitive to new physics
and hence a better place (by about a factor of 200) to search for a deviation
from the SM expectation.  Of course, strong and electroweak contributions to
$a_{\mu}$ are also enhanced by $m_{\mu}^{2}/m_{e}^{2}$ relative
to $a_{e}$; so, they must be evaluated much more precisely in any meaningful
comparison of $a_{\mu}^{\rm SM}$ with Eq.~(\ref{eq:aeav}). Fortunately, the 
recent experimental progress in $a_{\mu}^{\rm exp}$ has stimulated much 
theoretical improvement of $a_{\mu}^{\rm SM}$, uncovering errors and 
inspiring new computational approaches along the way, among these the use 
of hadronic $\tau$ decays.

It is convenient to separate the SM prediction for the
anomalous magnetic moment of the muon into its different contributions,
\beq
\label{eq:amu}
    \amu^{\rm SM} \:=\: \amu^{\rm QED} + \amu^{\rm weak} + \amuhad~,
\eeq
where  $\amu^{\rm QED}=(11\,658\,472.0\pm0.2)\tmten$ is 
the pure electromagnetic contribution (see~\cite{hughes,cm} and references 
therein),\footnote
{
	An improved calculation including all mass-dependent $\alpha^4$ QED
  	contributions has been published recently with 
  	a slightly different result~\cite{Kinoshita:2004}:
  	$\amu^{\rm QED}=116\,584\,719.58(0.02)(1.15)(0.85)\times 10^{-11}$.
  	Here, $0.02$ and $1.15$ are uncertainties in the $\alpha^4$ and
  	$\alpha^5$ terms, and $0.85$ is from the uncertainty in $\alpha$ 
  	measured by atom interferometry.
}
$\amu^{\rm weak}=(15.4\pm0.1\pm0.2)\tmten$,
with the first error being the hadronic uncertainty and the second
due to the Higgs mass range, accounts for corrections due to
exchange of the weakly interacting bosons up to two loops~\cite{amuweak}
(see third column in Fig.~\ref{fig:g-2feyn}).
The term \amuhad can be further decomposed into
\beq
\label{eq:amuhad}
 \amuhad = \amuhadLO + \amuhadHO + \amuhadLBL~,
\eeq
where \amuhadLO is the lowest-order contribution from hadronic 
vacuum polarization and \amuhadHO the corresponding higher-order 
part (Section~\ref{sec:hadho}). At the 3-loop level in $\alpha$, 
the so-called hadronic light-by-light (LBL) scattering contributions, 
\amuhadLBL, must be estimated in a model-dependent
approach.  Those estimates have been plagued by errors, which now seem to be
sorted out (Section~\ref{sec:hadho}).
Nevertheless, the remaining overall uncertainties 
from hadronic vacuum polarization and LBL scattering in \amuhad
represent the main theoretical error in $a_{\mu}^{\rm SM}$.
% We follow~\cite{Davier:2004} and use the value
% $\amuhadLBL=(12.0\pm3.5)\tmten$ from the latest 
% evaluation~\cite{lbl_mv}, slightly corrected for the missing 
% contribution from (mainly) the pion box.

Owing to unitarity and to the analyticity of the vacuum-polarization 
function\footnote
{
	The photon vacuum polarization function is the spin-one
	component of the time-ordered product of two electromagnetic currents, 
	$\Pi_\gamma(q^2)$ (\cf\   Eq.~(\ref{eq:correlator})).
	Due to Ward identities, all contributions from self energy 
	and vertex correction graphs to $\Pi_\gamma(q^2)$ cancel, and 
	only vacuum polarization modifies the charge of an elementary 
	particle. One can therefore write for the running electron 
	charge (charge screening) at energy scale $s$
	\beqns
		\alpha(s) = \frac{\alpha(0)}{1-\Delta\alpha(s)}~,
	\eeqns
	where $\alpha(0)$ is the fine structure constant
	in the long-wavelength Thomson limit and
	$\Delta\alpha(s)=\Delta\alpha_{\rm lep}(s)+\Delta\alpha_{\rm had}(s)=
        -4\pi\alpha\Re[\Pi_\gamma(s)-\Pi_\gamma(0)]$.
	The leptonic part $\Delta\alpha_{\rm lep}(s)$ is calculable 
	within QED and known up to three loops~\cite{steinhauser}.
	However, quark loops are modified by long-distance hadronic 
	physics that cannot be calculated within QCD. Instead, the 
	optical	theorem
	\beqns
		12\pi\Im\Pi_\gamma(s) =
		\frac{\sigma(\ee\to{\rm hadrons})}{\sigma(\ee\to\mup\mun)}
		\equiv R(s)~, 
	\eeqns
	and, owing to the analyticity of $\Pi_\gamma$, the dispersion relation
	\beqns
		\Pi_\gamma(s) - \Pi_\gamma(0) =
		\frac{s}{\pi}\intl_0^\infty ds^\prime
			\frac{\Im\Pi_\gamma(s^\prime)}
                             {s^\prime(s^\prime-s)-i\varepsilon}~,
	\eeqns
	lead to
	\beqns
		\Delta\alpha_{\rm had}(s)=
		-\frac{\alpha(0)s}{3\pi}\Re\intl_0^\infty ds^\prime
			\frac{R(s^\prime)}
                             {s^\prime(s^\prime-s)-i\varepsilon}~,
	\eeqns
	Its r.h.s.\ can be obtained with the use of experimental 
	data for $R(s)$. Similarly, one obtains Eq.~(\ref{eq:int_amu})
	for the hadronic vacuum polarization contribution to $\amu$.
}, 
the lowest-order hadronic vacuum polarization contribution to $\amu$ 
can be computed \via\ the dispersion integral~\cite{rafael}
\beq
\label{eq:int_amu}
    \amuhadLO \:=\: 
           \frac{1}{3}\left(\frac{\alpha}{\pi}\right)^{\!2}\!
           \intl_{m_\pi^2}^\infty\!\!ds\,\frac{K(s)}{s}R^{(0)}(s)~,
\eeq
where $K(s)$ is the QED kernel~\cite{rafael2},
\beq
      K(s) \:=\: x^2\left(1-\frac{x^2}{2}\right) \,+\,
                 (1+x)^2\left(1+\frac{1}{x^2}\right)
                      \left[{\rm ln}(1+x)-x+\frac{x^2}{2}\right] \,+\,
                 x^2\,{\rm ln}x\frac{1+x}{1-x}~,
\eeq
with $x=(1-\beta_\mu)/(1+\beta_\mu)$ and $\beta_\mu=(1-4m_\mu^2/s)^{1/2}$.
In Eq.~(\ref{eq:int_amu}), $R^{(0)}(s)$ denotes the ratio of the ``bare'' 
cross section for \epem annihilation into hadrons to the pointlike 
muon-pair cross section at center-of-mass energy $\sqrt{s}$. 
The bare cross section is defined as the measured cross section
corrected for initial-state radiation, electron-vertex loop contributions
and vacuum-polarization effects in the photon propagator. As the only
QED effect, photon radiation in the final state is to be included.
The reason for using the bare (\ie, lowest-order) 
cross section is that a full treatment of higher orders is anyhow 
needed at the level of $\amu$, so that the use of the ``dressed''
cross section would entail the risk of double-counting some of the 
higher-order contributions.

The function $K(s)\sim1/s$ in Eq.~(\ref{eq:int_amu}) gives a strong 
weight to the low-energy part of the integral. About 91\% of the 
total contribution to \amuhadLO  is accumulated at center-of-mass
energies $\sqrt{s}$ below $1.8\gev$ and 73\% of \amuhadLO is covered 
by the $\pi\pi$ final state, which is dominated by the $\rho(770)$ 
resonance. 

\subsection{Input data and their treatment}

The information used for the evaluation of the integral~(\ref{eq:int_amu}) 
comes mainly from direct measurements of the cross sections in \ee 
annihilation and \via\ CVC from $\tau$ \sfs. 
In addition to the data with the two- and four-pion final states shown
in Section~\ref{sec:cvceetau}, other final states and
at different energies are also used (References for experimental data
can be found in~\cite{Davier:2003a}). 
However, due to the higher hadron multiplicity at energies 
above $\sim2.5\gev$, the exclusive measurement of the many hadronic 
final states is not practicable. Consequently, the experiments at 
the high-energy colliders ADONE, SPEAR, DORIS, PETRA, PEP, VEPP-4, 
CESR and BEPC have measured the total inclusive cross section ratio $R$. 
In some cases measurements are incomplete, as for example 
$\ee\to K\Kb\pi\pi$ or $\ee\to6\pi$, and one has to rely on 
isospin symmetry to estimate or bound the unmeasured cross 
sections~\cite{Davier:2003a}.
To obtain the corresponding \ee annihilation cross sections using 
the $\tau$ data, the isospin rotations~(\ref{eq:cvc}, \ref{eq:cvc_4pi}) 
are applied, and all identified sources of isospin breaking are corrected.

To exploit the maximum information from available data, 
weighted measurements of different experiments at a given energy 
should be combined instead of calculating the integrals for every 
experiment and finally averaging them. Different averaging methods are 
used in the literature varying from classical $\chi^2$ minimization 
methods~\cite{adh, Davier:2003a,menke}, to the so-called clustering 
method~\cite{teubner} in which a cluster of arbitrary size is predefined 
and precise data are used to rescale imprecise measurements
within the (correlated) systematic errors of the imprecise data set.

To overcome the lack of precise data at threshold energies, 
%and to benefit from the analyticity property of the pion form factor, 
a phenomenological expansion in powers of $s$ is used: 
\beq\label{eq:taylor}
|F^0_{\pi}| =
      1 + \frac{1}{6}\rpisq\,s + c_1\,s^2 +c_2\,s^3~.
%      O(s^4)~.
\eeq
This interpolating function satisfies the known form factor constraints
at $s=0$. Above the $2\pi$ threshold, where a branch cut develops in the
form factor, such a parameterization has been derived within the framework
of Chiral Perturbation Theory~\cite{colangelo_finke}.
Exploiting precise results from space-like data~\cite{space_like}, the 
pion charge radius-squared is constrained to
\beq
\label{eq:pionchargeradius}
	\rpisq=(0.439\pm0.008)~{\rm fm}^2~,
\eeq 
and the two parameters $c_{1,2}$ are fit to the data in the range 
$[2m_\pi,\,0.6\gev]$. Good agreement is observed  
in the low energy region 
%where the expansion should be 
%reliable (\cf\  Fig.~\ref{fig:taylor}). Since 
%the fits incorporate unquestionable constraints from first principles, 
and this parameterization is used for evaluating the integrals in
the range up to $0.5\gev$~\cite{Davier:2003a}. 
\begin{figure}[t]
  \centerline{\epsfxsize9.3cm\epsffile{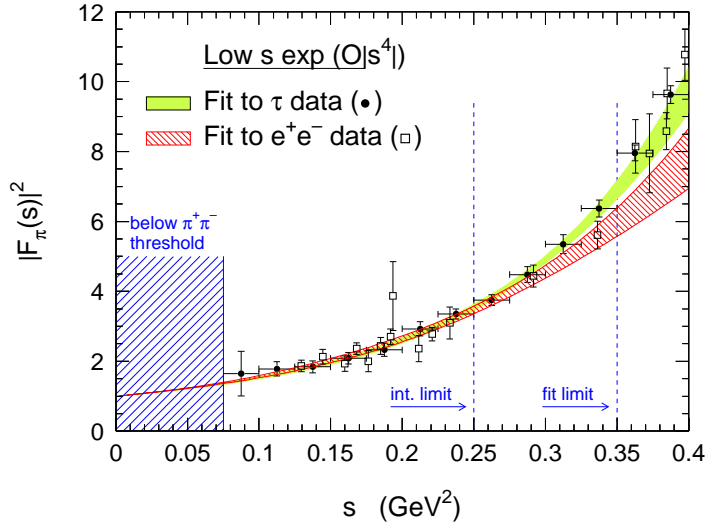}}
  \vspace{-0.3cm}
  \caption[.]{Fit of the pion form factor from $4 m_\pi^2$ to
	$0.35\gev^2$ using a third-order expansion with 
	the constraints at $s=0$ and the measured
    	pion r.m.s. charge radius from space-like data~\cite{space_like}.
	The result of the fit is integrated only up to $0.25\gev^2$.
	The \epem data do not yet include the recent SND 
	measurement~\cite{snd:2005}. }
\label{fig:taylor}
\end{figure}

\subsection{QCD for the high energy contributions}

Since problems still remain regarding the \sfs\ from low-energy data, 
the prediction from essentially perturbative QCD is used above an energy 
of $1.8$ and $5\gev$, respectively. The details of the calculation 
can be found in~\cite{dh97,dh98} and in the references therein.

The perturbative QCD prediction uses a next-to-next-to-leading order
$O(\as^3)$ expansion of the Adler $D$-function~\cite{adler1,adler2}, 
with second-order quark mass corrections included~\cite{kuhnmass}. 
$R(s)$ is obtained by evaluating numerically 
a contour integral in the complex $s$ plane (\cf\  the detailed discussion
of this technique in Section~\ref{sec:qcd}). Nonperturbative effects
are considered through the Operator Product Expansion, giving small
power corrections controlled by gluon and quark condensates. The value
$\as(M^2_{Z}) =  0.1193 \pm0.0026$, used for the evaluation 
of the perturbative part, is taken as the average of the results from
the analyses of $\tau$ decays~\cite{aleph_asf} (see Section~\ref{sec:qcd})
and of the $Z$ width in the global electroweak fit~\cite{lepewwg}. 
Uncertainties on the QCD prediction are taken to be equal 
to the common error on $\as(M^2_{Z})$, to half of the 
quark mass corrections and to the full nonperturbative contributions.
A local test of the QCD prediction can be performed in the energy range between
1.8 and $3.7\gev$. The contribution to \amuhadLO in this region is computed
to be $(33.87\pm0.46)\tmten$ using QCD, to be compared with the 
result, $(34.9\pm1.8)\tmten$ from the data. The two values agree
within the $5\%$ accuracy of the measurements.

The evaluation of \amuhadLO was shown to be improved by applying QCD 
sum rules~\cite{dh98,groote}. 
The latest full reanalyses~\cite{Davier:2003a, Davier:2003b} 
however did not consider this possibility since the main problem 
at energies below $2\gev$ was the inconsistency between the \ee and 
$\tau$ input data, which had to be resolved with priority. Also
the improvement provided by the use of QCD sum rules results from 
a balance between the experimental accuracy of the data and the 
theoretical uncertainties. With an increase in the experimental 
precision, the gain from the theoretical constraint would be smaller 
than before.

\subsection{Results for the lowest order hadronic vacuum polarization} 

Many calculations of the hadronic vacuum polarization contribution have 
been carried out in the past, taking advantage of the \ee data available 
at that time and benefiting from a more complete treatment of isospin-breaking 
corrections~\cite{ecker,ecker2}. 
In this review, we will follow the latest published analysis 
by~\cite{Davier:2003b} (and the update~\cite{hoecker_ichep04}
including data from KLOE, see also the result in~\cite{yndurain}), which 
considers input from both $\tau$ and \ee data.
%Comparison with other independent approaches are also given.

\begin{figure}
\epsfxsize12.5cm
\centerline{\epsffile{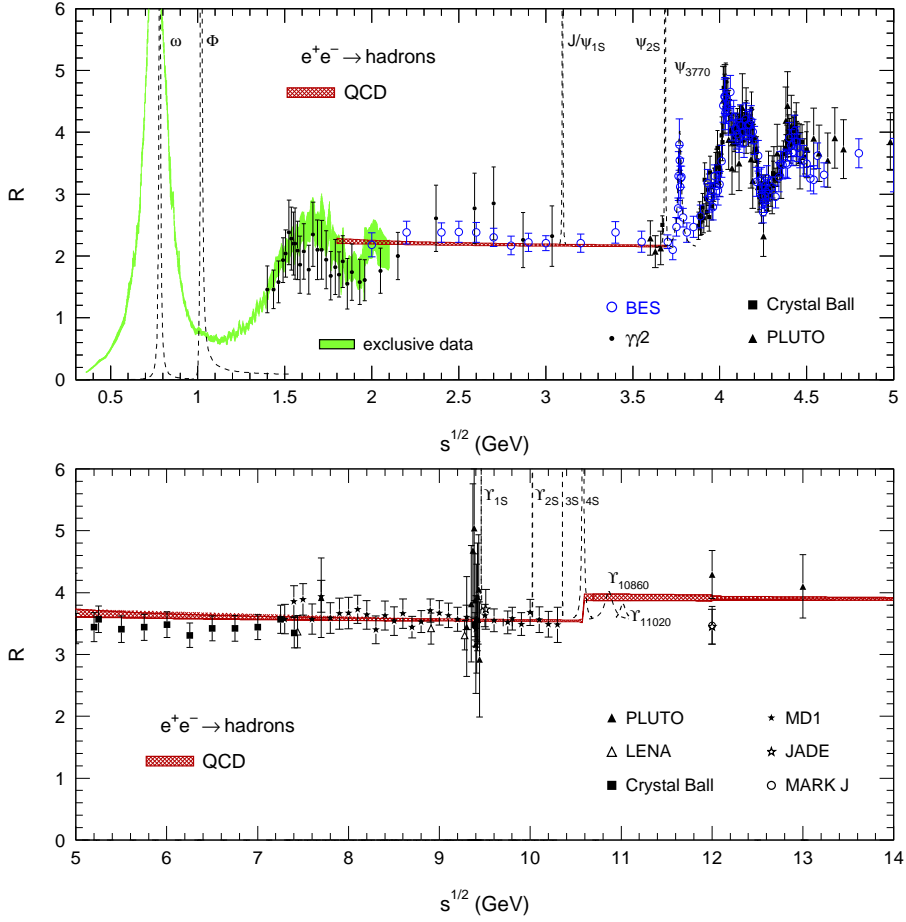}}
\vspace{0.1cm}
\caption[.]{Compilation of the data contributing to \amuhadLO. 
        Shown is the total hadronic over muonic cross section ratio $R$.
        The shaded band below $2\gev$ represents the sum of the exclusively
        measured channels, with the exception of the contributions from the 
	narrow resonances that are given as dashed lines.
        All data points shown correspond to inclusive measurements. 
	The cross-hatched band gives the prediction
        from (essentially) perturbative QCD (see text).}
\label{fig_ree_all}
\end{figure}
Figure~\ref{fig_ree_all} gives a panoramic view of the \ee data in the 
relevant energy range. The shaded band below $2\gev$ represents the sum 
of the exclusive channels. 
The QCD prediction is indicated by the cross-hatched band. Note that the 
QCD band is plotted taking into account the thresholds for open flavor 
$B$ states, in order to facilitate the comparison with the data in the 
continuum. However, for the evaluation of the integral, the $b\bbar$ threshold 
is taken at twice the pole mass of the $b$ quark, so that the contribution 
includes the narrow $\Upsilon$ resonances, according to global 
quark-hadron duality.

Compiling all contributions, the \epem-based result (including
the KLOE data, but not yet SND) for the lowest-order hadronic contribution 
is~\cite{hoecker_ichep04}
\beq
  \amuhadLO = (693.4\pm5.3\pm3.5_{\rm rad})\tmten ~,
\eeq
where the second error is due to the treatment of (potentially) 
missing radiative corrections in the older data~\cite{Davier:2003a}. 
The new KLOE data decreased the contribution by about $3\times10^{-10}$.
For a comparison we also give the results without KLOE for the
\ee  and $\tau$-based analyses~\cite{Davier:2003b}:
$\amuhadLO[\ee$-${\rm based}]=(696.3\pm6.2_{\rm exp}\pm3.6_{\rm rad})\tmten$
and
$\amuhadLO[\tau$-${\rm based}]=(711.0\pm5.0_{\rm exp}\pm0.8_{\rm rad}\pm2.8_{\rm SU(2)})\tmten$.

\subsection{Hadronic three-loop effects}
\label{sec:hadho}

The three-loop hadronic contributions to $a_{\mu}^{\rm SM}$ involve one 
hadronic vacuum polarization insertion with an additional loop 
(either photonic or another leptonic or hadronic vacuum polarization). 
They can be evaluated~\cite{krause} with the use of the 
same $e^+e^-\to{\rm hadrons}$ data sets used for \amuhadLO.
Denoting that subset of ${\cal O}(\alpha / \pi)^3$ hadronic contributions
\amuhadNLO, we quote here the result of a recent 
analysis~\cite{teubner},
\beq
\label{eq:had}
  \amuhadNLO= (-9.8\pm 0.1)\tmten~,
\eeq
which is consistent with earlier studies~\cite{krause,adh}.  
It would change by about $-0.3 \tmten$ if the $\tau$ data 
were also used.

More controversial are the hadronic light-by-light scattering 
contributions illustrated in the lower diagram of the second column in
Fig.~\ref{fig:g-2feyn}.  Since it invokes a four-point correlation
function, a dispersion relation approach using data is not possible and 
a first-principles calculation (\eg, lattice gauge theory~\cite{blum}) 
has so far not been carried out. Instead, calculations involving pole 
insertions, short distance quark loops~\cite{kino_light,bij_light} 
and charged-pion loops have been individually performed in a large $N_C$ 
QCD approach. The pseudoscalar poles ($\piz, \eta$ and $\etapr$) dominate 
such a calculation.  Unfortunately, in early studies the sign of the
contribution was incorrect. Its 
correction~\cite{knecht_light,knecht_lightb,blokland,kino_light_cor,bij_light_cor} 
led to a large shift in the $a_{\mu}^{\rm SM}$ prediction.  A 
representative estimate~\cite{Davier:2003a} of the LBL contribution, which 
includes $\pi,\eta$ and $\eta'$ poles as well as other resonances, 
charged pion loops and quark loops, currently gives
$\amuhadLBL\simeq (8.6 \pm 3.5)\tmten$ (representative).

A new analysis of LBL scattering~\cite{lbl_mv} takes into account the 
proper matching of the asymptotic short distance behavior of pseudoscalar 
and axial-vector contributions with the free quark loop behavior. It 
yields the somewhat larger result $\amuhadLBL= (13.6 \pm 2.5)\tmten$.

As pointed out in~\cite{Davier:2004}, several small but (likely) negative 
contributions such as charged pion loops and scalar resonances were not 
included in~\cite{lbl_mv} and could reduce somewhat the magnitude of 
the result. The authors~\cite{lbl_mv} provide a consistency check on their result 
much in the spirit of the EW hadronic triangle diagram study~\cite{amuweak,vain} 
discussed in~\cite{Davier:2004}. Using constituent quark masses in the 
LBL diagram combined with a pion pole contribution that properly 
accounts for the chiral properties of massless QCD and 
avoiding short distance double counting, they find 
$\amuhadLBL\simeq 12.0 \tmten$ with about half of the 
contribution coming from quark diagrams and the other half from 
the pion pole. Using that result along with an inflated error 
assigned so that the representative result and the one obtained
in~\cite{lbl_mv} overlap within their errors, 
\cite{Davier:2004} suggest the value
\beq
\label{eq:2had}
  \amuhadLBL\simeq (12.0 \pm 3.5)\tmten~.
\eeq
Further resolution of the LBL scattering contribution is very 
important. 

Combining Eqs.~(\ref{eq:had}) and (\ref{eq:2had}), one finds
\beq
\label{eq:fmd}
  a_{\mu}^{\rm had,3-loop} = (2.2 \pm 3.5)\tmten~,
\eeq
which we identify with \amuhadHO in Eq.~(\ref{eq:amuhad}).

\subsection{Comparing $\amu$ between theory and experiment}
\label{sec:g-2com}

Summing the results from the previous sections on $a_{\mu}^{\rm QED}$, 
$a_{\mu}^{\rm EW}$, \amuhadLO, \amuhadNLO, and \amuhadLBL, 
one obtains the SM prediction for $\amu$. The newest
\ee-based result reads~\cite{hoecker_ichep04}
\beq
\label{eq:smres}
  a_{\mu}^{\rm SM} = (11\,659\,182.8
                     \pm 6.3_{\rm had,LO+NLO}
                     \pm\, 3.5_{\rm had,LBL} 
                     \pm 0.3_{\rm QED+EW})\tmten~.
\eeq

This value can be compared to the present measurement~(\ref{eq:aeav});
adding all errors in quadrature, the difference between experiment
and theory is
\begin{figure}[t]
  \centerline{\epsfxsize8.6cm\epsffile{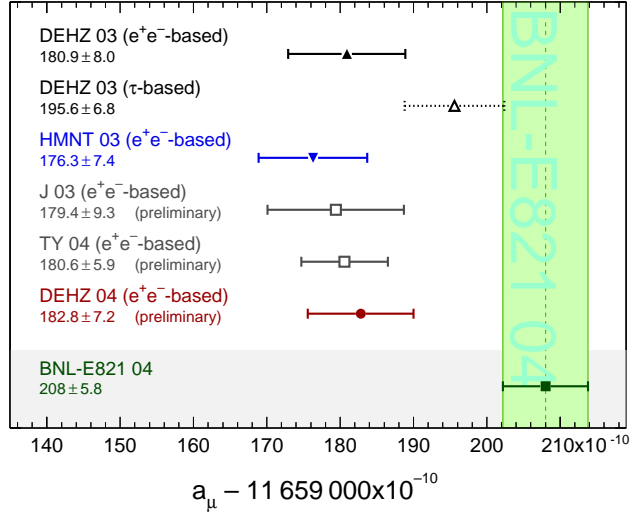}}
  \vspace{0.2cm}
  \caption[.]{
	Comparison of the result~(\ref{eq:smres})~\cite{hoecker_ichep04} 
        labelled DEHZ~04 with the 
	BNL measurement~\cite{bnl_2004}. Also given
	are the previous estimate~\cite{Davier:2003b}, where the 
	triangle with the dotted error bar indicates the 
	$\tau$-based result, as well as the estimates 
        from~\cite{teubner,yndurain,jegerlehner}, not yet including 
	the KLOE data.}
\label{fig:results}
\end{figure}
\beq
\label{eq:diffbnltheo}
  \amu^{\rm exp}-\amu^{\rm SM} = (25.2\pm9.2)\tmten~,
\eeq
which corresponds to 2.7 ``standard deviations'' (to be interpreted
with care due to the dominance of experimental and theoretical 
systematic errors in the SM prediction).
A graphical comparison of the result~(\ref{eq:smres}) with 
previous evaluations (also those containing $\tau$ data) and the 
experimental value is given in Fig.~\ref{fig:results}. 

Whereas the evaluation based on the \ee data only disagrees with
the measurement, the evaluation including the tau data is consistent
with it. The dominant contribution to the discrepancy between the two 
evaluations stems from the $\pi\pi$ channel with a difference of 
$(-11.9\pm6.4_{\rm exp}\pm2.4_{\rm rad}\pm2.6_{\rm SU(2)}
\,(\pm7.3_{\rm total}))\tmten$, and a more significant energy-dependent
deviation\footnote
{
	The systematic problem between $\tau$ and \epem data is more 
	noticeable when comparing the $\taum\to\pim\piz\nut$ branching 
	fraction with the prediction obtained from integrating the 
	corresponding isospin-breaking-corrected \epem
	spectral function (\cf\  Section~\ref{sec:cvc}).
}.
As a consequence, during the previous evaluations of \amuhadLO, 
the results using respectively the $\tau$ and \epem  data were 
quoted individually, but on the same footing since the \epem-based 
evaluation was dominated by the data from a single experiment (CMD-2).

The seeming confirmation of the $\ee$ data by KLOE could lead to
the conclusion that the $\tau$-based result be discredited for the use 
in the dispersion integral~\cite{hoecker_ichep04}. However, the newest
SND data~\cite{snd:2005} alter this picture in favor of the $\tau$
data, along with prompting doubts on the validity of the KLOE results
(see discussion in Section~\ref{sec:comp_eetau_2pi}). Comparing 
the SND and CMD-2 data in the overlapping energy region between 
$0.61\gev$ and $0.96\gev$, the SND-based evaluation of \amuhadNLO
is found to be larger by $(9.1\pm6.3)\times10^{-10}$. However, once 
these two experiments are averaged using the trapezoidal rule, the 
increase over the CMD-2 only evaluation is reduced to $2\times10^{-10}$
due to the much smaller systematic errors and the higher density of 
the CMD-2 measurements.

The proper response to the problems in the LO hadronic contribution 
is to further improve the experimental precision by a renewed 
experimental effort, and to continue to improve theory.  
In this respect, new $\epem\to{\rm hadrons}$ data and further 
study of LBL scattering could potentially reduce the overall 
theoretical uncertainty. The excitement that was caused by the 
deviation~(\ref{eq:diffbnltheo}) stems from the expectation that 
new physics could cause an effect of similar magnitude,
some examples of which are discussed in~\cite{Davier:2004,marciano2}.

\subsection{Conclusions and perspectives}

After considerable effort, the experiment E821 at Brookhaven has improved the
determination of $\amu$ by about a factor of 14 relative to the classic
CERN results of the 1970's~\cite{bailey,cern}. The result is still 
statistics limited and could be improved by another factor of 2 or so 
(to a precision of $3.0 \tmten$) before systematic effects become 
a limitation. Pushing the experiment to that level seems to be an obvious 
goal for the near term. In the longer term, a new experiment 
with improved muon acceptance and magnetic fields could potentially 
reach $0.6 \tmten$~\cite{newexp,newexpa}.

In spite of the new and precise data on the two-pion spectral function 
from the KLOE and SND collaborations, the lowest-order hadronic 
vacuum-polarization contribution remains the most critical component 
in the SM prediction of $a_\mu$. 
The publication of the KLOE results led to an outward confirmation of 
the discrepancy between the $\tau$ data and \epem  annihilation observed 
in the $\pip\pim$ channel~\cite{Davier:2003b}. Including the KLOE data
in the hadronic evaluation, the \epem-based SM prediction 
of $a_\mu$ was found to differ from the experimental value by 2.7 
``standard deviations''.

An opposite point view is advocated in~\cite{maltman_2005}. 
A QCD sum-rule analysis of low-energy \ee data results in the value 
$\asZ= 0.1150 \pm 0.0024$, which is somewhat lower than the world 
average~\cite{Eidelman:2004}, $0.1187 \pm 0.0020$, and even less 
consistent with the average $0.1200 \pm 0.0020$, which excludes 
$\tau$ data (for the sake of comparison) and heavy quarkonium 
results (the accuracy of which has been questioned in~\cite{brambilla_04}). 
The same analysis using the $\tau$ vector spectral function provides
agreement, with a value of $\asZ=0.1201 \pm 0.0021$. Although
the disagreement between the \ee data and other measurements 
is not yet significant, the observed trend is worth more 
investigations.
%
% removed after comment from Maltman
%\footnote
%{ 
%	We note that the rather high spectral moments 
%	used in~\cite{maltman_2005} should be inspected for the 
%	possibility of systematic deviations from the Operator 
%	Product Expansion, which---if present---could affect the 
%	reliability of the result.
%}. 
It is supported by the newest measurement from SND that tend
to amend the compatibility with the $\tau$ data in the 2 pion channel.

The further improvement of the \ee data is the better route and one is 
looking forward to the forthcoming results on the 
low- and high-energy two-pion spectral function from the 
CMD-2 collaboration. These data will help to significantly
reduce the systematic uncertainty due to the corrective 
treatment of radiative effects, often omitted by part by 
the previous experiments. The initial-state-radiation program 
of the \babar\  collaboration has already proved its performance 
by publishing the spectral functions for $\pip\pim\piz$~\cite{babar_3pi} 
and $2\pip 2\pim$~\cite{babar_4pi}, while results for the two-pion 
final state are expected. All sources considered, and assuming all 
discrepancies to be resolved, it appears however difficult to reach 
an uncertainty much better than $3 \tmten$ in this sector\footnote
{
   	We also point out that lattice gauge theories with 
	dynamical fermions can in principle provide 
	a determination of \amuhadLO~\cite{blum}. 
}.

The $30 \%$ uncertainty in the hadronic LBL contribution
($3.5 \tmten$) is largely model-dependent. Here one could conceive
that further work following the approach of~\cite{lbl_mv} or lattice 
calculations could help to reduce the error. Doing much better will
be difficult, but an improvement in the precision by about a factor 
of 2 would well match short-term experimental capabilities.

If the above experimental and theoretical improvements do occur, 
they will lead to a total uncertainty in 
$\Delta \amu=\amu^{\rm exp}-\amu^{\rm SM}$ of about $5.0 \tmten$,
\ie, a factor of 2 reduction with respect to today. This would 
provide an important step towards a better test of the SM.
The result could be used in conjunction with future collider 
discoveries to sort out the properties of new physics (\eg, 
the size of $\tan\! \beta$ in SUSY) or to further constrain 
possible extensions of the SM.

\section{HADRONIC TAU DECAYS AND QCD}	
\label{sec:qcd}

Tests of Quantum Chromodynamics and the precise measurement of the 
strong coupling constant \as at the $\tau$ mass 
scale~\cite{narisonpich:1988,braaten:1989,bnp,pichledib}, 
carried out for the first time by the ALEPH~\cite{aleph_as} and 
CLEO~\cite{cleo_as} collaborations have triggered many theoretical
developments. They concern primarily the perturbative expansion
for which different optimized rules have been suggested. Among these 
are contour-improved (resummed) fixed-order perturbation 
theory~\cite{pert,pivov1,pivov2}, effective charge and minimal 
sensitivity schemes~\cite{grunberg1,grunberg2,dhar1,dhar2,pms}, the
large-$\beta_0$ expansion~\cite{beneke,altarelli,neubert}, as well as
combinations of these approaches. They mainly distinguish themselves
in how they deal with the fact that the perturbative
series is truncated at an order where the missing part is not
expected to be small. Also, in the discussion of the so-called 
{\em contour improved} approach we will point out the importance 
to fully retain the complete information from the renormalization 
group when computing integrals of perturbatively expanded quantities.
With the publication of the full vector and axial-vector spectral 
functions by ALEPH~\cite{aleph_vsf,aleph_asf} and OPAL~\cite{opal_vasf}
it became possible to directly study the nonperturbative properties of 
QCD through vector minus axial-vector sum rules. These analyses are 
described in Section~\ref{sec:sumrules}. 

One could wonder how $\tau$ decays may at all allow us to learn something about 
perturbative QCD. The hadronic decay of the $\tau$ is dominated by resonant 
single particle final states. The corresponding QCD interactions that bind 
the quarks and gluons into these hadrons necessarily involve small momentum
transfers, which are outside the domain of perturbation theory. On the 
other hand, perturbative QCD describes interactions of quarks and gluons
with large momentum transfer. Indeed, it is the inclusive character of 
the sum of all hadronic $\tau$ decays that allows us to probe fundamental
short distance physics. Inclusive observables like
the total hadronic $\tau$ decay rate \Rtau~(\ref{eq:rtau}) can be accurately 
predicted as function of \asm using perturbative QCD, and including
small nonperturbative contributions within the framework of the Operator 
Product Expansion (OPE)~\cite{svz}. In effect, \Rtau is a doubly inclusive 
observable since it is the result of an integration over all hadronic final 
states at a given invariant mass and further over all masses between 
$m_\pi$ and $m_\tau$.
Hadronic $\tau$ decays are more inclusive than \ee annihilation data, 
since vector and axial-vector contributions are of approximately
equal size, while the isospin-zero component in \ee data is three 
times smaller than the isovector part, which is (about) equal to the 
vector contribution in $\tau$ decays.
The scale $m_\tau$ lies in a compromise region 
where \asm is large enough so that \Rtau is sensitive to its value, 
yet still small enough so that the perturbative expansion converges 
safely and nonperturbative power terms are small.

If strong and electroweak radiative corrections are neglected, the 
theoretical parton level prediction for $SU_C(N_C)$, $N_C=3$ reads
\beq
\label{eq:parton}
   R_\tau =
     N_C\left(|V_{ud}|^2 + |V_{us}|^2\right) = 3~,
\eeq
so that we can estimate a perturbative correction to this value
of approximately 21\% to obtain~(\ref{eq:rtau}), assuming other sources 
to be small. One realizes the increase in sensitivity to \as 
compared to the $Z$ hadronic width, where due to the 
three times smaller \asZ the perturbative QCD correction
reaches only about 4\%.

The nonstrange inclusive observable \Rtau can be theoretically separated 
into contributions from specific quark currents, namely vector ($V$) and 
axial-vector ($A$) $\ubar d$ and $\ubar s$ quark currents. It is therefore 
appropriate to decompose
\beq
\label{eq:rtausum}
   \Rtau = \RtauV + \RtauA + \RtauS~,
\eeq
where for the strange hadronic width $\RtauS$ vector and axial-vector
contributions are not separated so far due to the lack of the
corresponding experimental information for the Cabibbo-suppressed modes.
Parton-level and perturbative terms do not distinguish vector and
axial-vector currents (for massless partons). Thus the corresponding 
predictions become 
$R_{\tau,V/A} = (N_C/2)|V_{ud}|^2$ and $\RtauS = N_C|V_{us}|^2$,
which add up to Eq.~(\ref{eq:parton}).

A crucial issue of the QCD analysis at the $\tau$ mass scale concerns the 
reliability of the theoretical description, \ie, the use of the OPE to 
organize the perturbative and nonperturbative expansions, and the control
of unknown higher-order terms in these series. A reasonable stability test 
is to continuously vary $m_\tau$ to lower values $\sqrt{s_0}\le m_\tau$ for 
both theoretical prediction and measurement, which is possible since the 
shape of the full $\tau$ spectral function is available. The kinematic 
factor that takes into account the $\tau$ phase space suppression 
at masses near to $m_\tau$ is correspondingly modified so that 
$\sqrt{s_0}$ represents the new mass of the $\tau$. 

%In the following subsections we proceed with a fairly detailed discussion
%of the various approaches to the perturbative prediction of $\Rtau$. 
%Improvements over the traditional fixed-order perturbative expansion 
%have been developed in the past

%
% ----------------------------- Tools -------------------------------
%
\subsection{Renormalization group equations}
\label{sec:qcd_RGEs}

Like in QED, the subtraction of divergences in QCD is equivalent to the 
renormalization of the coupling ($\as\equiv g_s^2/4\pi$), the 
quark masses ($m_q$), etc, and the fields in the bare  (superscript $B$) 
Lagrangian such as $\as^B=s^{\varepsilon}Z_{\as}\as$, 
$m_q^B=s^{\varepsilon}Z_mm_q$, etc.
Here $s$ is the renormalization scale, $\varepsilon$ the 
dimensional regularization parameter, and $Z$ denotes a series of 
renormalization constants obtained from the generating functional 
of the bare Green's function.
The renormalization procedure introduces an energy scale, $s$, 
which represents the point at which the subtraction to remove 
the divergences is actually performed.
This leads to the differential {\em Renormalization Group Equations} (RGE)
\beqn
\label{eq:betafun}
	\frac{d a_s}{d \ln s}
	&=& 
	\beta(a_s) = -a_s^2\sum_n \beta_n a_s^n~, \\
\label{eq:gammafun}
   	\frac{1}{m_q}\frac{d m_q}{d \ln s}
     	&=& 
   	\gamma(\as)= -a_s\sum_n \gamma_n a_s^n~,
\eeqn
with $a_s=\as/\pi$. Expressed in the {\em modified minimal subtraction 
renormalization scheme} (\MSbar)~\cite{msbar1,msbar2} and for three 
active quark flavors at the $\tau$ mass scale, the perturbative coefficients, 
known to four loops~\cite{betafourloop,gammafourloop}, are\footnote
{
	The full expressions for an arbitrary number of quark flavors ($n_f$)
	are~\cite{betafourloop,gammafourloop}
	\newcommand\hshere{\hspace{0.5cm}}
	\beqns
	   \beta_0 = \frac{1}{4}\left(11 - \frac{2}{3}n_f\right)~,  \hshere
	   \beta_1 = \frac{1}{16}\left(102 - \frac{38}{3}n_f\right)~,  \hshere
	   \beta_2 = \frac{1}{64}\left(\frac{2857}{2} - \frac{5033}{18}n_f 
                                  + \frac{325}{54}n_f^2\right)~, \\
	   \beta_3 = \frac{1}{256}\left[
	       \frac{149753}{6} + 3564\,\zeta_3 -
	       \left(\frac{1078361}{162} + \frac{6508}{27}\,\zeta_3\right)n_f 
	       + \left(\frac{50065}{162} + \frac{6472}{81}\,\zeta_3\right)n_f^2 +
	       \frac{1093}{729}n_f^3\right]~,
	\eeqns
	and
	\beqns
	   \gamma_0 = 1~, \hshere
	   \gamma_1 = \frac{1}{16}\left(\frac{202}{3} - \frac{20}{9}n_f\right)~,  \hshere
	   \gamma_2 = \frac{1}{64}\left[1249 - 
                                  \left(\frac{2216}{27}+\frac{160}{3}\,\zeta_3\right)n_f 
                                  - \frac{140}{81}n_f^2\right]~, \\
	   \gamma_3 = \frac{1}{256}
	       \left[\frac{4603055}{162} + \frac{135680}{27}\,\zeta_3 - 8800\,\zeta_5
	       + \left(-\frac{91723}{27} - \frac{34192}{9}\,\zeta_3 + 880\,\zeta_4 
               + \frac{18400}{9}\,\zeta_5\right)n_f\right.
	        \\
		\left.
	       + \left(\frac{5242}{243} + \frac{800}{9}\,\zeta_3 
			- \frac{160}{3}\,\zeta_4\right)n_f^2 +
	       \left(-\frac{332}{243} + \frac{64}{27}\,\zeta_3\right)n_f^3\right]~,
	\eeqns
	where the $\zeta_{i=\{3,4,5\}}=\{1.2020569,\pi^4/90,1.0369278\}$ are the Riemann 
	$\zeta$-functions.
}
\newcommand\hshere{\hspace{0.5cm}}
\beqn
\label{eq:beta}
 	\beta_0=2.25~, 			\hshere
	\beta_1=4~,   			\hshere
	\beta_2\approx 10.05989~,	\hshere 
	\beta_3\approx 47.22804, \\
\label{eq:gamma}
 	\gamma_0=1~, 			\hshere 
	\gamma_1=3.79166~, 		\hshere 
	\gamma_2\approx 12.42018~, 	\hshere 
	\gamma_3\approx 44.26278~.
\eeqn
The solutions for the evolutions~(\ref{eq:betafun}) and (\ref{eq:gammafun}) 
are obtained from integration\footnote
{
	Note that the integration is only defined for a non-vanishing
	derivative $\beta(a_s)\ne0$. Vanishing $\beta(a_s)$ may indeed 
	occur for example in effective charge schemes with large $K_4$ 
	or $K_5$ coefficients (see later in this section).
%	An iterative solution of the four-loop $\beta$ function has been
%	computed in~\cite{chet1}, which we give here for completeness
%	\beqns
%		a_s(s) &=&
%		\frac{1}{\beta_0 L} - \frac{b_1\ln L}{(\beta_0 L)^2}
%		+ \frac{1}{(\beta_0 L)^3}\left[b_1^2\left(\ln^2L-\ln L -1\right) + b_2\right]
%		\nonumber\\
%		&&
%		+\: \frac{1}{(\beta_0 L)^4}\left[b_1^3\left(-\ln^3L+\frac{5}{2}\ln^2L
%				+ 2\ln L - \frac{1}{2}\right)
%			- 3b_1 b_2\ln L + \frac{b_3}{2}\right]~,
%	\eeqns
%	where $L=\ln(s/\Lambda^2)$, $b_i=\beta_i/\beta_0$ and
%	terms of order $\mathcal{O}(L^{-5})$ have been neglected.
}~\cite{pms,chet1} 
\beqn
\label{eq:betafunint}
	\ln\frac{s}{\Lambda^2} &=& 
	\intl_0^{a_s(s)} da_s^\prime \left(\frac{1}{\beta(a_s^\prime)}
		     + \frac{1}{\beta_0(a_s^\prime)^2+\beta_1(a_s^\prime)^3}\right)
         +\intl_{a_s(s)}^\infty da_s^\prime 
		\frac{1}{\beta_0(a_s^\prime)^2+\beta_1(a_s^\prime)^3}~, \\
%	&\approx&
%	\frac{1}{\beta_0}\left[\frac{1}{a_s} + \frac{\beta_1}{\beta_0}\ln a_s
%	+ \left(\frac{\beta_2}{\beta_0}-\frac{\beta_1^2}{\beta_0^2}\right)a_s
%	+\left(\frac{\beta_3}{2\beta_0} - \beta_1\beta_2 
%		+ \frac{\beta_1^3}{2\beta_0^3}\right)a_s^2\right]
%	+ C~, \\
\label{eq:gammafunint}
	\ln\frac{m_q(s)}{m_q(s_0)} &=& \intl^{a_s(s)}_{a_s(s_0)}
			           d a_s^\prime \frac{\gamma(a_s^\prime)}{\beta(a_s^\prime)}~,
\eeqn
where the \MSbar asymptotic scale parameter $\Lambda$ and the RG-invariant
quark mass $\hat m_q$ (which appears after evaluating the 
integral~(\ref{eq:gammafunint})) are fixed by referring to known values 
$a_s(s_0)$ and $m_q(s_0)$, respectively, at scale $s_0$. 
% The explicit 
% solution of the integral~(\ref{eq:betafunint}) implies the integration 
% constant $C=(\beta_1/\beta_0^2)\ln\beta_0$, which is chosen to suppress 
% terms $\propto \ln(s/\Lambda^2)$~\cite{msbar2,furmanski} (the reader is 
% referred to~\cite{chet1} for a discussion of other suitable choices).

For most practical purposes, it is however unnecessary to explicitly
perform the integrations~(\ref{eq:betafunint}, \ref{eq:gammafunint}), since 
the evolution of the differential RGEs can be solved numerically by means 
of {\em single-step integration} using a modified Euler method. 
It consists of Taylor-developing, say, Eq.~(\ref{eq:betafun}) around the 
reference scale $s_0$ in powers of $\eta\equiv\ln(s/s_0)$, and reordering 
the perturbative series in powers of $a_s\equiv a_s(s_0)$,
\beqn
\label{eq:astaylor}
	a_s(s) &=& 
  		a_s - 
		\beta_0 \eta a_s^2 + 
		\left(-\beta_1\eta + \beta_0^2 \eta^2\right) a_s^3 +
    		\left(-\beta_2\eta  + \frac{5}{2} \beta_0\beta_1 \eta^2 - \beta_0^3 \eta^3
				\right) a_s^4 
		\nonumber \\
		&&   +\: 
    		\left(-\beta_3\eta  + \frac{3}{2} \beta_1^2\eta^2  + 3 \beta_0 \beta_2\eta^2  
						- \frac{13}{3} \beta_0^2 \beta_1 \eta^3
                				+ \beta_0^4 \eta^4
				\right) a_s^5 
		\\
		&&   +\: 
    		\left(-\beta_4\eta  + \frac{7}{2} \beta_1 \beta_2\eta^2 
						+ \frac{7}{2} \beta_0 \beta_3\eta^2  
						- \frac{35}{6} \beta_0 \beta_1^2 \eta^3 
                				- 6 \beta_0^2 \beta_2 \eta^3
						+ \frac{77}{12} \beta_0^3 \beta_1 \eta^4
						- \beta_0^5 \eta^5
				\right)a_s^6 
		\;+\;\mathcal{O}(a_s^7)~. \nonumber
\eeqn
For later use, we have
expressed the expansion up to order $a_s^6$, which involves the 
unknown five-loop coefficient $\beta_4$. An efficient integration 
of the RGEs can also be obtained using the CERNLIB routine
``RKSTP'', which performs a $4^{\rm th}$ order Runge-Kutta single 
step approximation with excellent accuracy.

\subsection{Theoretical prediction of $\Rtau$}\label{sec:rtau_th}

According to Eq.~(\ref{eq:imv}) the absorptive parts of the vector 
and axial-vector two-point correlation functions 
$\Pi^{(J)}_{\ubar d,V/A}(s)$, with the spin $J$ of the hadronic 
system, are proportional to the $\tau$ hadronic \sfs\  with 
corresponding quantum numbers. The nonstrange ratio \RtauVpA
can be written as an integral of these \sfs\  over the 
invariant mass-squared $s$ of the final state hadrons~\cite{bnp}
\beq
\label{eq:rtauth1}
   \RtauVpA(s_0) =
	12\pi \Sew\intl_0^{s_0}
		\frac{ds}{s_0}\left(1-\frac{s}{s_0}
                                    \right)^{\!\!2}
     \left[\left(1+2\frac{s}{s_0}\right){\rm Im}\Pi^{(1)}(s+i\e)
      \,+\,{\rm Im}\Pi^{(0)}(s+i\e)\right]~,
\eeq
where $\Pi^{(J)}$ can be decomposed as
$\Pi^{(J)}=|V_{ud}|^2\left(\Pi_{ud,V}^{(J)}+\Pi_{ud,A}^{(J)}\right)$.
The lower integration limit is zero because the pion pole is at 
zero mass in the chiral limit.

The correlation function $\Pi^{(J)}$ is analytic in the complex $s$ plane 
everywhere except on the positive real axis where singularities exist.
Hence by Cauchy's theorem, the imaginary part of $\Pi^{(J)}$ is 
proportional to the discontinuity across the positive real axis
\begin{figure}[t]  
  \centerline{\epsfysize6.5cm\epsffile{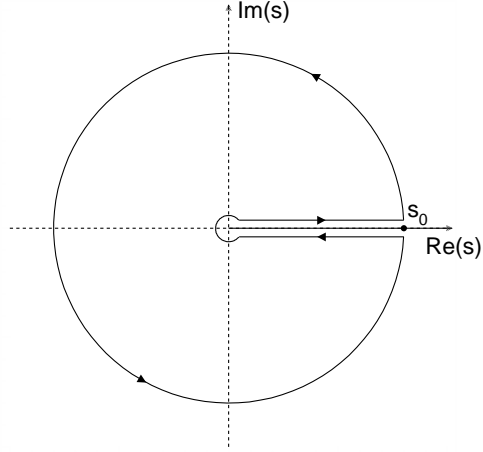}}
  \caption[.]{\label{fig:contour}
              Integration contour for the r.h.s. in Eq.~(\ref{eq:contour}).}
\end{figure} 
\beq
\label{eq:contour}
   \frac{1}{\pi}\intl_0^{s_0}ds\,w(s){\rm Im}\Pi(s) =
   -\frac{1}{2\pi i}\hm\ointl_{|s|=s_0}\hm\hm\hm ds\,w(s)\Pi(s)~,
\eeq
where $w(s)$ is an arbitrary analytic function, and the contour 
integral runs counter-clockwise around the circle from $s=s_0+i\e$ to 
$s=s_0-i\e$ as indicated in Fig.~\ref{fig:contour}. 

The energy scale $s_0= m_\tau^2$ is large enough that contributions 
from nonperturbative effects are expected to be subdominant and the use 
of the OPE is appropriate. The kinematic factor $(1-s/s_0)^2$ suppresses 
the contribution from the region near the positive real axis where 
$\Pi^{(J)}(s)$ has a branch cut and the OPE validity is restricted 
due to large possible quark-hadron duality 
violations~\cite{quinnetal,braaten88}. 

The theoretical prediction of the vector and axial-vector
ratio \RtauVA can hence be written as
\beq
\label{eq:delta}
   \RtauVA =
     \frac{3}{2}|V_{ud}|^2\Sew\left(1 + \delta^{(0)} + 
     \delta^\prime_{\rm EW} + \delta^{(2,m_q)}_{ud,V/A} + 
     \hm\hm\sum_{D=4,6,\dots}\hm\hm\hm\hm\delta_{ud,V/A}^{(D)}\right)~,
\eeq
with the massless perturbative contribution $\delta^{(0)}$,
the residual non-logarithmic electroweak correction 
$\delta^\prime_{\rm EW}=0.0010$~\cite{braaten} (\cf\  the discussion 
on radiative corrections in Section~\ref{sec:cvc_isobreak}), and the 
dimension $D=2$ {\em perturbative} contribution $\delta^{(2,m_q)}_{ud,V/A}$ 
from quark masses, which is lower than $0.1\%$ for $u,d$ quarks. 
The term $\delta^{(D)}$ denotes the OPE contributions of mass
dimension $D$
\beq
\label{eq:ope}
    \delta_{ud,V/A}^{(D)} =
       \hm\hm\hm\sum_{{\rm dim}{\cal O}=D}\hm\hm\hm C_{ud,V/A}(s,\mu)
            \frac{\langle{\cal O}_{ud}(\mu)\rangle_{V/A}}
                 {(-\sqrt{s_0})^{D}}~,
\eeq
where the scale parameter $\mu$ separates the long-distance 
nonperturbative effects, absorbed into the vacuum expectation 
elements $\langle{\cal O}_{ud}(\mu)\rangle$, from the short-distance 
effects that are included in the Wilson coefficients 
$C_{ud,V/A}(s,\mu)$~\cite{wilson}. Note that 
$\delta_{ud,V+A}^{(D)}=(\delta_{ud,V}^{(D)}+\delta_{ud,A}^{(D)})/2$.

\subsubsection{The Perturbative Prediction}\label{sec:pert}

The perturbative prediction used by the experiments follows the
work of~\cite{pert}. The perturbative contribution is 
given in the chiral limit. Effects from quark masses have been 
calculated in~\cite{pertmass} and are found to be well below
1\% for the light quarks. As a consequence, the contributions from 
vector and axial-vector currents coincide to any given order of 
perturbation theory and the results are flavor independent.

For the evaluation of the perturbative series, it is convenient to 
introduce the analytic Adler function~\cite{adler}
\beq
\label{eq:adler}
   D(s) \;\equiv\;
      - s\frac{d\Pi(s)}{ds} = 
            \frac{s}{\pi}\intl_0^\infty ds^\prime\,
            \frac{{\rm Im}\Pi(s^\prime)}{(s^\prime-s)^2}~,
\eeq
where the second identity is the dispersion relation (\cf\  Section~\ref{anomaly}). 
The derivative of the correlator avoids extra subtractions (renormalization) 
on the r.h.s. of Eq.~(\ref{eq:adler}), which are unrelated to QCD dynamics.
The function $D(s)$ calculated in perturbative QCD within the \MSbar 
renormalization scheme depends on a non-physical parameter $\mu$ 
occurring as $\ln(\mu^2/s)$. Furthermore it is a function of \as. 
On the other hand, since $D(s)$ is connected to a physical 
quantity, the \sf\  $\Im\Pi(s)$, it cannot depend on the subjective 
choice of $\mu$. This can be achieved if \as\  becomes a function of 
$\mu$ providing independence of $D(s)$ of the choice of $\mu$. 
Nevertheless, in the realistic case of a truncated series, some $\mu$ 
dependence remains and represents an irreducible systematic uncertainty. 
$D(s$) is then a function of $\mu^2/s$, $\as$ 
and can be understood as an effective charge (see below) that obeys the 
RGE. Choosing $\mu^2=s$, the perturbative series reads
\beq
\label{eq:pertAdler}
   D(\as(s)) = \hm\hm\hm
      \sum_{n=0}^{\infty}r_n a_s^{n}(s)~,
\eeq
with renormalization scheme dependent coefficients $r_n$.

To introduce the Adler function in Eqs.~(\ref{eq:rtauth1}) and 
(\ref{eq:contour}) one uses partial integration
\beqn
\label{eq:identity}
   \ointl_{|s|=s_0}\hm\hm\hm ds\,g(s)\Pi(s) 
	&=&
     -\hm\hm\hm\ointl_{|s|=s_0}\hm\hm
     \frac{ds}{s}\left(G(s) - G(s_0)
                 \right)\,s\frac{d\Pi(s)}{ds}~, 
\eeqn
with the kernel $g(s)=(1-s/s_0)^2(1+2s/s_0)/s_0$ and
$G(s)=\int_0^s ds^\prime\,g(s^\prime)$. This gives~\cite{pert}
\beq
\label{eq:rtauadler}
   1+\delta^{(0)} = 
      -2\pi i\hm\hm\hm\ointl_{|s|=s_0}\hm\hm\frac{ds}{s}
       \left[1-2\frac{s}{s_0} + 2\left(\frac{s}{s_0}\right)^{\!\!3}
             - \left(\frac{s}{s_0}\right)^{\!\!4}\right] D(s)~.
\eeq
The perturbative expansion of the Adler function can be inferred 
from the third-order calculation of the \ee  inclusive cross section ratio 
$R_{\ee}(s)=\sigma(\ee\to{\rm hadrons}\,(\gamma))/\sigma(\ee\to\mu^+\mu^-\,(\gamma))$~\cite{adler1,adler2,2loop,loopbis,loopbisbis,pert}
\beq
\label{eq:adlerpert}
   D(s) =
     \frac{1}{4\pi^2}\sum_{n=0}^\infty \tilde{K}_n(\xi)a_s^{n}(-\xi s)~,
\eeq
with the $\tilde{K}_n(\xi)$ functions up to order $n=5$~\cite{pert}
\beqn
	&&
   \tilde{K}_0(\xi) = K_0~, 	\hspace{1cm}
   \tilde{K}_1(\xi) = K_1~, 	\hspace{1cm}
   \tilde{K}_2(\xi) = K_2 + L~, \nonumber\\
	&&
   \tilde{K}_3(\xi) = K_3 + \left(b_1 + 2\,K_2\right)L + L^2 \nonumber\\
	&&
\label{eq:kfun}
   \tilde{K}_4(\xi) = K_4 + \left(b_2 + 2\,b_1 K_2 + 3\,K_3\right)L
 		            + \left(\frac{5}{2}\,b_1 + 3\,K_2\right)L^2 + L^3~, \\
	&&
   \tilde{K}_5(\xi) = K_5 + \left(b_3 + 2\,b_2 K_2 + 3\,b_1 K_3 
                                    + 4\, K_4\right)L
                        + \left(\frac{3}{2}\,b_1^2 + 3\,b_2 + 7\,b_1 K_2 
                                + 6\, K_3\right)L \nonumber\\
	&& \hspace{1.7cm}
 		        +\: \left(\frac{13}{3}\,b_1 + 4\,K_2\right)L^3 + L^4~, 
	\nonumber
\eeqn
where $b_i=\beta_i/\beta_0$ and $L=\beta_0\ln\xi$. The factor $\xi=s_0/s$ in 
Eq.~(\ref{eq:adlerpert}) represents the arbitrary renormalization 
scale ambiguity of which the Adler function~(\ref{eq:adler}) is independent. 
The $\xi$ dependence of the $K_i(\xi)$ coefficients is 
determined by inserting the expansion~(\ref{eq:astaylor}) into 
(\ref{eq:adlerpert}), and equating the two series obtained for $\xi=1$ 
and arbitrary $\xi$ at each order in $a_s$.\footnote
{
	Note that this procedure corresponds to a fixed-order
	perturbation theory rule since known higher order
	pieces of the expansion~(\ref{eq:astaylor}) are neglected
	when solving 
	\beqns
	\frac{\partial}{\partial\xi}\sum_{n=0}^N \tilde{K}_n(\xi)a_s^{n}(-\xi s)
	=\frac{\beta(a_s^{n}(-\xi s))}{\xi}
		\sum_{n=0}^N (n+1)\tilde{K}_n(\xi)a_s^{n}(-\xi s)
	 +      \sum_{n=0}^N a_s^{n+1}(-\xi s)\frac{\partial \tilde{K}_n(\xi)}{\partial\xi}
	 \sim \mathcal{O}(a_s^{N+1})~,
	\eeqns
	for a truncated series with maximum known order $N$. 
	A resummation of the subleading logarithms has been proposed
	in~\cite{kleiss}.
} The 
coefficients $K_n$ are known up to third order in $\as^3$. 
For $n\ge2$ they depend on the renormalization scheme used, while 
the first two coefficients are universal
\beq
\label{eq:kf1}
    K_0 = 1~,	\hspace{0.5cm}
    K_1 = 1~,	\hspace{0.5cm}
    K_2(\MSbar) = F_3(\MSbar)~, \hspace{0.5cm}
    K_3(\MSbar) = F_4(\MSbar) 
      \,+\, \frac{1}{3}\pi^2\beta_0^2~,
\eeq
where $F_3(\MSbar) = 1.9857 - 0.1153\,n_f$ and
$F_4(\MSbar) = -6.6368 - 1.2001\,n_f - 0.0052\,n_f^2$ are the 
coefficients of the perturbative series of $R_{\ee}$. For 
$n_f=3$ one has $K_2=1.640$ and $K_3=6.371$. With the
series~(\ref{eq:adlerpert}), inserted in the r.h.s of 
Eq.~(\ref{eq:rtauadler}), one obtains the perturbative expansion
\beq 
\label{eq:knan}
   \delta^{(0)} = 
       \sum_{n=1}^5 \tilde{K}_n(\xi) A^{(n)}(a_s)~,
\eeq
with the functions
\beqn
\label{eq:an}
   A^{(n)}(a_s) 
	&=&
      \frac{1}{2\pi i}\hm\ointl_{|s|=s_0}\hm\hm
      \frac{ds}{s}
       \left[1-2\frac{s}{s_0} + 2\left(\frac{s}{s_0}\right)^{\!\!3}
             - \left(\frac{s}{s_0}\right)^{\!\!4}
       \right]a_s^{n}(-\xi s) \nonumber\\
	&=&
  	\frac{1}{2\pi} \intl_{-\pi}^{\pi} d\varphi
       	\left[1 + 2e^{i\varphi} - 2e^{i3\varphi} - e^{i4\varphi}
       	\right] a_s^{n}(\xi s_0e^{i\varphi})~,
\eeqn
where $s=-s_0e^{i\varphi}$ has been substituted in the second line and
the integral path proceeds according to Fig.~\ref{fig:contour}. 
%Formally, 
%the integrals~(\ref{eq:an}) also obey an RGE~\cite{pert}
%\beq
%	\xi\frac{\partial A^{(n)}\left(a_s(-\xi s)\right)}
%                {\partial\xi} =
%               n\sum_{k=1}\beta_kA^{(n+k)}\left(a_s(-\xi s)\right)~.
%\eeq

\subsubsection{Fixed-order perturbation theory (FOPT)}

Inserting the series~(\ref{eq:astaylor}) in Eq.~(\ref{eq:knan}), 
evaluating the contour integral, and collecting the terms with 
equal powers in $a_s$ leads to the familiar expression~\cite{pert}
\beq 
\label{eq:kngn}
   \delta^{(0)} = 
       \sum_{n} \left[\tilde{K}_n(\xi) + g_n(\xi)\right]
       \left(\frac{\as(\xi s_0)}{\pi}\right)^{\!\!n}~,
\eeq
where the $g_n$ are functions of $\tilde{K}_{m<n}$ and $\beta_{m<n-1}$, and of
elementary integrals with logarithms of power $m<n$ in the integrand. Setting 
$\xi=1$ and replacing all known $\beta_i$ and $K_i$ coefficients
by their numerical values, Eq.~(\ref{eq:kngn}) simplifies to
\beqn
\label{eq:delta0exp}
   \delta^{(0)}
   &=&
      a_s(s_0)
      + (1.6398 +  3.5625)\,a_s^{2}(s_0)
      + (6.371  + 19.995)\,a_s^{3}(s_0) \nonumber\\
   &&
      +\: (K_4 + 78.003)\,a_s^{4}(s_0) \nonumber\\
   &&
      +\: (K_5 + 14.250\,K_4 - 391.54)\,a_s^{5}(s_0)\nonumber\\
   &&
      +\: (K_6 + 17.813\,K_5 + 45.112\,K_4 + 1.58333\,\beta_4 - 8062.1)\,a_s^{6}(s_0)~,
\eeqn
where for the purpose of later studies we have kept terms up to sixth order.
When only two numbers are given in the parentheses, the first number 
corresponds to $K_n$, and the second to $g_n$. 

The FOPT series is truncated at given order 
despite the fact that parts of the higher coefficients $g_{n>4}(\xi)$ are 
known to all orders and could be resummed. These known parts are the 
higher (up to infinite) order terms of the expansion~(\ref{eq:astaylor})
that are functions of $\beta_{n\le3}$ and $K_{n\le3}$ only. 
In effect, beyond the use of the perturbative expansion
of the Adler function~(\ref{eq:adlerpert}), two approximations 
have been used to obtain the FOPT series~(\ref{eq:delta0exp}):
{\em (i)} the RGE~(\ref{eq:betafun}) has been Taylor-expanded 
and terms higher than the given FOPT order have been truncated,
and {\em (ii)} this Taylor expansion is used to predict $a_s(-s)$ on
the entire $|s|=s_0$ contour.

\subsubsection{Contour-improved fixed-order perturbation theory (\FOPTCI)}
\label{sec:cipt}

A more promising approach to the solution of the contour integral~(\ref{eq:an}) 
is to perform a direct numerical evaluation by means of single-step
integration and using the solution of the RGE to four loops as 
input for the running $a_s(-\xi s)$ at each integration 
step~\cite{pivov1,pivov2,pert}.
It implicitly provides a partial resummation of the (known) higher 
order logarithmic integrals and improves the convergence of the 
perturbative series. While for instance the third order 
term in the expansion~(\ref{eq:delta0exp}) contributes with $17\%$ 
to the total (truncated) perturbative prediction, the corresponding 
term of the numerical solution amounts only to 
%zz $7\%$ 
$6\%$ 
(assuming $\as(m_\tau^2)=0.35$). This numerical solution of 
Eq.~(\ref{eq:knan}) will be referred to as {\it contour-improved} 
fixed-order perturbation theory (\FOPTCI) in the following.

\begin{figure}[t]  
  \centerline{\epsfysize6.3cm\epsffile{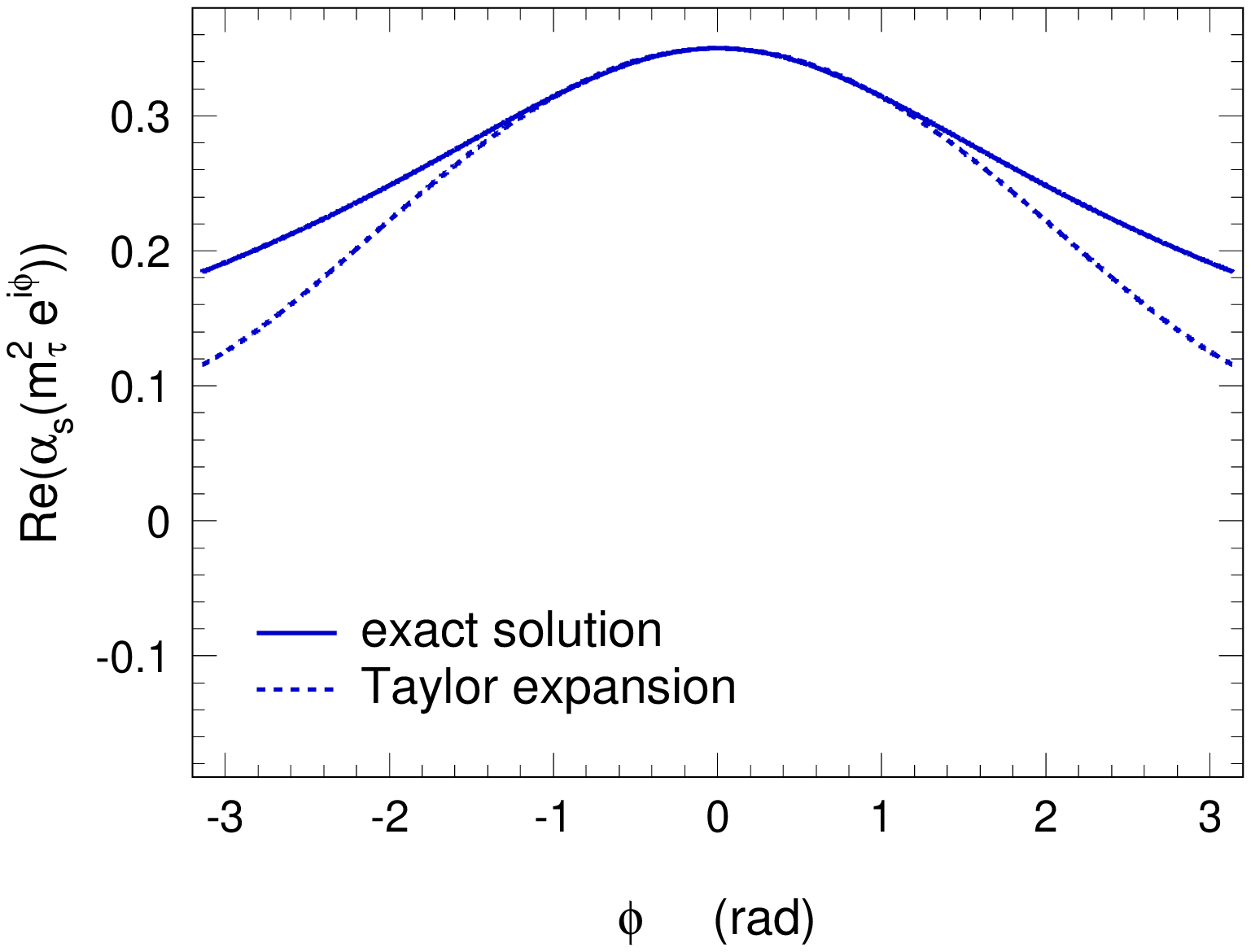}\hspace{0.1cm}
              \epsfysize6.3cm\epsffile{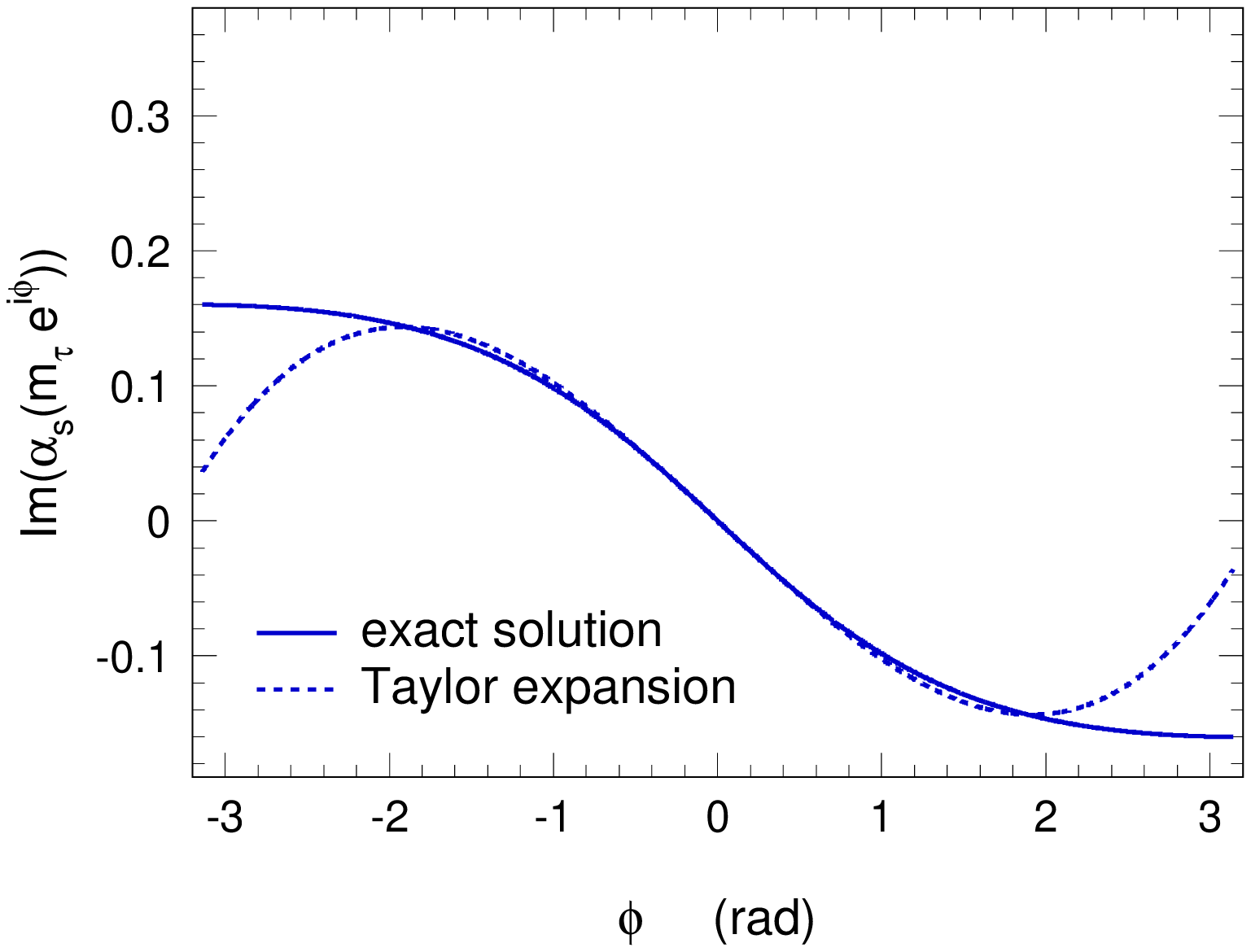}}
  \vspace{-0.1cm}
  \caption[.]{\label{fig:taylor_test}
	Comparison between the iteratively evolved 
	$\as(m_\tau^2e^{i\varphi})$ (exact solution) and the fourth-order
	Taylor expansion~(\ref{eq:astaylor}) on the circle 
	$\varphi=-\pi\dots\pi$~\cite{menke_phd}. This example
	uses $\as(m_\tau^2)=0.35$.}
\end{figure} 
Single-step integration also avoids the Taylor approximation of the RGE on 
the entire contour, since $a_s$ is iteratively computed from the previous 
step using the full known RGE.\footnote
{
	It has been tested that the single-step integration, \ie,
	$a_s(\xi s_0 e^{i(\varphi+\Delta\varphi)})$ is computed from
	$a_s(\xi s_0 e^{i\varphi})$, using a modified Euler method 
	(that is, a Taylor expansion of $a_s$ in each integration step),
	or using explicit $4^{\rm th}$ order Runge-Kutta integration,
	leads to the same result, as far as the step size is chosen
	to be small enough (1000 integration steps on the contour
	are found to be largely sufficient).
}
Figure~\ref{fig:taylor_test} shows the comparison between 
the iteratively evolved $\as(m_\tau^2e^{i\varphi})$ (referred
to as the exact solution), and the fourth-order Taylor expansion~(\ref{eq:astaylor}) 
on the circle $\varphi=-\pi\dots\pi$~\cite{menke_phd}. The differences
appear to be significant far away from the reference value at $(\varphi=0)$.
However, as shown below, the numerical effect of this discrepancy
on the contour integral is small compared to the main difference 
between FOPT and \FOPTCI, which is the truncation of the perturbative 
series after integration.

\subsubsection{Effective charge perturbation theory (ECPT)}
\label{sec:ecpt}

Inspired by the pioneering work in~\cite{grunberg1,grunberg2,dhar1,dhar2,pms} 
the use of effective charges to approach the perturbative prediction
of \Rtau has become quite popular in the recent 
past~\cite{maxwellrs1,raczka,maxwellrs2,pivoveffch}. 
The advocated advantage of this technique is that the perturbative 
prediction of the effective charge is renormalization scheme (RS) and
scale invariant since it is a physical observable. The effective $\tau$ 
charge is defined by
\beq
	a_\tau = \delta^{(0)}~,
\eeq
and obeys the RGE
\beqn
\label{eq:betataufun}
	\frac{d a_\tau}{d \ln s}
	&=& 
	\beta_\tau(a_\tau) = -a_\tau^2\sum_{n=0}^\infty \beta_{\tau,n} a_\tau^n~.
\eeqn
The first two terms in the expansion of $\beta_\tau$ are universal
(\ie, RS-independent), while the higher order terms must be 
computed using perturbation theory. The prescription for this 
computation is obtained from the relation between the renormalization 
scheme and scale used to obtain the FOPT coefficients in the 
series~(\ref{eq:kngn}), and the requirement of RS and scale 
invariance. Since $a_\tau$ is scheme and scale invariant, its 
evolution coefficients $\beta_{\tau,i}$ have the same property.
``{\em Another way of looking at this is to say that the method of 
effective charges involves a specific choice of renormalization
scheme and scale (\dots), so that the energy evolution of the 
observable is identical to the beta-function evolution of the 
coupling}''~\cite{maxwellagain}. 

The function $\beta_\tau$ is obtained from the equation
\beq
	\frac{d a_\tau}{d a_s}
	=
	\frac{d \beta_\tau(a_\tau)}{d \beta(a_s)}~,
\eeq
which, setting $\xi=1$, can be rewritten in the form 
\beq
\label{eq:effchargerelation}
	\beta_\tau(a_\tau) 
	= \frac{d a_\tau}{d\ln s}
	= \frac{d a_\tau}{d a_s}
	  \frac{d a_s}{d\ln s}\bigg|_{a_s=a_s(a_\tau)}
	= \frac{d a_\tau}{d a_s}
	  \beta(a_s)\bigg|_{a_s=a_s(a_\tau)}~.
\eeq
Inserting Eq.~(\ref{eq:kngn}) for $a_\tau$ on the l.h.s. and 
the r.h.s. of Eq.~(\ref{eq:effchargerelation}), and equating
the coefficients for each order in $a_s$ leads to the coefficients\footnote
{
	For completeness we also give the sixth-order coefficient, which 
	is later needed for extrapolation tests
	\beqns
	\beta_{\tau,5} &=& 
                        \beta_5 - 4 \,\beta_4 c_2
                        + 2\,\beta_3 (4 \,c_2^2 - c_3)
                        + 4\,\beta_2 (-2 \,c_2^3 + c_2 c_3)
			+ \beta_1 (-6 \,c_2^4 + 16 \,c_2^2 c_3 - 3 \,c_3^2 - 
   		              8 \,c_2 c_4 + 2 \,c_5) \\
			&&
			+\: 4\,\beta_0 (12 \,c_2^5 - 30 \,c_2^3 c_3 + 12 \,c_2 c_3^2 + 
			   14 \,c_2^2 c_4 - 4 \,c_3 c_4 - 5 \,c_2 c_5 + c_6)~.
	\eeqns
}
\beqn
	\beta_{\tau,2} &=& \beta_2 - \beta_1 c_2 + \beta_0 (-c_2^2 + c_3)~, \\
	\beta_{\tau,3} &=& \beta_3 - 2 \,\beta_2 c_2 + \beta_1 c_2^2 
                           + 2 \,\beta_0 \left( 2\,c_2^3 - 3 \,c_2 c_3 + c_4\right)~, \\
	\beta_{\tau,4} &=& \beta_4 - 3 \,\beta_3 c_2 
                           + \beta_2 (4 \,c_2^2 - c_3)
                           + \beta_1 (-2 \,c_2 c_3 + c_4) \nonumber\\
			&&
                           +\: \beta_0 (-14 \,c_2^4 + 28 \,c_2^2 c_3 - 5 \,c_3^2 
                                 - 12 \,c_2 c_4 + 3 \,c_5)~,
\eeqn
where $c_i=K_i + g_i(1)$ (\cf\  Eq.~\ref{eq:kngn}) and
$\beta_{\tau,0}=\beta_0$, $\beta_{\tau,1}=\beta_1$ because $K_0=K_1=1$. 
Inserting the known $\beta_i$ and $c_i$ coefficients, one finds~\cite{pivoveffch}
\beqn
	\beta_{\tau,2} &=& -12.320~, \nonumber\\
	\beta_{\tau,3} &=& -182.72 + 4.5\,K_4~, \nonumber\\
\label{eq:betataunum}
	\beta_{\tau,4} &=& -236.2 + \beta_4 - 40.27\,K_4 + 6.75\,K_5~,  \\
	\beta_{\tau,5} &=& 16427 - 20.80\,\beta_4 + \beta_5 - 
                           521.6\,K_4 - 65.79\,K_5 + 9\,K_6~,\nonumber
\eeqn
with a noticeable sign alteration for $\beta_{\tau,2}$ (and also for 
$\beta_{\tau,3}$ if $K_4<40$). Neglecting nonperturbative contributions, 
one has
\beq
	a_\tau(s_0)=\RtauVpA(s_0)/(3\,|V_{ud}|^2)-1~, 
\eeq
which using Eq.~(\ref{eq:rtauvpa}) and the $|V_{ud}|$ value given in   
Section~\ref{sec:tauspecfun} yields $a_\tau(m_\tau^2)=0.2219\pm0.0055$.

\subsubsection{Estimating unknown higher order perturbative coefficients}
\label{sec:estimhigherorder}

Compared with the \MSbar RGE~(\ref{eq:betafun}) the $\beta_\tau(a_\tau)$ 
series is significantly less convergent, taking into account that 
$a_\tau(m_\tau^2)\simeq1.8\times a_s(m_\tau^2)$. This property has been 
used in the past to estimate an effective $K_4^{\rm eff}$ coefficient,
which was found to be $K_4^{\rm eff} \sim 27$~\cite{k4_pms}. It 
approximates the true value of $K_4$ in the limit of vanishing 
higher order contributions. In effect, the derivative of 
$a_\tau(m_\tau^2)$ with respect to $K_4$ is large\footnote
{
	One can obtain the solution for $a_\tau(m_\tau^2)$ as a function 
	of the $\beta_\tau$ coefficients without explicitly 
	integrating the RGE~(\ref{eq:betataufun}), by using as fixed point 
        of the differential equation the one-loop solution at $s\to\infty$,
	$\Lambda^2=s e^{-1/(\beta_0 a_\tau(s))}$, so that the four-loop
	$a_\tau(\infty\to m_\tau^2)$ evolution reproduces the experimental 
	value. This gives $\Lambda=0.41\gev$.
}
\beq
\label{eq:k4estimate}
 	\frac{d a_\tau(m_\tau^2)}{dK_4}\bigg|_{a_\tau(m_\tau^2)=0.22,\,K_4=25}
	\simeq 0.0020~,
\eeq
so that the precise experimental error on $a_\tau(m_\tau^2)$ allows us to
predict $K_4$ with an accuracy of a few units, if we assume that the 
subleading contributions can be neglected. However, this assumption
is unfounded since, taking naively the unknown higher order coefficients 
to grow like a geometric series, \ie, $z_n\sim z_{n-1}^2/z_{n-2}$, with $z=\beta,K$, 
and using the estimate $K_4=25$ so that $K_5\sim98$, $\beta_{\tau,4}\sim-360$ 
(with $\beta_4\sim222$), one finds
\beq
\label{eq:k5estimate}
 	\frac{d a_\tau(m_\tau^2)}{dK_5}\bigg|_{a_\tau(m_\tau^2),K_4,K_5,\beta_4\dots}
	\simeq 0.0004~,
\eeq
which is not small with respect to~(\ref{eq:k4estimate}), in particular considering
the coarseness of the assumptions that went into the estimate~(\ref{eq:k5estimate}).
It is found that even the sixth-order coefficient is not negligible. Assuming the 
same geometric growth, one would expect $K_6\sim395$ with a derivative of 
$d a_\tau(m_\tau^2)/dK_6\sim5\times10^{-5}$, which appears small. However,
considering that the uncertainty on $K_6$ is at least of the same size as
its estimated value (which is probably still optimistic) generates an 
uncertainty in $a_\tau(m_\tau^2)$ that is equivalent to $\sigma(K_4)\sim10$. 
Taking correspondingly
the fifth-order uncertainty of the order of the estimated value of $K_5$,
we conclude that the systematic uncertainty on $K_4$ obtained with the 
effective charge scheme is at least of the order of $\sigma(K_4)\sim23$.

Another method to estimate $K_4^{\rm eff}$ has been suggested 
in~\cite{k4_fld}. It consists of increasing the sensitivity of the \FOPTCI
prediction of $\delta^{(0)}$ on $K_4$ by reducing the renormalization scale
$\xi$ in Eq.~(\ref{eq:knan}). The result $27\pm5$ is consistent with the one 
obtained from ECPT. However, these methods are not independent as can be 
seen in the following. Using the integral~(\ref{eq:betafunint}), and taking
$s\to\infty$ so that $a_\tau,a_s\to 0$, we can write~\cite{pms}
\beqn
	&&
	\ln\frac{s}{\Lambda^2}-\ln\frac{s}{\Lambda_\tau^2}
	=
	2\ln\frac{\Lambda_\tau}{\Lambda}
	= 
	\frac{1}{\beta_0a_s(s)} - \frac{1}{\beta_0a_\tau(s)} + \mathcal{O}(a_s)
	= 
	c_2 + \mathcal{O}(a_s)~, \\
	&&\hspace{1cm}
	\Longleftrightarrow\hspace{1cm}
	\Lambda_\tau \simeq \Lambda e^{c_2/\beta_0} \approx 3.2\Lambda~.
\eeqn
Since the asymptotic scale parameters are reciprocal to the scale,
transforming FOPT into ECPT is equivalent to reducing the renormalization 
scale, at lowest order. 

\begin{figure}[t]  
  \centerline{\epsfysize6.3cm\epsffile{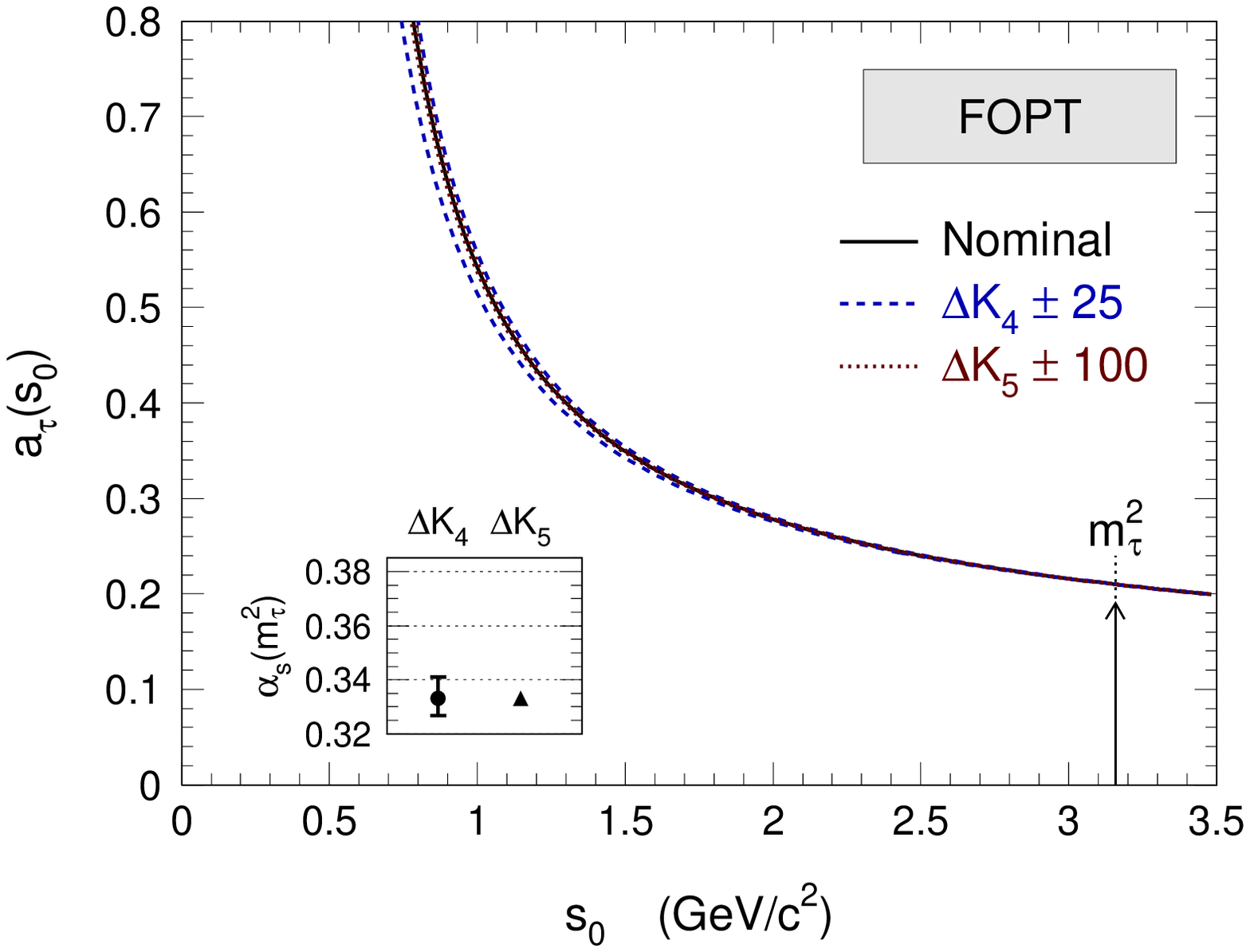}
              \epsfysize6.3cm\epsffile{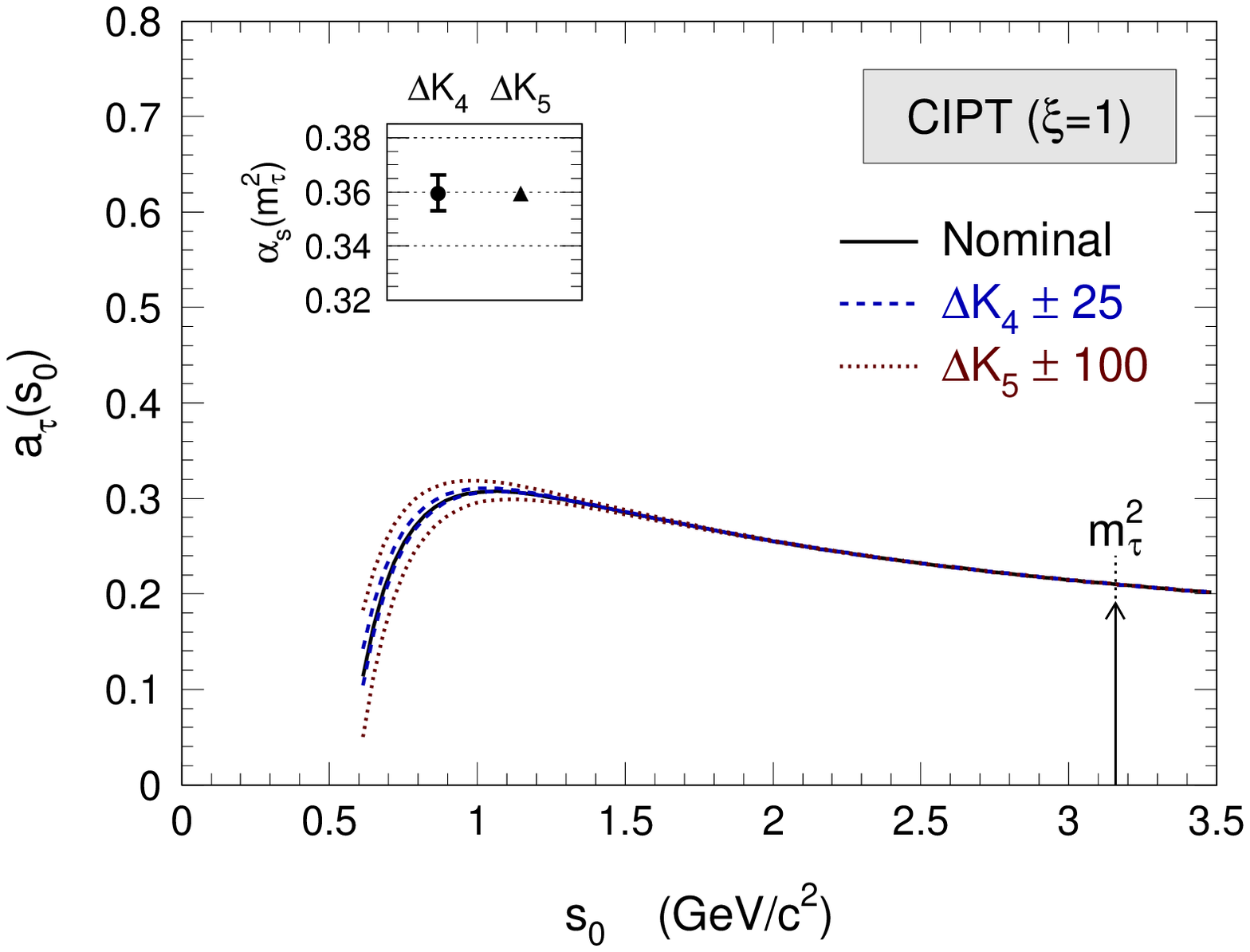}}
  \centerline{\epsfysize6.3cm\epsffile{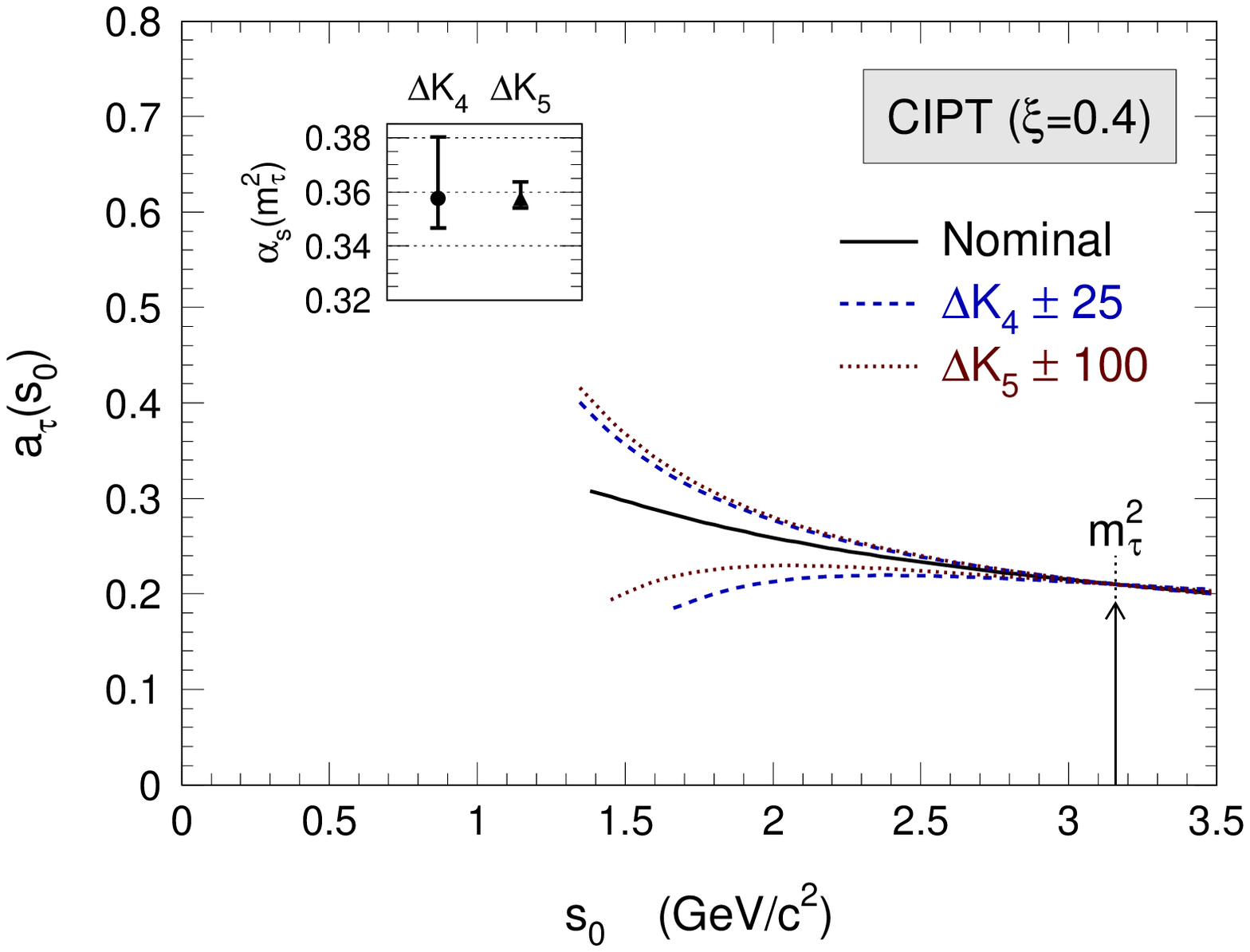}
              \epsfysize6.3cm\epsffile{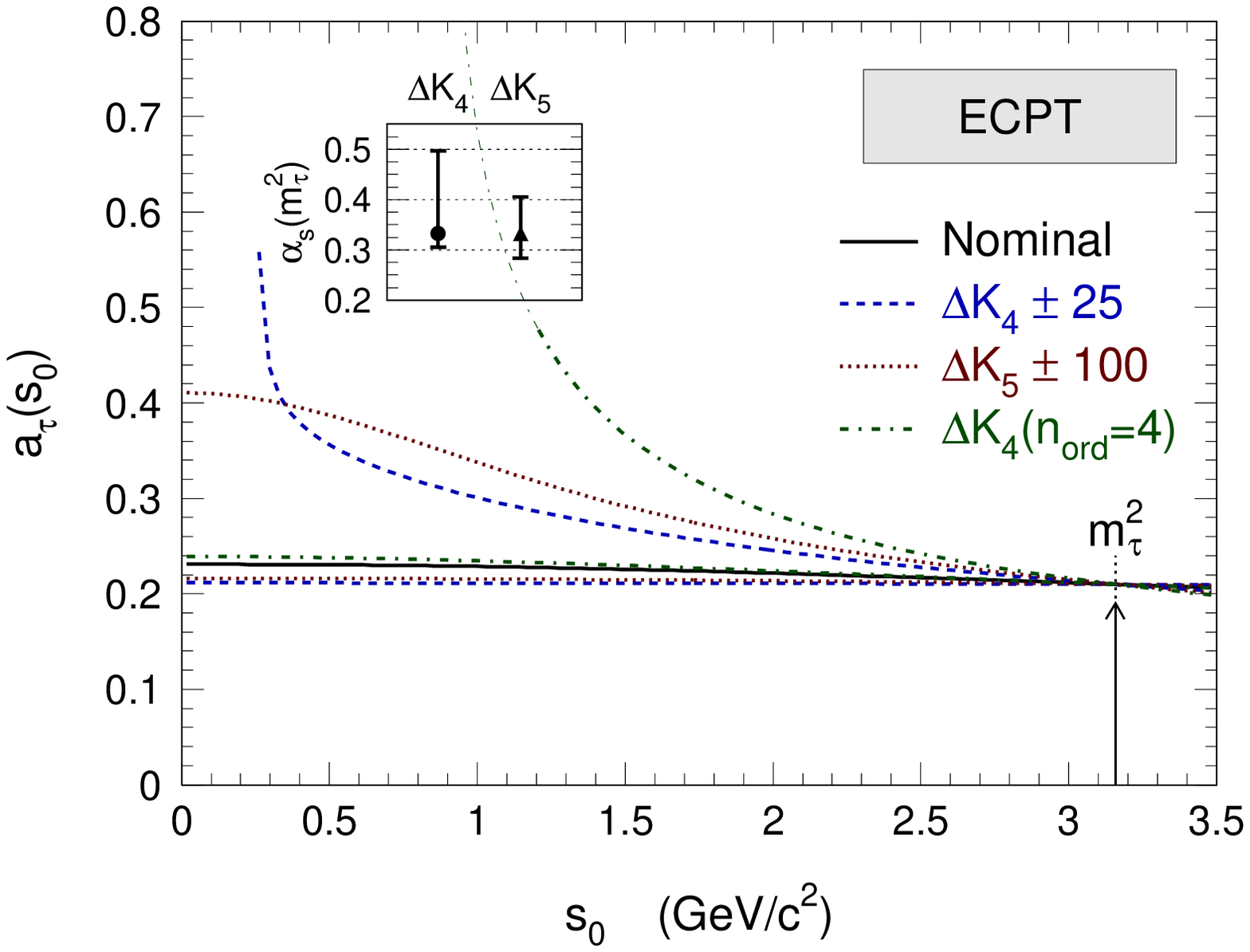}}
  \vspace{-0.1cm}
  \caption[.]{\label{fig:atau_evolution}
     	Evolution of $a_\tau(s_0)$ for the different perturbative methods, 
        and for variations of the unknown higher order perturbative
	coefficients $K_4$ (dashed) and $K_5$ (dotted). All methods 
	are normalized to $a_\tau(m_\tau^2)=0.22$. The inserted frames
	give the $\as(m_\tau^2)$ values derived from this $a_\tau$
	value in the respective methods. The error bars show
	the uncertainties, added in quadrature, from the $K_4$ and $K_5$ 
	variations quoted on the plots.}	
\end{figure} 
In terms of the measurement of $\as(m_\tau^2)$ the three perturbative
methods FOPT, \FOPTCI and ECPT use similar information. However their 
dependence on the unknown higher order coefficients $K_{n\ge4}$ can be 
dramatically different. Figure~\ref{fig:atau_evolution} shows the evolution 
of $a_\tau(s_0)$ for FOPT, \FOPTCI with renormalization scales $\xi=1$ 
and $\xi=0.4$, and ECPT, all normalized to $a_\tau(m_\tau^2)=0.22$.
The solid lines correspond to fifth order perturbation theory with
assumed $K_4=25$, $K_5=98$, and for variations of these (dashed and
dotted lines). One observes relatively small dependence for FOPT and 
\FOPTCI with $\xi=1$, while large and very large effects are found
with \FOPTCI ($\xi=0.4$) and ECPT. The latter effect provides the
sensitivity to estimate the unknown coefficients. To better analyze this
sensitivity, we have compared the $\Delta K_4$ variation in fifth-order
ECPT to the fourth-order effect (dashed-dotted line in the lower right
hand plot of Fig.~\ref{fig:atau_evolution}). Including the bold estimate
for the next order significantly reduces the dependence on the $K_4$
term. This cancellation becomes obvious when looking at the numerical
expansion~(\ref{eq:betataunum}): the $K_4$ coefficient enters 
$\beta_{\tau,3}$ and $\beta_{\tau,4}$ with different signs, and 
with a larger coefficient at fifth order to compensate for the $a_\tau$ 
suppression. A similar behavior is observed for $K_5$ when turning on the 
sixth-order term. We conclude that as for the $K_4$ term, the strong $K_5$ 
dependence observed in ECPT is (by part) an artifact of the truncation of 
the perturbative series, and it is therefore dangerous to use it for an 
estimate of the unknown coefficients.

The embedded frames in Fig.~\ref{fig:atau_evolution} give the 
$\as(m_\tau^2)$ values derived from the given $a_\tau$ value in 
the respective methods. The error bars indicate the uncertainty, 
added in quadrature, from the $K_4$ and $K_5$ variations quoted
on the figures. One observes
that the difference in $\as$ between FOPT and \FOPTCI is not covered
by the uncertainties on $K_4$ and $K_5$. This is because the main
contribution to the difference in these methods is due to the 
truncation of the {\em known} higher order terms of the Taylor
expansion after the contour integration, rather than due to 
the unknown coefficients. Due to the good convergence of these
series, the dependence on the $K_5$ coefficient is marginal.

It should be mentioned that a large computational effort is underway
with the goal to calculate the $K_4$ coefficient. The large
number of five-loop diagrams needed to calculate the two-point current
correlator at this order is somewhat discouraging, however the results 
on two gauge invariant subsets are already available. The subset of order 
${\cal O}(\alpha_s^4n_f^3)$ was evaluated long ago through the summation of 
renormalon chains~\cite{beneke1993}, while the much harder subset 
${\cal O}(\alpha_s^4n_f^2)$ was recently calculated~\cite{baikov_a4}.
The known parts amount to about half of the expected $K_4$ coefficient
assuming a geometric growth. 
Using the results of~\cite{k4_pms} for different $n_f$ obtained with the
effective charge method, it is possible to compare the $n_f^2$ estimated 
term at the $\alpha_s^4$ order of the Adler function expansion 
($3.64 n_f^2$), with the newly computed one ($1.875 n_f^2$). The difference
provides a conservative estimate for the uncertainty of the ECPT procedure
to predict the full $K_4$ term, as one could expect the sum of all
$\alpha_s^4n_f^i$ terms to be closer to the true value than 
the term-by-term comparison.
A successful test of the procedure has been performed for
$K_3$, where all the $\alpha_s^3n_f^i$ terms are exactly known. The two 
unknown remaining terms ${\cal O}(\alpha_s^4n_f^1)$ and 
${\cal O}(\alpha_s^4n_f^0)$, the direct calculation of which 
is prevented by the present computing power available to this project, 
have been estimated~\cite{k4_cal} to be
\beq
     K_4 = -0.01009\, n_f^3 + 1.875\, n_f^2 - 31.8\, n_f +105.7~,
\eeq
where the first two terms are known exactly. The calculation for
$n_f=3$ yields $K_4\simeq 27$ with an uncertainty estimated to $\pm16$
in~\cite{k4_cal}.

\subsubsection{Infrared behavior of the effective charge}
\label{sec:infraredbehavior}

In the effective charge scheme, the physical coupling should have a finite 
infrared behavior and hence must not suffer from the 
Landau-pole that lets \MSbar (but also momentum subtraction schemes) 
break down below a scale $\sqrt{s}\sim1\gev$. 
The universal two-loop function $\beta_\tau(a_\tau)$ 
does however suffer from a singularity at $\sqrt{s}\sim1.1\gev$ 
(for $a_\tau(m_\tau^2)=0.22$). Only adding the negative three-loop coefficient
regularizes $\beta_\tau(a_\tau)$ 
and leads to a freezing $a_\tau(s\to0)\simeq0.62$. Adding higher orders does
not qualitatively alter this behavior, however the freezing value is 
very unstable against variations of the unknown coefficients (which again 
exhibits the mediocre convergence of the ECPT series). The lower right 
hand plot of Fig.~\ref{fig:atau_evolution} shows the evolution of $a_\tau(s)$ 
for different orders and assumptions upon the unknown higher order coefficients
(\cf\  the discussion in~\cite{brodsky}).

\subsubsection{Renormalons and the large-$\beta_0$ expansion}
\label{sec:renormalons}

Perturbative series in quantum field theory\footnote
{
	The suppression of the constant term in the definition of 
	Eq.~(\ref{eq:ren_series}) is convenient to avoid $1/\alpha$
	terms in the definition of the Borel integral~(\ref{eq:borelint}).
	It is without consequence for the following discussion.
}
\beq
\label{eq:ren_series}
	R_N = \sum_{n=0}^N r_n\alpha^{n+1}~,
\eeq
be they abelian or not, are divergent at $N\to\infty$ for any 
value of $\alpha$.
Because the number of Feynman graphs increases at each order $n$, 
one expects a factorial growth of the perturbative coefficients of 
the corresponding $S$-matrix series
\beq
\label{eq:rnfactgrowth}
	r_n \sim K a^n n! n^\gamma~,
\eeq
with constant $K$, $a$, $\gamma$. The divergent behavior of perturbative
expansions often has physical significance: it indicates the 
nonperturbative structure of the vacuum and its excitations.
To be a useful approximation of the true value $R$, the 
series~(\ref{eq:ren_series}) must be {\em asymptotic to $R$} for 
given $\alpha$, \ie, it exists a number $K_N$ so that\footnote
{
	Our discussion of renormalons has greatly benefited from
	the comprehensive review given in~\cite{benekephysrp}.
}
$	|R - R_N| < K_{N+1}\alpha^{N+2}$.
If $r_n\sim K_n$, the best approximation of $R$ is achieved when
the series is truncated at its minimal (absolute) term and the 
truncation error can be roughly estimated by the size of the minimal 
term. The accuracy in the point of minimal sensitivity depends on
the strength of the perturbative expansion parameter (the coupling).
The larger it is, the lower the order when minimal sensitivity is 
reached, and the worse is the approximation. Note that as long as 
the computed order is below the order of minimal sensitivity, there
is no actual difference between a divergent or a convergent series.

A convenient way to deal with this divergence is to consider the Borel 
transform 
\beq
   B[R](t) \equiv
      \sum_{n=0}^\infty\frac{r_n}{n!}t^n~,
\eeq
which is believed to have a finite radius of convergence in the $t$-plane~\cite{mueller}.
The $n^{\rm th}$ fixed order perturbation coefficient is then generated by the 
$n^{\rm th}$ derivative 
\beq
\label{eq:generator}
   r_n = \frac{d^nB[R](t)}{dt^n}\bigg|_{t=0}~.
\eeq
The explicit factor $n!$ in the denominator of $B[R](t)$ makes the 
Borel-transformed series much better behaved. Summing up all orders 
leads to the integral representation\footnote
{
	The Borel integral is easily derived by using the integral
	representation $n!=\int_0^\infty dv\exp(-v)v^n$, and exchanging
	summation and integration in Eq.~(\ref{eq:ren_series})
	(see, \eg, \cite{maxwellwhoelse}). Note however that this interchange
	is formally not justified in presence of the Landau pole for the 
	coupling constant.
 }
\beq
\label{eq:borelint}
   R(\alpha) = \intl_0^\infty dt\,e^{-t/\alpha} B[R](t)~.
\eeq
If the integral exists, it gives the Borel sum of the original divergent
series. Existence requires that $B[R](t)$ 
has no singularities in the integration range. However, for a factorial
behavior akin to Eq.~(\ref{eq:rnfactgrowth}) (non-alternating 
series), the Borel transform is given by
$B[R](t) \propto (1-a t)^{-1-\gamma}$. Hence, if $\gamma$ is positive 
there exists a singularity $t=1/a$ so that the integral~(\ref{eq:borelint}) 
does not exist. Indeed, ``{\em the divergent behavior of the original series 
is encoded in the singularities of its Borel transform}''~\cite{benekephysrp}.
One notices that the larger $a$, \ie, the faster diverging the series is, 
the closer to zero moves the singularity. 

\begin{figure}[t]
  \epsfxsize8cm
  \centerline{\epsffile{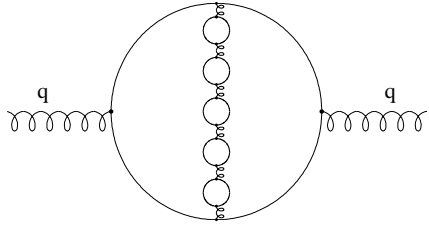}}
  \vspace{-0.1cm}
  \caption[.]{\label{fig:renormalons}
               Multi-fermion loop insertion (renormalons) into a
               fermion anti-fermion vacuum polarization diagram.}
\end{figure} 
Let us consider a renormalized fermion-loop insertion into a quark-antiquark
vacuum polarization diagram (\cf\  one bubble in Fig.~\ref{fig:renormalons}).
Assuming dominance of the $(\beta_0 a_s(-s))^n$ term (denoted as the 
{\em large-$\beta_0$ expansion}), one can resum this chain with a 
large number of $n$ bubbles. If one neglets higher order terms of the 
$\beta$-function, it is thus possible to derive estimates for the FOPT 
coefficients of a given perturbative series at all orders. This 
assumption is supported empirically by the observation that 
the $\beta_0$ term dominates second order radiative corrections for many 
observables in the \MSbar  scheme~\cite{beneke}. (We shall however 
anticipate here that the accuracy of these estimates is insufficient 
for the purpose of a precise perturbative prediction of $\Rtau$, and 
difficult to control.) The procedure provides a {\it naive non-abelianization} 
of the theory since lowest order radiative corrections do not include gluon 
self-coupling. 
%At sufficiently large orders of $n$ vacuum polarization 
%bubbles, the coefficients diverge as 
%\beq
%\label{eq:appr}
%	r_n \sim C_k n!\,n^{\gamma_k}(\beta_0/k)^n~,
%\eeq
%with constant $C_k$, $\gamma_k$, $k$.

\begin{figure}[t]
  \epsfxsize10.6cm
  \centerline{\epsffile{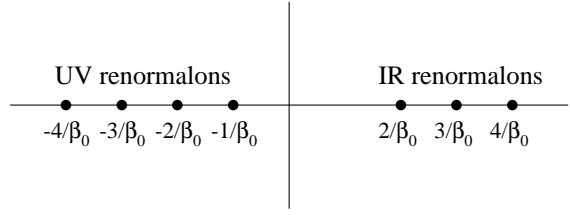}}
  \vspace{-0.3cm}
  \caption[.]{\label{fig:singularities}
              Ultraviolet (UV) and infrared (IR) renormalons
              in the Borel plane of the Adler function.}
\end{figure}

Following this line, the Adler function can be expanded as~\cite{benekebraun}
\beq
\label{eq:adlerlargebeta}
    D(s) = 1 + a_s\sum_{n=0}^N a_s^n\left(d_n\beta_0^{n} + \delta_n\right)~,
\eeq
where the coefficients $d_n$ are computed in terms of fermion bubble 
diagrams\footnote
{
	One identifies the coefficients $d_n$ in Eq.~(\ref{eq:adlerlargebeta})
	with their leading-$n_f$ pieces $d_n^{[n]}$ in the expression
	$d_n = d_n^{[n]}n_f^k + \dots + d_k^{[0]}$. It has been shown 
	in~\cite{broadhurst} that the $d_n^{[n]}$ in \MSbar
	are given by ($s=\mu^2$) 
	\beqns
		d_n^{[n]} = 2^{1-n} n! 
                \sum_{k=0}^n \left(-\frac{5}{9}\right)^{\!\!k}
                             \frac{\Psi_{n+2-k}^{[n+2-k]}}
                                  {k!(n-k)!}~,
	\eeqns
	with the generating function
	\beqns
		\Psi_m^{[m]}=
			\frac{3^{2-m}}{2}\left(\frac{d}{dx}\right)^{\!\!m-2}
  			\!\!\!\!\!P(x)\bigg|_{x=1}~,
	\hspace{0.5cm}\mbox{and}\hspace{0.5cm}
		P(x) = \frac{32}{3(1+x)}
                       \sum_{k=2}^{\infty}\frac{(-1)^kk}{(k^2-x^2)^2}~.
	\eeqns
} 
(see Fig.~\ref{fig:renormalons}). The
\MSbar Borel transform of Eq.~(\ref{eq:adlerlargebeta}) is given 
by~\cite{broadhurst,beneke1993}
\beq
\label{eq:boreladler}
      B[D](u) = \frac{32\pi}{3}e^{\frac{5}{3}u}\frac{u}{1-(1-u)^2}
                \sum_{k=2}^\infty\frac{(-1)^kk}{(k^2-(1-u)^2)^2}~,
\eeq
where $u=-\beta_0 t$, and where we have chosen the renormalization scale 
to be equal to the momentum scale. Neglecting the corrections
$\delta_n$, the series~(\ref{eq:adlerlargebeta}) leads to the
large-$\beta_0$ expansion of $D(s)$. The first elements of the 
series are
\beqn
\label{eq:dnlargebeta}
      d_0          = 1~,\hspace{1cm}
      d_1\beta_0   = 0.496~,\hspace{1cm}
      d_2\beta_0^2 = 15.71~,\hspace{1cm} \nonumber\\
      d_3\beta_0^3 = 24.83~,\hspace{1cm}
      d_4\beta_0^4 = 787.8~,\hspace{1cm}
      d_5\beta_0^5 = -1991 ~,\hspace{1cm}
\eeqn
which compare only coarsely with the exact terms~(\ref{eq:kf1})
where these are known.
The Borel transform $B[D](u)$ has singularities on the real axis, which  
can be distinguished in infrared (IR renormalons) for small virtuality,
and ultraviolet singularities (UV renormalons) for high virtuality of 
the exchanged gluon. One finds an infinite sequence of IR (UV) renormalons
at positive (negative) integer $u$, with the exception of $u=1$
(see Fig.~\ref{fig:singularities}).

The integral~(\ref{eq:borelint}) may still be solvable in presence of
renormalons by acquiring an imaginary part from moving the integration 
contour above or below the real axis. The associated ambiguity is 
exponentially small in the expansion parameter $\alpha$, and is 
therefore of nonperturbative origin. Correspondingly, the IR poles 
give rise to ambiguities when one uses the 
generators~(\ref{eq:generator}) to recompute $D(s)$ from 
its Borel transform. Using in first order 
\beq
\label{eq:alphaslog}
   a_s(s) = \frac{1}{\beta_0{\rm ln}(s/\Lambda^2)}~,
\eeq
one obtains for the associated ambiguity for $u=2,3,\dots$
\beqn
   \Delta D(s) 
   \sim
      e^{-u/(\beta_0a_s)} =
      \left(\frac{\Lambda^2}{s}\right)^{\!\!u}~.
\eeqn
These IR renormalons proportional to $(\Lambda^2/s)^u$ are reabsorbed 
into the nonperturbative terms of the OPE.
%\footnote
%{
%	Infrared renormalons are not connected to the existence
%	of the Landau pole. While the former always exist
%	(their location does not depend on the higher $\beta_n$
%	coefficients~\cite{beneke}), the Landau pole depends on the $\beta$
%	function. For instance, we have seen in Section~\ref{sec:infraredbehavior} 
%	that the ECPT scheme is IR finite.
%}. 
The absence of a $u=1$ IR 
renormalon is thereby related to the impossibility to build a gauge 
invariant operator of dimension $D=2$, which is an important property 
that is exploited for the determination of $\as$ from $\Rtau$.
The lowest IR renormalon is hence of the order $(\Lambda^2/s)^2$.
On the contrary to the UV renormalons, this term is scale independent
and cannot be decreased~\cite{keq1}. It should therefore have physical 
meaning and can be identified with the gluon condensate 
$\langle 0|G_{\mu\nu}G^{\mu\nu}|0\rangle$ as the leading infrared 
contribution to the Adler function. However, while IR renormalons
lead to nonperturbative operators for power corrections, it does
not necessarily contain all of them. For example, at $D=4$ one 
misses the quark condensate $\langle0|q \qbar|0\rangle$, which is
generated by chiral symmetry breaking that does not occur in 
perturbation theory.

Due to asymptotic freedom, governed by the negative sign in front
of the RGE $\beta$-function, the UV renormalons arise on the negative 
real axis\footnote
{
	In QCD, the presence of IR renormalons implies that 
	nonperturbative corrections should be added to define the theory 
	unambiguously. The same is true for UV renormalons in QED.
	This corresponds to the observation of large coupling 
	constants in the respective energy regimes.
} 
(on the contrary to QED where they occur on the positive 
side) so that they are outside the integration range of~(\ref{eq:borelint}) 
and insofar harmless, that is, Borel-summable~\cite{mueller}. Let us 
return for a moment to the series~(\ref{eq:ren_series}) in the 
large-$\beta_0$ expansion. For large $n$, the factorial growth of the 
perturbation series is dominated by the contribution of the leading UV 
renormalon $k=-1$ with alternating coefficients 
$r_n \sim n! n^\gamma(-\beta_0)^n$. A guess at which 
order $N_{\rm min}$ minimal sensitivity is achieved can be obtained from the 
argument that the series is convergent, \ie, reliable at order $n+1$ 
if $\alpha (r_{n+1}/r_n<1)$. Considering leading IR and UV renormalons only, 
one finds from the asymptotic behavior
$r_{n+1}/r_n\sim n\beta_0/2$ (IR) and $r_{n+1}/r_n\sim -n\beta_0$ (UV).
Hence the breakdown of convergence is first caused by the UV renormalon
and $N_{\rm min}\sim1/(\beta_0 a_s)$ is of order one. A series truncated 
at finite order 
brings an intrinsic limitation of accuracy along with it. The associated 
error is reasonably estimated with the magnitude of the order 
$N=N_{\rm min}$ term of the perturbative series. That gives using 
for the leading UV renormalon 
\beq
  |r_{N}|a_s^{N} 
   \sim
      N!N^\gamma(-\beta_0 a_s)^{N} \nonumber \\
%   &\simeq&
%      N^{N} e^{-N}
%     \sqrt{2\pi N}N^\gamma (-\beta_0a_s)^{N}~~~~~{\rm (Stirling~formula)}
%     \nonumber \\
%   &\sim& 
%     e^{-1/(\beta_0a_s)}(\beta_0a_s)^{-\gamma}
%     \sqrt{2\pi/(\beta_0\as)} \nonumber \\
%   &\sim&
	\sim
     \frac{\Lambda^2}{s} \times {\rm logarithms}~,
\eeq
where the r.h.s. is obtained with the use of the Stirling formula, 
and where we have fixed the renormalization scale at $\mu^2=s$.	
The truncation of the perturbative series is accompanied
by an uncertainty which scales like $1/s$, \ie, with apparent 
dimension $D=2$. This contribution is however not to be confounded
with nonperturbative IR renormalon ambiguities\footnote
{
	It has been shown in~\cite{keq1} that the truncation 
	uncertainty at $N_{\rm min}$ scales actually as $\Lambda^2s/\mu^4$ 
	once the renormalization scheme dependence of $\Lambda$ is taken into
	account.
}. 
Similarly to Eq.~(\ref{eq:adlerlargebeta}) the $\Rtau$ FOPT 
series~(\ref{eq:kngn}) can be written as
\beq
\label{eq:rtaulargebeta}
    \delta^{(0)}(s) = 1 + a_s\sum_{n=0}^Na_s^n\left(d_n^\tau\beta_0^{n} 
                                                 + \delta_n^\tau\right)~,
\eeq
where $d_n^\tau\beta_0^{n}+\delta_n^\tau=K_{n+1}+g_{n+1}(1)$.
Inserting Eq.~(\ref{eq:boreladler}) into (\ref{eq:rtauadler}), and taking 
advantage of the factorized $s$ dependence in the large-$\beta_0$ expansion, 
the Borel transform of~(\ref{eq:rtaulargebeta}) is~\cite{beneke1993}
\beq
      B[\Rtau](u)=B[D](u)\frac{\sin(\pi u)}{\pi}
                 \left(\frac{1}{u} + \frac{2}{1-u} + \frac{3}{3-u} 
                       \frac{1}{4-u}\right)~.
\eeq
The first elements of the series are
\beqn
\label{eq:dntau}
      d_0^\tau          = 1~,\hspace{1cm}
      d_1^\tau\beta_0   = 5.119~,\hspace{1cm}
      d_2^\tau\beta_0^2 = 28.78~,\hspace{1cm} \nonumber\\
      d_3^\tau\beta_0^3 = 156.6~,\hspace{1cm}
      d_4^\tau\beta_0^4 = 900.9~,\hspace{1cm}
      d_5^\tau\beta_0^5 = 4867 ~.\hspace{1cm}
\eeqn
They compare reasonably well with the exact terms~(\ref{eq:delta0exp})
where these are known.
In this approach, the unknown fixed-order fourth-order coefficient
is estimated to be $K_4\sim79$, which is significantly larger
than the estimate based on ECPT, and also larger than the one 
obtained from the large-$\beta_0$ expansion of the Adler 
function~(\ref{eq:dnlargebeta}).

Apart from the conceptual interest in the study of renormalons, the
main question regarding the concrete use of the large-$\beta_0$ expansion
certainly is: how large are the uncertainties? From Eq.~(\ref{eq:dntau})
we see that the series is positive and that minimal sensitivity is 
reached at $n\sim5$. On the other hand, the resummation of the known $g_n(\xi)$
coefficients~(\ref{eq:delta0exp}) shows that large negative coefficients
occur at higher order FOPT so that one would need a far larger than
a geometric growth of the $K_n$ coefficients to keep the series 
positive. A direct comparison between the large-$\beta_0$ expansion 
coefficients for the Adler function~(\ref{eq:dnlargebeta}) with the 
exact calculation~(\ref{eq:kf1}) reveals
an oscillating behavior of the large-$\beta_0$ series, \ie, small $d_n$ 
coefficients for $n$-even and large $d_n$ for $n$-odd. For example,
the large-$\beta_0$ $d_3$ coefficient is substantially overestimated.
One also has to worry about important contributions from the $d_5$ 
term, which is found to be $788$. As a consequence, our conclusion 
is that the uncertainties associated with the large-$\beta_0$ expansion 
are not under sufficient control to improve the perturbative prediction 
of $\Rtau$. It is also an unreliable estimator of the uncertainty of
the perturbative series.

\subsubsection{Comparison of the perturbative methods}
\label{sec:pert_comp}

Some comparisons between the fixed-order PT methods have already been 
given for the discussion of the estimate of the unknown higher order 
coefficients in Section~\ref{sec:estimhigherorder}. The slow convergence 
of ECPT (and CIPT at small renormalization scales, $\xi\ll1$) has been
pointed out in this respect. To further study the convergence of the 
perturbative series, we give in Table~\ref{tab:intsol} the contributions 
of the different orders in PT to $\delta^{(0)}$ for the various approaches
and using $\as(m_\tau^2)=0.35$. We assumed here $K_4=25$ and a geometric
growth of all other unknown PT and RGE coefficients. In the case of CIPT 
the results are given for the various techniques used to evolve 
$\as(s_0e^{i\varphi})$: the truncated Taylor expansion~(\ref{eq:astaylor}),
Runge-Kutta integration of the RGE with $\xi=1$, and Runge-Kutta integration
with $\xi=0.4$. 

\begin{table}[t]
  \caption[.]{\label{tab:intsol}
              Massless perturbative contribution to $\Rtau(m_\tau^2)$ 
	      for the various methods considered, and at orders 
              $n\ge1$ with $\asm=0.35$. The value of $K_4$ is 
              set to 25, while all unknown higher order $K_{n>4}$ and 
	      $\beta_{n>3}$ coefficients are assumed to follow a geometric 
	      growth.}
\begin{center}
\setlength{\tabcolsep}{0.0pc}
\begin{tabular*}{\textwidth}{@{\extracolsep{\fill}}lrrrrrrrr} 
\hline\noalign{\smallskip}
  	& \mc{7}{c}{$\delta^0$} \\
Pert. Method	
        & $n=1$	 & $n=2$  & $n=3$   & $(n=4)$ & $(n=5)$ & $(n=6)$   & $\sum_{n=1}^4$ & $\sum_{n=1}^6$ \\
\noalign{\smallskip}\hline\noalign{\smallskip}
FOPT ($\xi=1$)
	& 0.1114 & 0.0646 & 0.0365  & 0.0159  & 0.0010  & $-0.0086$ & 0.2283      & 0.2208 \\
\FOPTCI (Taylor RGE, $\xi=1$)         	
	& 0.1573 & 0.0317 & 0.0126  & 0.0042  & 0.0011  & 0.0001    & 0.2058      & 0.2070 \\
\FOPTCI (full RGE, $\xi=1$)           	
	& 0.1524 & 0.0311 & 0.0129  & 0.0046  & 0.0013  & 0.0002    & 0.2009      & 0.2025 \\
\FOPTCI (full RGE, $\xi=0.4$)           	
	& 0.2166 &$-0.0133$&0.0006  &$-0.0007$& 0.0010  &$-0.0007$  & 0.2032      & 0.2048 \\
ECPT
	& 0.1442 & 0.2187 &$-0.1195$&$-0.0344$&$-0.0160$&$-0.0120$  & 0.2090      & 0.1810 \\
Large-$\beta_0$ expansion
	& 0.1114 & 0.0635 & 0.0398  & 0.0241  & 0.0155  & 0.0093    & 0.2388      & 0.2636 \\
\noalign{\smallskip}\hline
\end{tabular*}
  \end{center}
\end{table}
As advocated in~\cite{pert}, faster convergence is observed for CIPT
compared to FOPT yielding a significantly smaller error associated with 
the renormalization scale ambiguity (while, somewhat counter intuitive, 
the uncertainty due to the unknown $K_4$ and higher order coefficients 
is similar in both approaches). Our coarse extrapolation
of the higher order coefficients could indicate that minimal sensitivity 
is reached at $n\sim5$ for FOPT, while the series further converges
for CIPT. Although the Taylor expansion in the CIPT integral exhibits 
significant deviations from the exact solution on the integration circle 
(\cf\  Fig.~\ref{fig:taylor_test}), the actual numerical effect from this 
on $\delta^{(0)}$ is small (\cf\  second and third column in 
Table~\ref{tab:intsol}). As discussed in Section~\ref{sec:ecpt}, the 
convergence of the ECPT series is much worse than for FOPT and CIPT.
Consequently, the difference between truncation at $n=4$ and $n=6$ is 
expected to be significant. A similar behavior is observed for the 
large-$\beta_0$ expansion.

The CIPT series is found to be better behaved than FOPT (as well as ECPT)
and is therefore to be preferred for the numerical analysis of the $\tau$
hadronic width. As a matter
of fact, the difference in the result observed when using a Taylor 
expansion and truncating the perturbative series after integrating
along the contour (FOPT) with the exact result at given order (CIPT) 
is exhibiting the incompleteness of the perturbative series. However, 
it is even worse than that since large known coefficients are neglected
in FOPT so that the difference between CIPT and FOPT is actually overstating
the perturbative truncation uncertainty. This can be verified by studying 
the behavior of this difference for the various orders in perturbation 
theory given in Table~\ref{tab:intsol}. The CIPT-vs.-FOPT discrepancy 
increases with the addition of each order, up to order four where a 
maximum is reached. Adding the fifth order does not reduce the effect,
and only beyond fifth order the two evaluations become asymptotic
to each other.
As a consequence varying the unknown higher order coefficients {\em and} using 
the difference between FOPT and CIPT as indicator of the theoretical 
uncertainties overemphasizes the truncation effect. 

\subsubsection{Quark-mass and nonperturbative contributions}\label{sec:nonpert}

Following SVZ~\cite{svz}, the first contribution to \Rtau\  beyond the 
$D=0$ perturbative expansion is the non-dynamical quark mass correction 
of dimension\footnote
{
	Corrections of dimension $D=2$ refer to the mass dependence
	of the perturbative OPE coefficients of the unit operator.
} 
$D=2$, \ie, corrections in powers of $1/s_0$. 
The leading $D=2$ corrections induced by the light-quark masses are
computed using the running quark masses evaluated at the two-loop level.
Evaluation of the contour integral in FOPT leads to~\cite{bnp}
\beqn
   \delta^{(2-\rm mass)}_{ud,V/A} 
  &=&
      -\,8\left(1+\frac{16}{3}a_s(s_0)\right)
      \frac{m_u^2(s_0) + m_d^2(s_0)}{s_0} \nonumber \\
  & & \pm\,
       4\left(1+\frac{25}{3}a_s(s_0)\right)
      \frac{m_u(s_0) m_d(s_0)}{s_0}~,
\eeqn
where $m_i(s_0)$ are the running quark masses evaluated at the scale 
$s_0$ using the RGE $\gamma$-function~(\ref{eq:gammafun}).
The following values are used by ALEPH for the renormalization group 
invariant quark mass parameters $\hat{m}_i$ defined in~(\ref{eq:gammafunint})
\beq
\label{eq:qmasses}	
	\hat{m}_u = (8.7\pm1.5)\mev~,~~ \hat{m}_d = (15.4\pm1.5)\mev~,
	~~\hat{m}_s = (270\pm30)\mev~.
\eeq

The dimension $D=4$ operators have dynamical contributions from the gluon 
condensate $\GG$ and light $u,d$ quark condensates 
$\langle m_iq_iq_j\rangle$, which are the matrix elements of the gluon field 
strength-squared and the scalar quark densities, respectively. Remaining $D=4$ 
operators are running quark masses to the fourth power. Inserting the Wilson 
coefficients of these operators~\cite{becchi,generalis,broadhurst1981,gluonterm}
in the contour integral one obtains~\cite{bnp}
\beqn
\label{eq:nondel4}
   \delta_{ud,V/A}^{(4)} 
   &=&
       \frac{11}{4}\pi^2 a_s^2(s_0)
       \frac{\GG}{s_0^2} \nonumber\\
   & & +\, 16\pi^2
           \left[1
             +\frac{9}{2} a_s^2(s_0)
           \right]\frac{\langle(m_u \mp m_d)
                        \left(\ubar u \mp \dbar d\right)\rangle}
                       {s_0^2} \nonumber \\
   & & -\, 18\pi^2 a_s^2(s_0)
           \left[\frac{\langle m_u\ubar u + m_d\dbar d\rangle}
                      {s_0^2}
                 + \frac{4}{9}\sum_{k=u,d,s}
                 \frac{\langle m_k\qbar_kq_k\rangle}
                      {s_0^2}
           \right] \nonumber \\
   & & -\, \left[\frac{48}{7}\frac{1}{a_s(s_0)}
                 - \frac{22}{7}
           \right]\frac{\left(m_u(s_0) \mp m_d(s_0)\right)
                        \left(m_u^3(s_0) \mp m_d^3(s_0)\right)}
                       {s_0^2} \nonumber \\
   & & \pm\, 6\frac{m_u(s_0)m_d(s_0)\left(m_u(s_0) \mp m_d(s_0)\right)^2}
                  {s_0^2}
             +36\frac{m_u^2(s_0)m_d^2(s_0)}{s_0^2}~.
\eeqn
Where two signs are given the upper (lower) one is for $V$ ($A$). 
The gluon condensate vanishes in first order \assz. However, there appear 
second order terms in the Wilson coefficients due to the $s$ dependence of 
$a_s$, which after integration becomes $a_s^2$.

The contributions from dimension $D=6$ operators are complex. 
The most important operators arise from four-quark dynamical effects of 
the form $\qbar_i\Gamma_1q_j\qbar_k\Gamma_2q_l$. Other operators, such 
as the triple gluon condensate whose Wilson coefficient vanishes at order 
\as, or those which are suppressed by powers of quark masses, are neglected 
in the evaluation of the contour integrals performed in~\cite{bnp}.
The large number of independent operators of the four-quark type occurring 
in the $D=6$ term can be reduced by means of the {\em factorization} 
(or {\em vacuum saturation}) assumption~\cite{svz} to leading order \as. 
The operators are then expressed as products of scale dependent two-quark 
condensates $\as(\mu)\langle\qbar_iq_i(\mu)\rangle\langle\qbar_jq_j(\mu)\rangle$.
To take into account possible deviations from the vacuum saturation assumption, 
one can introduce an effective scale independent operator 
$\rho \as\langle\qbar q\rangle^2$ that replaces the above product. The 
effective $D=6$ term obtained in this way is~\cite{bnp}
\beqn
\label{eq:nondel6}
   \delta_{ud,V/A}^{(6)} 
   &\simeq& \bigg(\!\begin{array}{c}7\\[-0.2cm]-11\end{array}\!\bigg)
   \frac{256\pi^4}{27}\frac{\rho\as\langle\qbar q\rangle^2}
                           {s_0^3}~,
\eeqn
predicting a different sign and hence a partial cancellation between 
vector and axial-vector contributions\footnote
{
	It has been pointed out in~\cite{pertmass} that 
	Eq.~(\ref{eq:nondel6}) assumes that the same $\rho$
	parameter can be used for both vector and axial-vector 
	contributions. 
}

The dimension $D=8$ contribution has a structure of nontrivial quark-quark, 
quark-gluon and four-gluon condensates whose explicit form is given 
in~\cite{dimeight}. For the theoretical prediction of \Rtau\  it is custom
to absorb the complete long and short distance part into the scale 
invariant phenomenological $D=8$ operator $\langle{\cal O}_8\rangle$,
which is fit simultaneously with \as\  and the other unknown 
nonperturbative operators to date.

Higher order contributions from $D\ge10$ operators are expected to be small 
as, equivalent to the gluon condensate, constant terms and terms in leading 
order \as vanish in Eq.~(\ref{eq:contour}) after integrating over the contour.

\subsection{Results} \label{sec:moment}
It was shown in~\cite{pichledib} that one can exploit the shape of the 
\sfs\  to obtain additional constraints on \assz and---more 
importantly---on the nonperturbative effective operators. The 
{\em $\tau$ spectral moments} at $s_0=m_\tau^2$ are defined by
\beq
\label{eq:moments}
   R_{\tau,V/A}^{k\l} =
       \intl_0^{m_\tau^2} ds\,\left(1-\frac{s}{m_\tau^2}\right)^{\!\!k}\!
                              \left(\frac{s}{m_\tau^2}\right)^{\!\!\l}
       \frac{dR_{\tau,V/A}}{ds}~,
\eeq
where $R_{\tau,V/A}^{00}=R_{\tau,V/A}$. The factor $(1-s/m_\tau^2)^k$ 
suppresses the integrand at the crossing of the positive real axis where the 
validity of the OPE less certain~\cite{braaten88} and the experimental accuracy 
is statistically limited. Its counterpart $(s/m_\tau^2)^\l$ projects upon
higher energies. The spectral information is used to fit simultaneously 
\asm  and the effective operators $\GG$, 
$\rho\as\langle\qbar q\rangle^2$ and $\langle{\cal O}_D\rangle$ for dimension
$D=4$, $6$ and $8$, respectively. Due to the intrinsic experimental correlations 
only five moments are used as input to the fit.

In analogy to \Rtau, the contributions to the moments originating from 
perturbative and nonperturbative QCD are decomposed through the OPE
\beq
\label{eq:rtaumom}
   R_{\tau,V/A}^{k\l} = 
       \frac{3}{2}|V_{ud}|^2\Sew
       \left(1 + \delta^{(0,k\l)} + 
             \delta^\prime_{\rm EW} + 
             \delta^{(2-{\rm mass},k\l)}_{ud,V/A} + 
             \hm\hm\sum_{D=4,6,\dots}\hm\hm\delta_{ud,V/A}^{(D,k\l)}
       \right)~.
\eeq
The prediction of the perturbative contribution takes the form
\beq 
   \delta^{(0,k\l)} = 
       \sum_{n=1}^3 \tilde{K}_n(\xi) A^{(n,k\l)}(a_s)~,
\eeq
with the functions
\beqn
\label{eq:anmom}
   A^{(n,k\l)}(a_s) 
   &=&
      \frac{1}{2\pi i}\hm\ointl_{|s|=m_\tau^2}\hm\hm
      \frac{ds}{s}
       \Bigg[2\Gamma(3 + k)
             \left(\frac{\Gamma(1 + \l)}{\Gamma(4 + k + \l)} +
                   2\frac{\Gamma(2 + \l)}{\Gamma(5 + k + \l)} 
             \right) \nonumber\\[0.2cm]
   &&\hspace{1.4cm}
             -\; I\left(\frac{s}{s_0},1+\l,3+k\right) 
             - 2 I\left(\frac{s}{s_0},2+\l,3+k\right)
            \Bigg]
	     a_s^n(-\xi s)~,
\eeqn
which make use of the elementary integrals 
$I(\gamma,a,b)=\int_0^\gamma t^{a-1}(1-t)^{b-1}dt$.
The contour integrals are numerically solved for the running 
$a_s(-\xi s)$ (CIPT).

In the chiral limit\footnote
{
	In the chiral limit, vector and axial-vector currents are 
	conserved so that $s\Pi_V^{(0)}=s\Pi_A^{(0)}=0$.		
} 
and neglecting the small logarithmic $s$ dependence of the 
Wilson coefficients, the dimension $D$ nonperturbative contributions 
in Expression~(\ref{eq:rtaumom}) read
\beq
\label{eq:array}
   \delta_{ud,V/A}^{(D,k\l)} =
      8 \pi^2 \left({\scriptsize\begin{array}{cccccc} 
               (D=2) & (D=4) & (D=6) & (D=8) & (D=10)& (k,\l) \\
                  1  &  0    & -3    &  -2   &   0   & (0,0) \\
                  1  &  1    & -3    &  -5   &  -2   & (1,0) \\
                  0  & -1    & -1    &   3   &   5   & (1,1) \\
                  0  &  0    &  1    &   1   &  -3   & (1,2) \\
                  0  &  0    &  0    &  -1   &  -1   & (1,3)
              \end{array}}\right)
      \sum_{{\rm dim}{\cal O}=D}\hm\hm C^{(1+0)}(\mu)
            \frac{\langle{\cal O}_D(\mu)\rangle_{V/A}}
                 {m_\tau^D}~,
\eeq
where the matrix is defined by the choice of the coefficients for the 
moments $k=1$, $\l=0,1,2,3$ and the corresponding dimension $D$. For 
completeness, we also give the coefficients for dimension $D=10$, which 
is not considered by the experiments since they do not contribute to 
$R_{\tau,V/A}^{00}$ at leading order. With increasing weight $\l$ the 
contributions from low dimensional operators vanish. For example, the 
only nonperturbative contribution to the moment $R_{\tau,V/A}^{13}$ stems 
from dimension $D=8$ and beyond. Hence, in a fit using spectral 
moments, $D=8$ will be determined by $R_{\tau,V/A}^{13}$. This 
observation is in some sense a ``paradox'', as higher moments project 
upon higher masses on the contrary to \Rtau\ and the spirit of the OPE, 
where the higher dimension terms are enhanced at small $s_0$. 

For practical purpose it is more convenient to define moments 
that are normalized to the corresponding \RtauVA\ in order 
to decouple the normalization from the shape of the $\tau$ \sfs
\beq
\label{eq:dkl}
   D_{\tau,V/A}^{k\l} =
     \frac{R_{\tau,V/A}^{k\l}}{R_{\tau,V/A}}~.
\eeq
The two sets of experimentally almost uncorrelated 
observables---\RtauVA\  on one hand and the spectral moments on the other 
hand---yield independent constraints on \asm and thus provide an important 
test of consistency. The correlation between these observables is completely
negligible in the $V+A$ case where \RtauVpA\ is calculated from the 
difference $R_\tau-R_{\tau,S}$, which has no correlation with the hadronic 
invariant mass spectrum. One experimentally obtains the $D_{\tau,V/A}^{k\l}$ 
by integrating the weighed normalized invariant mass-squared spectrum. 
The corresponding theoretical predictions are easily modified. 

%
% ----------------- Fit pf alpha_s ==== results ------------------------
%

\subsubsection{The ALEPH determination of $\as(m_\tau^2)$ and 
               nonperturbative contributions}
\label{sec:qcd_as_fit}

Combined fits to experimental spectral moments and the extraction of
$\as(m_\tau^2)$ together with the leading nonperturbative operators have 
been performed by ALEPH, CLEO and 
OPAL~\cite{aleph_as,aleph_asf,aleph_taubr,cleo_as,opal_vasf} using 
similar strategies and inputs.
Let us follow in this report the most recent analysis performed by the
ALEPH collaboration~\cite{aleph_taubr}.

Since the determination of $\as$ should be model-independent, the experiments
proceed by fitting simultaneously the nonperturbative operators, which 
is possible since the correlations between these and $\as$ turn out to be 
small enough. The theoretical framework and its intrinsic uncertainties were 
the subject of the previous section. Minor additional contributions originate 
from the CKM matrix element $|V_{ud}|$, the electroweak radiative correction 
factor $\Sew$, the light quark masses $m_{u,d}$ and the quark condensates 
(Section~\ref{sec:nonpert}). The largest contributions
have their origin in the truncation of the perturbative expansion.
Although it introduces some double-counting for the systematic error, the 
procedure used in~\cite{aleph_taubr} considers separate variations of 
the unknown higher order coefficient $K_4$ and the renormalization scale. 
The renormalization scale is varied around $m_\tau$ from 1.1 to $2.5\gev$
with the variation over half of the range taken as systematic uncertainty.
Taking advantage of the new theoretical developments discussed above, 
the value $K_4 = 25 \pm 25$ is used in \cite{aleph_taubr}.

\begin{table}[t]
  \caption[.]{\label{tab_asresults}
	Fit results~\cite{aleph_taubr} for \asm and the nonperturbative 
	contributions for vector, axial-vector and $V+A$ combined
	fits using the corresponding experimental spectral moments 
	as input parameters. Where two errors are given the first is
	experimental and the second theoretical. The $\delta^{(2)}$ term 
	is theoretical only with quark masses varying within their 
	allowed ranges (see text). The quark condensates in the $\delta^{(4)}$ 
	term are obtained from PCAC, while the gluon condensate is 
	determined by the fit. The total nonperturbative contribution is 
	the sum $\delta_{\rm NP}=\delta^{(4)}+\delta^{(6)}+\delta^{(8)}$.
	Full results are listed only for the \FOPTCI\
	perturbative prescriptions, except for \asm\ where both \FOPTCI\
        and FOPT results are given.}
  \begin{center}
{\small
\setlength{\tabcolsep}{0.0pc}
\begin{tabular*}{\textwidth}{@{\extracolsep{\fill}}lccc} 
\hline\noalign{\smallskip}
  Parameter     &Vector ($V$) &Axial-Vector ($A$)&  $V\,+\,A$\\
\noalign{\smallskip}\hline\noalign{\smallskip}
 \asm  (\FOPTCI) &  $0.355\pm0.008\pm0.009$  
                 &  $0.333\pm0.009\pm0.009$   
                 &  $0.350\pm0.005\pm0.009$   
\\
 \asm  (FOPT)    &  $0.331\pm0.006\pm0.012$  
                 &  $0.327\pm0.007\pm0.012$   
                 &  $0.331\pm0.004\pm0.012$ \\
\noalign{\smallskip}\hline\noalign{\smallskip}
  $\delta^{(2)}$ (\FOPTCI)    & $(-3.3\pm3.0) \times 10^{-4}$
                              & $(-5.1\pm3.0) \times 10^{-4}$
                              & $(-4.4\pm2.0) \times 10^{-4}$
 \\
%  $\delta^{(2)}$ (FOPT)       & $(-3.0\pm3.0) \times 10^{-4}$
%                              & $(-5.0\pm3.0) \times 10^{-4}$
%                              & $(-4.0\pm2.0) \times 10^{-4}$ \\
\noalign{\smallskip}\hline\noalign{\smallskip}
 $\GG$ ($\gev^4$) (\FOPTCI)   &  $(0.4\pm0.3) \times 10^{-2}$  
                                &  $(-1.3\pm0.4) \times 10^{-2}$   
                                &  $(-0.5\pm0.3) \times 10^{-2}$   
\\ 
% $\GG$ ($\gev^4$) (FOPT)    &  $(1.5\pm0.3) \times 10^{-2}$  
%                              &  $(-0.2\pm0.4) \times 10^{-2}$   
%                              &  $(\ph{-}0.6\pm0.2) \times 10^{-2}$   \\
\noalign{\smallskip}\hline\noalign{\smallskip}
  $\delta^{(4)}$ (\FOPTCI)    & $(4.1\pm1.2) \times 10^{-4}$
                              & $(-5.7\pm0.1) \times 10^{-3}$
                              & $(-2.7\pm0.1) \times 10^{-3}$
 \\      
%  $\delta^{(4)}$ (FOPT)       & $(6.8\pm1.0) \times 10^{-4}$
%                              & $(-5.3\pm0.1) \times 10^{-3}$
%                              & $(-2.4\pm0.1) \times 10^{-3}$ \\
\noalign{\smallskip}\hline\noalign{\smallskip}
 $\delta^{(6)}$ (\FOPTCI) &  $(2.85\pm0.22) \times 10^{-2}$  
                          &  $(-3.23\pm0.26) \times 10^{-2}$   
                          &  $(-2.1\pm2.2) \times 10^{-3}$   
\\
% $\delta^{(6)}$ (FOPT)    &  $(2.70\pm0.25) \times 10^{-2}$  
%                          &  $(-2.96\pm0.31) \times 10^{-2}$   
%                          &  $(-1.6\pm2.5) \times 10^{-3}$     \\
\noalign{\smallskip}\hline\noalign{\smallskip}
 $\delta^{(8)}$ (\FOPTCI) &  $(-9.0\pm0.5) \times 10^{-3}$  
                          &  $(8.9\pm0.6) \times 10^{-3}$   
                          &  $(-0.3\pm4.8) \times 10^{-4}$   
\\
% $\delta^{(8)}$ (FOPT)    &  $(-8.6\pm0.6) \times 10^{-3}$  
%                          &  $(8.6\pm0.6) \times 10^{-3}$   
%                          &  $(\ph{-}1.2\pm5.2) \times 10^{-4}$     \\
\noalign{\smallskip}\hline\noalign{\smallskip}
  Total $\delta_{\rm NP}$ (\FOPTCI)    & $(1.99\pm0.27) \times 10^{-2}$  
                                       & $(-2.91\pm0.20) \times 10^{-2}$
                                       & $(-4.8\pm1.7) \times 10^{-3}$
 \\
%  Total $\delta_{\rm NP}$ (FOPT)       & $(1.91\pm0.31) \times 10^{-2}$  
%                                       & $(-2.63\pm0.25) \times 10^{-2}$
%                                       & $(-3.9\pm2.0) \times 10^{-3}$ \\
\noalign{\smallskip}\hline\noalign{\smallskip}
 $\chi^2/$DF (\FOPTCI)       & 0.52            & 4.97            & 3.66    
 \\ 
% $\chi^2/$DF (FOPT)          & 0.01            & 0.63            & 0.11  \\
\noalign{\smallskip}\hline
  \end{tabular*}
}
  \end{center}
\end{table}

%\begin{table}[t]
%  \caption[.]{\label{tab_rescorr}
%        Correlation matrices for the fits given in Table~\ref{tab_asresults} 
%	for vector (left table), axial-vector (middle) and $V+A$ (right table) 
%	using \FOPTCI~\cite{aleph_taubr}. }
%{\footnotesize
%\begin{center}
%\setlength{\tabcolsep}{0.6pc}
%\begin{tabular*}{\textwidth}{@{\extracolsep{\fill}}l|ccc|ccc|ccc} 
%\hline\noalign{\smallskip}
%Parameter &  
%$\GG_V$ &  $\delta^{(6)}_V$ & $\delta^{(8)}_V$ &
%$\GG_A$ &  $\delta^{(6)}_A$ & $\delta^{(8)}_A$ &
%$\GG_{V+A}$
%                     &  $\delta^{(6)}_{V+A}$ & $\delta^{(8)}_{V+A}$ \\[0.15cm]
%\hline
%&&&&&&&&& \\[-0.3cm]
%\asm          & $-0.39$  & $-0.28$  & $-0.34$
%              & $-0.57$  & $\ph{-}0.52$  & $-0.55$   
%              & $-0.37$  & $\ph{-}0.38$  & $-0.45$ \\
%$\GG$
%              &  $   1$  & $\ph{-}0.44$  &  $\ph{-}0.46$
%              &  $   1$  & $-0.81$  &  $\ph{-}0.80$   
%              &  $   1$  & $-0.65$  &  $\ph{-}0.65$ \\
%
%$\delta^{(6)}$ 
%              &     --   &  $   1$  & $-0.98$
%              &     --   &  $   1$  & $-0.99$   
%              &     --   &  $   1$  & $-0.98$ \\
%$\delta^{(8)}$ 
%              &     --   &     --   &  $   1$
%              &     --   &     --   &  $   1$
%              &     --   &     --   &  $   1$ \\
%\noalign{\smallskip}\hline
%  \end{tabular*}
%\end{center}
%}
%\end{table}
The fit minimizes the $\chi^2$ of the differences between measured 
and adjusted quantities contracted with the inverse of the sum of the 
experimental and theoretical covariance matrices. 
The results of~\cite{aleph_taubr} are listed in Table~\ref{tab_asresults}. 
% Table~\ref{tab_rescorr} gives the corresponding correlation matrices
% of the fit parameters. 
%The limited number of observables
%and the strong correlations between the spectral moments explain the
%large correlations observed, especially between the
%nonperturbative operators, which are solely determined by the shapes 
%of the spectral functions. 
The precision of \asm  obtained with the 
two perturbative methods employed is comparable, however their central
values differ by  0.01--0.02, which is larger than the theoretical errors
assigned. The $\delta^{(2)}$ term is theoretical only with quark masses 
varying within their allowed ranges~(\ref{eq:qmasses}). The quark 
condensates in the $\delta^{(4)}$ term are obtained from PCAC, 
while the gluon condensate is determined by the fit. The main contributions
to the theoretical uncertainty on \asm are listed in Table~\ref{tab_theoerr}.
The fact that the $\chi^2$ value of the axial-vector fit is better
with FOPT than with CIPT should not---in our view--- affect the 
conclusion in favor of the latter method, reached in 
Sections~\ref{sec:pert_comp} and \ref{sec:cipt} on the basis of
a better perturbative behavior, but it may require further studies.
\begin{table}[t]
  \caption[.]{\label{tab_theoerr}
              Sources of theoretical uncertainties and their impact 
              on $R_{\tau,V+A}$ and \asm~\cite{aleph_taubr}.}
  \begin{center}
\setlength{\tabcolsep}{0.0pc}
\begin{tabular*}{\textwidth}{@{\extracolsep{\fill}}lccccc} 
\hline\noalign{\smallskip}
 &     & \mc{2}{c}{$\Delta^{\rm th} R_{\tau,V+A}$} 
       & \mc{2}{c}{$\Delta^{\rm th}$\asm} \\
\rs{Error source}  & \rs{Value\pms$\Delta^{\rm th}$}
       & \FOPTCI & FOPT
       & \FOPTCI & FOPT \\ 
\noalign{\smallskip}\hline\noalign{\smallskip}
 $\Sew$          & $1.0198\pm0.0006$              & \mc{2}{c}{0.002}
                                                    & \mc{2}{c}{0.001} \\
 $|V_{ud}|$      & $0.9745\pm0.0004$              & \mc{2}{c}{0.002} 
                                                    & \mc{2}{c}{0.001} \\
 $K_4$           & $25\pm25$                      & 0.015 & 0.009
                                                    & 0.007 & 0.003 \\
 R-scale $\mu$ & $m_\tau^2 \pm 2\gev^2$           & 0.012 & 0.035
                                                    & 0.006 & 0.011 \\
\noalign{\smallskip}\hline\noalign{\smallskip}
 \mc{2}{c}{Total errors}                        & 0.019 & 0.036
                                                      & 0.009 & 0.012\\ 
\noalign{\smallskip}\hline
  \end{tabular*}
  \end{center}
\end{table}

The final result on \asm  given by~\cite{aleph_taubr} is taken as the arithmetic 
average of the \FOPTCI\  and FOPT values given in Table~\ref{tab_asresults}, 
with half of their difference added as additional theoretical error
$\as(m_\tau^2) = 0.340 \pm 0.005_{\rm exp} \pm 0.014_{\rm th}$.
The first error accounts for the experimental uncertainty and the 
second error gives the uncertainty related to the theoretical prediction 
of the spectral moments including the ambiguity between \FOPTCI\  and FOPT.

There is a remarkable agreement within statistical errors between the 
\asm  determinations using the vector and axial-vector data.
This provides an important consistency check of the results, since
the two corresponding spectral functions are experimentally independent and
manifest a quite different resonant behavior. 
%The results are displayed
%in Fig.~\ref{alphas_res}.
%\begin{figure}[t]
%   \centerline{\psfig{file=figures/alphas_res.eps,width=78mm}}
%  \caption[.]{\label{alphas_res}
%              Results for \asm from the fits of $R_{\tau,V,A,V+A}$ and 
%              the moments $D^{k\l}_{V,A,V+A}$ using the \FOPTCI and FOPT
%              perturbative expansions. The measurements are correlated due 
%              to the theoretical errors (see Table~\ref{tab_asresults}).}
%\end{figure}

The advantage of 
separating the vector and axial-vector channels and comparing to
the inclusive $V+A$ fit becomes obvious in the adjustment of the 
leading nonperturbative contributions of dimension $D=6$ and $D=8$, 
which approximately cancel in the inclusive sum. This cancellation of the 
nonperturbative terms increases the confidence in the \asm determination 
from the inclusive $V+A$ observables. The gluon condensate is 
determined by the first $k=1$, $\l=0,1$ moments, which receive 
lowest order contributions. It is observed that the values obtained in 
the $V$ and $A$ fits are not very consistent, which could indicate 
problems in the validity of the OPE approach used once the 
nonperturbative terms become significant. Taking the value obtained in 
the $V+A$ fit, where nonperturbative effects are small, 
and adding as systematic uncertainties half of the difference between 
the vector and axial-vector fits as well as between 
the \FOPTCI\ and FOPT results, ALEPH measures the gluon condensate to be
\beq
\label{eq:gluon_cond}
	\GG=(0.001\pm0.012)\gev^4.
\eeq
This result does not provide evidence for a nonzero gluon condensate, but it 
is consistent with and has comparable accuracy to the independent value 
obtained using charmonium sum rules and \ee data in the charm 
region, $(0.011\pm0.009)\gev^4$ in a combined determination with 
the $c$ quark mass~\cite{ioffe}. 

The $D=6,8$ nonperturbative contributions are obtained after averaging the
FOPT and \FOPTCI\ values:
\beq
\begin{array}{rclrcl}
  \delta^{(6)}_V	&=& (2.8 \pm 0.3) \times 10^{-2}~, &
  \delta^{(8)}_V 	&=& (-8.8 \pm 0.6) \times 10^{-3}~,\\
  \delta^{(6)}_A 	&=& (-3.1 \pm 0.3) \times 10^{-2}~, & 
  \delta^{(8)}_A 	&=& (8.7 \pm 0.6) \times 10^{-3}~,  \\
  \delta^{(6)}_{V+A} 	&=& (-1.8 \pm 2.4) \times 10^{-3}~,  &
  \delta^{(8)}_{V+A} 	&=& (0.5 \pm 5.1) \times 10^{-4}~. 
\end{array}
\eeq
The remarkable feature is the approximate cancellation of these contributions
in the $V+A$ case, both for $D=6$ and $D=8$. This property was 
predicted~\cite{bnp} for $D=6$ using the simplifying assumption of vacuum
saturation for the matrix elements of four-quark operators, yielding
$\delta^{(6)}_V /\delta^{(6)}_A =-7/11=-0.64$, in fair agreement with
the above results which reads $-0.90\pm0.18$. The estimate~\cite{bnp} for 
$\delta^{(6)}_V = (2.5 \pm 1.3) \times 10^{-2}$ agrees well with 
the experimental result. 

The total nonperturbative $V+A$ correction, 
$\delta_{{\rm NP},V+A}=(-4.3\pm1.9) \times 10^{-3}$, is an
order-of-magnitude smaller than the corresponding values in the $V$ and 
$A$ components, $\delta_{{\rm NP},V}=(2.0\pm0.3) \times 10^{-2}$ and 
$\delta_{{\rm NP},A}=(-2.8\pm0.3) \times 10^{-2}$.

\subsubsection{The OPAL results}

The OPAL analysis~\cite{opal_vasf} makes use of the spectral functions 
shown in Fig.~\ref{fig:vasf} and provides also a complete description 
in terms of QCD of their vector and axial-vector data. It proceeds 
along very similar lines as the ALEPH analysis, with the difference 
that OPAL has chosen to quote their results separately for three 
prescriptions for the perturbative expansion, respectively
\FOPTCI, FOPT and with renormalon chains, {\it i.e.}
\beq
\label{qcd_opal}
\asm =\left\{\begin{array}{ll}
0.348\pm 0.009_{\rm exp}\pm 0.019_{\rm th} & \hspace{5mm}{\rm CIPT}~, \\
0.324\pm 0.006_{\rm exp}\pm 0.013_{\rm th} & \hspace{5mm}{\rm FOPT}~, \\
0.306\pm 0.005_{\rm exp}\pm 0.011_{\rm th} & \hspace{5mm}{\rm renormalon}~.
\end{array}\right.
\eeq

The OPAL results using \FOPTCI and FOPT agree with those from ALEPH.
The quoted theoretical uncertainties are consistently treated between 
the two analyses , with the important difference that the more recent
ALEPH analysis~\cite{aleph_taubr} uses a reduced uncertainty range for 
the $K_4$ parameter ($K_4=25\pm25$) as compared to $K_4=50\pm50$ in the 1998 
analysis~\cite{aleph_asf} and $K_4=25\pm50$ in OPAL. The justification
of the smaller $K_4$ uncertainty was provided in Section~\ref{sec:pert}.
As for the value obtained with the renormalon chains, we refer to our
discussion in Sections~\ref{sec:renormalons} and \ref{sec:pert_comp} 
where we concluded that however interesting this approach is, its
performance for a reliable and precise determination of \asm is
not adequate. OPAL also derives values for the gluon condensate: averaging
the results of the \FOPTCI and FOPT analyses, one obtains
$ \GG=(0.005\pm0.017)\gev^4$, in agreement with ALEPH.

\subsubsection{Running of $\as(s)$ below $m_\tau^2$}
%\subsubsection{Measuring the running of $\as(\mu^2)$ for $\mu<m_\tau$}
\label{sec:qcd_as_running}
Using the \sfs, one can simulate the physics of a hypothetical 
$\tau$ lepton with a mass $\sqrt{s_0}$ smaller than $m_\tau$
through Eq.~(\ref{eq:rtauth1}). Assuming quark-hadron duality, 
the evolution of $R_\tau(s_0)$ provides a direct test of the running 
of $\as(s_0)$, governed by the RGE $\beta$-function. On the other 
hand, it is also a test of the stability of the OPE approach 
in $\tau$ decays. The studies performed in this section employ only 
\FOPTCI. Results obtained with FOPT are similar\footnote
{
	As already indicated by the smaller $\chi^2$ value of the 
	spectral moments fit (see Table~\ref{tab_asresults}), the 
	agreement between the FOPT evolution computation and the 
	data is better than it is when using \FOPTCI.
} 
and differ mostly in the central \asm value.
\begin{figure}[t]
   \centerline{
        \epsfxsize8.3cm\epsffile{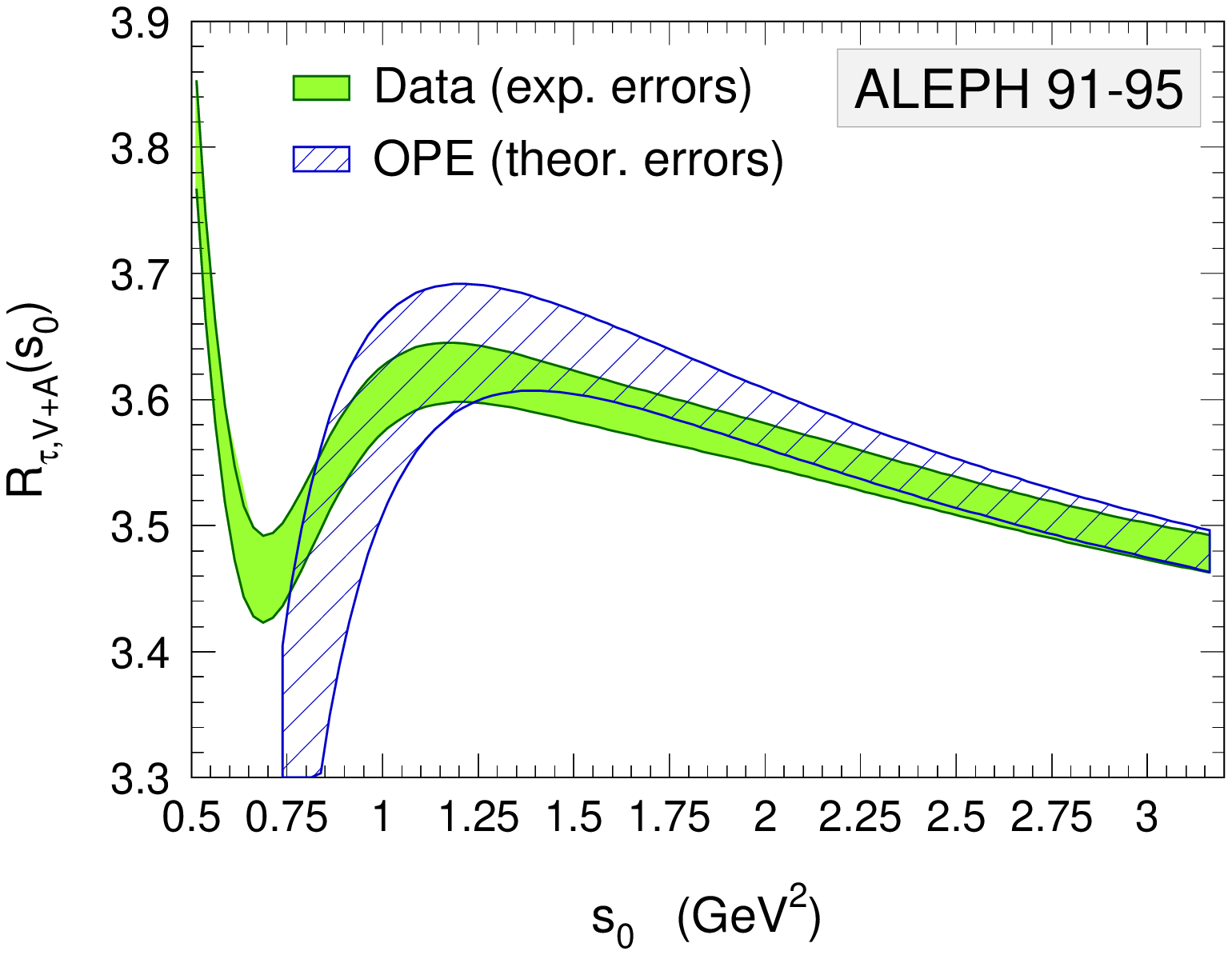}
        \epsfxsize8.3cm\epsffile{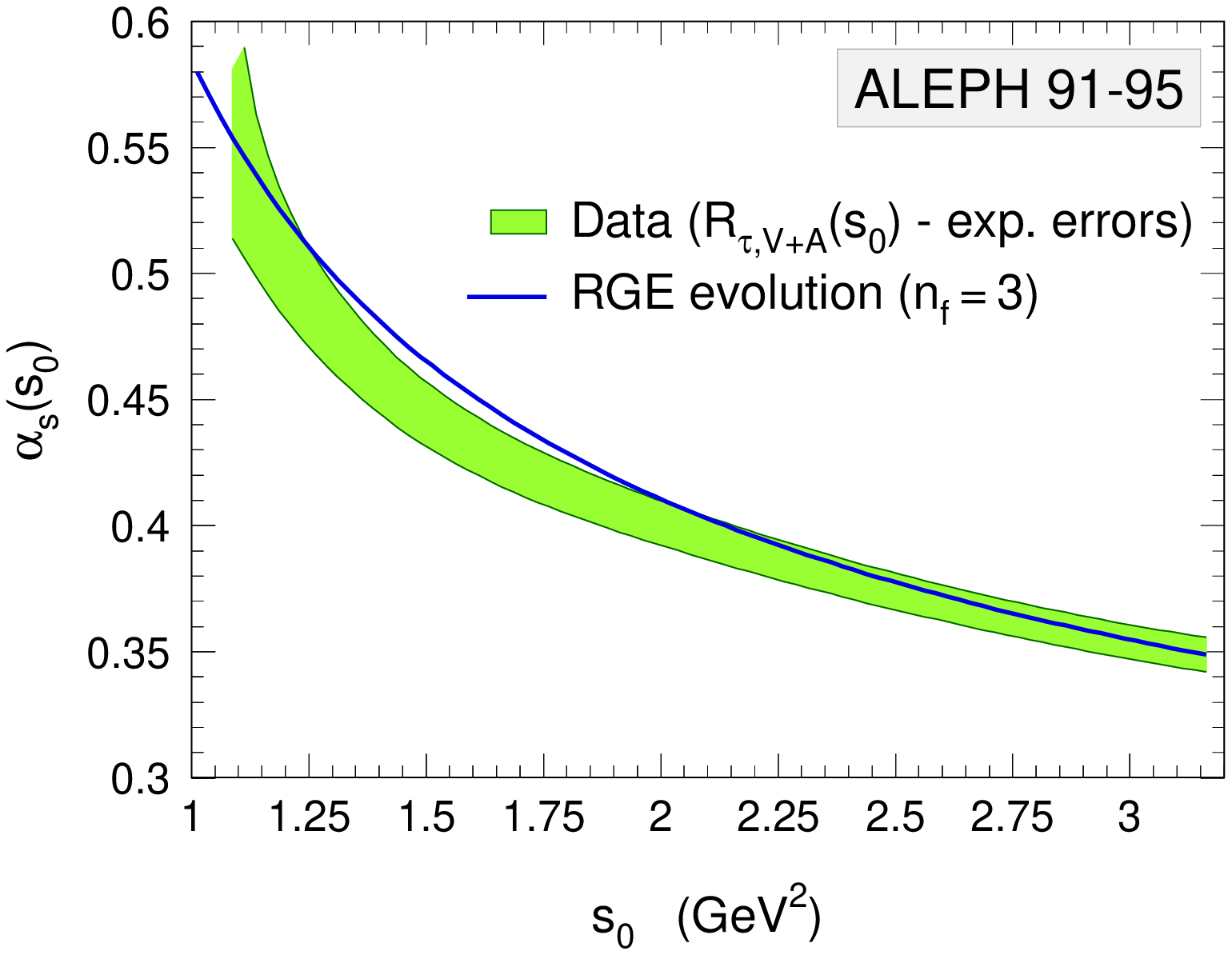}}
  \vspace{-0.2cm}
  \caption[.]{\label{vpa_runrtauas}
        \underline{Left:} 
	The ratio $R_{\tau,V+A}$ versus the square ``$\tau$ mass'' $s_0$.
      	The curves are plotted as error bands to emphasize their 
       	strong point-to-point correlations in $s_0$. Also 
      	shown is the theoretical prediction using \FOPTCI\ and
       	the results for $R_{\tau,V+A}$ and the nonperturbative 
       	terms from Table~\rm\ref{tab_asresults}.
        \underline{Right:} 
	The running of $\as(s_0)$ obtained from the 
       	fit of the theoretical prediction to $R_{\tau,V+A}(s_0)$ using
       	CIPT. The shaded band shows the data including only experimental
       	errors. The curve gives the four-loop RGE evolution 
       	for three flavors.}
\end{figure}

The functional dependence of $R_{\tau,V+A}(s_0)$ is plotted in the left
hand plot of Fig.~\ref{vpa_runrtauas} together with the theoretical 
prediction using the results of Table~\ref{tab_asresults}. 
The spreads due to uncertainties are shown as error bands. 
The correlations between two adjacent points in $s_0$ are large as 
the only new information is provided by the small mass difference 
between the two points and the slightly modified weight functions. 
Moreover they are reinforced by the original experimental and theoretical 
correlations. Below $1\gev^2$ the error of the theoretical 
prediction of $R_{\tau,V+A}(s_0)$ starts to blow up due to the 
increasing uncertainty from the unknown $K_4$ perturbative term;
errors of the nonperturbative contributions are {\it not} contained 
in the theoretical error band. Figure~\ref{vpa_runrtauas} (right) shows the plot 
corresponding to Fig.~\ref{vpa_runrtauas} (left), translated into the running 
of $\as(s_0)$, \ie, the experimental value for $\as(s_0)$ 
has been individually determined at every $s_0$ from the comparison 
of data and theory. Also plotted is the four-loop RGE evolution using 
three quark flavors.

It is remarkable that the theoretical prediction using the parameters 
determined at the $\tau$ mass and $R_{\tau,V+A}(s_0)$ extracted from 
the measured
$V+A$ \sf\ agree down to $s_0 \sim 0.8\gev^2$. The agreement is good to
about 2\% at $1\gev^2$. This result, even more directly illustrated by the
right hand plot of Fig.~\ref{vpa_runrtauas}, demonstrates the validity of 
the perturbative approach
down to masses around $1\gev$, well below the $\tau$ mass scale. The
agreement with the expected scale evolution between 1 and $1.8\gev$ is an
interesting result, considering the relatively low mass range, where
$\as$ is seen to decrease by a factor of 1.6 and reaches rather 
large values $\sim 0.55$ at the lowest masses. This behavior provides
confidence that the \asm  measurement is on solid phenomenological ground.

\subsubsection{Final assessment on the $\as(m_\tau^2)$ determination}
\label{sec:assess}

Although the recent ALEPH evaluation of $\as(m_\tau^2)$ represents the
state-of-the art, several remarks can be made:
\begin{itemize}

\item 	The analysis is based on the ALEPH spectral functions and branching
	fractions, ensuring a good consistency between all the observables, 
        but not exploiting the full experimental information currently 
        available from other experiments. Since the result on $\as(m_\tau^2)$ 
        is limited by theoretical uncertainties, one should expect only a 
        small improvement of the final error in this way, however it can 
        influence the central value.

\item 	One example for this is the evaluation of the strange component. Some 
	discrepancy is observed between the ALEPH measurement of the 
	$(K \pi \pi)^- \nu$ mode and the CLEO and OPAL results, as discussed 
	in Section~\ref{sec:brs_k}. Although this could still be the
	result of a statistical fluctuation, their average provides a 
	significant shift in the central value compared to using the ALEPH 
	number alone.
	Another improvement is the substitution of the measured branching
	fraction for the $\Km\nu$ mode by the more precise value predicted from
	$\tau$--$\mu$ universality (see Eq.~(\ref{k_uni})). Both operations 
	have the effect to increase the strange $\RtauS$ ratio from  
	$0.1603 \pm 0.0064$, as obtained by ALEPH, to $0.1686 \pm 0.0047$
	for the world average.

\item 	One can likewise substitute the world average value for the 
	universality-consolidated value of the electronic branching fraction 
        given in Eq.~(\ref{eq:uni_be}), $\BR_e^{\rm uni}=(17.818 \pm 0.032)\%$,
        to the corresponding
	ALEPH result, $\BR_e=(17.810 \pm 0.039)\%$, with little effect
	on the central value, but some improvement in the precision.

\item 	Most of the theoretical uncertainty originates from the limited 
	knowledge of the perturbative expansion, only predicted to third order.
	This problem was discussed in detail in Section~\ref{sec:pert} and, 
	among the standard FOPT and \FOPTCI\  approaches, strong arguments 
	have been presented in favor of the latter one. 
        We therefore take the result from the CIPT expansion, 
        not introducing any additional 
	uncertainty spanning the difference between FOPT and CIPT results. The 
	dominant theoretical errors are from the uncertainty in $K_4$ and from 
	the renormalization scale dependence, both covering the effect of 
	truncating the series after the estimated fourth order.

\end{itemize}

From this analysis, one finds the new value for the nonstrange ratio,
\beqn
\label{eq:rtau_new}
  R_{\tau,V+A} 	&=& R_\tau - R_{\tau,S} \nonumber \\
       	        &=& (3.640 \pm 0.010)~-~(0.1686 \pm 0.0047) \nonumber \\
                &=& 3.471 \pm 0.011~, 
\eeqn
to be compared to the ALEPH value of $3.482 \pm 0.014$. The 
result~(\ref{eq:rtau_new}) translates into an updated determination 
of $\as(m_\tau^2)$ from the inclusive $V+A$ component using the \FOPTCI\  
approach
\beq
\label{astau:final}
  \as(m_\tau^2) = 0.345 \pm 0.004_{\rm exp} \pm 0.009_{\rm th}~,
\eeq
with improved experimental and theoretical precision over the ALEPH
result.

\subsubsection{Evolution to $M_Z^2$}
\label{sec:qcd_as_evolution}

It is customary to compare \as values, obtained at different 
renormalization scales, at the scale of the $Z$-boson mass. 
To do this, one evolves the \as result using the RGE~(\ref{eq:betafun}).
A difficulty arises from the quark thresholds: 
in the MS scheme, or any of its modifications such as \MSbar,
the beta function governing the running of the strong coupling constant
is independent of quark masses.
This simplifies the calculation of QCD corrections beyond the one-loop 
level. On the other hand, decoupling of heavy quarks~\cite{decoupling} 
is not manifest in each order of perturbation theory~\cite{bernr}. 
A heavy quark decoupling is realized in momentum-space 
dependent renormalization schemes (MO). The RGE in an MO scheme 
involves the scaling-function 
$\beta_{\rm MO}=\beta(\as^\prime,m_i^\prime,a^\prime)$
that depends on the coupling $\as^\prime(\mu^2)$, the quark masses 
$m_i^\prime(\mu^2)$ and a gauge parameter $a^\prime(\mu^2)$, where the 
primes stand for the scheme dependence of the parameters. Quark-loop 
calculations in this scheme appear rather complicated, however the 
mass dependence of the $\beta$-function provides a suppression of 
heavy-quark effects at scales much smaller than the masses of these 
quarks, \ie, it decouples heavy quarks from the light-particle Green's
function to each order in perturbation theory.

To obtain decoupling in MS schemes, one builds in the decoupling region, 
$\mu\ll \mbar_q^{(n_f)}(\mu^2)$, where $\mbar^{(n_f)}(\mu^2)$ is the 
running \MSbar mass of the 
heavy quark with flavor $f$, an effective field theory that behaves as 
if only light quarks up to flavor $n_f-1$ were present. Matching conditions
connect the parameters of the low energy effective Lagrangian to the 
full theory. The coupling constant of the effective theory can then 
be developed in a power series of the coupling constant of the full 
theory with coefficients that depend on $x=\ln(\mu^2/\mbar_q^2)$.
Doing so one obtains for the matching of $a_s$ and the light quark 
masses between the ``light'' flavor $n_f-1$ and the heavy-quark flavor 
$n_f$~\cite{betafourloop,chet1,chet2,wetzel1,wetzel2,pichsanta}
\beqn
   a_s^{(n_f)}(\mu^2) 
     &=&
        a_s^{(n_f-1)}(\mu^2)
        \left[1 + \sum_{k=1}^\infty C_k(x)\left(a_{s}^{(n_f-1)}(\mu^2)\right)^k\right]~, \nonumber\\
   \mbar_q^{(n_f)}(\mu^2) 
     &=&
\label{eq:CHfunctions}
        \mbar_q^{(n_f-1)}(\mu^2)
        \left[1 + \sum_{k=1}^\infty H_k(x)\left(a_{s}^{(n_f-1)}(\mu^2)\right)^k\right]~.
\eeqn
There is no straightforward choice of the matching scale $\mu$. However,
to acquire good convergence of the perturbative expansion, one should
satisfy $\mu/\mbar_q\sim \mathcal{O}(1)$. Since vacuum polarization of
quark pairs modifies the coupling constant, we believe that the scale
$\mu=2\mbar_q$ is quite meaningful. The effect from this scale ambiguity 
is part of the systematic error assigned to the RGE evolution (see below).

The functions $C_k(x)$ and $H_k(x)$ are
derived by inserting the RGE-based Taylor expansions (Eq.~(\ref{eq:astaylor}) 
and equivalent for the quark masses)
into Eq.~(\ref{eq:CHfunctions}), and 
solving for each order in $a_s^{(n_f-1)}$ the corresponding coupled
differential equations~\cite{pichsanta}
\beqn
	&&
	C_1=\frac{1}{6}\,x~,\hspace{0.5cm
	C_2 = c_{2,0} + \frac{19}{24}\,x + \frac{1}{36}\,x^2~,} \\
	&&
	C_3 = c_{3,0} + \left[\frac{241}{54} + \frac{13}{4}\,c_{2,0}
		- \left(\frac{325}{1728} + \frac{1}{6}\,c_{2,0}\right)n_f\right]x
		+ \frac{511}{576}\,x^2 + \frac{1}{216}\,x^3~,\\
	&&
	H_1 = 0~,\hspace{0.5cm}
	H_2 = d_{2,0} + \frac{5}{36}\,x - \frac{1}{12}\,x^2~, \\
	&&
	H_3 = d_{3,0} + \left[\frac{1627}{1296} - c_{2,0} + \frac{35}{6}\,d_{2,0}
		+ \left(\frac{35}{648} - \frac{1}{3}\,d_{2,0}\right)n_f
		+ \frac{5}{6}\,\zeta_3\right]x \nonumber\\
	&&\hspace{2.4cm}
		-\: \frac{299}{432}\,x^2 
		- \left(\frac{37}{216} - \frac{1}{108}\,n_f\right)x^3~,
\eeqn
where the integration coefficients $c_{i,0}$, $d_{i,0}$ are computed
in the \MSbar scheme at the scale of the quark masses 
(\ie, $m_q=\mbar_q(m_q)$)~\cite{wetzel1,wetzel2,chet1,pichsanta}
\beq
	c_{1,0}=d_{1,0}=0~,\hspace{0.3cm}
	c_{2,0}=-\frac{11}{72}~, \hspace{0.3cm}
	c_{3,0}= \frac{82043}{27648}\,\zeta_3 - \frac{575263}{124416}
			+ \frac{2633}{31104}\,n_f~, \hspace{0.3cm}
	d_{2,0}=-\frac{89}{432}~.
\eeq

The evolution of the \asm measurement from the inclusive $V+A$ 
observables given in Eq.~(\ref{astau:final}), based on Runge-Kutta 
integration of the RGE~(\ref{eq:betafun}) to N$^3$LO 
(see Section~\ref{sec:qcd_RGEs}), and three-loop quark-flavor 
matching~(\ref{eq:CHfunctions}), gives
\beqn
\label{eq:asres_mz}
   \as(M_Z^2) &=& 0.1215 \pm 0.0004_{\rm exp} 
                                       \pm 0.0010_{\rm th} 
                                       \pm 0.0005_{\rm evol}~, \nonumber \\
                         &=& 0.1215 \pm 0.0012~.
\eeqn
The first two errors originate from the \asm determination given
in Eq.~(\ref{astau:final}), and the last error stands for
ambiguities in the evolution due to uncertainties in the 
matching scales of the quark thresholds~\cite{pichsanta}. This
evolution error receives contributions from the uncertainties in 
the $c$-quark mass (0.00020, $m_c$ varied by $\pm0.1\gev$) 
and the $b$-quark mass (0.00005, $m_b$ varied by $\pm0.1\gev$), the 
matching scale (0.00023, $\mu$ varied between $0.7\,m_q$ 
and $3.0\,m_q$), the three-loop truncation in the matching 
expansion (0.00026) and the four-loop truncation in the RGE 
equation (0.00031), where we used for the last two errors 
the size of the highest known perturbative term as systematic 
uncertainty. These errors have been added in quadrature.
The result~(\ref{eq:asres_mz}) is a determination of the strong coupling
at the $Z$ mass scale with a precision of 1\%.

\begin{figure}[t]
%
%...kumac is in: ~hoecker/alpha_s/evolution/eval.kumac
  \centerline{\epsfxsize12.0cm\epsffile{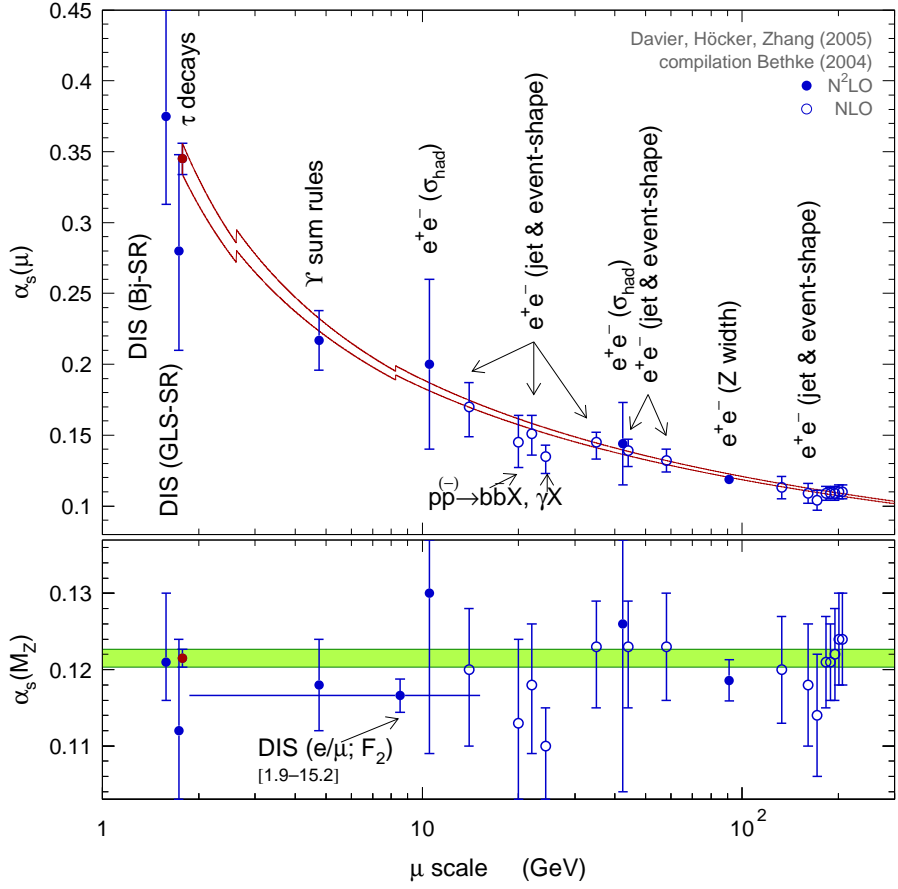}}
  \vspace{-0.7cm}
  \caption[.]{\label{fig:evolution}
      	\underline{Top}: The evolution of \asm~(\ref{astau:final}) to 
	higher scales $\mu$ using the four-loop RGE and the 3-loop 
	matching conditions applied at the heavy quark-pair thresholds 
	(hence the discontinuities at $2m_c$ and $2m_b$). The
      	evolution is compared with other independent measurements (see
      	text) covering scales varying over more than two orders magnitude. 
	The experimental values are briefly discussed in the text; they are 
	mostly taken from the compilation~\cite{bethke04}.
        \underline{Bottom}: The corresponding extrapolated $\as$ values
       	at $M_Z$. The shaded band displays the $\tau$ decay result within 
	errors.
	}
\end{figure}

The evolution path of \asm is shown in the upper plot 
of Fig.~\ref{fig:evolution}.
The two discontinuities are due to the quark-flavor matching
at $\mu=2\mbar_q$. One could prefer to avoid the discontinuities by 
choosing $\mu=\mbar_q$ so that the logarithms in Eq.~(\ref{eq:CHfunctions})
vanish, and the matching becomes (almost) smooth. However, in this case,
one must first  evolve from $m_\tau$ down to $\mbar_c$ to match the 
$c$-quark flavor, before evolving to $\mbar_b$. 
The effect on \asZ from this ambiguity is within the assigned 
systematic uncertainty for the evolution.

\subsubsection{Comparison with other determinations of $\as(M_Z^2)$}
\label{sec:qcd_as_comp}

The evolution of \asm in Fig.~\ref{fig:evolution}
is compared with other independent measurements compiled 
in~\cite{bethke04}. Listed below are some outstanding measurements made 
over a large energy range in a variety of processes with different 
experimental and theoretical precisions:
\begin{description}

\item[DIS (Bj-SR):] $\as(1.58\gev)=0.375^{+0.062}_{-0.081}$~\cite{ek95}
     was obtained using the polarized structure function data 
     for protons $g^p_1(x)$ and neutrons $g^n_1(x)$ 
     from the CERN SMC~\cite{aea94} and SLAC E142~\cite{e143:95} experiments,
     based on the Bjorken sum rule~\cite{bj70} at 
     N$^2$LO~\cite{ltv91,lv91}
     \begin{equation}
        \intl^1_0\left[g^p_1(x)-g^n_1(x)\right]dx
	=\frac{1}{3}\left|\frac{g_A}{g_V}\right|
		\left[1-\frac{\as}{\pi}
			-3.58\left(\frac{\as}{\pi}\right)^{\!2}
			-20.2\left(\frac{\as}{\pi}\right)^{\!3}
		\right]\,.
     \end{equation}
     Here $g_V$ and $g_A$ are the vector and axial-vector coupling 
     constants of the neutron decay, $|g_A/g_V|=-1.2573\pm 0.0028$ (determined
     assuming CVC), and the last two coefficients are
     calculated for three active flavors. 
     No explicit corrections for nonperturbative 
     higher twist effects\footnote
     {
         Interactions involving more than one parton,
         like secondary interactions of quarks with the target remnant,
         give rise to additional contributions to the structure functions.
         These ``higher twist'' terms are classified according to their $Q^2$
         dependence $1/Q^{2n}$, where $n=0$ is leading twist (twist two),
         $n=1$ is twist four, etc.
     } 
     were applied to
     derive this result. However an estimate of the size of 
     the perturbative ${\cal O}(\alpha^4_s)$ term was taken into account.

\item[DIS (GLS-SR):] 
     $\as(1.73\gev)=0.280\pm 0.061_{\rm exp}\left.^{+0.035}_{-0.030}\right|_{\rm th}$ is an update of 
     the analysis using structure function $F_3$ data in the $Q^2$ range 
     from $1$ to $15.5\gev^2$ performed by the CCFR 
     collaboration~\cite{ccfr98}. It uses the Gross-Llewellyn-Smith (GLS) 
     sum rule~\cite{gls69} at N$^2$LO~\cite{ck92,ltv91,lv91}
     \begin{equation}
        \intl^1_0 F_3(x,Q^2)dx
		\equiv3\left[1-\frac{\as}{\pi}
				-3.58\left(\frac{\as}{\pi}\right)^{\!2}
				-19.0\left(\frac{\as}{\pi}\right)^{\!3}
			\right]\,,
     \end{equation}
      given for three active flavors.
      The systematic uncertainty is dominated by the extrapolation of 
      the GLS integral to the region $x<0.01$ where no measurements exist, 
      and to $x>0.5$, which is substituted by the structure function $F_2$ from 
      SLAC data~\cite{withlow:1992}. 
      The theoretical error is dominated by uncertainties 
      in the higher twist corrections.

\item[\boldmath $\Upsilon$ sum rules:] 
     $\as(m_b)|_{m_b=4.75\gev}=0.217\pm 0.021$
     was found together with the bottom quark pole mass $m_b$ 
     from sum rules for the $\Upsilon$ system in N$^2$LO, resumming all 
     ${\cal O}\left(\as^2,\as v, v^2\right)$ terms with $v$ being 
     the velocity of the heavy quark~\cite{pp98}. The precision of 
     this result on $\as(m_b)$ was not confirmed by a similar 
     analysis~\cite{Hoang}, where the theoretical parameter space
     has been scanned, which leads to a more conservative estimate 
     of the theory uncertainty.

\item[\boldmath $e^+e^- (\sigma_{\rm had})$:] 
     $\as(10.52\gev)=0.20\pm 0.06$
     was determined in N$^2$LO by CLEO~\cite{cleo98} using the total 
     $e^+e^-$ hadronic cross section according to
     \begin{equation}
     R=\frac{\sigma(e^+e^-\to {\rm hadrons})}
     	{\sigma(e^+e^-\to \mu^+\mu^-)}=3\sum_i Q^2_i
     		\left[1+\frac{\as}{\pi}
			+1.52\left(\frac{\as}{\pi}\right)^2
			-11.5\left(\frac{\as}{\pi}\right)^3
		\right]~,
     \end{equation}
     where $Q_i$ are the electric charges of quark flavors $i$ 
     that are produced at center-of-mass energy just below the 
     $\FourS$ resonance in the $e^+e^-$ continuum.
     Another measurement $\as(42.4\gev)=0.175\pm 0.028$ 
     was obtained with the use of data from PETRA and TRISTAN in the 
     center-of-mass energy range from $20$ to $65\gev$~\cite{dh95,bethke00}. 

\item[\boldmath $e^+e^-$ (jet \& event-shape):] Using data from PETRA, 
     TRISTAN, SLC and LEP1,2 over a large energy range,
     several determinations of $\as$ were obtained
     based on the jet rate and event-shape variables that are sensitive 
     to gluon radiation governed by the strong coupling constant.
     The observables are calculated in resummed NLO (\ie, NLO matched 
     with next-to-leading order logarithms). 
     In all these determinations, the theoretical uncertainty from the
     renormalization scale uncertainty is the dominant
     error source. At low PETRA energy the hadronization uncertainty 
     is also important. See references in~\cite{bethke00}.

\item[\boldmath $p\overline{p}\to b\bbar X$:]
     $\as(20\gev)=0.145\left.^{+0.012}_{-0.010}\right|_{\rm
     exp}\left.^{+0.013}_{-0.016}\right|_{\rm th}$ was obtained by UA1~\cite{ua1:96}
     from a measurement of the cross section of the process 
     $p\overline{p}\to b\bbar X$ for which NLO QCD predictions 
     exist. The theoretical error includes uncertainties due to 
     different sets of structure functions, renormalization/factorization 
     scale uncertainties and the $b$-quark mass.

\item[\boldmath $p\overline{p},pp\to \gamma X$:]
     $\as(24.3\gev)=0.135\pm0.006_{\rm exp}\left.^{+0.011}_{-0.005}\right|_{\rm th}$ 
     was determined by UA6~\cite{ua6:99} in NLO from a measurement of 
     the cross section difference 
     $\sigma(p\overline{p}\to \gamma X)-\sigma(pp\to \gamma X)$
     such that the poorly known contributions of the sea quarks and
     gluon distributions in the proton cancel. 
     The theoretical error includes uncertainties from the scale choice 
     and from the variation of the parton distribution functions.

\item[\boldmath $e^+e^-$ ($Z$ width):]
     $\as(M_Z)_{Z\,{\rm width}}=0.1186\pm 0.0027$ was determined from
     the hadronic width at the $Z^0$ resonance by a global fit to all 
     electroweak data in N$^2$LO~\cite{ewfit}.

\end{description}

A comparison is best achieved by extrapolating these measurements 
to $M_Z$ as shown in the lower plot in Fig.~\ref{fig:evolution}.
The most precise result at the $M_Z$ scale stems from $\tau$ 
decays~(\ref{eq:asres_mz}), which is a 1.1\% determination limited in 
accuracy by theoretical uncertainties in the perturbative expansion. 
The significant improvement in precision compared to the previous ALEPH 
result~\cite{aleph_asf}, 
$0.1202 \pm 0.0008_{\rm exp} \pm 0.0024_{\rm th} \pm 0.0010_{\rm evol}$, 
is due to the higher statistics and the more detailed experimental analysis,
but mostly because of the smaller theory uncertainty assigned to 
the perturbative expansion and our conclusion to lift the 
ambiguity between \FOPTCI\  and FOPT.

Another precise determination (DIS ($e/\mu; F_2$) [$1.9-15.2$]\gev),
$\as(M_Z^2)=0.1166\pm 0.0009_{\rm stat}\pm 0.0020_{\rm syst}$~\cite{sy01,bethke03},
obtained using the structure function $F_2$ from deep inelastic electron 
($e$) and muon ($\mu$) scattering (DIS) data in the $Q^2$ range between 
$3.5\gev^2$ and $230\gev^2$ is also shown. The analyses use N$^2$LO
calculations wherever available. The systematic error is mostly theoretical
including higher twist effects and an estimate of the N$^3$LO corrections, 
and is doubled to account for missing contributions associated with the
renormalization scale and scheme uncertainties.

In addition to the measurements shown in Fig.~\ref{fig:evolution}, 
there is now a rather precise determination from lattice 
QCD simulations~\cite{davis04}: $\alpha_s(M^2_Z)=0.121\pm 0.003$. This is
the first lattice determination with realistic quark vacuum polarization
and an ${\cal O}(a^2)$ improved staggered-quark discretization of the
light-quark action.

\subsubsection{A measure of asymptotic freedom between $m_\tau^2$ and $M_Z^2$}
\label{sec:qcd_as_runningness}

The $\tau$-decay and $Z$-width determinations have comparable accuracies, 
which are however very different in nature. The $\tau$ value is 
dominated by theoretical uncertainties, whereas the determination 
at the $Z$ resonance, benefiting from the much larger energy scale 
and the correspondingly small uncertainties from the truncated 
perturbative expansion, is limited by the experimental precision 
on the electroweak observables, essentially the ratio of leptonic 
to hadronic peak cross sections. The consistency between the two results
provides the most 
powerful present test of the evolution of the strong interaction 
coupling, over a range of $s$ spanning more than three 
orders of magnitude, as it is predicted by the nonabelian nature of 
the QCD gauge theory. The difference between the extrapolated
$\tau$-decay value and the measurement at the $Z$ is:
\beq
 \as^\tau(M_Z^2)-\as^Z(M_Z^2) = 0.0029 \pm 0.0010_\tau \pm 0.0027_Z~,
\eeq
which agrees with zero with a relative precision of 2.4\%.

In fact, the comparison of these two values is valuable since they are among
the most precise single measurements and they are widely spaced in energy 
scale. Thus it allows one to perform an accurate test of asymptotic freedom.
Let us consider the following evolution estimator~\cite{delphi_run} 
for the inverse of $\as(s)$,
\beq
 r(s_1,s_2) = 2\cdot\frac {\as^{-1}(s_1)-\as^{-1}(s_2)} 
                           {\ln s_1 -\ln s_2}~,
\eeq
which reduces to the logarithmic derivative of $\as^{-1}(s)$ 
when $s_1 \rightarrow s_2$, 
\beqn
\frac {d\as^{-1}}{d\ln \sqrt{s}} &=& -\frac {2\pi\beta(s)}{\as^2}~,\nonumber \\
 &=&  \frac {2\beta_0}{\pi}~(1 + \frac {\beta_1} {\beta_0} \frac {\as}{\pi} +\cdots~)~, 
\eeqn
with the notations of Eq.~(\ref{eq:betafun}). 
At first order, the logarithmic derivative is driven by $\beta_0$.

The $\tau$ and $Z$ experimental determinations of $\as(s)$ yield the value
\beq
 r_{\rm exp}(m_\tau^2,M_Z^2) = 1.405 \pm 0.053~,
\eeq
which agrees with the QCD prediction using the RGE to N$^3$LO, 
and three-loop quark-flavor matching~(\ref{eq:CHfunctions}), as discussed 
in Section~\ref{sec:qcd_as_evolution},
\beq
 r_{\rm QCD}(m_\tau^2,M_Z^2) = 1.353 \pm 0.006~.
\eeq
This is to our knowledge the most precise experimental test of the
asymptotic freedom property of QCD at present. It can be compared to 
an independent  determination~\cite{delphi_run}, using an analysis of event 
shape observables at LEP between the $Z$ energy and 207~\gev,
$r(M_Z^2,(207~\gev)^2) = 1.11 \pm 0.21$, for a QCD expectation
of 1.27.

\section{SPECTRAL FUNCTIONS OF TAU FINAL STATES WITH STRANGENESS}
\label{sec:strangesf}
%\section{Strange Quark Mass from the Invariant Mass Distribution 
%	of Cabibbo-Suppressed Tau Decays}
%
%\subsection{Introduction}
%
Complete experimental studies of the Cabibbo-suppressed
decays of the $\tau$ became available at LEP~\cite{ALEPH:1999},
allowing to initiate systematic QCD 
studies~\cite{CDH99a,CDH99b,CDH99c,CDH99d,DA:00} of the effect
induced by the strange quark mass in the $\tau$ decay 
width~\cite{bnp,pertmass,ChK97a,ChK97b,MA98,PP98a,PP98b,PP98c,PP99,KM00,CKP98,KKP00}. 
Indeed, from the separate measurement of the strangeness $S=0$ and $S=-1$
$\tau$ decay widths, it is possible to pin down the ${\rm SU}(3)$ breaking
effects and to obtain information on the strange quark mass.
Recent experimental information~\cite{CLEOK99,OPAL04} can now be 
included in addition to the ALEPH data. In this section the current situation
on the physics of $\tau$ decays with net strangeness is reviewed, with 
emphasis on the determination of the strange quark mass and the CKM 
element $|V_{us}|$. We anticipate here that while the determination 
of $|V_{us}|$ through QCD sum rules is found to be robust, a reliable 
and competitive determination of the strange quark mass $m_s$ is 
difficult because of poor convergence of the mass-dependent
perturbative series, and lack of sufficient inclusiveness of the
higher-order spectral moments that are most sensitive to $m_s$. 

\subsection{Strange spectral functions}

The experimental determination of nonstrange and strange (subscript $S$)
spectral functions has been presented and discussed in Section~\ref{sec:tauspecfun}.
As shown in Section~\ref{sec:moment}, one
uses the spectral functions, or equivalently the fully corrected 
mass-squared distributions, to construct the spectral moments
\beq
\label{eq_moments}
   R_{\tau,(S)}^{k\l} = 
       \intl_0^{m_\tau^2} ds\,\left(1-\frac{s}{m_\tau^2}\right)^{\!\!k}
                              \left(\frac{s}{m_\tau^2}\right)^{\!\!\l}
       \frac{dR_{\tau,(S)}}{ds}~,
\eeq
where $R_{\tau,(S)}^{00}=R_{\tau,(S)}$. 

\begin{figure}[t]
\epsfxsize9.1cm
\centerline{\epsffile{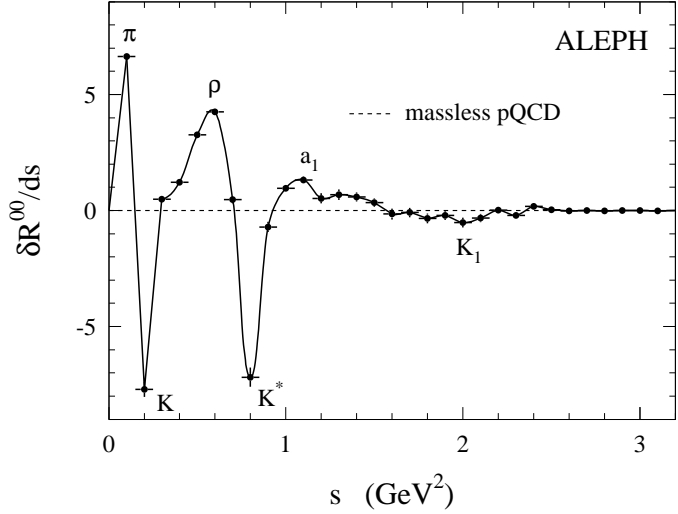}}
\vspace{-0.1cm}
\caption{Integrand of Eq.~(\ref{eq_moments}) with $(k=0, \l=0)$ for the 
	 difference~(\ref{eq_dmoments}) of the Cabibbo-corrected nonstrange 
	 and strange invariant mass spectra from~\cite{ALEPH:1999}. The 
         contribution from massless perturbative QCD (pQCD) vanishes.
	 To guide the eye, the solid line interpolates between 
	 the bins of constant $0.1\gev^2$ width.}
\label{diffmoms}       % Give a unique label
\end{figure}

To reduce the theoretical uncertainties and emphasize the
SU(3) breaking component originating from the larger $s$ quark mass, it
is convenient to consider the difference between 
nonstrange and strange spectral moments, properly normalized with their 
respective CKM matrix elements~\cite{ALEPH:1999}
\beq
\label{eq_dmoments}
   \delta R_\tau^{k\l} =
     \frac{1}{|V_{ud}|^2}R_{\tau,S=0}^{k\l} - 
     \frac{1}{|V_{us}|^2}R_{\tau,S=-1}^{k\l}~,
\eeq
for which the massless perturbative contribution vanishes.
The contributions from the various decay modes to $\delta R_\tau^{00}$
are visualized in Fig.~\ref{diffmoms}. 

\subsection{Theoretical analysis of $R_{\tau,S}$}

The inclusive $\tau$ decay ratio into strange hadronic final states
\beq
     R_{\tau,S} =
\frac{\Gamma(\tau^-\to{\rm hadrons}_{S=-1}^-\,\nut)}
                   {\Gamma(\tau^-\to e^-\,\nueb\nut)}~,
\eeq
can be used through the framework of the OPE to determine \mss\ at the 
scale $s=m_\tau^2$~\cite{CDH99a,CDH99b,CDH99c,CDH99d}. However, the 
perturbative expansion computed for the mass term in~\cite{pertmass} 
was shown to be incorrect~\cite{MA98}. After correction the series
shows a problematic convergence behavior~\cite{MA98,PP98a,PP98b,PP98c,CKP98},
which is discussed in detail below.

Following~\cite{bnp} we recall the theoretical prediction for the 
inclusive vector plus axial-vector hadronic decay rate, given by
\beq
\label{eq_rtau}
   R_\tau(m_\tau^2) =
     12\pi \Sew\intl_0^{m_\tau^2}
     \frac{ds}{m_\tau^2}\left(1-\frac{s}{m_\tau^2}\right)^{\!\!2}
     \left[\left(1+2\frac{s}{m_\tau^2}\right){\rm Im}\Pi^{(1+0)}(s+i\e)
      \,-\,2\frac{s}{m_\tau^2}{\rm Im}\Pi^{(0)}(s+i\e)\right]~,
\eeq
with the two-point correlation functions
$\Pi^{(J)}=|V_{ud}|^2\Pi_{ud,V+A}^{(J)}+|V_{us}|^2\Pi_{us,V+A}^{(J)}$
for the hadronic final state of spin $J$. The choice of the particular
spin combination for the correlators taken in Eq.~(\ref{eq_rtau}) is
justified in~\cite{bnp}. Using the OPE~\cite{svz}, 
Eq.~(\ref{eq_rtau}) for the strange part can be decomposed as
\beq
\label{eq_delta}
   R_{\tau,S} =
     3|V_{us}|^2\Sew\left(1 + \delta^{(0)} + 
     \delta^\prime_{\rm EW} +
     \delta^{(2,m_q)}_S + 
     \hm\hm\sum_{D=4,6,\dots}\hm\hm\hm\hm\delta^{(D)}_S
     \right)~.
\eeq
The $\delta^{(0)}$ term is the perturbative part 
of mass dimension $D=0$, known to third order in the expansion with 
$a_s=\as(m_\tau^2)/\pi$, and discussed in detail in Section~\ref{sec:pert}.
The residual non-logarithmic 
electroweak correction $\delta^\prime_{\rm EW}\simeq0.0010$~\cite{braaten}
is neglected in the following. Throughout this section, all QCD observables
are expressed in the \MSbar renormalization scheme.

%\subsubsection{Evidence for the effect of a massive strange quark}
%
From the result obtained in Section~\ref{sec:assess} for the world average 
branching fraction for inclusive $\tau$ decays to strange hadronic final 
states (improved using lepton universality for the $K\nu$ mode),
one obtains $R_{\tau,S} = 0.1686 \pm 0.0047$, 
using the value for $\BR^{\rm uni}_e$ given in Section~\ref{sec:brs_lepton}.
This result can be compared with the QCD prediction for a massless strange 
quark neglecting the nonperturbative contributions,
  $ R^{(0)}_{\tau,S} = 3|V_{us}|^2\Sew\left(1+\delta^{(0)}\right)
   = 0.1847 \pm 0.0029$, 
where the quoted error reflects the uncertainties on mostly $|V_{us}|$ and 
to a lesser extent $\as(m^2_\tau)$. The two values differ by $2.9 \sigma$, 
in the direction predicted for a non-zero $m_s$ value. 
%
%\subsubsection{Convergence of the perturbative series}

The $ \delta^{(2,m_q)}_S$ term in Eq.~(\ref{eq_delta}) is the mass contribution 
of dimension $D=2$, which is in practice only relevant for the strange 
quark mass. Fixed-order perturbation theory (FOPT) gives~\cite{pertmass,MA98}
\beq
\label{eq_delta2}
   \delta^{(2,m_s)}_S =-\,8\frac{m^2_s(m_\tau^2)}{m_\tau^2}
     \left[ 1 + \frac{16}{3}a_s + 46.00\,a_s^2
+ \left(283.6 + \frac{3}{4}\,x_3^{(1+0)}\right)a_s^3 +\cdots \right]~.
\eeq
The third order coefficient $x_3^{(1+0)}$ occurring in the expansion of 
the massive $J=1+0$ correlator has been recently calculated~\cite{baikov},
while the $J=0$ correlator was already known up to third order.
Previous analyses used an estimate for $x_3^{(1+0)}$, however it turns 
out to be of minor numerical importance for the result on $m_s$.
The true value~\cite{baikov} is $x_3^{(1+0)}=202.3$, while the naive
estimate, assuming a geometric growth of the perturbative coefficients,
gives $x_3^{(1+0)} \simeq x_2^{(1+0)}(x_2^{(1+0)}/x_1^{(1+0)})\simeq 165$,
which was taken with a 200\% uncertainty.
Setting $\as(m_\tau^2)=0.334$, one finds that Eq.~(\ref{eq_delta2})
does not exhibit good convergence.
%\beq
%\label{eq_convfopt}
%   \delta^{(2,m_s)}_S = 
%     -\,8\frac{m^2_s(m_\tau^2)}{m_\tau^2}
%     \left(1 + 0.57 + 0.52 + 0.52 +\cdots \right)~,
%\eeq
%which converges badly.

A contour-improved analysis (CIPT) for the dimension $D=2$ contribution 
has been used in~\cite{PP98a,PP98b,PP98c,CKP98}. 
It consists of a direct numerical evaluation of the contour integral 
derived from Eq.~(\ref{eq_rtau}), using the solution of the 
renormalization group equation (RGE) to four loops as 
input to the running $\as(s)$ and $m_s(s)$. This provides a resummation of 
all known higher order logarithmic integrals and has been found to improve 
the convergence of the massless perturbative series. 
This is also the case for Eq.~(\ref{eq_delta2}), however not solving
the convergence problem.
%With the above value for $\as$,
%the contour-improved evaluation of the perturbative series reads
%\beq
%\label{eq_convfoptci}
%   \delta^{(2,m_s)}_{S,{\rm CIPT}} = 
%     -\,8\frac{m^2_s(m_\tau^2)}{m_\tau^2}
%     \left(0.97 + 0.49 + 0.38 + 0.33 +\cdots \right)~,
%\eeq
%with somewhat improved but still unsatisfactory convergence.
Independently of whether FOPT or CIPT is used, the origin of the
convergence problem is found in the $J=0$ component as defined 
in Eq.~(\ref{eq_rtau}). 
%Using FOPT the series reads
%\beq
%\label{eq_cfoptci0}
%   \delta^{(2,m_s)}_S(J=0) = 
%     -\,8\frac{m^2_s(m_\tau^2)}{m_\tau^2}
%     \left(0.41 + 0.32 + 0.32 + 0.37 +\cdots \right)~,
%\eeq
%while the $J=1+0$ part converges well,
%\beq
%\label{eq_cfoptci10}
%   \delta^{(2,m_s)}_S(J=1+0) = 
%     -\,8\frac{m^2_s(m_\tau^2)}{m_\tau^2}
%     \left(0.56 + 0.16 + 0.05 - 0.05 +\cdots \right)~.
%\eeq
%Similar but worse convergence problems occur in the mass corrections
%to the $(k,\l)$ $J=0$ moments.

%\subsubsection{Phenomenological analysis}
%\label{analysis}

The phenomenological analysis is performed using the OPE framework
including the light quark masses and the $D=4$ quark-mass corrections.
Higher dimension terms are assumed to be numerically 
insignificant compared to the present experimental uncertainty 
so that for all practical purposes the strange quark mass can be
obtained from the relation
\beq
\label{mass}
m_s^2(m_\tau^2) \simeq \frac{m_\tau^2} {\Delta_{k\l}(a_s)}
 \left[ \frac{\delta R_\tau^{k\l}}{24 \Sew}
  + 2 \pi^2 \frac{\langle\delta O_4(m_\tau^2)\rangle}
		{m_\tau^4} Q_{k\l}(a_s)\right] \, , 
\eeq
where $\Delta_{k\l}(a_s)$ and $Q_{k\l}(a_s)$ are the
pQCD series, defined in~\cite{PP99}, associated with 
the $D=2$ and $D=4$ contributions to $\delta R_\tau^{k\l}$, and
\beqn\label{dim4}
\langle\delta O_4(m_\tau^2)\rangle &=& \langle 0| m_s\, \sbar s
  - m_d \,\dbar  d | 0 \rangle (m_\tau^2) \nonumber\\
&\simeq& - (1.5 \pm 0.4) \times 10^{-3}\gev^4 \, . 
\eeqn

As discussed above, the QCD series for 
$\Delta_{k\l}=\Delta_{k\l}^{(1+0)}+\Delta_{k\l}^{(0)}$ are
problematic, since they exhibit bad convergence originating from 
the longitudinal component 
($\Delta_{k\l}^{(0)}$)~\cite{MA98,PP98a,PP98b,PP98c,PP99}. 
The value of the strong coupling constant is taken to be
$\as(m_\tau^2)=0.334\pm0.022$, from the analyses of the nonstrange 
$V+A$ moments $R_\tau^{k\l}$. 
Correlations between $\as(m_\tau^2)$ and
the moments $\delta R_{k\l}$ are negligible since the 
uncertainty on the former value is dominated by theory and on the latter 
one by the measurement of the strange component.
Using these inputs, the pQCD series 
$\Delta_{k\l}$ can be displayed up to third order 
using CIPT, separating the $(1+0)$ and $(0)$ series
($a_s=0.106$):
%\beqn
%\label{pertseries}
%\Delta_{00}(a_s) &=& 0.9734 + 0.4811 + 0.3718 + 0.3371 +\dots
%+ \dots \nonumber \\ 
%\Delta_{10}(a_s) &=& 1.0390 + 0.5576 + 0.4820 + 0.4771 +\dots
%+ \dots  \nonumber \\
%\Delta_{20}(a_s) &=& 1.1154 + 0.6432 + 0.6082 + 0.6470 +\dots
%+ \dots  \nonumber \\
%\Delta_{30}(a_s) &=& 1.1990 + 0.7374 + 0.7516 + 0.8507 +\dots
%+ \dots  \nonumber \\
%\Delta_{40}(a_s) &=& 1.2880 + 0.8404 + 0.9142 + 1.0928 +\dots
%+ \dots \nonumber\\
%\eeqn
\beqn
\label{pertseries}
\Delta^{(1+0)}_{00}(a_s) &=& 0.565 + 0.161 + 0.049 - 0.038 
+ \cdots \nonumber \\ 
\Delta^{(1+0)}_{10}(a_s) &=& 0.684 + 0.251 + 0.144 + 0.042 
+ \cdots  \nonumber \\
\Delta^{(1+0)}_{20}(a_s) &=& 0.791 + 0.338 + 0.248 + 0.142 
+ \cdots  \nonumber \\
\Delta^{(1+0)}_{30}(a_s) &=& 0.893 + 0.428 + 0.363 + 0.264 
+ \cdots  \nonumber \\
\Delta^{(1+0)}_{40}(a_s) &=& 0.993 + 0.523 + 0.493 + 0.413 
+ \cdots \nonumber \\ \nonumber \\
\Delta^{(0)}_{00}(a_s) &=& 0.408 + 0.320 + 0.323 + 0.375 
+ \cdots \nonumber \\ 
\Delta^{(0)}_{10}(a_s) &=& 0.355 + 0.307 + 0.338 + 0.435 
+ \cdots  \nonumber \\
\Delta^{(0)}_{20}(a_s) &=& 0.324 + 0.305 + 0.360 + 0.505 
+ \cdots  \nonumber \\
\Delta^{(0)}_{30}(a_s) &=& 0.306 + 0.309 + 0.389 + 0.587 
+ \cdots  \nonumber \\
\Delta^{(0)}_{40}(a_s) &=& 0.295 + 0.317 + 0.421 + 0.680 
+ \cdots~.
\eeqn
The exhibited behavior is that of asymptotic series close to 
their point of minimum sensitivity and a prescription is needed 
to evaluate the expansions and to make a reasonable estimate of their
uncertainties. 

\subsection{Several approaches}

The bad convergence properties of the perturbative QCD expansion can 
be alleviated in several ways.

%\subsubsection{Summing up the full known terms}
%\label{fullFOPTCI}

For instance, one could try to work with the full series up to the highest 
known order (third), observing that the terms in the overall expansion show
a reasonable rate of decrease, at least in the CIPT approach. 
Fourth order coefficients can be estimated, using for instance the principle 
of minimal sensitivity~\cite{pms} or the simplistic geometric growth rule. 
The latter method was used in early analyses~\cite{CDH99a,CDH99b,CDH99c,CDH99d}. 

%\subsubsection{Optimal truncation of the series}
%\label{truncatedFOPTCI}

Examination of examples of asymptotic series suggests that a reasonable 
procedure is to truncate the expansion where the terms reach their minimum 
value. The precise prescription---cutting at the minimum, or one order 
before or after, including the full last term or only a fraction of it---is 
arbitrary and this ambiguity must be reflected by a specific uncertainty
attached to the procedure. In~\cite{DA:00,CDGHP2} the adopted rule 
was to keep all terms up to (and including) the minimal one and to assign 
as a systematic uncertainty the value of the last term retained. 
It follows that the $\Delta_{k0}$ series are summed up to third 
order for $k=0,1$, second order for $k=2$ and first order for $k=3,4$. 

%\subsubsection{Removing the longitudinal parts}
%\label{Lremoved}
%\subsubsection*{The kaon pole and other spin-0 components}
%\label{spin0_direct}

Since the culprit is the longitudinal expansion, it could be wise to remove 
the longitudinal piece beforehand from the inclusive data and to apply the 
QCD analysis only to the well-behaved $(1+0)$ part\footnote
{
	A drawback of this procedure is that it reduces the inclusiveness 
	of the data and hence makes the phenomenological analysis more
	vulnerable to quark-hadron duality violations.
}. 
This method was already used in~\cite{ALEPH:1999}, where it was
noted that the dominant contribution to the strange $J=0$ component is 
given by the kaon pole.
Other longitudinal contributions can be identified and estimated, giving
%{\it (i)} off-shell vector $\Kstar$ resonances can generate
%a $0^+$ component~\cite{finkescalar}, however too small to have 
%a significant effect in the analysis, and {\it (ii)} production of the
%scalar $\Kstar_0(1430)$ resonance. 
%The $\Kstar_0(1430)$ state decays almost exclusively into $K \pi$ with a
%branching fraction of $(93 \pm 10) \%$~\cite{Eidelman:2004}. A fit to the
%$K^- \pi^0$ mass distribution measured by ALEPH~\cite{ALEPH:1999} 
%results in the branching fraction 
%$\BR(\tau \to \Kstarb_0(1430)\nut)= (0.0 \pm 2.5) \times 10^{-4}$.
%Hence $0^+$ contributions are suppressed
%($< 1.6 \times 10^{-2}$) compared to the dominant $J^P=1^-$ production. 
%This is expected from chiral symmetry breaking, with $J=0$ contributions 
%reduced by $\sim (m_K / m_\tau)^4$ as discussed in~\cite{finkescalar}, not
%including the $\tau$ kinematic factor, which suppresses higher mass
%states.
%
%No direct estimate of extra $0^-$ components beyond the single $K$ can
%be made from the present data in the $\Kb \pi \pi$ modes. This is
%due to the lack of statistics and to the background subtraction, 
%especially in the high mass region where some evidence for 
%a broad pseudoscalar state at $1460\mev$ exists~\cite{Eidelman:2004}.
%A contribution (upper limit) equal to the $0^+$ one is assumed in this 
%case.
%
%The ALEPH evaluation of the $J=0$ contributions to $R_{\tau,S}$ gives 
\beqn
\label{eq_spin0K}
         R_{\tau,S}^{(0)}(K)&=& -0.00615 \pm 0.00026~, \\
\label{eq_spin0rest}
         R_{\tau,S}^{(0)}({\rm other}~0^-,0^+)&=& -0.0015 \pm 0.0015~,
\eeqn
leading to the result
\beq
\label{eq_delta000}
         \delta R_\tau^{00}(J=0) = 0.147 \pm 0.031~,
\eeq
where the $J=0$ contributions in the nonstrange part are assumed to be
saturated by the pion pole and have been computed
to be $-0.0078$. Similarly, one estimates the $J=0$ contributions 
for all ($k,\l$) moments and subtracts them from the measured moments. 

%\subsubsection*{Other spin-0 components: phenomenological parameterizations}
%\label{spin0_ph}

The more recent analyses rely on phenomenological parameterizations of
the additional scalar and pseudoscalar contributions~\cite{kambor01,jop02}. 
The $0^\pm$ spectral functions constructed in this way rely on not so
well-established resonances with poorly known couplings, however they 
probably provide a realistic level for these contributions. 
Even with their intrinsically large uncertainties they are an order 
of magnitude more precise than the experimental
estimate from ALEPH (Eq.(\ref{eq_spin0rest})). 
In the analysis of~\cite{GJP2S} 
the corresponding contribution to $R_{\tau,s}$ yields 
\beq
\label{eq_delta000_ph}
         \delta R_\tau^{00}(J=0,{\rm ph}) = 0.155 \pm 0.005~,
\eeq
where ``ph'' stands for phenomenological parameterizations. The quoted 
uncertainty in this method is clearly much smaller than in 
Eq.~(\ref{eq_delta000}). It reflects the fact that pieces of the spin-0
spectral functions can be reconstructed using known resonances.

\subsubsection{Verifying the stability of the results}
\label{sec:ms_running}
\begin{figure*}[t]
\epsfxsize12cm
\centerline{\epsffile{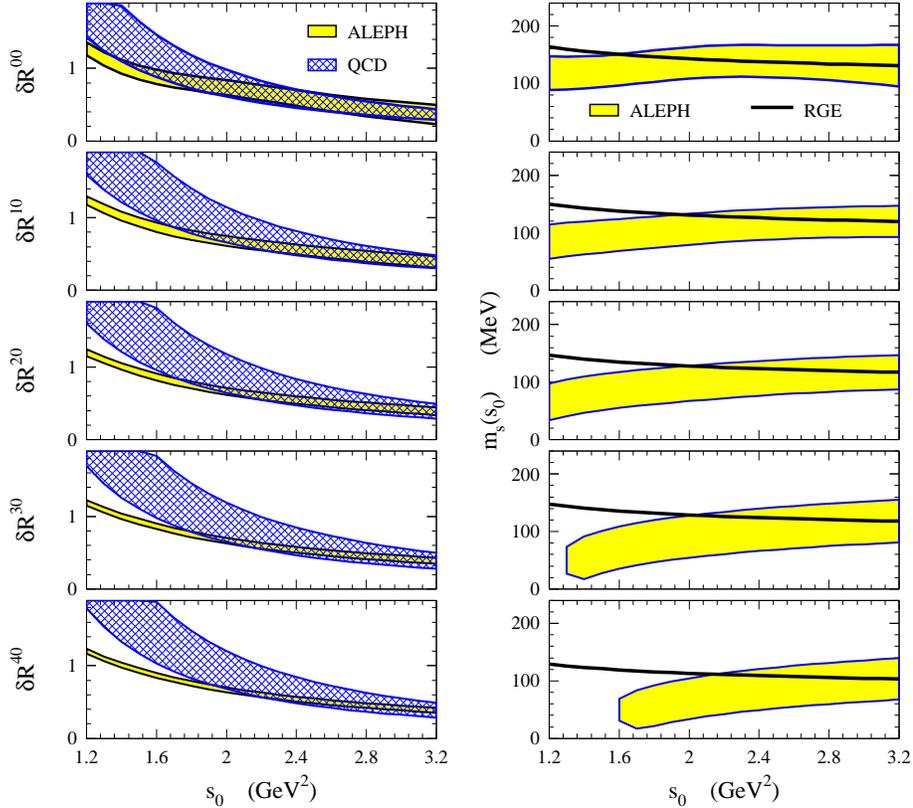}}
\vspace{-0.1cm}
\caption{The observables $\delta R^{k0}_\tau (s_0)$ as a function of
       	the ``$\tau$-mass''-squared $s_0$ confronted with the QCD 
	predictions (left plots), and the derived $m_s(s_0)$ compared 
	with the QCD RGE running (right plots). The bands correspond 
	to the experimental and theoretical uncertainties (left plots),
	and experimental errors only (right plots).
       	By construction, data and theory agree at $s_0=m_\tau^2$.
	The figure is taken from~\cite{CDGHP2}.
	}
\label{fig_running}
\end{figure*}

As shown in the analysis of the nonstrange $\tau$ 
decays in~\cite{aleph_asf}, one can test the validity of the QCD analysis
by simulating the physics of a hypothetical $\tau$ lepton of lower mass,
$\sqrt{s_0}\le m_\tau$. This is obtained by replacing $m_\tau^2$ by $s_0$ 
everywhere in Eqs.~(\ref{eq_moments}, \ref{eq_rtau}), and
correcting the latter equation for the modified kinematic 
factor. Under the assumption of quark-hadron
duality, the evaluation of the observables as function of $s_0$
represents a test of the OPE approach, since the energy dependence
of the theoretical predictions is determined once the parameters 
of the theory are fixed.
The results of this exercise are given in Fig.~\ref{fig_running}, showing
the variation with $s_0$ of the first four $\delta R^{k0}_\tau$ moments,
and the value for $m_s(s_0)$ derived from each moment. The bands indicate
the experimental and theoretical uncertainties. The agreement between data
and theory, perfect at $s_0=m_\tau^2$ by construction, remains acceptable
for lower $s_0$ values, down to $1.6\;(2.4)\gev^2$ for $k=0\;(4)$. For the 
first three moments used in the final determination, the running observed
in data follows the RGE evolution down to about $2\gev^2$. The 
validity of using higher moments thus appears more questionable. 
%It should 
%be pointed out that the $s_0$ dependence of the theoretical prediction is 
%obtained following the truncation method
%and applied at $s_0=m_\tau^2$. If the same rule had been consistently
%used at each $s_0$ point a better scaling agreement would have been 
%found, at the price of introducing steps in the prediction, corresponding 
%to dropped terms in the expansion (according to the truncation prescription). 
%This observation provides another consistency test of the procedure, with
%however some warning concerning the use of the high moments.

\subsubsection{Results and discussion}
\label{others}

Several analyses of the ALEPH strange spectral function have been performed
in order to extract $m_s(m_\tau^2)$. More recent analyses take advantage of 
new experimental input on the strange spectral function from~\cite{OPAL04} 
and improvements in branching fractions for some strange channels 
(see Table~\ref{brs}).
\bei
\item	In~\cite{ALEPH:1999}, 
	a rather conservative road was followed, using only the $1+0$ part 
	and a prudent estimate of the removed longitudinal part directly 
        from data. A price was paid in the 
	experimental sensitivity, leading to large errors.
	Also, the experimental moments were fit, not only to the strange quark 
	mass, but also to the values of the nonperturbative operators 
	$\langle\delta O_{6}\rangle$ and $\langle\delta O_{8}\rangle$.
	The value found by ALEPH is 
	$m_s(m_\tau^2) = (176 \left.^{+37}_{-48}\right|_{\rm exp}\,
		\left.^{+24}_{-28}\right|_{\rm th}
                  \pm 14_{\rm meth})\mev$.

\item  Using the truncation method with the $k=0,1,2$ moments~\cite{CDGHP2},
       the value $m_s(m_\tau^2) = (120 \pm 11_{\rm exp} 
       \pm 8_{|V_{us}|} \pm 19_{\rm th})\mev$ is found.

\item	The analysis of~\cite{KKP00} uses the ALEPH $\delta R^{00}_\tau$ moment
	and obtains $m_s(m_\tau^2) = (130 \pm 27_{\rm exp} \pm 9_{\rm th})\mev$. 
	It advocates contour-improved resummation and employs 
	an effective charge as well as effective masses absorbing the 
	higher perturbative terms. The small quoted theoretical uncertainty
	is not exhaustively motivated in their paper, and apparently no
	uncertainty is included from $|V_{us}|$, which should be of the 
	order of $\pm 13\mev$.

\item	Another analysis~\cite{KM00} using the ALEPH data makes use of
	weight functions multiplying the correlators in Eq.~(\ref{eq_rtau}). 
	These weights are designed to improve the convergence of the 
	perturbative series, while suppressing the less accurate high-mass 
	part of the strange spectral function. This latter feature is 
	similar to using higher moments in $k$ and it is not entirely 
	clear to what extent this procedure provides stable results and 
	hence how reliable the answer is. 
	Nevertheless that result is in good agreement with other results,
 	with a smaller theoretical uncertainty, but apparently not including 
	the effect from the renormalization scale ambiguity. 

\item	In~\cite{GJP2S}, the longitudinal part is subtracted 
	using a phenomenological prescription. The result
	$m_s(m_\tau^2) = (107 \pm 18)\mev$ is obtained by taking the
	weighted average of the $m_s$ values from $k=0,\dots,4$ and $\l=0$ moments.
	This procedure is statistically hardly justified as the experimental
	moments are strongly correlated. The result is more precise than 
	other determinations, however a large systematic
	trend is observed for different $k$ values: the $m_s$ central values 
	decrease monotonically from 151 for $k=0$ to 91\mev for $k=4$. 
	No additional systematic uncertainty is included to account for this
	effect, which is larger than the final quoted error, dominated by the
	experimental error and the uncertainty on $|V_{us}|$. The updated
        value using the OPAL strange spectral function with world-average
        branching ratios yields~\cite{GJP2S_v} a value 
        $m_s(m_\tau^2) = (84 \pm 23)\mev$.

\item	Another analysis~\cite{baikov} uses the full perturbative series for
	the $1+0$ correlator up to third order and an estimate of the fourth 
	order contributions, while the longitudinal part is subtracted using
	the same phenomenological model as in~\cite{GJP2S}. 
	The result, based on the $k=3$ $\l=0$ moment, is 
	$m_s(m_\tau^2) = \left(100\left.^{\,+5}_{\,-3}\right|_{\rm th}\,
		\left.^{\,+17}_{\,-19}\right|_{\rm rest}\right)\mev$. 
	Here the first (quite small) error shall cover the uncertainties 
	on the perturbative treatment, while the remaining error is dominated
	by the experimental and $|V_{us}|$ uncertainties. The same
	systematic trend is observed with $m_s(m_\tau^2)$ values ranging from 
	126\mev for $k=2$, to 82\mev for $k=4$.
\eei

The determinations of $m_s$ from $\tau$ decays compare reasonably well to 
the results of analyses of the divergence of the vector and axial-vector 
current two-point function correlators~\cite{gasser,jamin1,chetyrkin1,becchi,dominguez1,dominguez2,jamin2,kataev,colangelo,narison1,chetyrkin2}. 
In these approaches the phenomenological information on the associated 
scalar and pseudoscalar spectral functions is reconstructed from 
phase-shift resonance analyses, which are yet incomplete over the 
considered mass range and need to be supplemented by other assumed 
ingredients, in particular the description of the continuum.

Another approach~\cite{narison3} considers the difference between
isovector and hypercharge vector current correlators as related
to the $I=1$ and $I=0$ spectral functions accessible in $e^+e^-$
annihilation into hadrons at low energy. The author obtains
$m_s(1\gev^2) = (198 \pm 29)\mev$. This approach was criticized 
in~\cite{maltman2}, pointing out the possibility of large isospin
breaking leading to a significant deviation for the extracted
$m_s$ value. Subsequently in~\cite{narison4}, new ${\rm SU}(3)$-breaking
sum rules much less affected by ${\rm SU}(2)$ breaking were studied, 
with the result $m_s(1\gev^2) = (178 \pm 33)\mev$.

Finally, full lattice QCD calculations of $m_s$ including dynamical quark
actions have become available recently. A pioneering analysis using the 
so-called {\em staggered} fermion approach~\cite{lattice:aubin04} finds in 
the \MSbar scheme $m_s(4\gev^2)=(76\pm8)\mev$. The quoted error is dominated 
by the uncertainty in the perturbative theory, needed to obtain the 
\MSbar mass from the pole mass, itself calculated from the lattice 
bare mass. The included systematic uncertainty from the simulation 
on the lattice is quoted to be 3\mev, while the statistical 
error is negligible. Other computations using $N_f=2$ Wilson fermions
find the significantly larger values of 
$m_s(4\gev^2)=(119\pm10)\mev$~\cite{gockeler} and
$m_s(4\gev^2)=(116\pm6)\mev$~\cite{becirevic}.

\begin{figure}[t]
\epsfxsize8.5cm
\centerline{\epsffile{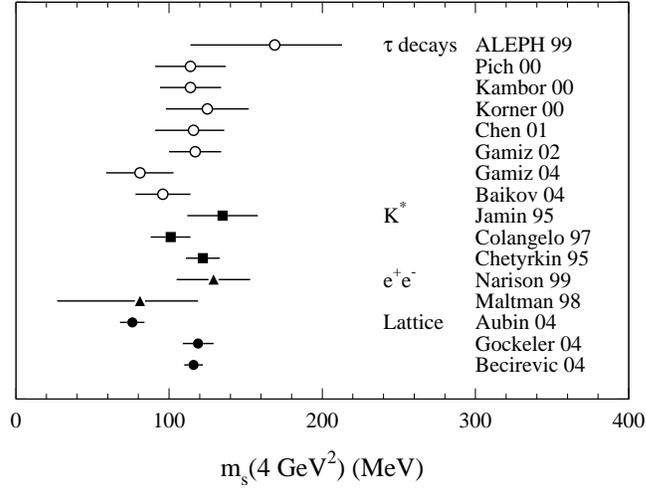}}
\vspace{-0.1cm}
\caption{Determinations of the \MSbar mass $\mbar_s(4\gev^2)$ based on 
	$\tau$ spectral functions and on other 
	approaches: $K^*$ and \ee sum rules, lattice calculations.}
\label{ms_res} 
\end{figure}

Available results on $m_s$ are displayed in Fig.~\ref{ms_res}. They
are scaled for convenience at a mass of $2\gev$, using 
the four-loop RGE $\gamma$-function~\cite{ritmass}. One observes a
scatter among the values that exceeds the quoted uncertainties
in some cases.

\subsection{Sensitivity to $|V_{us}|$ and concluding remarks on the $m_s$ 
determinations}

Before concluding the discussion of strange $\tau$ decays, it is useful 
to combine the available experimental information. Taking detector-corrected 
experimental mass spectra from~\cite{ALEPH:1999}, the world average
branching fractions for the strange final states (Table~\ref{brs}) and
the value~(\ref{k_uni}) for the $\taum \to \Km \nut$ branching 
ratio, an improved strange spectral function can be constructed.
We shall use it in this last section.

\begin{figure}[t]
\epsfxsize9.0cm
\centerline{\epsffile{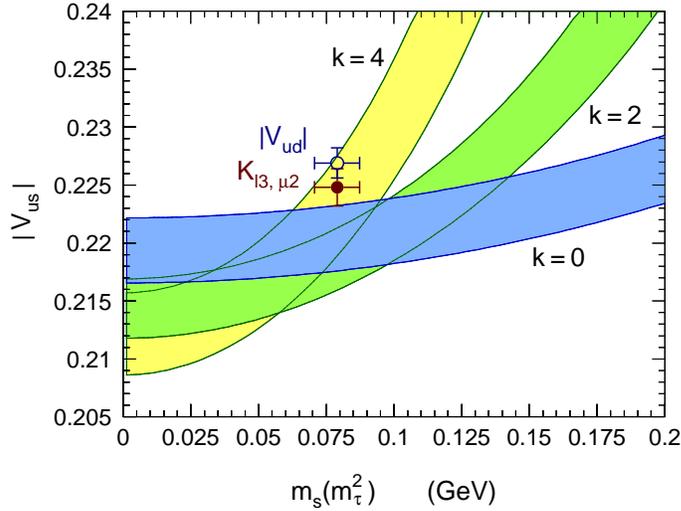}}
\vspace{-0.3cm}
\caption{The values of $|V_{us}|$ and $m_s(m_\tau^2)$ derived from the experimental
	strange and nonstrange moments obtained from the ALEPH 
        spectral functions normalized to the world average branching
        ratios. The various curves correspond to the different
	$k,\l=0$ moments with $k$ from 0 to 4. The width of the bands 
	indicates the present experimental accuracy. The longitudinal 
	components are subtracted using the $K$ and $\pi$ branching 
	fractions together with $\tau$--$\mu$ universality, and the 
	phenomenological contributions estimated in~\cite{GJP2S}. 
	The data points indicate the $|V_{us}|$ and $m_s(m_\tau^2)$ 
	results from outside $\tau$ physics. The two $|V_{us}|$ values
	shown correspond to the average of the $K_{\l3}$ and $K_{\mu2}$ 
	results (full circle), and to the $\sqrt{1-|V_{ud}|^2}$ 
	unitarity-based number (open circle)~\cite{isidori_ckm05}.}
\label{vus_ms} 
\end{figure}

Besides the strange quark mass, the detailed analysis of the strange
$\tau$ width can bring independent information on the CKM matrix element
$|V_{us}|$. In fact the present uncertainty on $|V_{us}|$ is a major
contribution to the final error on $m_s$. One can turn the argument around 
and consider the possibility that data on strange $\tau$ decays could 
ultimately provide a powerful determination of $|V_{us}|$~\cite{GJP2S,GJP2S_v}.
Figure~\ref{vus_ms} illustrates the method: the bands show the regions 
in the $|V_{us}|$--vs.--$m_s(m_\tau^2)$ plane which are allowed by the different
moments $\delta R_\tau^{k0}$ for $J=1+0$ (\ie, longitudinal part 
excluded) within their total (experimental and theoretical) errors. 
It is found that the various moments are sensitive to $|V_{us}|$ and $m_s$ 
in a different way. While higher-$k$ moments exhibit a strong $m_s$
dependence, $\delta R_\tau^{00}$ is particularly sensitive to
$|V_{us}|$, almost independently of $m_s$. This property can be exploited 
to achieve a determination of $|V_{us}|$ from $\tau$ decays. Since the
$\tau$ $SU(3)$ breaking relation~(\ref{eq_dmoments}) involves both 
$|V_{ud}|$ and $|V_{us}|$, unitarity of the CKM matrix is used 
($|V_{ub}|$ being negligible) and only $|V_{us}|$ is kept as a variable. 

Using as an independent determination the $m_s$ value from recent lattice
calculations~\cite{lattice:aubin04} evolved to the $\tau$ scale,
$m_s(m_\tau^2)=(79\pm8)\mev$, one gets from $\delta R_\tau^{00}$ alone
\beq
\label{eq:ourVus}
 |V_{us}|=0.2204 \pm 0.0028_{\rm exp} \pm 0.0003_{\rm th} \pm 0.0001_{m_s}~, 
\eeq
where the quoted errors are from the $\tau$ data (dominantly from the
strange channels), the theory-at-large ($\Sew$, $\delta O_4$, $f_\pi$, 
$f_K$, \as, $\m_\tau$, the phenomenological longitudinal part), and $m_s$.
The theoretical error is strongly reduced because the $(1+0)$ expansion
of the (0,0) moment is well behaved, in contrast to the higher moments.
Thus the uncertainty on the result is dominated by the
experimental error on the branching ratios for strange hadronic final
states.

The present result is not yet competitive with the best determinations
of $|V_{us}|$, but it is approaching these. Although it could be a statistical
fluctuation, the fact that the result~(\ref{eq:ourVus}) is on the low side 
of the recent results from $K_{\l3}$ and $K_{\mu2}$ decays may be worrisome. 
It could indicate that not all $\tau$ strange final states have been identified.
In this context it is important to revisit in particular the branching ratio 
for the $\nut K^- \pi^+ \pi^-$ mode where the agreement is marginal 
between ALEPH on the low side, and CLEO and OPAL on the high side 
(Section~\ref{sec:brs_k}).

One could forecast a significant improvement with the analyses of high
statistics $\tau$ samples available at the \B Factories. The detectors at 
these facilities (\babar\ and Belle) are equipped with powerful Cherenkov 
counters, well suited to separate kaons and pions in a large momentum 
domain, and one should expect significant improvement in the measurement
of the small, statistics-limited strange branching fractions.
Then it is conceivable that a competitive determination of $|V_{us}|$ can
be obtained. 

\begin{figure}[t]
\epsfxsize9.0cm
\centerline{\epsffile{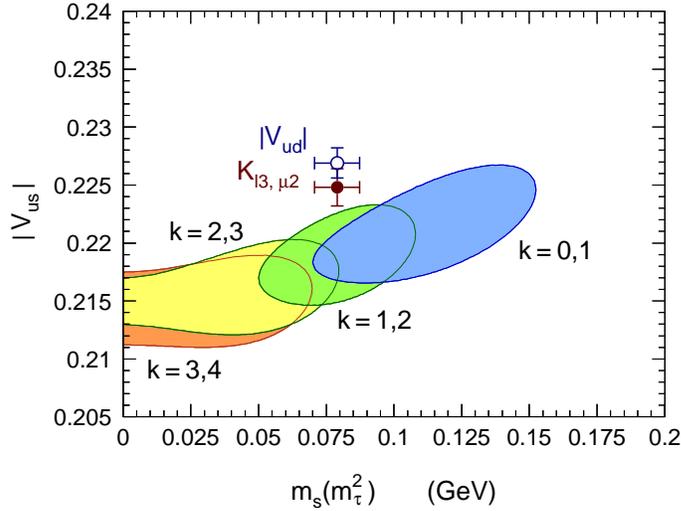}}
\vspace{-0.3cm}
\caption{The 68\% CL contours in the $|V_{us}|$--vs.--$m_s(m_\tau^2)$ plane from 
        the fit of different pairs of moments derived from the experimental
	strange and nonstrange ALEPH 
        spectral functions normalized to the world average branching
        ratios. The longitudinal components are subtracted 
        using the $\tau$--$\mu$ universality-improved $K$ and $\pi$ 
	branching fractions, and the phenomenological contributions estimated 
        in~\cite{GJP2S}. The results exhibit a systematic trend with the orders
        of the pair chosen. The data points show the best present knowledge
        of $|V_{us}|$ (the open circle determined using unitarity) and 
	$m_s(m_\tau^2)$ from outside $\tau$ physics.The two $|V_{us}|$ values
	shown correspond to the average of the $K_{\l3}$ and $K_{\mu2}$ 
	results (full circle), and to the $\sqrt{1-|V_{ud}|^2}$ 
        unitarity-based number (open circle)~\cite{isidori_ckm05}.}
\label{vus_ms_contours} 
\end{figure}

We now turn back to the $m_s$ determination. In fact, owing to the
availability of the different moments, it is possible to fit them
simultaneously and to derive both $|V_{us}|$ and $m_s$. In doing so it is
mandatory to take into account the strong correlations between the
moments. With the present best data it seems not possible to achieve 
a consistent description of all five moments
($k=0-4$) with a unique solution for $|V_{us}|$ and $m_s$. The situation
is best illustrated in Fig.~\ref{vus_ms_contours} showing the 68\% CL
contours when only pairs of moments are considered in the fit. Typical 
examples are given with the following ``adjacent'' $k$-pairs: going from 
$k=0,1$ through $k=3,4$ pairs clearly selects different regions in the 
plane. The situation is worse than it looks at first sight since the 
moments are highly correlated. Indeed, the correlation coefficient ranges 
from 0.68 ($k=0-4$) to 0.99 ($k=2-3$, $3-4$). A fit of the three $k=0,2,4$ 
moments (adding more does not extract any new information from the data) 
yields a $\chi^2$ of 6.4 for 1 DF, corresponding to a goodness-of-fit 
probability of only 1\%.

We are afraid that the pattern emerging from Fig.~\ref{vus_ms_contours}
undermines the reliability in the previous findings of $m_s$ from $\tau$ 
decays, at least within their quoted errors. The more recent 
determinations~\cite{GJP2S_v,baikov} have focused on the higher $k=2,3,4$
moments (only the $k=3$ moment is used in~\cite{baikov}) since they are 
more sensitive to $m_s$ and they used a world average 
value for $|V_{us}|$. However, Fig.~\ref{vus_ms_contours} shows that the 
data prefer a lower value, not consistent with the world average. This
might be the result of the worse behavior of the perturbative 
expansion of the higher moments as detailed in Eq.~(\ref{pertseries}),
even for the $J=1+0$ part. Another sign of potential problems is found in 
the $m_s$ running study as function of $s_0$ in Section~\ref{sec:ms_running}, 
where the stability of the results for the higher moments could not be safely
established.

It may be that some of these alarming features will disappear with better
experimental data, especially if systematic deviations are found. More 
precise data from the \B Factories on the multi-hadronic strange final
states will certainly add valuable information to this
problem. As for the moment, whereas the $|V_{us}|$ measurement from
the $k=0$ SU(3) breaking moment appears reasonably safe, the $m_s$ 
determination is unfortunately plagued with systematic effects and 
does not appear to be sufficiently robust for a precise and 
model-independent measurement.

\section{CHIRAL SYMMETRY AND QCD SUM RULES}
\label{sec:sumrules}

In the limit of $n_f$ massless quarks ($m_i=0$, $i=u,d,\dots$), the QCD 
Lagrangian possesses a global ${\rm SU}_L(n_f)\times {\rm SU}_R(n_f)$ chiral symmetry 
between the left- and right-handed quarks in flavor space. The associated 
conserved Noether currents are the vector ($V^\mu$) and axial-vector ($A^\mu$) 
quark currents. The vector and axial-vector charges
\beq
   Q_{V/A} = \int d^3x\,V^0/A^0(x)~,
\eeq
are the generators of the symmetry group. For a state $|\phi\rangle$, which 
is symmetric under ${\rm SU}(3)_L\times {\rm SU}(3)_R$, one then must 
have~\cite{gasser}\footnote
{  
   The conserved right- ($R=V+A$) and left-handed ($L=V-A$) 
   currents under ${\rm SU}(3)_L\times {\rm SU}(3)_R$ transform like (\underline{8},
   \underline{1}) and (\underline{1}, \underline{8}), respectively.
   If $|\phi\rangle$ is invariant then $\langle\phi|RL|\phi\rangle=0$
   $\Rightarrow$ $\langle\phi|V^2-A^2|\phi\rangle=0$, which 
   is Eq.~(\ref{eq:gasser}).
}
\beq
\label{eq:gasser}
   \langle\phi|A_\mu^a(x)A_\mu^b(y)|\phi\rangle =
   \langle\phi|V_\mu^a(x)V_\mu^b(y)|\phi\rangle~~~~~~(a,b=,1,\dots,8)~,
\eeq
which requires that for every contribution on the r.h.s.\ of Eq.~(\ref{eq:gasser})
($J^P=0^+$ or $1^-$) there exists a mass degenerate partner on the l.h.s.\
($J^P=0^-$ or $1^+$). However, chiral symmetry, which should be a good symmetry 
for the light $u,d,s$ quarks, is not observed in the low energy hadronic spectra 
of \ee annihilation or $\tau$ \sfs. In order to be consistent with this 
experimental fact, the chiral flavor group ${\rm SU}(3)_L\times {\rm SU}(3)_R$ is 
assumed to be spontaneously broken down to ${\rm SU}(3)_{L+R}$ where the vacuum 
expectation values are symmetric
\beq
   \langle \ubar u \rangle = 
   \langle \dbar d \rangle = 
   \langle \sbar s \rangle \ne 0~.
\eeq

The spontaneously breaking of the axial charge symmetry generates an octet 
of massless pseudoscalar Goldstone mesons~\cite{goldstone1,goldstone2,nambu}, 
which is identified with the 8 lightest hadronic states: $\pi^+$, $\pi^0$, 
$\pi^-$, $\eta$, $K^+$, $K^-$, $\Kz$ and $\Kzb$. For non-zero quark masses 
the axial-vector current is not conserved and its divergence reads
\beq
\label{eq:mpi1}
   \partial_\mu A^\mu_{ij} =
        (m_i+m_j)\qbar_i(i\gamma_5)q_j~,
\eeq
which is associated with the decay constant $f_{\!P}$ of a pseudoscalar meson 
$P$ with four momentum $q_\mu$, defined as
\beq
\label{eq:deffpi}
   \langle0|\partial_\mu A^\mu_P|P\rangle =
       \sqrt{2}f_{\!P}m_P^2~~~~{\rm or:}~~~
   \langle0|A^\mu_P|P\rangle =
       \sqrt{2}f_{\!P}q^\mu.
\eeq
The non-vanishing physical masses of the pseudoscalars reflect the fact that
chiral symmetry is not exact in QCD. True zero $u,d,s$ quark masses would
generate massless pseudoscalar mesons. Explicit mass relations between the
pseudoscalars can be deduced from Chiral Perturbation Theory 
(\cf\ Section~\ref{sec:invmomentssumrules}).
There is no indication of spontaneously breaking of the vector 
charge symmetry (the vector charge operators annihilate the vacuum): no light 
scalars have been found experimentally, and vector mesons can be 
classified in degenerate multiplets, which are ${\rm SU}(3)_V$ representations. 
Hence, the longitudinal part of the vector correlator in Eq.~(\ref{eq:correlator}) 
vanishes in the chiral limit, $\Pi^{(0)}_{ij,V}(q^2)=0$.

Reducing chiral symmetry to the nonstrange sector, \ie, to 
${\rm SU}(2)_L\times {\rm SU}(2)_R$, one has the Lorentz-decomposed two-point 
correlation function\beqn
\label{eq:lrcorrelator}
   \Pi_{{\ubar}d,LR}^{\mu\nu}(q) &\equiv &
      i\int d^4x\,e^{iqx}
      \langle 0|T(L^\mu(x)R^\nu(0)^\dag)|0\rangle \nonumber\\
                          &  =    &
      (-g^{\mu\nu}q^2 + q^\mu q^\nu)\,\Pi^{(1)}_{ud,LR}(q^2)
      \,+\,q^\mu q^\nu\,\Pi^{(0)}_{ud,LR}(q^2)~,
\eeqn
where $L^\mu$ and $R^\mu$ are the left- and right-handed quark currents
\beq
\label{eq:lrcurrents}
    L^\mu = \ubar\gamma^\mu(1-\gamma_5)d~,~~~~~
    R^\mu = \ubar\gamma^\mu(1+\gamma_5)d~.
\eeq
%zz SpectralChiral
With Eqs.~(\ref{eq:lrcorrelator}) and (\ref{eq:lrcurrents}) 
and the correlators for vector and axial-vector currents~(\ref{eq:correlator}) 
one finds
\beq
   \Pi_{\ubar d,LR}^{\mu\nu}(q) = 
       \Pi_{\ubar d,V}^{\mu\nu}(q) \,-\, 
           \Pi_{\ubar d,A}^{\mu\nu}(q)~.
\eeq
In the chiral limit and for $q^2\rightarrow\infty$, the correlator 
$\Pi_{ud,LR}^{\mu\nu}(q)$ vanishes (which again implies the 
degeneracy of Eq.~(\ref{eq:gasser})). With the use of a dispersion 
relation and from the comparison of the $q^\mu q^\nu$ and the $q^2$ 
terms in Eq.~(\ref{eq:lrcorrelator}), one obtains the two Weinberg 
sum rules (WSR) for $u,d$ quark correlators~\cite{wsr}
\beqn
\label{eq:wsr1}
 {\rm 1.~WSR:} & &  \intl_{0}^{s_0\to\infty}ds\,
 {\rm Im}\left[\Pi^{(1)}_{\ubar d,V}(s) - \Pi^{(1)}_{\ubar d,A}(s) 
               + \Pi^{(0)}_{\ubar d,V}(s) - \Pi^{(0)}_{\ubar d,A}(s)
         \right] = 0~, \\
\label{eq:wsr2}
 {\rm 2.~WSR:} & &  \intl_{0}^{s_0\to\infty}ds\,s\,
 {\rm Im}\left[\Pi^{(1)}_{\ubar d,V}(s) - \Pi^{(1)}_{\ubar d,A}(s)
         \right] = 0~.
\eeqn
The upper integration bound $s_0$ is finite when integrating
over experimental spectral functions, and the missing saturation of 
the integral at the cut-off is a source of systematic uncertainty.
The first WSR can be simplified using the pion decay constant \fpi
defined in Eq.~(\ref{eq:deffpi}), which fixes the integral 
$\int\! ds\,{\rm Im}\Pi^{(0)}_{\ubar d,A}(s)=\fpi^2$, and the 
fact the longitudinal vector correlator vanishes. Defining
\beq
	\rhoVmA(s) = v_1(s) - a_1(s)~,
\eeq
the two Weinberg sum rules read
\beqn
\label{eq:wsr1bis}
 {\rm 1.~WSR:} & &  \frac{1}{4\pi^2}\hsm\intl_{0}^{s_0\to\infty}ds\,\rhoVmA(s) 
			= \fpi^2~, \\
\label{eq:wsr2bis}
 {\rm 2.~WSR:} & &  \frac{1}{4\pi^2}\hsm\intl_{0}^{s_0\to\infty}ds\,s\,\rhoVmA(s) 
  = 0~.
\eeqn
In the presence of 
non-zero quark masses, only the first WSR remains valid while the second 
WSR breaks down due to contributions from the difference of non-conserved 
currents of order $m_q^2/s$, leading to a quadratic 
divergence of the integral~\cite{floratos}. However, convergence can be
recovered by considering a {\em Borel-transformed} (also denoted 
{\em Laplace-transformed})
version of the second WSR (see Section~\ref{sec:borelsumrules}), 
where the result is expressed as a function of the Borel parameter $M^2$ 
and the $u,d$ running quark masses at scale $M^2$ and quark condensates 
in powers of $1/M^0$, $1/M^2$, $\dots$~\cite{floratos,svz,peccei}.

Using a coarse narrow-width approximation for the resonances and a saturation 
of the corresponding vector, axial-vector \sfs\  by the $\pi$, $\rho(770)$ 
and $a_1(1260)$, Weinberg deduced from these sum rules the mass formula~\cite{wsr}
$m_{a_1}/m_\rho \simeq 1.41$, which compares not too badly with the 
experimental value of $1.60\pm0.05$~\cite{Eidelman:2004}.

Das, Mathur and Okubo (DMO)~\cite{dmo} showed that the derivative of the 
vector minus axial-vector correlator $\Pi_{\ubar d,V-A}(q^2)$ taken at $q^2=0$
is connected with the radiative $\pi^-\rightarrow \ell^-\nub_\ell\gamma$
decay axial-vector form factor $F_A$ via\footnote
{
   In cases where the radiated photon is real the differential
   decay rate can be written as
   \beq
      \frac{d^2\Gamma_{\pi\rightarrow \ell\nu\gamma}}
           {dE_\gamma dE_\ell} =
      \frac{d^2(\Gamma_{\rm IB}+\Gamma_{\rm SD}+\Gamma_{\rm INT})}
           {dE_\gamma dE_\ell}~,
   \eeq
   where $\Gamma_{\rm IB}$, $\Gamma_{\rm SD}$ and $\Gamma_{\rm INT}$
   are the contributions from inner bremsstrahlung, structure-dependent
   radiation and their interference. The structure-dependent term
   is parameterized by using vector and axial-vector form factors
   $F_{V/A}$ (for the complete expression see Ref.~\cite{Eidelman:2004}).
}
\beq
\label{eq:dmo}
   \frac{d(q^2\Pi_{\ubar d,V-A}(q^2))}{dq^2}\bigg|_{q^2=0} =
       \frac{1}{4\pi^2}\hsm\intl_0^{s_0\to\infty} ds\frac{1}{s}\,\rhoVmA(s)
         = \fpi^2\frac{\rpisq}{3} - F_A~,
\eeq
with the pion charge radius-squared $\rpisq$ given in 
Eq.~(\ref{eq:pionchargeradius}). 
This relation has been used in the past to perform a finite energy 
sum-rule determination of the electric polarizability\footnote
{
	The electric polarizability $\alpha_E$ of a physical system 
	can be classically understood as the proportionality constant 
	that governs the induction of the dipole moment $\bf p$ of a 
	system in presence of an external electric field vector 
	$\bf E$: ${\bf p}=\alpha_E \bf E$. The polarizability is an 
	important quantity to characterize a particle, \ie, in probing 
	its inner structure. In chiral perturbation theory, it can be
	shown~\cite{elpol1,elpol2} that the polarizability of the pion is given 
	by $\alpha_E=\alpha F_A/(m_\pi\fpi^2)$, which gives~\cite{CDH99a} 
	$\alpha_E = (2.86\pm0.33)\times 10^{-4}\;{\rm fm}^3$.
	Using the value~(\ref{eq:pionchargeradius}) for the pion charge 
	radius-squared, the analysis of a Borel-improved 
	DMO sum rule yields~\cite{CDH99a} 
	$\alpha_E = (2.96\pm0.32)\times 10^{-4}\;{\rm fm}^3$.
} 
of the pion~\cite{polarizability1,CDH99a,margvpol,menke_tau}. Equation~(\ref{eq:dmo})
is an example for an {\em inverse moment spectral sum rule}. It is special
because the integration kernel is singular at $s=0$ so that its residue 
is sensitive to the chiral properties of the spectral functions
(\cf\  Section~\ref{sec:invmomentssumrules}).

The electromagnetic mass splitting between the charged and the neutral 
pion has been calculated using current algebra techniques in chiral symmetry 
by Das \ea~\cite{das}. They derived the Das-Guralnik-Low-Mathur-Young (DGLMY) 
sum rule in the chiral limit
\beqn
\label{eq:das}
      16\pi^2 i\int\frac{d^4q}{(2\pi)^4}
      \frac{1}{q^2}\intl_{0}^{s_0\to\infty} ds\,s\,
      \frac{{\rm Im}\left[\Pi^{(1)}_{\ubar d,V}(s) 
                          - \Pi^{(1)}_{\ubar d,A}(s)\right]}
           {q^2+s-i\e}
	&=&
   \frac{1}{4\pi^2}\intl_0^{s_0\to\infty}\hsm ds\,s\,
            {\mathrm{ln}}\frac{s}{\Lambda^2}\,\rhoVmA(s)\nonumber\\
        &=&
   -\frac{4\pi f_0^2}{3\alpha}\left(m_{\pi^\pm}^2 \,-\, m_{\pi^0}^2\right)~,
\eeqn
where 
$f_0 = f_\pi(1-13m_\pi^2/(192\pi^2\fpi^2)-m_\pi^2\rpisq/6+{\cal O}(m_\pi^4))
=(87.6\pm0.3)\mev$
corresponds to $\fpi=(92.4\pm0.1\pm0.3)\mev$~\cite{Eidelman:2004}, 
expressed in the chiral limit ($m_u=m_d=0$, $m_s\ne0$)~\cite{moussallam}.
Making use of the KSFR relation~\cite{ksfr1,ksfr2} and lowest 
resonance saturation, they obtained the relation
\beq
   m_{\pi^\pm}^2 \,-\, m_{\pi^0}^2 =
      \frac{3\alpha m_\rho^2\,{\rm ln}2}{4\pi m_\pi} =
      5.1\mev~,
\eeq
in approximate agreement with the experimental value of
$4.59\mev$~\cite{Eidelman:2004}. Although apparently containing an 
arbitrary energy scale $\Lambda$, the sum rule~(\ref{eq:das}) is actually 
independent of $\Lambda$ by virtue of the second WSR. However, 
for finite $s_0$ a residual $\Lambda$ dependence is maintained. 

\subsection{Chiral sum rules with $\tau$ spectral functions}

\begin{figure}[t]  
  \centerline{\epsfysize6.3cm\epsffile{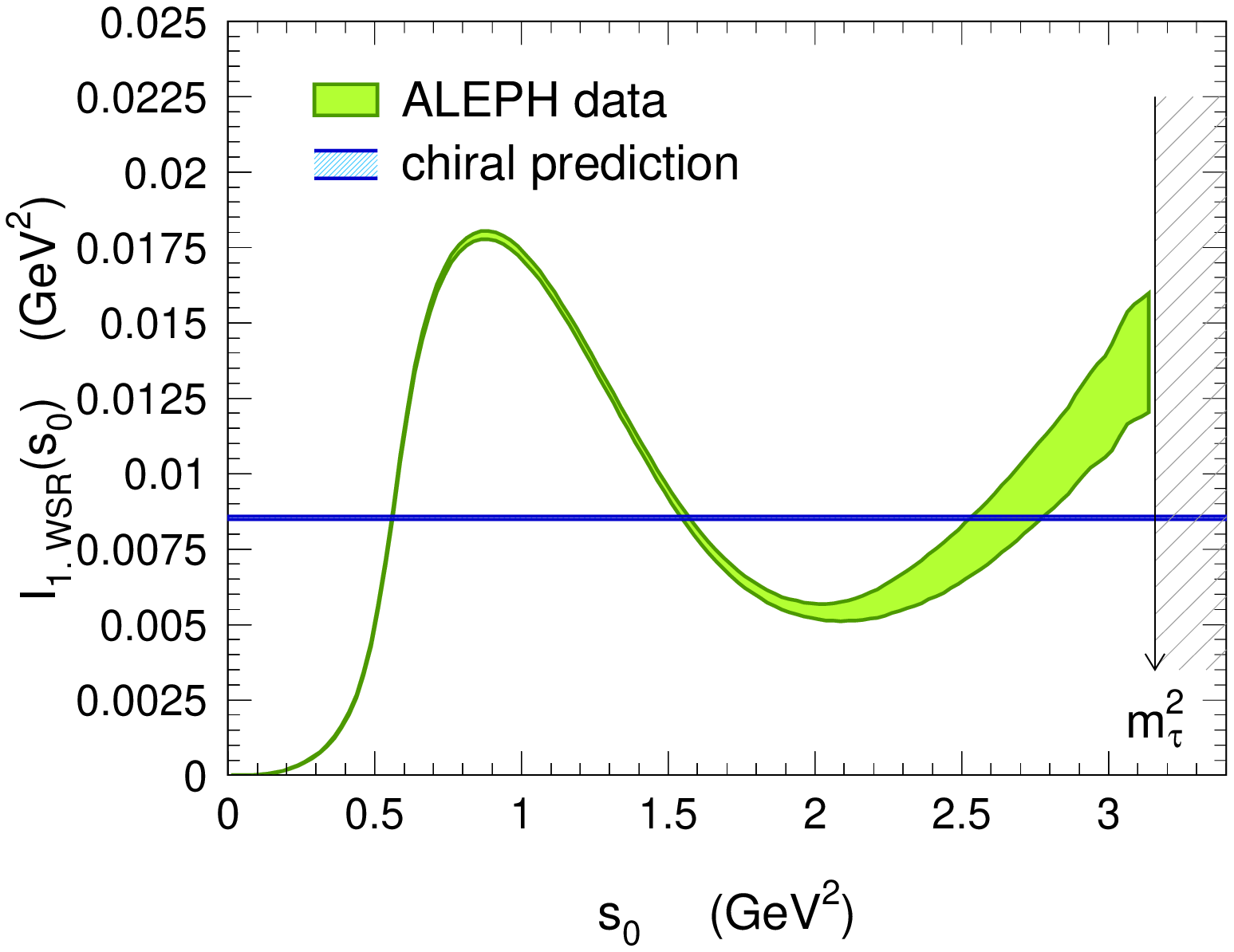}\hspace{0.5cm}
              \epsfysize6.3cm\epsffile{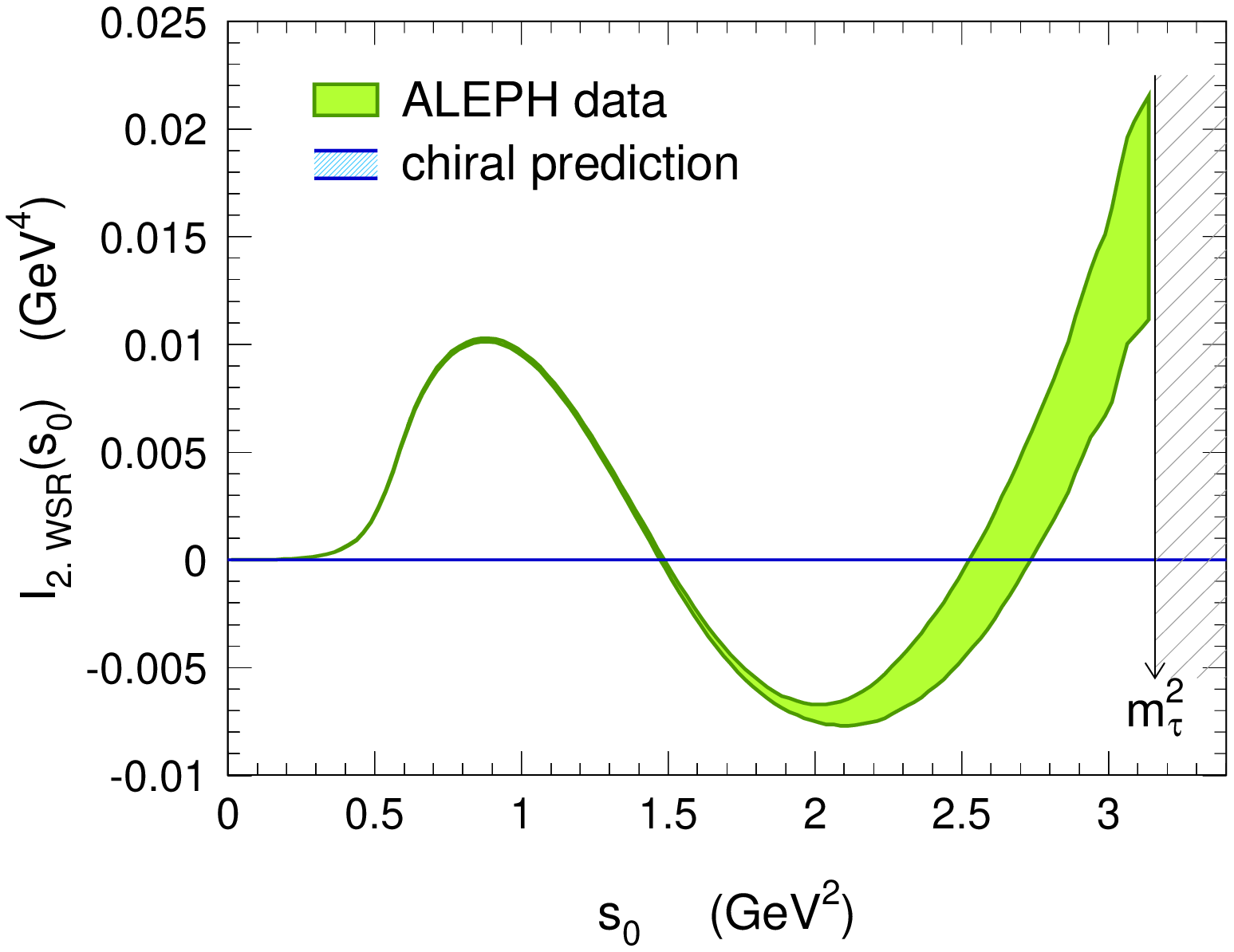}}
  \centerline{\epsfysize6.3cm\epsffile{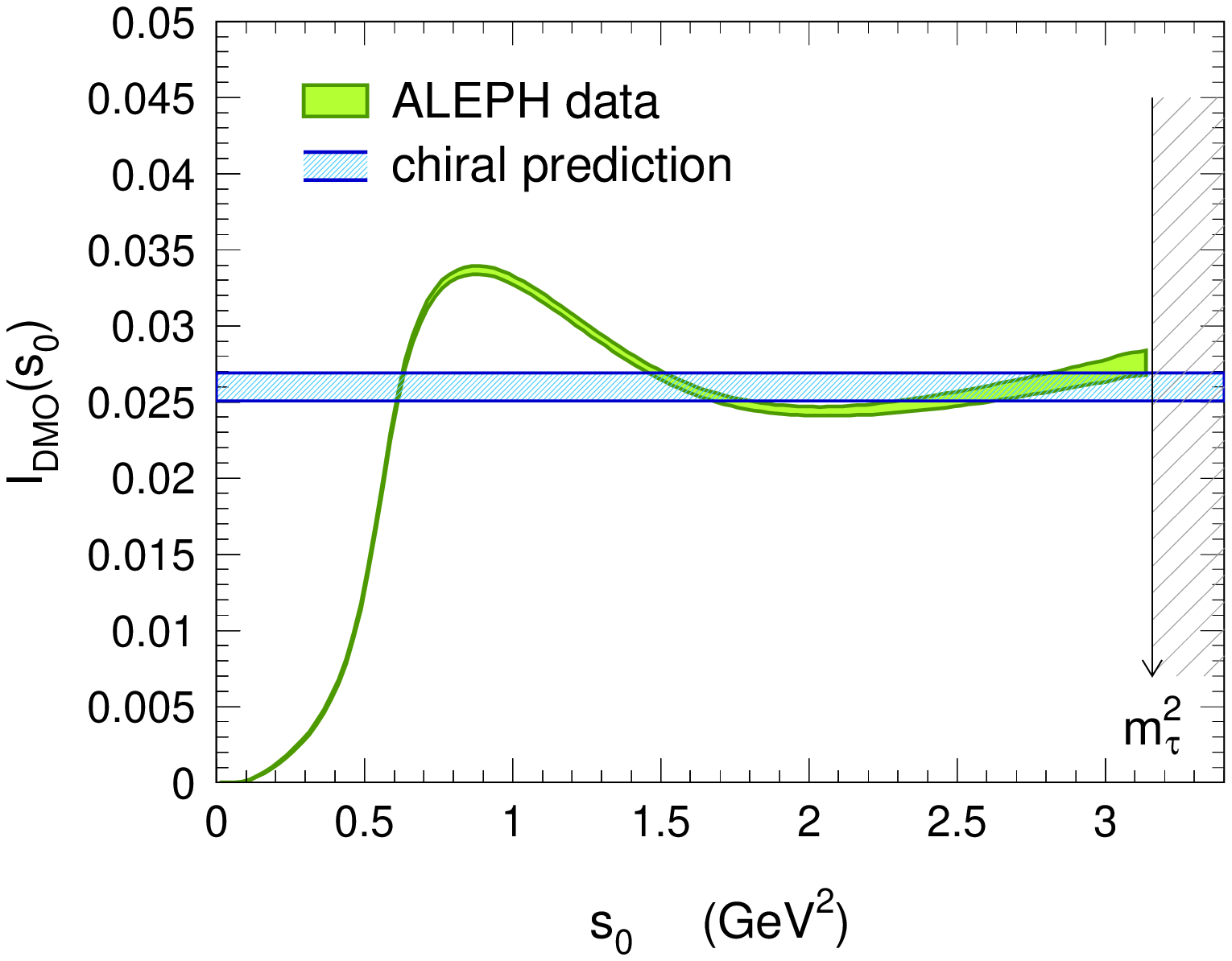}\hspace{0.5cm}
              \epsfysize6.3cm\epsffile{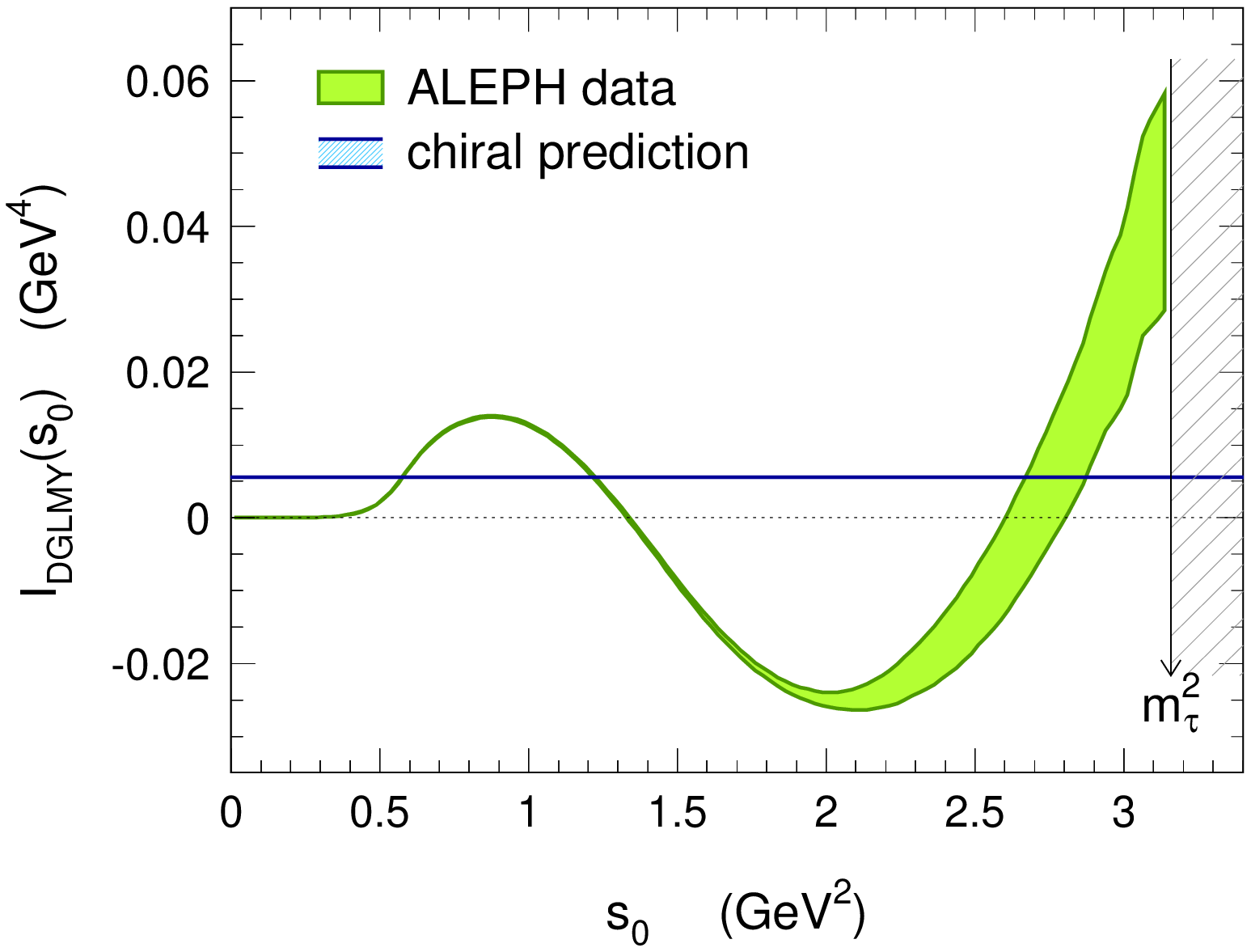}}
  \vspace{-0.2cm}
  \caption[.]{\label{fig:sumrules}
	The first and second Weinberg sum rules (upper left and right), 
	and the DMO and DGLMY sum rules (lower left and right)
	versus the upper integration cut-off ($s_0$). The horizontal 
	bands indicate the chiral predictions within errors, assuming
	full saturation of the integrals.}	
\end{figure} 
The spectral integrals~(\ref{eq:wsr1}) through (\ref{eq:das}) are
computed with variable upper integration bounds $s_0\le m_\tau^2$ 
using the experimental $\tau$ hadronic \sfs\  and their respective 
covariance matrices. The availability of the covariance matrices allows
one to perform a straightforward error propagation taking 
into account the strong bin-to-bin correlations of the \sfs.
%\footnote
%{
%	The spectral integrals with weight function $w(s)$ are evaluated as
%	\beqns
%	   I(s_0) =
%       		\intl_0^{s_0}\hm ds\,w(s)\left[v_1(s)-a_1(s)\right] 
%          	=
%       		\!\!\!\sum_{i=1}^{N_{\rm bin}(s_0)}\!\!\!
%			\Delta s_i w(s_i)
%			\left[v_{1,i}-a_{1,i}\right]~,
%	\eeqns
%	and their errors-squared read
%	\beqns
%   		\sigma^2 (I(s_0)) =
%       		\!\!\!\sum_{i,j=1}^{N_{\rm bin}(s_0)}\!\!\!
%		\Delta s_i\Delta s_j w(s_i)w(s_j)
%       		\left[C^{V}_{ij} + C^{A}_{ij}\right]~,
%	\eeqns
%	where the anti-correlations between $v_1$ and $a_1$ due to the
%	estimates of the vector/axial-vector parts of the $K\Kb\pi(\pi)$
%	final states, and also due to experimental feed through between 
%	modes with adjacent numbers of pions, must be reabsorbed in the 
%	respective covariance matrices $C^{V/A}$, so that $v_1$ and $a_1$ 
%	can be assumed to be uncorrelated.
%}.
%The right hand plot in Fig.~\ref{fig:vpmasf} shows the vector minus 
%axial-vector $\tau$ spectral functions measured by ALEPH~\cite{aleph_taubr} 
%and OPAL~\cite{opal_vasf}. 
The vector minus axial-vector $\tau$ spectral functions measured 
by ALEPH~\cite{aleph_taubr} and OPAL~\cite{opal_vasf} have been shown
in the right hand plot in Fig.~\ref{fig:vpmasf}.
Local saturation of the spectral functions at the 
integral end point $s_0$ would require convergence of central values and 
all derivatives at zero. This is not yet achieved at $s_0=m_\tau^2$.
We hence expect that the sum rules are not entirely satisfied for the 
$\tau$ spectral functions.
The sum rules~(\ref{eq:wsr1}) through (\ref{eq:das}) versus the upper 
integration cut-off $s_0\le m_\tau^2$ are plotted in Fig.~\ref{fig:sumrules}
for the ALEPH spectral functions. The horizontal band depicts the 
chiral predictions of the integrals, neglecting OPE power corrections.
One observes that only for the DMO sum rule (lower left plot), where 
contributions from higher masses are suppressed, saturation at the 
$\tau$ mass is approximately satisfied within errors. The other sum rules 
are not saturated at $m_\tau^2$, and no further conclusions can be drawn.

Many $V-A$ sum-rule analyses have been performed since the experimental 
spectral functions became 
available~\cite{dghs,ioffe01,zyablyuk,peris,bijnenssr,cgm1,cgm2,schilchersr1,schilchersr2,rojosr}. Some results of these will be discussed in the following.

\subsection{Operator product expansion}

% The conservation of currents in the 
% chiral limit implies 
% $s \left[\Pi^{(1)}_{\ubar d,V}(s)-\Pi^{(1)}_{\ubar d,A}(s)\right]=0$ 
% in momentum space. For large space-like momenta $Q^2=-s$ one 
% can expand the dispersion relation of the $V-A$ correlator, and with 
% the use of the first and the second WSR one obtains
% \beq
% \label{eq:proj1}
%    \frac{1}{4\pi^2}\intl_0^{s_0\to\infty} ds\,\rhoVmA(s)
%    \left[\frac{1}{s+Q^2} -
%          \left(\frac{s^2}{Q^6} - \frac{s^3}{Q^8} 
%                + \cdots
%          \right)
%    \right] \approx 0~.
% \eeq
% 
Nonperturbative effects give rise to nonperturbative OPE contributions
to the $V-A$ correlation function
\beq
\label{eq:proj2}
   \Pi^{(1)}_{\ubar d,V}(s)
             - \Pi^{(1)}_{\ubar d,A}(s)
       = 
   	\sum_{D=2,4,6,\dots}\frac{1}{(-s)^{D/2}} 
  	\sum_{{\rm dim}{\cal O}=D}C_{D,V-A}(s,\mu)
		                \langle{\cal O}_D(\mu)\rangle_{V-A}~,
\eeq
where we shall omit the $V-A$ subscripts in the following. 
The dimension $D=0$ contribution is of pure perturbative origin and 
is degenerate in all orders of perturbation theory for vector and 
axial-vector currents. To NLO in QCD one has
\beq
\label{eq:logOPE}
	C_D(s,\mu)\langle{\cal O}_D(\mu)\rangle
	= a_D(\mu) + b_D\ln\left(\frac{s}{\mu^2}\right)~.
\eeq
The explicit logarithmic dependence of the Wilson coefficients (the $b_D$
terms) is known for $D=2,4,6$, but not for higher dimensions. 
Useful relations between the OPE terms and the \sfs\ are obtained from 
projective sum rules~\cite{margv}. When neglecting the logarithmic power
terms in Eq.~(\ref{eq:logOPE}), one finds~\cite{cgm1}
\beqn
\label{eq:projsumrule}
         \frac{1}{4\pi^2}\intl_0^{s_0} ds\,
         \left(\frac{s}{s_0}\right)^{\!\!k}\,\rhoVmA(s)
	&=&
	-\frac{1}{2\pi i}\ointl_{|s|=s_0}\hm\hm\hm ds\,
		\frac{a_D}{(-s)^{D/2}}\left(\frac{s}{s_0}\right)^{\!\!k}\nonumber\\
	&=&
	(-1)^k\frac{a_D}{s_0^k}\delta_{k,(D/2)-1}~,
\eeqn
where $\delta_{m,n}$ has the value 1 if $m=n$, and 0 elsewhere.

The validity of the OPE depends on the size of the cut-off scale $s_0$.
Apart from the saturation of the spectral integral, this represents 
another source of systematic uncertainty related to the cut-off. The
insertion of appropriate weight functions $w(s)$ under the spectral
integral in~(\ref{eq:projsumrule}) may help to reduce the impact of 
these systematics (see also the discussion in~\cite{cgm1,cgm2}). While the 
impact of missing saturation can be reduced by suppressing the high-energy
tail of the $V-A$ spectral function, the breakdown of the OPE due to 
quark-hadron duality violations is unavoidable to some extent. 
In practice, this duality violation is assumed to be absorbed by the 
OPE power operators that are fit phenomenologically to the spectral
sum rules. The fit protects the physical observable (for instance
the strong coupling constant as discussed in Section~\ref{sec:qcd})
that one is aiming to extract. If the nonperturbative corrections 
are found to be large, this protection mechanism may not work
sufficiently well anymore, and duality violations can entail systematic 
errors. The optimization between saturation of the integral on one hand, 
and small duality violation on the other hand requires a compromise: 
if one aims at a minimum
saturation loss, \ie, small residual modulations of the spectral 
function in the vicinity of the cut-off $s_0$, one has to apply a 
weight function that suppresses the high-energy tail\footnote
{
	The suppression of the high-energy spectral
	function also reduces the experimental uncertainty on the 
	integral since the statistical errors are large at the kinematic
	endpoint of the $\tau$ phase space.
}.
However, the stronger the suppression the larger the quark-hadron
duality violation will be.

Using the formulae of~\cite{bnp} and~\cite{pertmass} for the nonperturbative 
power expansion of the correlators, one obtains for the $V-A$ correlators
the OPE~\cite{dghs}
\beqn
\label{eq:corr10}
     \Pi_{\ubar d, V-A}^{(0+1)}(-s) 
       &=& 
             -\, \frac{a_s(s)}{\pi^2}\frac{\hat{m}^2 (s)}{s}
            + \Biggl(\frac{8}{3}a_s(s) + \frac{59}{3}a_s^2(s)\Biggr)
              \frac{\hat{m}\langle\ubar u+\dbar d\rangle}{s^2} 
            - \frac{16}{7\pi^2}\frac{\hat{m}^4(s)}{s^2} \nonumber \\
       & &  -\, 8 \pi^2a_s(\mu^2)\Biggl[1 + \left(\frac{119}{24} - \frac{1}{2}
                                         L(s)\right) a_s(\mu^2)
                              \Biggr]\frac{\langle {\cal O}_6^{1}(\mu^2)
                                      \rangle}{s^3} \nonumber \\
       & &  +\, \frac{2 \pi^2}{3}\biggl[3 + 4 \, L(s) \biggr]a^2_s(\mu^2)
              \frac{\langle {\cal O}_6^2(\mu^2)\rangle}{s^3}
              + \frac{\langle {\cal O}_8\rangle}{s^4}~, \\[0.2cm]
\label{eq:corr0}
     \Pi_{\ubar d,V-A}^{(0)}(-s) 
       &=& 
            -\, \frac{3}{\pi^2}\Biggl[2a_s^{-1}(s)-5 + 
              \left( -\frac{21373}{2448} + \frac{75}{17} \zeta (3) 
              \right) a_s (s) \Biggr]\frac{\hat{m}^2(s)}{s}
            - 4\hat{C}(\mu^2)\frac{\hat m^2(\mu^2)}{s}
             \nonumber \\
       & &
            -\, 2\frac{\hat{m}\langle\ubar u+\dbar d\rangle}{s^2} 
            - \frac{1}{7\pi^2}\Biggl[\frac{53}{2} - 12a_s^{-1}(s)
              \Biggr]\frac{\hat{m}^4(s)}{s^2}~,
\eeqn
with $a_s(s)=\as(s)/\pi$, $L(s)=\ln(s/\mu^2)$ and the
dimension $D=6$ operators
\beqn
  {\cal O}_6^1 
        &\equiv& \ubar\gamma_{\mu} \gamma_5 T^a d \dbar \gamma^{\mu}
                  \gamma_5 T^a u - \ubar\gamma_{\mu}  T^a d \dbar 
                  \gamma^{\mu} T^a u \nonumber~, \\
\label{eq:dim6}
  {\cal O}_6^2 
        &\equiv& \ubar\gamma_\mu\dbar d\gamma^\mu u
                 -\ubar\gamma_\mu\gamma_5d\dbar\gamma^\mu\gamma_5u~,
\eeqn
and where the ${\rm SU}(3)$ generators $T^a$ are normalized so that 
${\rm tr}(T^a T^b )=\delta^{ab}/2$. The average mass $\hat{m}$ stands 
for $(m_u + m_d)/2$, \ie, ${\rm SU}(2)$ symmetry is assumed. The constant 
$\hat{C}(\mu^2)$ depends on the renormalization procedure and should 
not affect physical observables. Dimension $D=2$ mass terms are 
calculated perturbatively to order $\as^2$, which suffices for 
the light $u,d,$ quarks. The coefficient functions of the dimension 
$D=4$ operators for vector and axial-vector currents have been calculated 
to subleading order in~\cite{generalis,gluonterm}. Their vacuum 
expectation values are expressed in terms of the scale invariant gluon 
and quark condensates. Since the Wilson coefficients of the gluon condensate 
are symmetric for vector and axial-vector currents, they vanish in the 
difference. The expectation values of the dimension $D=6$ 
operators~(\ref{eq:dim6}) obey the inequalities 
$\langle {\cal O}_6^1(\mu^2)\rangle\ge0$ 
and $\langle {\cal O}_6^2(\mu^2)\rangle\le0$, which can be derived 
from first principles. The corresponding coefficient functions 
were calculated in the chiral limit~\cite{lanin}.
Since $\langle {\cal O}_6^2(\mu^2)\rangle$ is 
subleading in $a_s(\mu^2)$, it is neglected in most phenomenological
analyses. In leading order, one can use factorization of the matrix elements 
(that is, vacuum insertion) to rewrite $\langle {\cal O}_6^1\rangle$ 
in the form\footnote
{
	Following~\cite{zyablyuk}, factorization for the $D=6$ operators
	can be written as 
	$
	\langle (\ubar \lambda \Gamma_1 d)(\dbar \lambda \Gamma_2 u)\rangle
	=
	-(16/9){\rm Tr}
	[\langle u \otimes \ubar\rangle\Gamma_1
	 \langle d \otimes \dbar\rangle\Gamma_2]
	$, where $\Gamma_i$ are Dirac matrices (acting on color, flavor 
	and Lorentz indices), the factor $16/8$ includes
	the ${\rm SU}(3)$ color number, and the $4\times4$ matrix 
	$\langle q\otimes\qbar\rangle$ is proportional to the quark condensate 
	$\langle q\otimes\qbar\rangle=-(1/4)\langle0|q\qbar|0\rangle$.
} 
\beq
\label{eq:D6factorization}
	\langle {\cal O}_6^1\rangle \propto 
	-\frac{64}{9}\pi^2 a_s
	\langle 0|\ubar u|0\rangle\langle 0|\dbar d|0\rangle~.	 
\eeq
For the dimension $D=8$ operators no higher order calculations are 
available in the literature, so that it is customary in the QCD sum-rule 
analyses to neglect their logarithmic $s$ dependence. Again, the $J=0$ 
contribution vanishes in the chiral limit. A factorization result similar
to Eq.~(\ref{eq:D6factorization}) can be derived, which now involves a 
five-dimensional quark-gluon mixed condensate~\cite{grozin,ioffe01,zyablyuk}.
Higher OPE dimensions are neglected (see~\cite{zyablyuk} for an analysis 
that includes the relevant $D=10$ operators).

\subsection{Borel and pinched weight sum rules}
\label{sec:borelsumrules}

Many ways have been explored to improve the saturation
of the integrals at finite cut-off scale $s_0$, and to better control
the validity of the OPE. In general, the approaches distinguish
themselves by the definition of the weight $w(s)$ used under the 
integral
\beq
\label{eq:weighed_sumrule}
	\frac{1}{4\pi^2}\intl_0^{s_0}
            ds\,w(s)\rhoVmA(s)~.
\eeq

Since duality violations are expected to be largest in the vicinity of 
the timelike real axis~\cite{quinnetal}, some authors suggest to use
weights of the form $w_n(s)\propto(1-s/s_0)^n$~\cite{pichledib} 
(\cf\  Section~\ref{sec:moment}). The class of functions that satisfy 
$w(s_0)=0$ is introduced as ``pinched weights'' in~\cite{cgm1,cgm2}. As 
mentioned before, pinched weights improve the experimental accuracy of 
the sum rule, but also reduce the effective inclusiveness of the spectrum, 
and hence make it more sensitive to duality violations.

\begin{figure}[t]  
  \centerline{\epsfysize6.3cm\epsffile{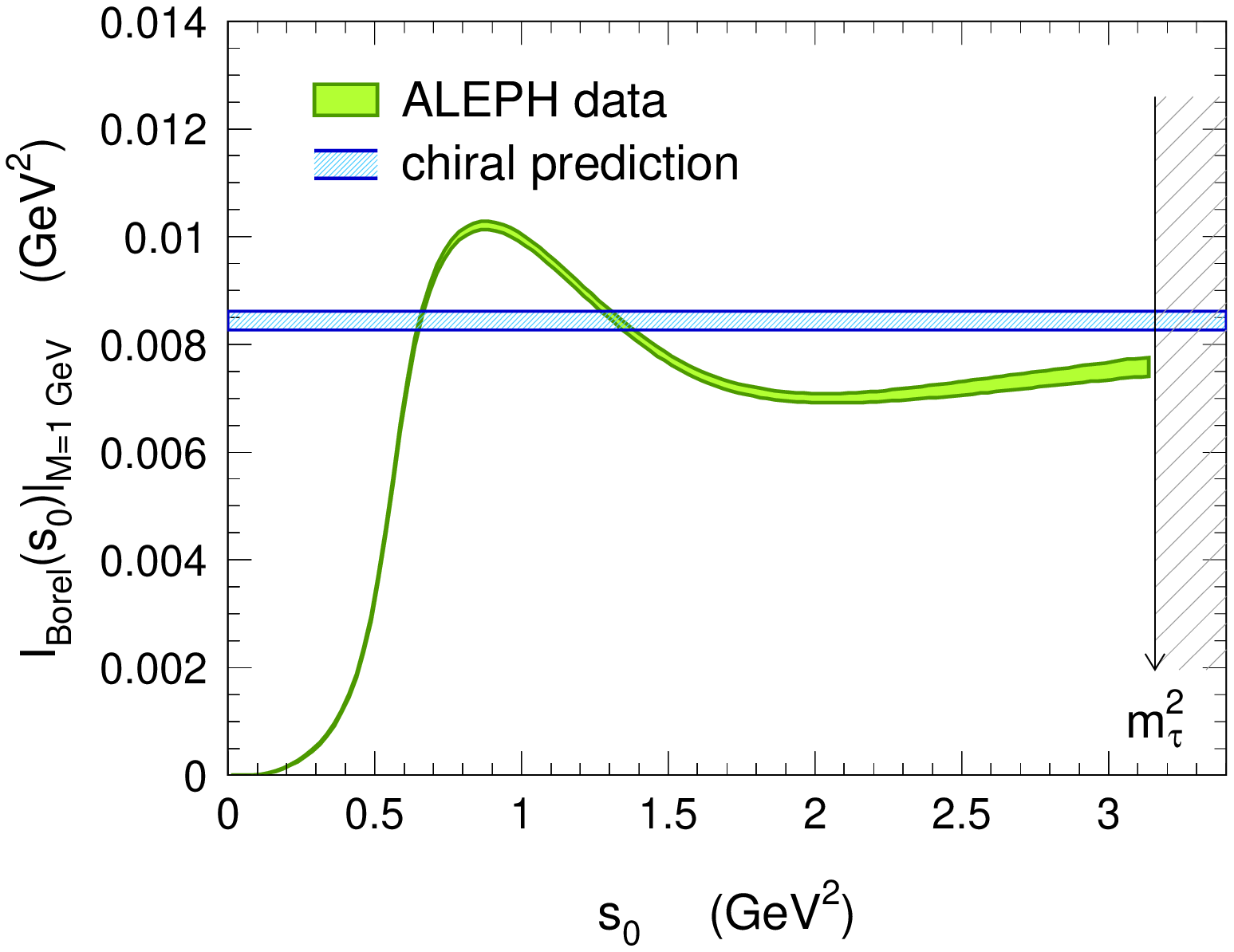}\hspace{0.5cm}
              \epsfysize6.3cm\epsffile{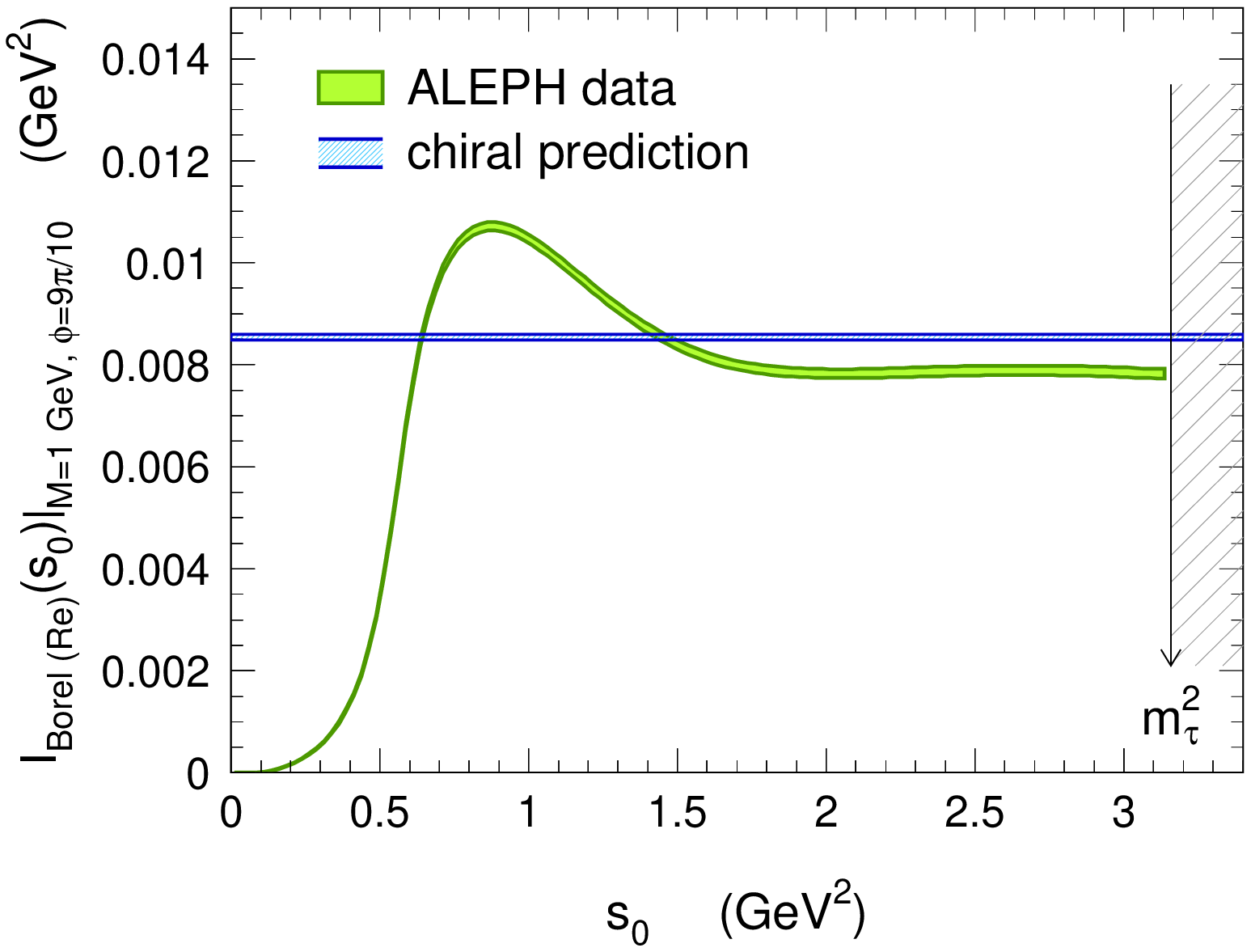}}
  \vspace{-0.2cm}
  \caption[.]{\label{fig:w1_borel}
	The Borel sum rule~(\ref{eq:w1_borel}) (left)
	and the real part of the complex Borel sum rule~(\ref{eq:borel_re})
	for $\phi=9\pi/10$ (right), at $M=1\gev$, and as a function of the 
	integral cut-off $s_0$.	The chiral prediction shown in the right 
	hand plot does not include OPE power corrections. }	
\end{figure}

Another way to improve the saturation of finite energy sum rules is 
to apply the Borel weight $w(s)=\exp(-s/M^2)$, emphasizing the lower 
resonances of the spectrum. Here, the OPE operators of dimension $D=n$
are suppressed as $\langle {\cal O}_{2n+2}\rangle/n!$, and all leading
terms of the OPE vanish. The Borel-transformed sum rules give rise to 
nonperturbative corrections from quark condensates and non-vanishing 
quark masses that scale with the Borel parameter $M^2$ and with 
$\as(M^2)$. For example, the Borel-transformed first WSR 
reads~\cite{floratos,pascual_sr,narison_sr,peccei}
\beqn
\label{eq:w1_borel}
\lefteqn{\frac{1}{4\pi^2}\hsm\intl_0^{s_0\to\infty}\hsm
            ds\,e^{-s/M^2}\rhoVmA(s)
        =  \fpi^2} \nonumber \\
   & &      +\; \frac{\as(M^2)}{2\pi}
          \left[-\frac{m_u(M^2)m_d(M^2)}{\pi^2} 
                - \frac{8}{3}\frac{\fpi^2m_\pi^2}{M^2}
                - \frac{32\pi}{9}\frac{\left(\langle0|\qbar q|0\rangle
                                       \right)^2}{M^4} + \cdots
          \right]~,
\eeqn
where the lowest-order current algebra Gell-Mann-Oakes-Renner relation
\beq
\label{eq:gmor}
	\fpi^2 m_\pi^2=-(m_u+m_d)\langle0|\ubar u|0\rangle~,
\eeq
obtained from Eqs.~(\ref{eq:mpi1}) and (\ref{eq:deffpi}) has been used~\cite{gmor}, 
and where $m_u(M^2)$ and $m_d(M^2)$ denote the running light quark masses
(see~\cite{jamin:2002} for a derivation of chiral corrections to 
Eq.~(\ref{eq:gmor}), which are found to be of the order of $-5\%$). 
The Borel sum rule is plotted in Fig.~\ref{fig:w1_borel} (left hand) for 
$M=1\gev$. The comparison with Fig.~\ref{fig:sumrules} 
(upper left hand plot) exhibits the gain in experimental precision and 
in the approach to the asymptotic regime.
Inserting the experimental values for the parameters on
the r.h.s.\ of Eq.~(\ref{eq:w1_borel}), and neglecting higher order 
nonperturbative contributions ${\cal O}(M^{-6})$ and contributions other 
than quark condensates, one obtains $(8.44\pm0.17)\times10^{-3}\gev^2$.
From Fig.~\ref{fig:w1_borel} we conclude that further nonperturbative 
power terms need to be added to satisfy the sum rule.

\begin{figure}[t]  
  \centerline{\epsfysize6.5cm\epsffile{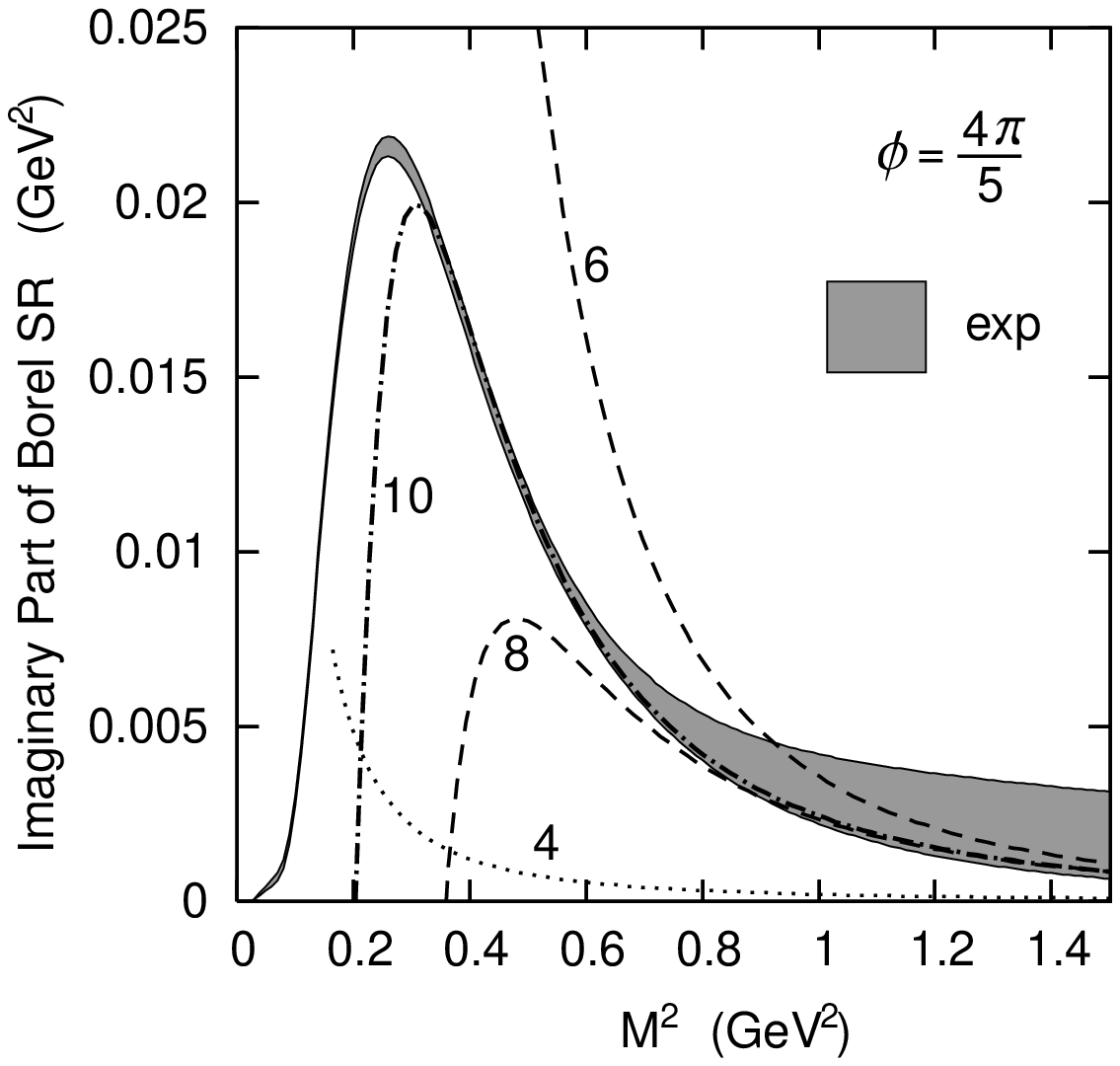}\hspace{1.0cm}
              \epsfysize6.5cm\epsffile{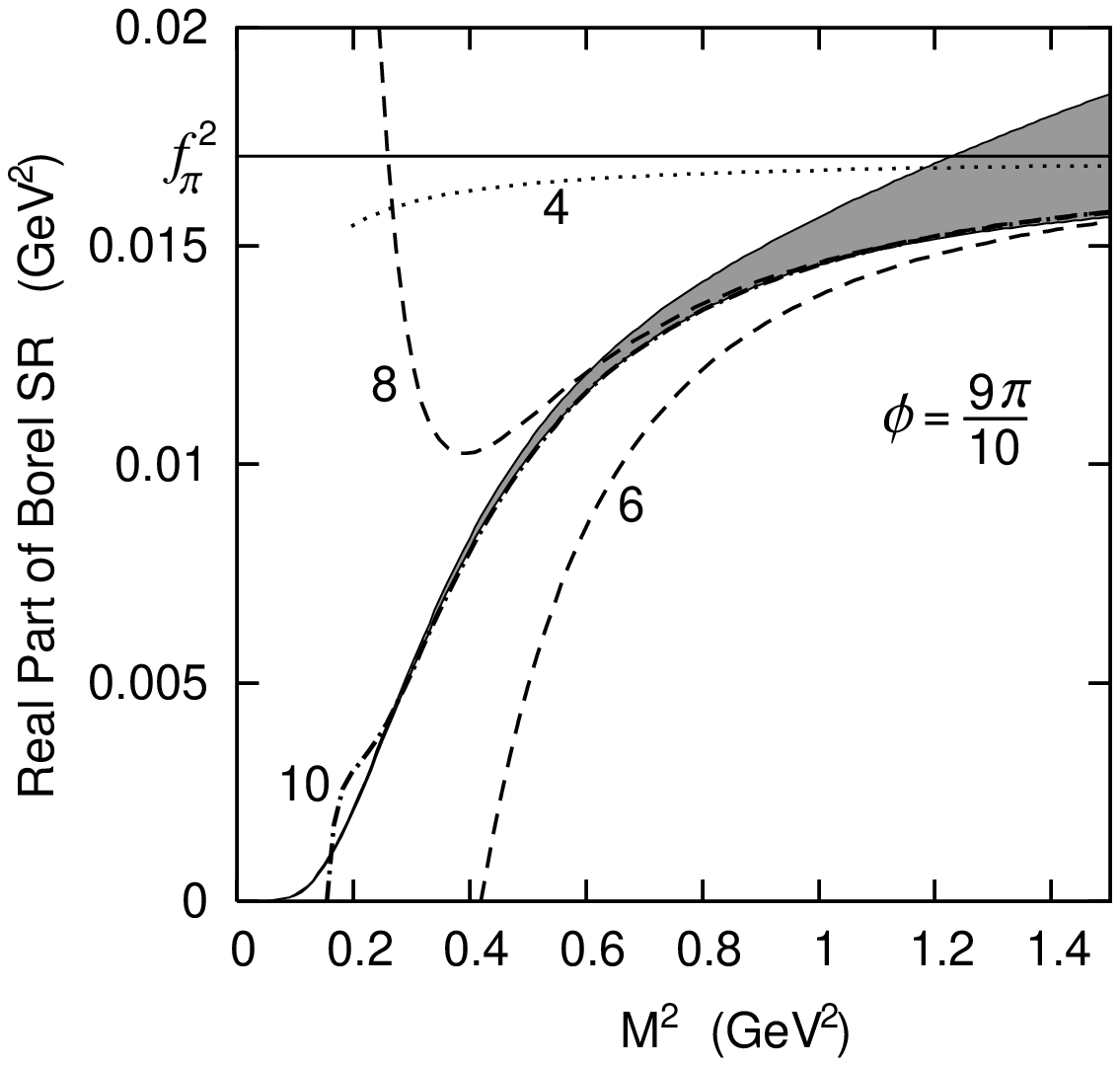}}
  \vspace{0.2cm}
  \caption[.]{\label{fig:zyablyuk}
	Real part~(\ref{eq:borel_re}) of the Borel transformation
	for $\phi=9\pi/10$ (left), and its imaginary part~(\ref{eq:borel_re}) for 
	$\phi=4\pi/5$ (right), as a function of the modulus $M^2$ of the Borel
	parameter. The leading order $D=12$ operator $a_{12}$ vanishes in 
	these sum rules. The shaded band shows the ALEPH 
	data~\cite{aleph_vsf,aleph_asf}, and the lines display the 
	leading order operator series, \ie, neglecting all logarithmic 
	terms. The numbers indicate the highest dimension considered
	in the operator sum. The figures are taken from~\cite{zyablyuk}.
}	
\end{figure}

To reduce the correlation between the different operators, one can
consider a Borel transformation in the complex plane of the Borel scale
parameter, that is $M^2 \to M^2 \exp[i(\pi-\phi)]$. Keeping only
the leading order coefficients $a_D$ in Eq.~(\ref{eq:logOPE}), one finds 
for the real and imaginary parts of the Borel transformation~\cite{zyablyuk} 
\beqn
\label{eq:borel_re}
	\frac{1}{4\pi^2}\!\!
	\intl_0^{s_0\to\infty}\!\!
			 e^{c_\phi}
			 \cos(s_\phi)\rhoVmA(s)
	&=& \fpi^2 + \sum_{k=1}^\infty(-1)^k\frac{\cos(k\phi)a_{2k+2}}{k!M^{2k}}~,
	\\
\label{eq:borel_im}
	\frac{1}{4\pi^2M^2}\!\!
	\intl_0^{s_0\to\infty}\!\!
			 e^{c_\phi}
			 \sin(s_\phi)\rhoVmA(s)
	&=& \sum_{k=1}^\infty(-1)^k\frac{\sin(k\phi)a_{2k+2}}{k!M^{2k+2}}~,
\eeqn
with $s_\phi=(s/M^2)\sin\phi$, $c_\phi=(s/M^2)\cos\phi$.
If the sum rule is applied to the $\tau$ spectral functions, the argument of 
the cosine function must be negative to suppress the unknown $s_0>m_\tau^2$ 
tail of the spectrum. In the case of $\phi=\pi(2k-1)/(2k)$ ($k=2,3,\dots$)
for the real part~(\ref{eq:borel_re}), and $\phi=\pi(k-1)/k$ ($k=3,4,\dots$) 
for the imaginary part~(\ref{eq:borel_im}), the sum rules become projective
with vanishing leading operators of dimension $D=2k+2$. The real part for
a particular working point chosen in~\cite{zyablyuk} as a function of the 
cut-off $s_0$ is shown in the right hand plot of Fig.~\ref{fig:w1_borel}. 
The chiral prediction given contains the pion pole, but no OPE power corrections.
The sum rule is well saturated at $s_0=m_\tau^2$. The remaining discrepancy
with the data is due to nonperturbative contributions.
An update~\cite{zyablyuk} of the numerical analysis performed in~\cite{ioffe01} 
finds for the leading $V-A$ operators using the ALEPH data~\cite{aleph_asf}
\beqn
\label{eq:borel_a6}
   a_6 &=&      -(7.2\pm1.2)\times10^{-3}\gev^6~, \\
\label{eq:borel_a8}
   a_8 &=& \ph{-}(7.8\pm2.5)\times10^{-3}\gev^8~. 
\eeqn
The author also estimates the leading $D=10$ operator to be
$a_{10}=-(4.4\pm2.8)\times10^{-3}\gev^{10}$, without however reporting the 
statistical correlations with $a_{6}$ and $a_{8}$. The imaginary 
part~(\ref{eq:borel_im}) of the Borel sum rule were used as input, as well 
as the angles $\phi=2\pi/3,3\pi/4,4\pi/5$ for which the operators 
$a_{8},a_{10}, a_{12}$ vanish, respectively. For the Borel parameter 
$M^2$ a range between 0.4 and $1.0\gev^2$ is used, which is integrated 
over to determine an average $\chi^2$ estimator~\cite{zyablyuk} that is 
minimized in the fit. The fit results for the real and imaginary parts
of the sum rules~(\ref{eq:borel_re}), (\ref{eq:borel_im}), and for two 
working points in the parameter $\phi$ are shown in Fig.~\ref{fig:zyablyuk}.
The contribution from $D=10$ is significant here.

Recent theoretical developments match the $D=6$ operator, derived
from the $V-A$ spectral sum rules, with the vacuum matrix elements
$\langle {\cal O}_1\rangle_\mu$ and $\langle {\cal O}_8\rangle_\mu$, 
which are related to the electroweak penguin operator ${\cal Q}_8$
that contributes to $K\to\pi\pi$ decays (see, \eg, Refs.~\cite{bijnenssr,cgm2}). 
Together with the QCD penguin ${\cal Q}_6$, these operators
generate direct $CP$ violation in neutral kaon decays to two pions,
the computation of which suffers from large uncertainties.

\subsection{Inverse moments sum rules and chiral perturbation theory}
\label{sec:invmomentssumrules}

We have seen that at sufficiently high energies nonperturbative effects 
are parameterized by vacuum condensates, following the OPE rules.
A particular role is played by the condensates that are order parameters of
the spontaneous breakdown of chiral symmetry (SB$\chi$S). They vanish
at all orders of perturbation theory and control the high energy
behavior of chiral correlation functions, such as the difference of vector
and axial-vector current two-point functions. On the  other hand, at low 
energies, SB$\chi$S  makes it possible to construct an effective theory
of QCD, the Chiral Perturbation Theory ($\chi$PT)~\cite{wein,gale1},
which uses  the Goldstone bosons as  fundamental fields and provides a
systematic expansion of QCD correlation functions in powers of momenta and
quark masses. Any missing information is then parameterized by low-energy 
coupling constants, which can be determined phenomenologically in low-energy 
experiments involving  pions and kaons. The fundamental parameters 
describing chiral symmetry breaking, the running quark masses and the 
quark condensates appear both in low-energy $\chi$PT and the high energy 
OPE expansion. For this reason it is useful to combine the two expansions 
in order to get a truly systematic approach to the chiral sum 
rules~\cite{donoghue}. Such a combined approach is illustrated in~\cite{dghs}
through the determination of the $L_{10}$ parameter of the chiral Lagrangian, 
including high-energy corrections coming from the OPE. The connection 
between the two domains is provided by the experimental $\tau$ spectral
function $\rhoVmA$.

At leading  order $\chi$PT, $L_{10}$ is obtained by the DMO
sum rule~(\ref{eq:dmo})
\beq
\label{eq:dmo_second}
       \frac{1}{4\pi^2}\hsm\intl_0^{s_0\to\infty} ds\frac{1}{s}\,\rhoVmA(s)
       \simeq -4L_{10}~.
\eeq
As it stands the DMO sum rule~(\ref{eq:dmo}) is subject to chiral
corrections due to non-vanishing quark masses~\cite{kagodmo}. On the 
other hand, the truncation of the integral at $s_0\leq m_{\tau}^2$ 
introduces an error that competes with the low-energy chiral 
corrections. Both types of corrections can be systematically included 
through $(i)$~the high-energy expansion in $\as(s_0)$ and in 
inverse powers of $s_0$, and $(ii)$~the low-energy expansion in powers 
of quark masses and of their logarithms.

\begin{figure}[t]  
%...kumac is on: ~hoecker/unfolding/data/srpap/srpap.kumac
  \centerline{\epsfysize6.5cm\epsffile{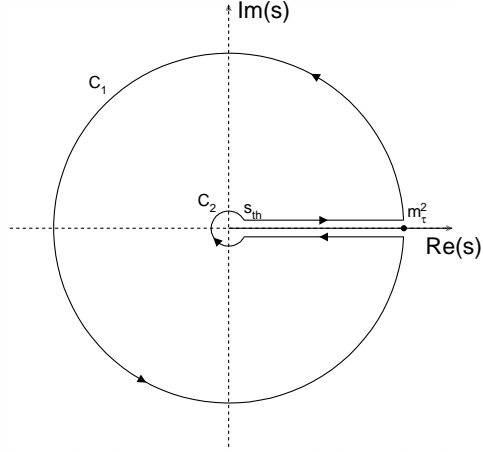}}
  \caption[.]{\label{fig:contour_imsr}
              Integration contour around the circles at $s=m_\tau^2$
              and $s=s_{\rm th}$.}
\end{figure}

In Section~\ref{sec:moment} we have reviewed the simultaneous
determination of the strong coupling constant and the OPE operators using 
the spectral weights 
\beq
\label{eq:moments_imsr}
	w^{(k,\l)}(s) = \left(1-\frac{s}{s_0}\right)^{\!\!k}
               \left(\frac{s}{s_0}\right)^{\!\!\l}~,
\eeq
for $k,\l\ge0$. The extension of the spectral moment analysis to negative 
integer values of $\l$ (``inverse moment sum rules'', (IMSR)~\cite{imsr})
requires, due to the pole at $s=0$, a modified contour of integration 
in the complex plane, as illustrated in Fig.~\ref{fig:contour_imsr}.
This is where $\chi$PT comes into play: along the small circle $C_2$ 
placed at the production threshold $s_{\rm th}=4m_\pi^2$ one can use 
$\chi$PT predictions for the two-point correlators. 
Generalizing the spectral moments~(\ref{eq:moments}) to IMSR's, one obtains 
for the Cauchy integral
\beq
\label{eq:momtheorycontour}
   R_{\tau,V/A}^{(k,\l)} 
         = 6 \pi i |V_{ud}|^2 \Sew
           \ointl_C \frac{ds}{s_0} w^{(k,\l)}(s)
                    \left[\left(1 + 2\frac{s}{s_0}
                          \right)\Pi^{(0+1)}_{V/A} (s) - 2
                          \frac{s}{s_0} \Pi^{(0)}_{A}(s) 
                    \right]~,
\eeq
where $C=C_1+C_2$ for the inverse moments and $C=C_1$ for the 
positive moments (see Fig.~\ref{fig:contour_imsr}).

The nonstrange correlators~(\ref{eq:correlator}) are known up to two 
loops~\cite{gale1,gale2,kago1,kago2} in standard $\chi$PT\footnote
{
	The authors of~\cite{dghs} use generalized $\chi$PT 
	(G$\chi$PT~\cite{gchpt}) to ${\cal O}(p^4)$ one-loop
	order only. Standard $\chi$PT assumes~\cite{weinberg77} the 
	validity of Eq.~(\ref{eq:gmor}) and 
	$r\simeq 2m_K^2/m_{\pi}^2-1\approx25.9$, whereas G$\chi$PT 
	admits lower values of these two 
	quantities~\cite{sterngchpt1,sterngchpt2,gchpt}.
	The alterations of the standard ${\cal O}(p^4)$ results 
	for nonstrange correlators introduced by G$\chi$PT are marginal. 
	They merely concern the symmetry breaking $J=0$ component of 
	the \sfs\ and most of them are absorbed into the renormalization 
	of $\fpi$.
} 

%For the Cabibbo-allowed channel one finds for the $\chi$PT correlation 
%functions~\cite{dghs}
%\beqn
%\label{eq:chpt1}
%   \Pi^{(0+1)}_{\ubar d,V}(s)       &=& 
%          4M^r_{KK}(s) + 8 M^r_{\pi\pi}(s) - 4(L_{10}^r + 2H_1^r)~,  \\
%   \Pi^{(0)}_{\ubar d,V} (s)   &=& 0~, \\
%   \Pi_{\ubar d,A}^{(0+1)} (s) &=& 
%          - \frac{2\fpi^2}{s-M_{\pi}^2} - 4(2H_1^r-L_{10}^r)~, \\
%   s\Pi_{\ubar d,A}^{(0)} (s)  &=& 
%          - \frac{2\fpi^2M_{\pi}^2}{s-M_{\pi}^2} + 8{\hat m}^2(H_{2,2}-2B_3)~,
%\eeqn
%where $M^r_{PP^\prime}(s)$ and $L_{PP^\prime}(s)$ are loop 
%integrals defined in~\cite{gale2}. The superscript $r$ refers to renormalized 
%quantities, which depend on the scale $\mu_{\chi{\rm{PT}}}$. The whole 
%expressions are $\mu_{\chi{\rm{PT}}}$ independent. The terms $H_{2,2}$ and 
%$B_3$ are found to be finite, and do not need renormalization. The coefficients 
%$H_1^r$ and $H_{2,2}$ multiply contact terms of the sources. They are counterterms 
%needed to renormalize the ultraviolet divergences of the Green functions and 
%do not appear in physical observables.  In the difference between vector and
%axial-vector correlators the constant $H_1^r$ and all massless perturbative 
%expressions cancel. 
%
The analysis described in~\cite{dghs} performs a fit of the OPE to the IMSR 
$R_{\tau,V-A}^{(1,-1)}$ 
and the spectral moments~(\ref{eq:moments}). The inner integration circle $C_2$ 
occurring in $R_{\tau,V-A}^{(1,-1)}$ introduces the sensitivity to the effective 
$\chi$PT parameter $\Leff$, which is determined by the fit. This quantity is 
a well defined observable that can be related to $L_{10}^r(\mu_{\chi{\rm{PT}}})$
at each loop in $\chi$PT.
%In the particular case of the one-loop G$\chi$PT 
%calculation, its expansion reads
%\beq \label{eq:l10eff}
%   \Leff
%        = L_{10}^r(\mu_{\chi{\rm{PT}}}) + \frac{1}{128 \pi^2} 
%          \left( \ln\frac{m_{\pi}^2}{\mu_{\chi{\rm{PT}}}^2}+1 \right) 
%          + \frac{1}{384 \pi^2} \ln\frac{m_K^2}{m_{\pi}^2} 
%          + \frac{2\hat m^2}{m_{\tau}^2}\left(2B_3-\hat H_{2,2}\right)~,
%\eeq
%which is independent of $\mu_{\chi{\rm{PT}}}$. 

The spectral information is used to simultaneously determine $\Leff$
and the nonperturbative phenomenological operators. For dimension $D=6$ the 
contribution from ${\cal O}_6^2$ is neglected since it is large-$N_C$ and 
$\as^2$ suppressed. The results of the fit are reported as~\cite{dghs}
\beqn
\label{eq:l10}
   \Leff &=& -(6.36\pm0.09_{\rm exp}\pm0.14_{\rm th}
             \pm0.07_{\rm fit} \pm0.06_{\rm OPE}) \times 10^{-3}~,\\
  \label{eq:o6}
   \langle {\cal O}_6 \rangle 
        &=&  (5.0\pm0.5_{\rm exp}\pm0.4_{\rm th}
             \pm0.2_{\rm fit}\pm1.1_{\rm OPE}) 
             \times 10^{-4}\gev^6~, \\
  \label{eq:o8}
   \langle {\cal O}_8\rangle          
        &=&  (8.7\pm1.0_{\rm exp}\pm0.1_{\rm th}
             \pm0.6_{\rm fit}\pm2.1_{\rm OPE}) \times 10^{-3}\gev^8~,
\eeqn
where ${\cal O}_6 ={\cal O}_6^1(m_\tau^2)$.
The fit errors are separated in experimental (first errors) and theoretical
(second errors) parts, and systematic fit uncertainties (third errors) are
added. The last errors estimate the uncertainties due to the OPE 
breakdown~\cite{aleph_asf}. They are small for $L_{10}^{\rm{eff}}$ and 
dominant for the nonperturbative operators. Correcting for the factor of 
about $-13$ between the definition of $\langle {\cal O}_6 \rangle$ and 
$a_6$ of Eq.~(\ref{eq:borel_a6}), the two estimates are in agreement. 
The results~(\ref{eq:o6}) and (\ref{eq:o8}) have been confirmed 
in~\cite{opal_vasf} using OPAL data. In general it is observed that 
while all available sum-rule analyses\footnote
{
	See also the detailed recent analysis~\cite{peris:2005} on duality 
	violations in $V-A$ sum rules, which came to late to be 
	included in this review.
} 
agree on the value for the $D=6$ contribution to the $V-A$ correlator, 
the $D=8$ contribution, due to its smallness, is less stable. It 
exhibits a model dependence that exceeds the experimental error 
in~(\ref{eq:o8}). 
%For example, the analyses performed in~\cite{cgm2,bijnenssr} find
%negative values for $\langle{\cal O}_8\rangle$. We refer to the detailed
%comparisons of the various fit results for the nonperturbative power
%terms using the $\tau$ spectral functions in~\cite{friot:2004,rojosr}.

A stability test of the results~(\ref{eq:l10}) through (\ref{eq:o8}) at 
$m_\tau^2$ has been performed in~\cite{dghs} by fitting with variable 
``$\tau$ masses'' $s_0\le m_\tau^2$~\cite{dghs}. The resulting convergence
was found satisfactory.

Expressing $L_{10}^{\rm eff}$ of Eq.~(\ref{eq:l10}) at the 
renormalization scale $\mu_{\chi\rm PT}=m_\rho\simeq770\mev$, one 
obtains~\cite{dghs} $L_{10}^r(m_\rho)=-(5.13\pm0.19)\times10^{-3}$,
where experimental and theoretical errors of different sources have 
been added in quadrature.
This result can be compared with the value of $L_{10}^r$ that is
obtained from the one-loop expression of the axial form factor, $F_A$, of 
$\pim \to e^-\nueb\gamma$ decays (see~\cite{pienugamma} for notations)
\beq
   F_A = \frac{4\sqrt{2}\,m_\pi}{\fpi}(L_9+L_{10})~.
\eeq
Using $L_9^r(m_{\rho})= ( 6.78 \pm 0.15 ) \times 10^{-3}$~\cite{dghs,gale2,daphne}
and $F_A=0.0116\pm0.0016$~\cite{Eidelman:2004}, one obtains
$L_9 + L_{10}  = (1.36 \pm 0.19 ) \times 10^{-3}$. This gives
$L_{10}^r(m_{\rho}) = -(5.42 \pm 0.24) \times 10^{-3}$,
in agreement with the $\tau$ result. 
%Note that the quoted errors 
%do not take into account uncertainties from higher order chiral 
%corrections. Nonetheless, the determination of $\Leff$ is independent 
%of any chiral expansion: its value can be directly used in a two-loop analysis,
%where it suffices to include higher order corrections in Eq.~(\ref{eq:l10eff}).
%
%\begin{figure}[t]
%  \epsfxsize8.8cm
%  \centerline{\epsffile{figures/running_imsr.eps}}
%  \caption[.]{\label{fig:running_imsr}
%              Fit results for $L_{10}^{\rm eff}$ and the 
%              nonperturbative operators as a function of the 
%              ``$\tau$ mass'' $s_0$. The bands depict the 
%              values~\rm(\ref{eq:l10})--(\ref{eq:o8}) within 
%              errors, obtained at $m_\tau^2$.}
%\end{figure}
%The total purely nonperturbative contribution to $R_{\tau,V-A}$ found in 
%the fit, taking into account the correlations between the operators, amounts 
%to~\cite{dghs} 
%\beq
%\label{eq:np_imsr}
% $      R_{\tau,V-A}  =  0.061 \pm 0.014$,
%\eeq
%which is in agreement with the measurement $0.092\pm0.023$ 
%(Eq.~(\ref{eq:rtauvma})).

%The results 
%of such fits are shown in Fig.~\ref{fig:running_imsr}.
%The horizontal bands give the results at $m_\tau^2$ within one standard 
%deviation (all errors added in quadrature). 
%All curves exhibit a convergent 
%behavior for $s_0\to m_\tau^2$. Any deviation from the fit values for 
%$s_0>m_\tau^2$ should be covered by the ``OPE'' errors assigned to the 
%results.

\section{CONCLUSIONS AND PERSPECTIVES}
\label{sec:conclusions}

% In this concluding section we summarize the main analyses discussed in this
% review, and give perspectives for future progress in the field of precision
% measurements using hadronic $\tau$ decays.

The physics of hadronic $\tau$ decays has been the subject of much
progress in the last decade, both at the experimental and the theoretical 
level. Somewhat unexpectedly, hadronic $\tau$ decays provide one
of the most powerful testing grounds for QCD. This situation results from
a number of favorable conditions:
\bei

\item 	The $\tau$ lepton is heavy enough to decay into a variety of hadrons, 
	with net strangeness 0 and $\pm1$.

\item 	$\tau$ leptons are copiously produced in pairs at \ee colliders, 
	leading to simple event topologies without extra background particles.

\item 	The experimental study of $\tau$ decays could be done with large
	data samples: at LEP with high efficiency and small 
	background; at the $\FourS$ energy with much increased statistics, 
        but irreducible backgrounds. The two experimental setups offer 
	complementary advantages.

\item 	As a consequence, $\tau$ decays are experimentally known to
	great details and their rates measured with high precision.

\item 	The theoretical description of hadronic $\tau$ decays is based on solid
	ground and found to be dominated by perturbative QCD, as the 
	result of several lucky circumstances, which shall be briefly 
	recalled below.

\eei 

The $\tau$ decay rates into hadrons are expressed through spectral functions
of different final states with specific quantum numbers. 
The spectral functions are the basic ingredients to the theoretical 
description of these decays, since they represent the probability 
to produce a given hadronic system from the vacuum,
as a function of its invariant mass-squared $s$. The spectral functions are
dominated by known resonances at low mass, and tend to approach the 
quark-level asymptotic regime towards $m_\tau^2$. They have been 
determined for individual hadronic modes,
however they are conveniently summed and separated into their inclusive vector 
and axial-vector components for the nonstrange part, while the 
Cabibbo-suppressed strange part cannot be fully separated at present,
due to the lack of necessary experimental information.

The nonstrange $\tau$ vector spectral functions can be compared to
the corresponding quantities obtained in \ee annihilation by the virtue of
isospin symmetry. The ${\cal O}(\%)$ precision reached makes it necessary 
to correct for isospin-symmetry breaking. Beyond a superficial
agreement, the detailed comparison unveils discrepancies with some 
\epem data sets. Since the vector spectral functions are the necessary
ingredients to compute vacuum polarization integrals, required for
the evaluation of the running of \aqed  or the anomalous magnetic moment of
the muon \amu, this disagreement leads to different results when using
$\tau$ or \ee spectral functions. While the \ee-based theoretical \amu value 
disagrees with the measurement by $2.7$ ``standard deviations'', possibly 
indicating a contribution from physics beyond the Standard Model, the 
$\tau$-based calculation is consistent with experiment. Although the use 
of \ee data is {\em a priori} more direct, the physics at stake is such 
that the present situation must evolve, requiring more experimental and 
theoretical cross-checks. Newest \epem results already indicate better 
agreement with the $\tau$ data.

The observation that hadronic $\tau$ spectral functions can provide 
precision information on perturbative QCD is surprising, considering 
the moderate energy scale involved in these decays. Hence it is useful 
to recall briefly the reasons behind this success:
\bei

\item 	The total semileptonic decay rate normalized to the leptonic width, \Rtau, is 
	obtained by integrating over $s$ the total spectral function weighed by a 
	known kinematic function originating from the $V-A$ leptonic tensor. 
	This integral can be transformed to a complex contour integral over 
	a circle at $|s|=m_\tau^2$, hence involving only large complex $s$ 
	values where perturbative QCD can be applied.

\item 	One factor in the weight function, $(1-s/m_\tau^2)^2$, chops off 
	the integrand in the vicinity of the real axis where poles of the current 
	correlator are located.
	
\item 	The invariant-mass spectra in the hadronic $\tau$ decays, dominated by 
	resonances, are thus related to the quark-level contributions expected at 
	large mass. This global quark-hadron duality is expected to work best in
	more inclusive situations. In the case of hadronic $\tau$ decays, this
	condition is particularly well met, with several resonance contributions
	in each of equal-importance vector and axial-vector components.
	
\item 	The perturbative expansion of the spectral function 
	is known to third order in \as, with some estimates of the fourth 
	order coefficient.

\item 	The perturbative and nonperturbative contributions can be treated 
	systematically using the operator product expansion (OPE), giving 
	corrections organized in inverse powers $D$ of $m_\tau$.

\item 	The {\em a priori} dominant nonperturbative term from the gluon condensate
	($D=4$) is strongly suppressed by the second factor of the weight function,
	$(1+2s/m_\tau^2)$. This accidental fact renders hadronic $\tau$ decays 
	particularly favorable since nonperturbative effects are expected to be 
	small.

\item 	The next $D=6$ term is subject to a partial cancellation between the
	vector $V$ and axial-vector $A$ contributions, due to the $(1-\gamma_5)$ 
	relative factor.

\eei 

These fortunate circumstances explain why hadronic $\tau$ decays can be globally
described by perturbative QCD with very small nonperturbative corrections.
This behavior has been verified model-independently using the measured spectral 
functions. From a simultaneous fit of the total hadronic rate, obtained with 
precision from the leptonic branching ratio, and spectral moments that are
weighted integrals of the mass-squared distributions, the total nonperturbative
contribution is found to be $\delta_{{\rm NP},V+A}=(-4.3\pm1.9) \times 10^{-3}$
relative to the $V+A$ hadronic width. This is almost a factor of 50 smaller than
the $\as$-dependent perturbative correction to $\Rtau$. The predicted 
near-cancellation between the $V$ and $A$ contributions has been also 
observed, as  $\delta_{{\rm NP},V}=(2.0\pm0.3) \times 10^{-2}$ and 
$\delta_{{\rm NP},A}=(-2.8\pm0.3) \times 10^{-2}$.

Several approaches are available to treat the perturbative expansion of
the $\tau$ hadronic width. In fixed-order perturbation theory (FOPT),
all contributions beyond the known orders of the Adler function are neglected,
even if some parts resulting from the running of $\alpha_s$ are calculable (and
calculated). On the contrary, the contour-improved FOPT variant (CIPT) keeps 
all known terms. It has been shown that a faster convergence is achieved
with CIPT and, since the FOPT expansion represents an unnecessary
approximation, the CIPT approach should be preferred. Analyses using ALEPH
and OPAL spectral functions give \asm values that are in agreement. The 
robustness of the result can be established by comparing results from 
separate fits based on the $V$ or $A$ spectral functions, which are less 
inclusive and include larger, yet still small, nonperturbative contributions: 
their results are in reasonable agreement. A further test is achieved by 
partial integration of the spectral function up to a variable hypothetical 
$\tau$ mass $s_0$ less than $m_\tau^2$: the derived $\as(s_0)$ values are 
in agreement with the expected behavior from the QCD renormalization group 
down to $s_0$ values as low as $1\gev^2$, where the OPE begins to break down. 
Therefore the moderately high mass of the $\tau$ lepton is shown to be 
fully adequate for a precision measurement of \as.

Using the final spectral functions from ALEPH, 
supplemented by the world average leptonic and total strange hadronic 
fractions, the most precise determination of the strong coupling 
at the $\tau$ mass scale gives
\beq
  \as(m_\tau^2) = 0.345 \pm 0.004_{\rm exp} \pm 0.009_{\rm th}~.
\eeq
Evolving this value to the $Z$ mass, and adding experimental and theoretical
errors in quadrature, gives $\as(M_Z^2)_{\tau}=0.1215 \pm 0.0012$.
Taken together with the measurement from the $Z$ width at LEP,
$\as(M_Z^2)_{Z\,{\rm width}}=0.1186\pm 0.0027$, the $\as(m_\tau^2)$
determination provides a powerful test of the running of \as\ as 
predicted by QCD. The property of asymptotic freedom can be conveniently 
tested using the estimator
\beq
 r(s_1,s_2) = \frac {\Delta \as^{-1}}{\Delta \ln \sqrt{s}}
            = 1.405 \pm 0.053~,
\eeq
with $s_1=m_\tau^2$ and $s_2=M_Z^2$, while QCD predicts $1.353 \pm 0.006$.

Further progress in the QCD analyses of $\tau$ decays will have to wait for a
full calculation of the fourth-order coefficient $K_4$ in the Adler function
expansion. Such a program is underway and will succeed with the availability 
of improved computing power. When $K_4$ is known, we expect the perturbative 
uncertainty to become really small: the errors on \asm from $K_5$, estimated 
assuming a geometric growth of the perturbative coefficients with 100\% 
uncertainty, and from the renormalization scale will be 0.002 and 0.002, 
respectively. Using the existing data, the total error on \asm will be reduced 
by a factor of 2 and reach a relative precision of 1.4\%. Hence the calculation 
of $K_4$ will be highly rewarding.

Further progress can be considered with better quality data. 
Whereas ameliorating the precision on the
branching ratios appears to be difficult, several areas could be improved with
facilities such as the $B$ Factories, affording very large statistics samples.
Examples include the study of the high mass part of the spectrum near the 
$\tau$-mass endpoint, limited at the moment by statistics. This will 
improve the knowledge of the spectral function in an interesting region and
permit the study of the approach to the asymptotic regime. Such topics can
also be covered at a $\tau$-charm factory operating near the $\tau$ 
threshold. Detailed physics simulations are required to quantitatively 
assess the potential of these measurements.

The strange component of the $\tau$ hadronic width gives access to the study
of flavor-symmetry breaking induced by the massive $s$ quark as compared to
the nearly massless $u$ and $d$ quarks. Although the expected shift in the
strange hadronic width is indeed found experimentally, it turns out that a 
competitive determination of the strange quark mass $m_s$ is
difficult because of poor convergence of the mass-dependent
perturbative series and lack of sufficient inclusiveness in the higher-order 
spectral moments that are most sensitive to $m_s$. Thus various analyses 
have given somewhat different results, depending on the 
moments used. The situation might improve with a better determined strange
spectral function. Contrary to the situation for $m_s$, the determination
of the CKM matrix element $|V_{us}|$ from the strange $\tau$ decays appears 
on safe grounds, since it relies directly on the non-weighted spectral 
function with maximum inclusiveness and a better control of the perturbative 
expansion. Present data yield 
\beq
 |V_{us}|=0.2204 \pm 0.0028_{\rm exp} \pm 0.0003_{\rm th} \pm 0.0001_{m_s}~, 
\eeq
where $m_s$ has been taken from lattice calculations to be 
$m_s(m_\tau^2)=(79\pm8)\mev$.

The study of strange $\tau$ decays can benefit significantly from the large 
statistical samples provided by the $B$ Factories, complemented by the 
excellent $K-\pi$ separation available with the \babar\ and Belle detectors.
The prospects are excellent for improving over the present situation. As
discussed above, the determination of $|V_{us}|$ is limited by the 
experimental error, while the theoretical uncertainty would permit the best
measurement among all of this quantity. We express our wish that these 
experiments measure the various strange decay channels in a systematic 
way towards achieving this goal. It could also occur that the present 
unsatisfactory situation on the determination of $m_s$ will improve 
during this process. In any case, better strange spectral functions are 
needed.

Another active field of research is the study of finite energy chiral sum 
rules, involving the nonstrange $V-A$ spectral function, which are of
intrinsically nonperturbative nature. 
The analyses are divided into spectral moments 
that are weighted integrals over the $V-A$ spectral function 
with some high-mass damping function to reduce the systematic uncertainty 
due to the finite energy integral cut-off, and inverse moment sum rules,
which, due to the singularity at $s=0$, give direct access to the chiral 
structure of the QCD Lagrangian. The former sum rules are useful to analyze 
the nonperturbative structure of the OPE ,since it allows one to measure the
higher dimensional operators of the expansion independently of the 
perturbative contributions. Several analyses give
agreeing results on the $D=6$ contribution, while the much smaller $D=8$
contribution exhibits some model dependence. Even 
$D=10$ has been considered and found to improve the fit, though not 
significantly. No sign of nonconvergence of the OPE has been detected
at the $\tau$-mass scale. Inverse moment sum rules have been traditionally
used through the Das-Mathur-Okubo sum rule to determine the polarizability 
of the pion. In a more evolved later analysis the parameter $L_{10}$ of 
the chiral Lagrangian has been determined, which at one-loop and expressed
at the Chiral Perturbation Theory ($\chi$PT) renormalization scale
$\mu_{\chi\rm PT}=m_\rho\simeq770\mev$, is found to be
\beq
   L_{10}^r(m_\rho)=-(5.13\pm0.19)\times10^{-3}~,
\eeq
where experimental and theoretical errors of different sources have been 
added in quadrature. This result is in agreement with, but more precise than 
corresponding determinations from the axial form factor of 
$\pim \to e^-\nueb\gamma$ decays.

While the study of hadronic $\tau$ decays has achieved a high status thanks
to precise and complete sets of measurements and the development of
the relevant theoretical framework, it is satisfying that advances are
at hand both at the experimental and theoretical levels.

\section*{Acknowledgments}

We are indebted to all our experimental and theoretical colleagues who have 
provided valuable contributions to the field of $\tau$ physics. 
During this work we benefited from helpful discussions with Georges Grunberg, 
Fran\c{c}ois Le Diberder, William Marciano, Sven Menke, Antonio Pich, 
Joachim Prades, Jan Stern, and Konstantin Zyablyuk. 
We thank Konstantin Chetyrkin, Matthias Jamin, Sven Menke, Antonio Pich,
Achim Stahl and Arkady Vainshtein for critically reading the document and 
providing helpful comments.
Special thanks to Shaomin Chen and Changzheng Yuan for the fruitful
collaboration on the spectral function measurements.

% \vfill\newpage

\bibliographystyle{apsrmp}

\end{document}